%% file: exotic_nicer.tex
\documentclass[dvipsnames,11pt]{article}
\usepackage{amsmath, amsfonts, setspace, url, hyperref, verbatim, xcolor, graphicx, todonotes, cancel, slashed, mathtools}
\usepackage{cite}
\usepackage[a4paper, total={5.5in, 8.5in}]{geometry}
\usepackage[T1]{fontenc}
\usepackage[utf8]{inputenc}
\usepackage{enumitem}
\usepackage{tikz}
\usetikzlibrary{arrows,positioning,fit,shapes,snakes,backgrounds,tikzmark}
\usepackage{pgfplots}
\usetikzlibrary{shapes.geometric,calc}
\input{overhead.tex}
\usepackage{multirow}
\usepackage{multicol}
\usepackage{accents}
\usepackage{tabulary}
\usepackage{booktabs}
\usepackage{caption}
\usepackage{float}
\usepackage{multirow}
\usepackage{afterpage}
\usepackage{stmaryrd}	

\newcommand{\refeqn}[1]{\eqref{#1}}

\newcommand{\refref}[1]{\cite{#1}}

\newcommand{\tbl}[2]{#2 \caption{#1}}
\newcommand{\sayappendix}{appendix{} }
\newcommand{\bea}{\begin{eqnarray}}
\newcommand{\eea}{\end{eqnarray}}
\newcommand{\beq}{\begin{align}}
\newcommand{\eeq}{\end{align}}
\newcommand{\diag}{\mathrm{diag}}

\newcommand{\dd}{\mathrm{d}}		
\newcommand{\HH}{\mathcal{H}}		
\newcommand{\MM}{\mathcal{M}}			
\newcommand{\LL}{\mathcal{L}}			
\newcommand{\DD}{\mathcal{D}}			
\newcommand{\FF}{\mathcal{F}}
\newcommand{\AAA}{\mathcal{A}}

\newcommand{\btheta}{{\bar{\theta}}}

\newcommand{\ta}{\tilde{a}}

\begin{document}

\title{
\begin{flushright}
\normalsize{QMUL-PH-20-13}
\end{flushright}
\bigskip
\bf
The Geometry, Branes and Applications of Exceptional Field Theory
}

\author{\sc David S. Berman${}^1$\footnote{\tt d.s.berman@qmul.ac.uk}, Chris D. A. Blair${}^2$\footnote{\tt cblair@vub.ac.be}{}}

\date{\it $^1$ Centre for Research in String Theory, School of Physics and Astronomy,
Queen Mary University of London, 327 Mile End Road, London E1 4NS, UK 
\\ 
$^2$ Theoretische Natuurkunde, Vrije Universiteit Brussel, and the International Solvay Institutes,  Pleinlaan 2, B-1050 Brussels, Belgium 
}

\maketitle

\begin{abstract}
This is a review of exceptional field theory: a generalisation of Kaluza-Klein theory that unifies the metric and $p$-form gauge field degrees of freedom of supergravity into a generalised or extended geometry, whose additional coordinates may be viewed as conjugate to brane winding modes.
This unifies the maximal supergravities, treating their previously-hidden exceptional Lie symmetries as a fundamental geometric symmetry. Duality orbits of solutions simplify into single objects, that in many cases have simple geometric interpretations, for instance as wave or monopole-type solutions.
It also provides a route to explore exotic or non-geometric aspects of M-theory, such as exotic branes, U-folds, and more novel sorts of non-Riemannian spaces.
\end{abstract}

\tableofcontents

\section{Introduction}

\input{sec1Introduction}


\input{sec2GeneralizingKK}

\input{sec3SUGRAAndDuality}

\input{sec4and5ExFTIntroduction}

\section{Branes in Double and Exceptional Field Theory}
\label{branes}

\input{sec6BranesInExFT}

\section{Non-geometry and exotic branes}
\label{nongeoexotic}
\input{sec7NonGeometryFromDuality}

\input{sec8ExoticBranes}

\section{Non-Riemannian geometry}
\label{nonrie}
\input{sec9NonRieGeometry}


\section{Conclusions and Discussion}
\input{sec10conclusions}



\vspace{1em}
\section*{Acknowledgments}

We would like to thank the huge number of colleagues over the years who have worked to develop Double and Exceptional Field Theory and have immersed themselves in this new field. We have personally benefited from the expertise and enthusiasm of (and apologies to all the deserving people missed off the list): Luigi Alfonsi, Alex Arvanitakis, Eric Bergshoeff, Martin Cederwall, Athanasios Chatzistavrakidis, the Godazgars, Olaf Hohm, Chris Hull, Gianluca Inverso, Larisa Jonke, Axel Kleinschmidt, Kanghoon Lee, Dieter L\"ust, Emanuel Malek, Diego Marqu\'es, Edvard Musaev, Carmen N\'u{\~n}ez, Ray Otsuki, Jakob Palmkvist, Jeong-Hyuck Park, Malcolm Perry, Fabio Riccioni, Alejandro Rosabal Rodriguez, Alasdair Routh, Felix Rudolph, Christian Saemann, Henning Samtleben,  Masaki Shigemori, Charles Strickland-Constable, Richard Szabo, Daniel Thompson, Paul Townsend, Dan Waldram, Yi-Nan Wang, Peter West and Barton Zwiebach. 
CB would also like to thank Sofia Zhidkova for comments on the manuscript.

DSB would like to thank the organisers of the conferences: ``String and M-Theory: The New Geometry of the 21st Century'', Singapore, where he gave an overview covering some of this material. A recording is available at:

\begin{center}
 \url{https://www.youtube.com/watch?v=NXsJ13qjwNo}
\end{center} 
Similarly, CB would like to thank the organisers of the conference ``Geometry and Duality'' at the Max Planck Institute for Gravitational Physics (Albert Einstein Institute), Potsdam for an invitation to give an overview talk on 	``Non-geometry and exotic branes'', which inspired some of the material covered in this review. Slides and a recording of this talk are available at: 
\begin{center}
\url{https://workshops.aei.mpg.de/geometryduality/program/}
\end{center}

\noindent DSB is supported by an STFC consolidated grant ST/L000415/1, String Theory, Gauge Theory and Duality.
CB is supported by an FWO-Vlaanderen Postdoctoral Fellowship, and also in part by the FWO-Vlaanderen through the project G006119N and by the Vrije Universiteit Brussel through the Strategic Research Program ``High-Energy Physics''.

\appendix

\input{sec_appendix}


\begingroup
\setstretch{0.9}
\small
\bibliography{CurrentBib}
\endgroup

\end{document}

%% file: overhead.tex
\newcommand{\fM}{\mathcal{M}}
\newcommand{\fN}{\mathcal{N}}
\newcommand{\fP}{\mathcal{P}}
\newcommand{\fQ}{\mathcal{Q}}
\newcommand{\fK}{\mathcal{K}}
\newcommand{\fL}{\mathcal{L}}
\newcommand{\fR}{\mathcal{R}}
\newcommand{\fS}{\mathcal{S}}
\newcommand{\fT}{\mathcal{T}}

\newcommand{\fA}{\mathcal{A}}
\newcommand{\fB}{\mathcal{B}}
\newcommand{\fC}{\mathcal{C}}
\newcommand{\fD}{\mathcal{D}}
\newcommand{\fE}{\mathcal{E}}
\newcommand{\fF}{\mathcal{F}}

\newcommand{\be}{\begin{equation}}
\newcommand{\ee}{\end{equation}}
\numberwithin{equation}{section}

\newcommand{\bk}{\bar{k}}

\newcommand{\barm}{\bar{m}}
\newcommand{\bn}{\bar{n}}

\newcommand{\hmu}{\hat{\mu}}
\newcommand{\hnu}{\hat{\nu}} 
\newcommand{\hrho}{\hat{\rho}}

\newcommand{\hsigma}{\hat{\sigma}}

\newcommand{\gM}{\mathcal{M}}
\newcommand{\cH}{\mathcal{H}}

\newcommand{\Cdef}{{C}}
\newcommand{\lambdadef}{{\lambda}}
\newcommand{\Fdef}{{F}} 
\newcommand{\Celeven}{\hat{C}}
\newcommand{\Feleven}{\hat{F}}

\newcommand{\Aa}{\mathcal{A}}
\newcommand{\Ab}{\mathcal{B}}
\newcommand{\Ac}{\mathcal{C}}
\newcommand{\Ad}{\mathcal{D}}

\newcommand{\p}{\bullet}
\newcommand{\RR}{R} 
\newcommand{\sing}{\mathbb{S}}
\newcommand{\D}{\mathcal{D}}

\newcommand{\Fa}{\mathcal{F}}
\newcommand{\Fb}{\mathcal{H}}
\newcommand{\Fc}{\mathcal{J}}
\newcommand{\Fd}{\mathcal{K}}

\newcommand{\Gtwo}{\mathrm{SL}(2) \times \mathbb{R}^+}  
\newcommand{\Gthree}{\mathrm{SL}(3) \times \mathrm{SL}(2)}
\newcommand{\Gfour}{\mathrm{SL}(5)}
\newcommand{\Gfive}{\mathrm{SO}(5,5)}
\newcommand{\Gsix}{E_{6(6)}}
\newcommand{\Gseven}{E_{7(7)}}
\newcommand{\Geight}{E_{8(8)}}
\newcommand{\Gnine}{E_{9(9)}}

\newcommand{\Geleven}{E_{11(11)}}
\newcommand{\Edd}{E_{d(d)}}

\newcommand{\Hthree}{\mathrm{SO}(3) \times \mathrm{SO}(2)}
\newcommand{\Hfour}{\mathrm{SO}(5)}
\newcommand{\Hfive}{\mathrm{SO}(5)\times \mathrm{SO}(5)}
\newcommand{\Hsix}{\mathrm{USp}(8)}
\newcommand{\Hseven}{\mathrm{SU}(8)}
\newcommand{\Height}{\mathrm{SO}(16)}

\newcommand{\HfourLor}{\mathrm{SO}(2,3)}

\newcommand{\gendil}{\mathbf{d}}

\definecolor{vub}{RGB}{0,52,154}
\definecolor{vubo}{RGB}{255,102,0}
\definecolor{redd}{RGB}{255,40,40}
\definecolor{r}{RGB}{228,32,20}
\definecolor{o}{RGB}{238,69,4}
\definecolor{y}{RGB}{253,228,1}
\definecolor{g}{RGB}{108,160,0}
\definecolor{b}{RGB}{0,162,203}
\definecolor{i}{RGB}{120,42,117}

\newcommand{\red}[1]{{\color{red} #1}}

\usepackage[utf8]{inputenc}
\usepackage[framemethod=tikz]{mdframed}
\usetikzlibrary{calc}
\usepackage{chngcntr}

\newcounter{ex}
\counterwithin{ex}{section}

\makeatletter
\def\mdf@@mynote{}
\define@key{mdf}{mynote}{\def\mdf@@mynote{#1}}

\mdfdefinestyle{mystate}{
skipabove={1\baselineskip},
linewidth=0.5pt,
innertopmargin=1\baselineskip,
roundcorner=10pt,
backgroundcolor=black!0,
linecolor=black,
singleextra={
  \node[xshift=10pt,draw=black,fill=white,rounded corners,anchor=west] at (P-|O) %
  {\strut{\itshape  }\ifdefempty{\mdf@@mynote}{}{\itshape\bfseries \mdf@@mynote}};
},
firstextra={
  \node[xshift=10pt,draw=black,fill=white,rounded corners,anchor=west] at (P-|O) %
  {\strut{\itshape  }\ifdefempty{\mdf@@mynote}{}{\itshape\bfseries \mdf@@mynote}};
}
}

\mdfdefinestyle{mystateb}{
skipabove={1\baselineskip},
linewidth=1pt,
innertopmargin=1\baselineskip,
roundcorner=10pt,
backgroundcolor=white!5,
linecolor=vub!50,
singleextra={
  \node[xshift=10pt,thick,draw=vub!50,fill=b!0,rounded corners,anchor=west] at (P-|O) %
  {\strut{\itshape  }\ifdefempty{\mdf@@mynote}{}{\bf\mdf@@mynote}};
},
firstextra={
  \node[xshift=10pt,thick,draw=vub!50,fill=b!0,rounded corners,anchor=west] at (P-|O) %
  {\strut{\itshape  }\ifdefempty{\mdf@@mynote}{}{\bf\mdf@@mynote}};
}
}

\mdfdefinestyle{mystater}{
skipabove={1\baselineskip},
linewidth=1pt,
innertopmargin=1\baselineskip,
roundcorner=10pt,
backgroundcolor=white!5,
linecolor=r!50,
singleextra={
  \node[xshift=10pt,thick,draw=r!50,fill=b!0,rounded corners,anchor=west] at (P-|O) %
  {\strut{\itshape  }\ifdefempty{\mdf@@mynote}{}{\bf\mdf@@mynote}};
},
firstextra={
  \node[xshift=10pt,thick,draw=r!50,fill=b!0,rounded corners,anchor=west] at (P-|O) %
  {\strut{\itshape  }\ifdefempty{\mdf@@mynote}{}{\bf\mdf@@mynote}};
}
}

\mdfdefinestyle{mystateg}{
skipabove={1\baselineskip},
linewidth=1pt,
innertopmargin=1\baselineskip,
roundcorner=10pt,
backgroundcolor=white!5,
linecolor=g!50,
singleextra={
  \node[xshift=10pt,thick,draw=g!50,fill=g!0,rounded corners,anchor=west] at (P-|O) %
  {\strut{\itshape  }\ifdefempty{\mdf@@mynote}{}{\bf\mdf@@mynote}};
},
firstextra={
  \node[xshift=10pt,thick,draw=g!50,fill=g!0,rounded corners,anchor=west] at (P-|O) %
  {\strut{\itshape  }\ifdefempty{\mdf@@mynote}{}{\bf\mdf@@mynote}};
}
}

\mdfdefinestyle{myprinc}{
skipabove={1\baselineskip},
linewidth=1pt,
innertopmargin=1\baselineskip,
roundcorner=10pt,
backgroundcolor=black!0,
linecolor=ForestGreen,
singleextra={
  \node[xshift=10pt,thick,draw=white!75,fill=orange!25,rounded corners,anchor=west] at (P-|O) %
  {\strut{\itshape  }\ifdefempty{\mdf@@mynote}{}{\itshape\bfseries\underline \mdf@@mynote}};
},
firstextra={
  \node[xshift=10pt,thick,draw=white!75,fill=orange!25,rounded corners,anchor=west] at (P-|O) %
  {\strut{\itshape  }\ifdefempty{\mdf@@mynote}{}{\itshape\bfseries\underline \mdf@@mynote}};
}
}

\mdfdefinestyle{mydefn}{
skipabove={1\baselineskip},
linewidth=1pt,
innertopmargin=1.5\baselineskip,
innerbottommargin=0.5\baselineskip,
roundcorner=10pt,
backgroundcolor=red!0,
linecolor=red!70!black,
singleextra={
    \node[xshift=10pt,thick,draw=white!75,fill=blue!0!red!10,rounded corners,anchor=west] at (P-|O) %
	{\strut{\itshape  }\ifdefempty{\mdf@@mynote}{}{\itshape\bfseries\underline \mdf@@mynote}};
},
firstextra={
    \node[xshift=10pt,thick,draw=white!75,fill=blue!0!red!10,rounded corners,anchor=west] at (P-|O) %
    {\strut{\itshape  }\ifdefempty{\mdf@@mynote}{}{\itshape\bfseries\underline \mdf@@mynote}};
}
}

\mdfdefinestyle{myexer}{
skipabove={1\baselineskip},
settings={\refstepcounter{ex}},
linewidth=1pt,
innertopmargin=1.5\baselineskip,
innerbottommargin=1\baselineskip,
roundcorner=10pt,
backgroundcolor=red!0,
linecolor=ForestGreen!40,
singleextra={
    \node[xshift=10pt,thick,draw=white,fill=ForestGreen!25,rounded corners,anchor=west] at (P-|O) %
	{\strut{\bf\underline{Exercise~\theex}}\ifdefempty{\mdf@@mynote}{}{~(\mdf@@mynote})};
},
firstextra={
    \node[xshift=10pt,thick,draw=white,fill=ForestGreen!25,rounded corners,anchor=west] at (P-|O) %
	{\strut{\bf\underline{Exercise~\theex}}\ifdefempty{\mdf@@mynote}{}{~(\mdf@@mynote})};
}
}

\newmdenv[style=mystate,nobreak=true]{state}
\newmdenv[style=mystater,nobreak=true]{stater}
\newmdenv[style=mystateg,nobreak=true]{stateg}
\newmdenv[style=mystateb,nobreak=true]{stateb}
\newmdenv[style=myprinc,nobreak=true]{principle}
\newmdenv[style=myprinc,nobreak=false]{principlebreak}
\newmdenv[style=mydefn,nobreak=true]{defin}
\newmdenv[style=myexer,nobreak=false]{exercise}




%% file: sec1Introduction.tex
\subsection{A brief general introduction}

A century ago, confronted with the problem of unifying gravity and electromagnetism, Theodor Kaluza made what he described as ``the otherwise extremely odd decision to ask for help from a new fifth dimension of the world'' \cite{Kaluza:1921tu}.
To string theorists, there is no longer anything particularly odd about asking for help from extra dimensions, and what is now known as Kaluza-Klein reduction {\cite{Kaluza:1921tu,Klein:1926tv} is an essential tool. 

Indeed, it was with the aid of an eleventh dimension \cite{Hull:1994ys, Witten:1995ex} that the Second Superstring Revolution succeeded at unifying the five ten-dimensional superstring theories.
The significance of this event (if it is not, like the French Revolution, ``too soon to say'') may be in how it emphasised the role played by \emph{duality} in string theory.

The interplay between geometry and duality is particularly important.
For a start, duality relates string theories in different geometries. A basic example is the T-duality between type IIA string theory on a circle of radius $R$ and type IIB string theory on a circle of radius $\tilde R = \alpha^\prime / R$.

Conversely, geometry -- as in Kaluza-Klein theory -- serves to unify.
The classic route to M-theory is the strong coupling limit of type IIA string theory, where an eleventh dimension emerges.
At low energies, this eleven-dimensional M-theory leads to eleven-dimensional supergravity, and Kaluza-Klein reduction leads back to the ten-dimensional type IIA supergravity.
The radius of the eleventh dimension is interpreted as the dilaton field in supergravity, which controls the string coupling.
This Kaluza-Klein perspective unifies the ten-dimensional fields and branes: for example, the type IIA RR one-form gauge potential becomes part of the eleven-dimensional metric, identified with the Kaluza-Klein vector, and momentum modes in the eleventh direction (charged under the Kaluza-Klein vector) become D0 branes, while fundamental strings and D2 branes combine into the membrane in eleven-dimensions.

The further compactification of the eleven-dimensional theory on a two-torus $T^2$ can then be mapped to the reduction of the ten-dimensional type IIB string on a circle.
In the type IIB theory, there is a non-perturbative $\mathrm{SL}(2)$ duality, which swaps fundamental strings with D1 branes. 
From the eleven-dimensional point of view, this $\mathrm{SL}(2)$ has a geometrical origin in the action of large diffeomorphisms on the M-theory two-torus. The scalar fields of type IIB supergravity, namely the dilaton, $\Phi$ and the RR 0-form, $C_0$, are mapped to the complex structure of the torus, $\tau = C_0 + i e^{-\Phi}$.

An alternative description of this $\mathrm{SL}(2)$ is provided by F-theory \cite{Vafa:1996xn}, where the ten-dimensional type IIB theory is extended to a 12-dimensional theory, with the extra two dimensions corresponding to an auxiliary $T^2$ (of zero area) with complex structure $\tau$.

In these examples, we see that the degrees of freedom and symmetries of the type II theories have an explanation in terms of a higher-dimensional geometry.
These are just the simplest cases of a wide range of situations where geometric features of one description of M-theory explains the physics of another, and gives rise to a rich spectrum of dual descriptions.

This extends to field theory.
One of M-theory's most interesting objects is the six-dimensional (0,2) theory associated to the world volume of the M-theory five-brane.
Its reduction on a torus produces $\mathcal{N}=4$ Yang-Mills in four-dimensions  whose $\mathrm{SL}(2)$ duality again comes from the mapping class group of the torus. 
This is an extraordinary result. The $\mathrm{SL}(2)$ symmetry in $\mathcal{N}=4$ Yang-Mills is a non-perturbative quantum result and it is remarkable that there can be a geometric explanation for it.
Similarly, the Seiberg-Witten curve that encodes the quantum exact, low energy effective theory for $\mathcal{N}=2$ Yang-Mills may also be given a geometric origin through the reduction of the six-dimensional (0,2) theory on a Riemann surface. 

All this shows the crucial idea of M-theory is that the existence of dualities indicates the existence of a unifying description in higher dimensions, where the duality symmetries have a geometric origin in terms of a bigger space. 

Of course, there are many dualities for which this does not seem to be the case. 
Although it provides a unifying framework up to 11-dimensions, the Kaluza-Klein reduction of string theory and M-theory also leads to symmetries which seem not to have an evident higher-dimensional origin.
The reduction of type II supergravity on a $d$-dimensional torus leads to a theory with a global $O(d,d;\mathbb{R})$ symmetry; the full quantum string theory on a torus has a corresponding $O(d,d;\mathbb{Z})$ T-duality symmetry.
Including non-perturbative dualities enlarges this to the U-duality symmetry of M-theory on a $T^d$ or type II string theory on a $T^{d-1}$, which corresponds to the exceptional groups $\Edd(\mathbb{Z})$. In the Kaluza-Klein reduction of 11-dimensional supergravity on $T^d$, one finds correspondingly an $\Edd(\mathbb{R})$ global symmetry.

A general Kaluza-Klein reduction would only have an $\mathrm{GL}(d)$ global symmetry, coming from large diffeomorphisms of the $T^d$ on which we reduce.
These enlarged $O(d,d)$ or $\Edd$ symmetries appear thanks to the metric and form field degrees of freedom rearranging themselves rather remarkably into multiplets of these groups.
This reflects what also happens to  the spectrum of extended objects of M-theory wrapping the torus.
Supergravity is the point particle limit of string theory and yet the presence of duality symmetries shows that it retains some memory of its stringy origins and that, fundamentally, these are theories of extended objects (with T- and U-dualities exchanging momentum for brane winding modes).

From a philosophical point of view, one might then wonder about the nature of geometry in string theory and M-theory.
Is the standard Riemannian geometry that physicists have lived in since Einstein the most convenient language to capture the features of the backgrounds of string theory and M-theory?
Is there a better organisational principle that takes into account the menagerie of $p$-form gauge fields and the branes to which they couple? 
In this review, we will try to answer these questions using \emph{exceptional field theory}.

In exceptional field theory (ExFT), an $\Edd$ symmetry is manifest acting on an extended or generalised geometry.
Depending on how one chooses to identify the physical geometry with the extended geometry of ExFT, for each $d$, the $\Edd$ ExFT is equivalent to the full 11- or 10-dimensional maximal supergravities. It therefore provides a higher-dimensional origin of U-duality, in which no reduction is assumed, and on identifying the  novel coordinates of the extended geometry as conjugate to brane winding modes, ExFT offers a glimpse towards the geometry of M-theory beyond supergravity.

Beyond the intrinsic interest in obtaining a higher-dimensional perspective on duality, ExFT is a very powerful tool in understanding the geometry of string and M-theory backgrounds, both with a view towards reductions -- where it offers a way to efficiently characterise the properties of geometries with flux, and leads to new methods to obtain consistent truncations to gauged supergravities in lower-dimensions-- and towards expansions -- as it can be employed to obtain complicated higher-derivative corrections in an efficient manner.
Underlying these successes is the fact that the geometry of ExFT treats the metric and form-fields of supergravity on the same footing, rearranging all degrees of freedom into multiplets of $\Edd$ (in a form which is perfectly adapted to general dimensional reductions but completely general so that it works regardless of background). It manages to do this while simultaneously implementing and side-stepping issues with complicated non-linear electromagnetic dualisations that inevitably accompany the realisation of $\Edd$ symmetry \cite{Cremmer:1997ct, Cremmer:1998px} -- including of the metric itself.  

Exceptional field theory is by now a well-developed field with numerous interesting applications and outcrops.
The selection of topics in this review is of course biased by the authors' own interests and ignorances and further by the limitations of space, for all of which we ask for the understanding and patience of the reader.
We hope to be able to describe the general concepts and technical tools needed to understand and make of ExFT.
After developing the general theory, we will discuss some of the applications mentioned above and refer to the literature for further details when necessary.

We will then also describe the exceptional geometric perspective on brane solutions, and discuss the connections with \emph{non-geometry} and \emph{exotic branes}. 
By ``non-geometry'' we refer to backgrounds of string theory and M-theory which cannot be described purely in supergravity because they are globally defined only up to duality transformations. You can think of this as taking two ordinary geometries and patching them together via a duality transformation. This should be allowed if duality is a genuine symmetry of M-theory. Such spaces include T-folds and U-folds, where T- and U-duality are used in their definition. 
Exotic branes, meanwhile, are predictions of duality, and are characterised by the fact that their backreacted supergravity solutions -- obtained by duality transformations -- turn out to be non-geometric in nature: encircling an exotic brane one returns to a dual description to the one you started with.
The natural home for such phenomena should be in an extended geometry that knows about duality.

Putting this picture together leads to the conclusion that Kaluza's call for help from extra dimensions will be seen to have opened up new vistas in geometry and provided a truly unified approach to M-theory.

\subsection{This review}

As this review is fairly lengthy, let us offer a detailed outline for the readers' orientation.

\vspace{1em}

In \textbf{section \ref{stringduality}} we give a brief reminder on \textbf{duality in string and M-theory}.

\vspace{1em}

In \textbf{section \ref{genKK}} we rethink the usual approach to the classical geometry of string and M-theory by going back to the basics and \textbf{generalising Kaluza-Klein theory}. 
\begin{itemize}[noitemsep]
\item In \textbf{section \ref{genKK_KK}}, we start with the classic Kaluza-Klein setup, describing the unification of electromagnetism with the spacetime metric (in, of course, a particular example of an Einstein-Maxwell theory with a scalar field describing the radius of the extra dimension). 
\item In \textbf{section \ref{genKK_DFT}}, we will then be faced with the key puzzle, that strings fundamentally couple to two-forms and so we will seek a Kaluza-Klein unification of a metric and a two-form, leading to generalised geometry and double field theory (DFT).
\item In \textbf{section \ref{genKK_EFT}}, we take the next step: incorporating a three-form with the metric, and discovering a (simplified version of) exceptional field theory (ExFT) and its exceptional geometry.
\end{itemize}

In \textbf{section \ref{SUGRAduality}}, we explore the structure of \textbf{supergravity decompositions and $\Edd$ multiplets}.

\begin{itemize}[noitemsep]
\item In \textbf{section \ref{KKsplit}}, we again revisit Kaluza-Klein theory but now without doing any reduction: we perform an $(n+d)$-dimensional splitting of the coordinates and fields in order to explore the symmetry structures that appear, which will then get uplifted to and extended in exceptional field theory.
\item In \textbf{section \ref{11red}}, we specialise to the bosonic part of the action of 11-dimensional supergravity, and explain its decomposition using section \ref{KKsplit}.
\item In \textbf{section \ref{counting}}, we highlight how the field content of 11-dimensional supergravity reorganises into multiplets of $\Edd$ under this decomposition, in a way that underlies and pre-empts the construction of exceptional field theory.
\end{itemize}

In \textbf{section \ref{exft}}, we systematically introduce \textbf{exceptional field theory}.
\begin{itemize}[noitemsep]
\item  In \textbf{section \ref{exgld}}, we define the exceptional generalised Lie derivative.
\item In \textbf{section \ref{thier}}, we introduce the \emph{tensor hierarchy} of generalised $p$-forms and their gauge transformations. 
\item In \textbf{section \ref{exsl5th}}, we look in more detail at the tensor hierarchy for the $\Gfour$ ExFT.
\item In \textbf{section \ref{compensator}}, we explain how the tensor hierarchy deals with the presence of dual graviton degrees of freedom by introducing additional \emph{constrained compensator} fields and symmetries.
\item In \textbf{section \ref{action}}, we describe the general form of the action of ExFT.
\item In \textbf{section \ref{action5}}, we specialise to the $\Gfour$ ExFT, write down the action and show how to precisely identify it with that of 11-dimensional supergravity, using the results of sections \ref{11red} and \ref{exsl5th}.
\item In \textbf{section \ref{actionE7}}, we describe the (pseudo-)action of the $\Gseven$ ExFT.
\item In \textbf{section \ref{action8}}, we describe the $\Geight$ ExFT, where the definition of generalised diffeomorphisms must be modified. 
\end{itemize}

In \textbf{section \ref{features}}, we go into detail on further \textbf{features of exceptional field theory}.
\begin{itemize}[noitemsep]
\item  In \textbf{section \ref{ornotIIB}}, we explain how ExFT describes both the 11-dimensional supergravity and 10-dimensional type IIB supergravity.
\item In \textbf{section \ref{genSS}}, we describe generalised Scherk-Schwarz reductions of ExFT. These lead to gauged supergravities in lower dimensions, and furthermore allow the section condition to be relaxed. As examples, we will discuss two deformations of ten-dimensional supergravity (the Romans IIA theory and the so-called generalised IIB theory) which can be described in ExFT by allowing the fields to have a controlled dependence on dual coordinates.
\item In \textbf{section \ref{susy}}, we briefly discuss the interplay between ExFT and (various amounts of) supersymmetry.
\item In \textbf{section \ref{O}}, we discuss the notion of a generalised orientifold, or ``O-fold'', of the exceptional geometry, and how this can be used to describe half-maximal supergravities.
\item In \textbf{section \ref{corrections}}, we discuss work on describing higher-derivative corrections to supergravity using extended geometry.
\end{itemize}

In \textbf{section \ref{branes}}, we discuss the form of \textbf{brane solutions in double and exceptional field theory}.
\begin{itemize}[noitemsep]
\item  In \textbf{section \ref{kksolns}}, we review the pp-wave and Kaluza-Klein monopole solutions of pure gravity, which realise electric and magnetically charged solutions after Kaluza-Klein reduction.
\item  In \textbf{section \ref{branesDFT}}, we embed brane solutions in the doubled geometry of DFT, where they become generalised wave and monopole solutions.
\item  In \textbf{section \ref{branesinexft}}, we discuss the analogous picutre in ExFT.
\item In \textbf{section \ref{salgebras}}, we describe the superalgebra of ExFT, with all central charges combining into generalised momenta. 
\item In \textbf{section \ref{windingcentral}}, we connect central charges, winding modes and dual coordinates.
\item  In \textbf{section \ref{branewvol}}, we briefly comment on approaches to brane worldvolume actions where the target space is the extended geometry of either DFT or ExFT.
\end{itemize}

In \textbf{section \ref{nongeoexotic}}, we explore \textbf{non-geometry and exotic branes}.
\begin{itemize}[noitemsep]
\item  In \textbf{section \ref{toyT}}, we introduce non-geometry via the T-duality chain connecting a three-torus with $H$-flux to both global and local T-folds. This provides a toy model for non-geometry.
\item  In \textbf{section \ref{toyU}}, we demonstrate the U-duality version of this toy model.
\item  In \textbf{section \ref{exoticbranes}}, the toy models are upgraded to genuine string and M-theory backgrounds corresponding to exotic branes.
\item  In \textbf{section \ref{exoticmult}}, we discuss the apperance of such exotic branes in U-duality multiplets, and the interpretation of such states in string theory.
\item  In \textbf{section \ref{stubes}}, we describe the realisation of exotic branes via the supertube effect, which implies they appear generically via spontaneous polarisation of normal brane configurations, and so are unavoidable in string theory. 
\item  In \textbf{section \ref{windingmode}}, we describe winding mode localisation effects. 
\item  In \textbf{section \ref{exoticexft}}, we comment on the exceptional field theory perspective on exotic branes, and discuss further developments.
\end{itemize}

In \textbf{section \ref{nonrie}}, we describe a different version of non-geometry, which we may call \textbf{non-Riemannian geometry}, which arises by considering parametrisations of the fields of DFT/ExFT which do not admit a standard spacetime interpretation involving a Riemannian metric.

\subsection{Duality in string and M-theory}
\label{stringduality}

In order to cover the basic concepts of duality that will re-occur throughout this review, we include here a brief review of the standard T-, S- and U-dualities of string theory and M-theory. 
In this section, we explicitly display the dimensionful quantities $\alpha^\prime$ and $l_p$. Elsewhere in this review, they are (mostly) suppressed.

\subsubsection{T-duality}

Consider string quantisation in flat spacetime, with one spatial direction a circle of radius $R$.
Along the circle the string has momentum states with quantised momentum $p = n/R$, $n \in \mathbb{Z}$, and winding states characterised by the winding number $w \in \mathbb{Z}$. 
The mass squared of a string state with momentum and winding (measured from the point of view of the non-compact directions) is
\be
M^2 = \left(\frac{n}{R}\right)^2 + \left( \frac{w R}{\alpha^\prime} \right)^2 + \text{oscillators}\,, 
\label{mass}
\ee
where $\alpha^\prime = l_s^2$ is the string length squared, related to the string tension by $T_{\text{F1}} = \frac{1}{2\pi \alpha^\prime}$.
The spectrum of the bosonic string is unchanged under the T-duality transformation:
\be
R \mapsto \tilde R =\frac{\alpha^\prime}{R} \,,
\quad (n,w) \mapsto (\tilde n, \tilde w ) = ( w,n) \,.
\label{basicT}
\ee
Momentum and winding modes are interchanged, and the radius of the circular direction is inverted.
In the worldsheet CFT, this extends to a map $Y = Y_L + Y_R \rightarrow \tilde Y = Y_L - Y_R$ exchanging the string coordinate $Y$ on the original circle for a \emph{dual coordinate} $\tilde Y$ on the dual circle, flipping the sign of the right-moving part.

We can repeat this for string on a $d$-dimensional torus.
In this case, we have $d$ winding numbers $w^i$ and $d$ momenta quantum numbers $n_i$, where $i=1,\dots,d$.
We combine these into a $2d$-dimensional vector which we write as $p_M = ( n_i, w^i)$ introducing a doubled index $M=1,\dots,2d$ (viewed as $i=1,\dots,d$ twice, once down and once up).
We can take the torus to have a (constant) metric $g_{ij}$, and a two-form field, $B_{ij}$.
Then the mass squared formula \refeqn{mass} generalises to:
\be
M^2 = \frac{1}{\alpha^\prime} p_M \mathcal{H}^{MN} p_N  + \text{oscillators}\,, 
\ee
where now all the geometrical information about the background is contained in the $2d \times 2d$ matrix
\be
\mathcal{H}^{MN} = \begin{pmatrix} \alpha^\prime g^{ij} & - g^{ik} B_{kj} \\ B_{ik} g^{kj} & \frac{1}{\alpha^\prime} (g_{ij} - B_{ik} g^{kl}B_{lj} )\end{pmatrix} \,.
\label{genmetdd}
\ee
The momenta and winding also appear in the level-matching condition:
\be
N_R - N_L = \frac{1}{2} \eta^{MN} p_M p_N \,,
\label{levelmatching}
\ee
where 
\be
\eta^{MN} = \begin{pmatrix} 0 & I \\ I & 0 \end{pmatrix} \,.
\label{defeta}
\ee
These formula are invariant under $O(d,d;\mathbb{Z})$ transformations acting such that
\be
p_M \rightarrow \tilde p_M =  (U^{-1})^K{}_M p_K
\,,\quad
\mathcal{H}^{MN} \rightarrow \tilde{\mathcal{H}}^{MN} = U^M{}_K U^N{}_L \mathcal{H}^{KL}
\,,
\ee
where the transformation matrix $U^M{}_N \in O(d,d;\mathbb{Z})$ leaves $\eta^{MN}$ invariant:
\be
\eta^{MN}\rightarrow \tilde \eta^{MN} =  U^M{}_K U^N{}_L \eta^{KL} = \eta^{MN} \,.
\ee
The inverse of $\eta^{MN}$ is denoted $\eta_{MN}$ and is what we will call the $O(d,d)$ metric or $O(d,d)$ structure. 
The restriction to integer valued entries comes about in order that $p_M$ remain integer valued.

The matrix given in \refeqn{genmetdd}, which encodes the background $d$-dimensional metric and $B$-field, obeys 
\be
\cH^{MK} \cH^{NL} \eta_{KL} = \eta^{MN} \,.
\ee
This means that $\cH_{MN}$ itself defines an element of $O(d,d;\mathbb{R})$, and its inverse is $\cH_{MN} = \eta_{MK} \eta_{NL} \cH^{KL}$ which has the same components as in \refeqn{genmetdd}, just with the blocks rearranged.
We will call $\cH_{MN}$ the \emph{generalised metric}.

The basic T-duality transformation \refeqn{basicT} which inverts the radius in one direction can be realised in the full $O(d,d;\mathbb{Z})$ group as follows.
Let $i=(\mu,y)$, with $\mu=1,\dots,d-1$.  
The transformation \refeqn{basicT} in the $y$ direction corresponds to:
\be
U^M{}_N = \begin{pmatrix}
\delta^\mu{}_\nu & 0 & 0 & 0 \\
0 & 0 & 0 & 1 \\
0 & 0 & \delta_\mu{}^\nu & 0 \\
0 & 1 & 0 & 0 
\end{pmatrix} \,.
\ee
The effect on the matrix $\cH^{MN}$ is to swap $\cH^{yy} \leftrightarrow \cH_{yy}$, $\cH_\mu{}^y \leftrightarrow \cH_{\mu y}$, $\cH^\mu{}_{y} \leftrightarrow \cH^{\mu y}$ while leaving $\cH^{\mu\nu}$, $\cH_{\mu}{}^{\nu}$ and $\cH_{\mu\nu}$ unchanged.
Denoting by $\tilde g_{ij}$ and $\tilde B_{ij}$ the dual metric and $B$-field, this produces:
\be
\begin{array}{cclccl}
\tilde g_{yy} & = & \frac{(\alpha^\prime)^2}{g_{yy}}\,,& &&  \\
\tilde g_{\mu y} & = & \alpha^\prime \frac{B_{\mu y}}{g_{yy}}\,, &\tilde B_{\mu y} & =& \alpha^\prime \frac{g_{\mu y}}{g_{yy}} \,,\\
\tilde g_{\mu\nu} & = & g_{\mu\nu} - \frac{1}{g_{yy}} ( g_{\mu y} g_{\nu y} - B_{\mu y} B_{\nu y} )\,, &\tilde B_{\mu\nu} & = & B_{\mu\nu} + \frac{1}{g_{yy}} (g_{\mu y} B_{\nu y} - g_{\nu y} B_{\mu y} ) \,.
\end{array}
\label{Buscher}
\ee
These are known as the \emph{Buscher rules} \cite{Buscher:1987sk,Buscher:1987qj}.
They are in fact applicable as a duality transformation of any background which has an isometry in the $y$ direction, with the background fields still depending on the remaining spacetime coordinates, as can be established (following the procedure of \refref{Buscher:1987sk,Buscher:1987qj}) using the worldsheet sigma model.

The massless states $g$ and $B$ are common to the bosonic and type II theories - in the latter they appear in the quantisation of the NSNS sector along with the scalar dilaton, which we will denote by $\Phi$.
The dilaton couples to the worldsheet Ricci scalar, and the string coupling is related to the (asymptotic value of) the dilaton by $g_s = e^\Phi$.
When we carry out the Buscher procedure at the level of string path integral, zero mode effects imply that the path integral is not in fact invariant but can be made so by simultaneously transforming the dilaton. 
The rule for the dilaton follows from the statement that the combination
\be
e^{-2 \Phi}  \sqrt{ \text{det}(g_{ij})} \, .
\ee
is invariant under $O(d,d;\mathbb{Z})$ T-duality transformations.

In both the bosonic and type II strings, compactification on a $d$-torus leads to this $O(d,d;\mathbb{Z})$ duality symmetry. In the latter case, T-duality transformations with determinant minus one act as maps exchanging the type IIA theory on one torus with the type IIB theory on a dual torus, while maps with unit determinant act as duality symmetries within either type IIA or type IIB.

In the low energy effective actions describing these string theories, i.e. in the type II supergravities for supersymmetric case, a classical version of the full stringy T-duality symmetry appears again on dimensional reduction on a $T^d$. In this case, the reduced supergravity has a global $O(d,d;\mathbb{R})$ symmetry.

\subsubsection{S-duality}

Next we consider a non-perturbative duality: the S-duality of type IIB string theory.
The basic statement of this duality is that type IIB at strong coupling is dual to type IIB at weak coupling; the theory is invariant under the map $g_s \rightarrow 1/ g_s$.

S-duality extends to an action of $\mathrm{SL}(2;\mathbb{Z})$ on the BPS states of the type IIB theory, which also acts accordingly on the massless states of theory.
In the type IIB supergravity, this appears as a classical global $\mathrm{SL}(2;\mathbb{R})$ symmetry.
Here, as well as the metric and NSNS 2-form, we have RR sector $p$-forms of even rank, $C_0$, $C_2$, $C_4$.
The dilaton and RR 0-form are combined into a complex scalar which transforms under $\mathrm{SL}(2)$ via a modular transformation:
\be
\tau
= C_0 + i e^{-\Phi}
\,,\quad
\tau  \rightarrow \frac{a \tau + b}{c \tau +d} \,,
\label{defmodular}
\ee
where the parameters of the transformation can be expressed as a unit determinant matrix
\be
U = \begin{pmatrix} a & b \\ c& d \end{pmatrix} \,,\quad ad-bc = 1 \,.
\label{defmodularmx}
\ee
For later use, we can also express this transformation in terms of a matrix
\be
\gM = \frac{1}{\text{Im}\,\tau} \begin{pmatrix} |\tau|^2 & \text{Re}\,\tau \\ \text{Re}\,\tau & 1 \end{pmatrix} = e^\Phi \begin{pmatrix} C_0^2 + e^{-2\Phi} & C_0 \\ C_0 & 1 \end{pmatrix} \,,
\ee
itself of unit determinant, transforming as $\mathcal{M} \rightarrow U \mathcal{M} U^T$.
The coupling constant inversion corresponds to $a=0=d$, $b=1$, $c=-1$.

The two two-forms form a doublet, $(C_2, B_2)$ transforming as $(C_2,B_2)^T \rightarrow U^{-T} (C_2,B_2)^T$.
The four-form is invariant, as is the \emph{Einstein frame} metric.
The latter is related to the string frame metric by $g_{\mu\nu}^{\text{string}} = e^{\Phi/2} g_{\mu\nu}^{\text{Einstein}}$.
It is the string frame metric which naturally appears in the string sigma model with no dilaton factors.
The Einstein frame metric is distinguished by the fact that in the supergravity action it corresponds to having the usual Einstein-Hilbert term, unmultiplied by any power of $e^\Phi$.

\subsubsection{U-duality}

S-duality and T-duality transformations do not commute.
Combining them, we generate a larger group of dualities of the type II theories. 
This is known as U-duality. It is a non-perturbative duality of the type II strings on a torus, and hence also a duality of M-theory.
The latter can be motivated by considering the strong coupling limit of the type IIA string.
As the IIA string coupling goes to infinity, an eleventh dimension decompactifies, and we are led to conjecture the existence of an 11-dimensional M-theory, which when compactified on a circle reduces to the IIA string in the zero radius limit.
The 11-dimensional radius $R_{11}$ and Planck length $l_p$ are related to the 10-dimensional string coupling constant $g_s$ and string length $l_s$ by \cite{Witten:1995ex}
\be
R_{11} = l_s g_s \,,\quad l_p = g_s^{1/3} l_s \,.
\label{M}
\ee
This corresponds to a reduction ansatz for the 11-dimensional metric with
\be
ds^2_{11}  = R_{11}^2 (dx^{11} + A)^2 + ds_{10}^2 \,,
\ee
where $A$ is a ten-dimensional 1-form.
In the simplest case, the radius $R_{11}$ is constant, but this is a special case of a more general Kaluza-Klein ansatz with
\be
\frac{1}{l_p^2} ds^2_{11} = e^{4\Phi/3} (dx^{11} + A)^2 + e^{-2\Phi/3} \frac{1}{l_s^2} ds_{10}^2 \,,
\ee
where the string coupling $g_s$ can be defined as the value of $e^\Phi$ at asymptotic infinity (with respect to the ten-dimensional string frame metric $ds_{10}^2$).

We can combine the relationships of \refeqn{M} with T-duality and S-duality transformations to generate 11-dimensional U-duality transformations.
The (string frame) rules we need are
\be
S : g_s \rightarrow \frac{1}{g_s} \,,\quad l_s^2 \rightarrow g_s l_s^2 \,,
\label{S}
\ee
for S-duality, and for a T-duality in the direction $i$ we have
\be
T_i : R_i \rightarrow \frac{l_s^2}{R_i} \,,\quad g_s \rightarrow g_s \frac{l_s}{R_i} \,.
\label{T}
\ee
As the relationships \refeqn{M} only hold for the IIA string variables, we should consider even numbers of T-duality transformations.
For instance, T-dualising on two directions labelled by $i$ and $j$, and then swapping the resulting directions (denote this via $\sigma: i \leftrightarrow j$) gives the transformation:
\be
(\sigma T)_{ij}: (R_i,R_j) \rightarrow \left(\frac{l_s^2}{R_j}, \frac{l_s^2}{R_i} \right)\,,\quad
g_s \rightarrow g_s \frac{l_s^2}{R_i R_j} \,.
\ee
This uplifts to a symmetrical U-duality transformation acting on \emph{three} directions and on the 11-dimensional Planck length as
\be
U_{IJK}: 
R_I \rightarrow \frac{l_p^3}{R_J R_K} \,,\quad l_p^3 \rightarrow \frac{l_p^6}{R_IR_JR_K} \,,
\label{UIJK}
\ee
where $I,J,K$ are distinct and can be any of the indices $(i,j,11)$.
Define the volume of the three-torus on which we are acting by $V = R_IR_JR_K$. 
Denoting the U-dual volume and Planck length by $\tilde V$ and $\tilde l_p$, we have $\tilde V/\tilde l_p^3 = l_p^3 / V$.
The transformation $U_{IJK}$ is therefore a volume inversion, generalising the radius inversion of the basic T-duality.

It can be more convenient to phrase the transformation in terms of the dimensionless quantities $R_I / l_p$. We have
\be
U_{IJK} : \frac{R_I}{l_p} \rightarrow \frac{ l_p R_I^{1/3}}{ (R_JR_K)^{2/3} } \,.
\label{UIJKdimless}
\ee
In general we will mostly want to drop explicit factors of the Planck length, for instance by setting it to one, in which case the transformation \refeqn{UIJKdimless} of the radii will appear in practice. In this case, we will also have the volume inverse in the form $\tilde V = 1/V$.

The full U-duality group is generated by these transformations, along with global coordinate transformation and shifts of the massless form fields appearing in M-theory.
Acting purely in three directions as above, this means shifting $C_{IJK} \rightarrow C_{IJK} + l_p^3 \Lambda_{IJK}$, with $\Lambda_{IJK}$ integer-valued.
The full U-duality group generated in this way is in fact $\mathrm{SL}(3;\mathbb{Z}) \times \mathrm{SL}(2;\mathbb{Z})$. 

Let's take a moment to describe the M-theory origin of such transformations more precisely.
First, take a step back and consider M-theory on a torus. The global symmetry group is given by $\mathrm{GL}(2;\mathbb{R})$ acting on the two-torus which can be viewed as the  $\mathrm{SL}(2;\mathbb{R})$ of volume preserving global coordinate transformations, plus an $\mathbb{R}^+$ acting as overall scalings. These global symmetries are symmetries of the equations of motion acting on the moduli of the reduction. They are solution generating in the sense that they map one vacuum solution, a torus with complex structure $\tau$ and size $V$ to another inequivalent one with different complex structure and volume.
A subgroup, $\mathrm{SL}(2;\mathbb{Z})$ are large diffeomorphisms (that is those that are not connected to the identity) that preserve the torus. Thus the torus itself is invariant even though its complex structure, $\tau$, is transformed. 
This group is then the duality group of equivalent vacua. 
The restriction to the arithmetic subgroup in that case came from Dirac quantisation of charges. 

When we include a further direction, i.e. M-theory on a three-torus, we now have a geometric $\mathrm{SL}(3;\mathbb{R})$ of volume preserving transformations, but the scalings get enhanced to an $\mathrm{SL}(2;\mathbb{R})$ acting on both the volume of the three-torus and the single component of the background three-form on the torus.
To realise this $\mathrm{SL}(2;\mathbb{R})$, we define
\be
\tau = \frac{C_{123}}{l_p^3} + i \frac{\sqrt{\det g}}{l_p^3} \,,
\ee
where $g_{ij}$ is the metric on the torus. 
This complex combination transforms under $\mathrm{SL}(2;\mathbb{R})$ according to the modular transformation rule \refeqn{defmodular}, with the volume scaling transformation corresponding to $a = d^{-1}$, $b=c=0$.
Classically these transformations are all real-valued, and this is what we encounter in supergravity, while in the full quantum theory we expect that only the integer-valued subgroup remains as the full U-duality group. 

We can characterise these M-theory $\Gthree$ transformations as acting on two unit determinant matrices, 
\be
\mathcal{M}_{ij} = \frac{1}{(\det g)^{1/3}} g_{ij} \,,\quad
\mathcal{M}_{\alpha \beta} = 
\frac{1}{(\det g)^{1/2}}
\begin{pmatrix}
l_p^{-3} ( C_{123}^2 + \det g) & C_{123} \\ 
C_{123} & l_p^3 
\end{pmatrix}
\label{threetwogm}
\ee
which have the interpretation as coset elements, $\gM_{ij}  \in \mathrm{SL}(3) / \mathrm{SO}(3)$, $\gM_{\alpha\beta} \in \mathrm{SL}(2) / \mathrm{SO}(2)$.
Suppose we define a composite index $M = i\alpha$.
Then we can combine these two matrices into a single six-by-six matrix
\be
\gM_{MN} \equiv \gM_{i\alpha,j\alpha} = \gM_{ij} \gM_{\alpha\beta} \,,
\ee
which can be written as
\be
\gM_{MN} = 
(\det g)^{1/6}
\begin{pmatrix}
l_p^{-3} ( 
g_{ij} + \frac{1}{2} C_{i pq} C_{j}{}^{pq} ) &  C_i{}^{kl} \\
C_k{}^{ij} & 2 l_p^3 g^{i[k} g^{l]j} 
\end{pmatrix} \,,
\label{genmet32}
\ee
having replaced the index $i2$ for a pair of antisymmetric indices, that is, $\gM_{k}{}^{ij} \equiv \frac{1}{2} \epsilon^{ijp} \gM_{k ,p2}$, and $\gM_{ij,kl} \equiv \frac{1}{4} \epsilon^{ijp} \epsilon^{klq} \gM_{p2,q2}$.
The quantity \refeqn{genmet32} is to be compared with the matrix \refeqn{genmetdd} encoding the NSNS metric and B-field on which T-duality transformations acted.

The action of U-duality on the backgrounds of type IIA string theory follow on applying the reduction rules \refeqn{M}.
Then we can further T-dualise to identify the corresponding transformations in type IIB.
In particular, the geometric $\mathrm{SL}(2)$ appearing when $d=2$, i.e. for M-theory on a two-torus, becomes the S-duality of type IIB. 
For M-theory on a three-torus, the type IIB S-duality is likewise embedded in the $\mathrm{SL}(3)$ factor of the full U-duality group.

Acting in more than three directions, there are further shift symmetries possible.
For $d$ sufficiently large, we also include shifts of the six-form electromagnetically dual to the three-form, and also shifts of further ``exotic'' dualisations including of the metric itself.
The U-duality group acting on a $d$-dimensional torus in M-theory is then determined to be $\Edd(\mathbb{Z})$. 
These groups are listed in table \ref{tab:eddinfo}.

This sequence of U-duality symmetries was first found in the context of reductions of eleven-dimensional supergravity \cite{Cremmer:1978km} on a torus \cite{Cremmer:1978ds,Cremmer:1979up,Julia:1980gr}. They are the global symmetries of maximal supergravity in $n$ dimensions. 
As mentioned before, in terms of the supergravity action, these global symmetry groups are real-valued.
When we take into account Dirac quantisation of brane charges, as we must in the full quantum theory, then group is broken to the integers so that charge quantisation is preserved.  
This is an important feature of duality: the reduced supergravity will have a moduli space and there will be a continuous set of symmetries acting on the moduli that take one vacuum into another inequivalent vacuum. However, an arithmetic subgroup will leave the reduction space invariant and this will coincide with the duality group when taking into account quantum charge preservation.
In general, we will not explicitly notate whether we are dealing with $\mathbb{R}$ or $\mathbb{Z}$, and will only reinstate the distinction when it is necessary to do so.

The sequence of Lie algebras shown in table \ref{tab:eddinfo} is defined by its Dynkin diagram, which has $d$ nodes, and by a choice of real form. The latter means a choice of allowed signature for the Killing form, or equivalently the number of compact and non-compact generators.
The general notation is $E_{d(K)}$ where $K$ is the number of non-compact minus the number of compact generators.
The particular sequence $\Edd$ is referred to as the split real form, which is the version of the group with the maximal number of non-compact generators. Table \ref{tab:eddinfo} also shows the maximal compact subgroup of $\Edd$, denoted by $H_d$: the scalar moduli of a toroidal reduction of supergravity parametrise the coset $\Edd/H_d$.

\begin{table}[ht]
	\renewcommand{\arraystretch}{1.5}
	\tbl{The exceptional sequence of rank $d$, their Dynkin diagrams, dimension, plus their maximal compact subgroups $H_d$ and their dimensions.}{
	\centering
	\begin{tabular}{lrcccc}
	$d$ & & $\Edd$ & dim $\Edd$ & $H_d$ & dim $H_d$ \\\hline
	2 &
	\begin{tikzpicture}
	\draw (0,0) circle (0.1cm);
	\end{tikzpicture}
	& $\Gtwo$ & 4 & $\mathrm{SO}(2)$  & 1 \\ 
	3 &
	\begin{tikzpicture}
	\foreach \x in {0,...,1}
	{
	\draw (-\x*0.4,0) circle (0.1cm);
	}
	\foreach \x in {0}
	\draw (-0.1-0.4*\x,0) -- (-0.3 -0.4*\x,0);

	\draw (-0.8,0.4) circle (0.1cm);
	\end{tikzpicture}
	& $\Gthree$ & 11 &  $\Hthree$ & 4 \\
	4 &
	\begin{tikzpicture}
	\foreach \x in {0,...,2}
	{
	\draw (-\x*0.4,0) circle (0.1cm);
	}
	\foreach \x in {0,...,1}
	\draw (-0.1-0.4*\x,0) -- (-0.3 -0.4*\x,0);

	\draw (-0.8,0.4) circle (0.1cm);
	\draw (-0.8,0.3) -- (-0.8,0.1); 
	\end{tikzpicture}
	& $\Gfour$ & 24 & $\Hfour$ & 10 \\
	5 &
	\begin{tikzpicture}
	\foreach \x in {0,...,3}
	{
	\draw (-\x*0.4,0) circle (0.1cm);
	}
	\foreach \x in {0,...,2}
	\draw (-0.1-0.4*\x,0) -- (-0.3 -0.4*\x,0);

	\draw (-0.8,0.4) circle (0.1cm);
	\draw (-0.8,0.3) -- (-0.8,0.1); 
	\end{tikzpicture}
	& $\Gfive$ &  45 & $\Hfive$  & 20  \\
	6 &
	\begin{tikzpicture}
	\foreach \x in {0,...,4}
	{
	\draw (-\x*0.4,0) circle (0.1cm);
	}
	\foreach \x in {0,...,3}
	\draw (-0.1-0.4*\x,0) -- (-0.3 -0.4*\x,0);

	\draw (-0.8,0.4) circle (0.1cm);
	\draw (-0.8,0.3) -- (-0.8,0.1); 
	\end{tikzpicture}
	& $\Gsix$ & 78 & $\Hsix$  & 36 \\
	7 &
	\begin{tikzpicture}
	\foreach \x in {0,...,5}
	{
	\draw (-\x*0.4,0) circle (0.1cm);
	}
	\foreach \x in {0,...,4}
	\draw (-0.1-0.4*\x,0) -- (-0.3 -0.4*\x,0);

	\draw (-0.8,0.4) circle (0.1cm);
	\draw (-0.8,0.3) -- (-0.8,0.1); 
	\end{tikzpicture}
	& $\Gseven$ & 133 & $\Hseven$ & 63 \\
	8 &
	\begin{tikzpicture}
	\foreach \x in {0,...,6}
	{
	\draw (-\x*0.4,0) circle (0.1cm);
	}
	\foreach \x in {0,...,5}
	\draw (-0.1-0.4*\x,0) -- (-0.3 -0.4*\x,0);

	\draw (-0.8,0.4) circle (0.1cm);
	\draw (-0.8,0.3) -- (-0.8,0.1); 
	\end{tikzpicture}
	& $\Geight$ & 248 & $\Height$  & 120  \\
	\end{tabular} }
\label{tab:eddinfo}	
\end{table}
	\renewcommand{\arraystretch}{1}

%% file: sec2GeneralizingKK.tex
\section{Generalising Kaluza-Klein theory}
\label{genKK}

\subsection{Kaluza-Klein theory}
\label{genKK_KK}

Our starting point is the classic version of Kaluza-Klein theory \cite{Kaluza:1921tu,Klein:1926tv}, applied to an Einstein-Maxwell theory in four dimensions with a scalar field.
This is a theory of a metric, $g_{\mu \nu}$, a one-form gauge field, $A_\mu$, and a scalar $\phi$. 
The goal is to understand these fields in terms of a higher-dimensional geometry.

The local symmetries of the four-dimensional theory are diffeomorphisms, under which all the fields transform tensorially, and 	gauge transformations of the one-form.
We describe the action of (infinitesimal) diffeomorphisms generated by a vector $\xi^\mu$ using the Lie derivative, defined on vectors by
\begin{equation}
L_\xi U ^\mu = \xi^\nu \partial_\nu U^\mu -  \partial_\nu \xi^\mu  U^\nu \,,
\label{originallie}
\end{equation} 
while gauge transformations are generated by scalars $\lambda$.
The symmetry transformations of the fields are thus:
\be
\delta g_{\mu\nu} = L_\xi g_{\mu\nu} \,,\quad
\delta A_\mu  = L_\xi A_\mu + \partial_\mu \lambda\,,\quad
\delta \phi  = L_\xi \phi\,.
\label{deltaKK}
\ee
The gauge invariant field strength of the one-form is of course:
\be
F_{\mu\nu} = 2\partial_{[\mu} A_{\nu]} \,.
\ee
To match with expressions that we will encounter later on, we note here that if one shifts $\lambda \rightarrow \lambda - A_\nu \xi^\nu$, then the transformation of the vector $A_\mu$ can also be written as
\be
\delta A_\mu = \xi^\nu F_{\nu\mu} + \partial_\mu \lambda \,.
\label{deltaAF}
\ee
So, in 	total we have five independent parameters generating local symmetries.
In Kaluza-Klein theory, we will treat these as the components of a five-dimensional vector, $\hat{\xi}^{\hat{\mu}} = (\xi^\mu, \lambda)$, where $\hat \mu$ denotes a five-dimensional spacetime index.

The number of independent field components also matches the decomposition of a five-dimensional metric, $\hat g_{\hmu\hnu} \rightarrow (g_{\mu\nu}, A_\mu, \phi)$. 
We therefore construct the five-dimensional theory by defining its field content to consist solely of a metric $\hat g_{\hmu\hnu}$, transforming under five-dimensional diffeomorphisms as:
\be
\delta_{\hat \xi} \hat g_{\hmu\hnu} = L_{\hat \xi} \hat g_{\hmu\hnu} = \hat\xi^{\hrho} \partial_{\hrho} \hat g_{\hmu \hnu} + 2 \partial_{(\hmu} \hat \xi^{\hrho} \hat g_{\hnu) \hrho} \,.
\label{kk5diff}
\ee
This combines the $\mathrm{GL}(4)$ diffeomorphisms and $\mathrm{U}(1)$ gauge transformations of the four-dimensional theory into $\mathrm{GL}(5)$ diffeomorphisms.
(Observe here the structure of the general $d$-dimensional Lie derivative acting on some tensor: first, the transport term involving the derivative of the tensor, then $\mathrm{GL}(d)$ transformations of its indices using the $\mathrm{GL}(d)$ matrix $\partial_\mu \xi^\nu$.) 
 
We now seek two things: firstly, a higher-dimensional action for the metric $\hat g_{\hmu\hnu}$ which is invariant under the symmetries \refeqn{kk5diff}, and secondly, a parametrisation of $\hat g_{\hmu\hnu}$ in terms of four-dimensional variables such that decomposing \refeqn{kk5diff} in components we recover the expected four-dimensional symmetries and dynamics.
The higher-dimensional action we are looking for must be the Einstein-Hilbert action:
\begin{align}
S= \int \dd^5x \sqrt{|\hat{g}|} R(\hat{g}) \, ,
\label{KKaction}
\end{align}
and an appropriate parametrisation of the five-dimensional metric can be easily written down:
\be
\hat g_{\hmu\hnu} = \begin{pmatrix} 
\phi^{-\tfrac{1}{2}} g_{\mu\nu} + \phi A_\mu A_\nu & \phi A_\mu \\
\phi A_\nu & \phi 
\end{pmatrix}\,.
\label{kkmetric}
\ee
The factor of $\phi^{-\tfrac{1}{2}}$ in front of $g_{\mu\nu}$ ensures that we will below obtain the canonically normalised four-dimensional Einstein-Hilbert term i.e. the four-dimensional Einstein frame action.
Evaluating the transformation \refeqn{kk5diff} on the metric \refeqn{kkmetric} we indeed recover the four-dimensional symmetries \refeqn{deltaKK} assuming that we restrict our fields to obey the Kaluza-Klein constraint:
\begin{align}
\partial_{x^5}  {} \, \,= 0  \, .
\label{KKconstraint}
\end{align}
(Normally we should think of this a truncation to the zero modes in a Fourier expansion in the $x^5$ direction.)
Similarly, we can work through the Kaluza-Klein reduction of the action \refeqn{KKaction} given the parametrisation \refeqn{kkmetric} and the constraint \refeqn{KKconstraint}.
This leads to the four-dimensional action:
\begin{equation}
S= \int \dd^4 x \sqrt{|g|}\left( R(g) - \frac{1}{4} \phi^{\tfrac{3}{2}} {F}_{\mu \nu} {F}^{\mu \nu} - \frac{3}{8} \partial_\mu \ln \phi\,\partial^\mu \ln \phi\right) \,.
\label{4dem}
\end{equation}
If we define the canonically normalised scalar $\varphi = \tfrac{\sqrt{3}}{2} \ln \phi$, then the action is:
\begin{equation}
S= \int \dd^4 x \sqrt{|g|}\left( R(g) - \frac{1}{4} e^{\sqrt{3} \varphi} {F}_{\mu \nu} {F}^{\mu \nu} - \frac{1}{2} \partial_\mu \varphi \,\partial^\mu \varphi \right) \,.
\label{4demcanonical}
\end{equation}
A famous lesson of Kaluza-Klein theory is that it is impossible to geometrise just the metric and one-form alone; the higher-dimensional metric inevitably produces the Kaluza-Klein scalar and it is in general inconsistent to truncate this. (If one started with the ansatz \refeqn{kkmetric} with $\phi = 0$, solutions of the equations of motion of the reduced theory would not generically correspond to solutions of the equations of motion of the higher-dimensional theory.)
A different slant on this is to note that the result \refeqn{4dem} implies that the precise numerical coefficients accompanying the scalar kinetic term, and its coupling to the one-form, are necessary  for this action to come from a reduction of the higher-dimensional theory \refeqn{KKaction} (modulo field redefinitions).

An interesting feature of the action \refeqn{4demcanonical} is that it is invariant under the following global transformation:
\be
\phi \rightarrow c^2 \phi \,,\quad A_\mu \rightarrow c^{-\tfrac{3}{2}} A_\mu \,, \quad c^2 \in \mathbb{R}^+\,,
\ee
under which the Einstein frame metric does not transform.
We can further combine this with a $\mathbb{Z}_2$ acting as $A_\mu \rightarrow - A_\mu$ to obtain a $\mathrm{GL}(1)$ symmetry.
This is a very simple example of a global symmetry corresponding to a change in the moduli of reduction (in this case the radius $R \sim \sqrt{\phi}$ of the Kaluza-Klein circle). Conversely, without this precise symmetry present in the Einstein-Maxwell-scalar theory, there would be no uplift to the five-dimensional theory.
Note that if one drops the Kaluza-Klein constraint, allowing the higher dimensional field to depend on the coordinate $x^5$ or including the tower of Kaluza-Klein modes with masses $m^2 \sim ( \tfrac{n}{R} )^2$, then the spectrum in the reduced theory is sensitive to the radius modulus. 

\subsection{Double field theory}
\label{genKK_DFT}

Can we repeat the Kaluza-Klein procedure for $p$-form gauge fields, with $p>1$? 
Let's start with a theory of a metric and a two-form, $B_{\mu\nu}$, which we will refer to as the $B$-field.
The $B$-field arises naturally in string theory, as it couples to the fundamental string itself.
We now take the dimension of spacetime to be arbitrary, and denote it by $d$.
The theory we will obtain will not correspond precisely to a Kaluza-Klein reduction in the usual sense, but will lead, via ideas of generalised geometry \cite{Gualtieri:2003dx,Hitchin:2004ut} to \emph{double field theory} (DFT) \cite{Siegel:1993th, Siegel:1993xq,Hull:2009mi,Hull:2009zb, 
Hohm:2010jy,Hohm:2010pp}.
This Kaluza-Klein approach to DFT was emphasized in \refref{Berman:2019biz} and a elaborated on mathematically in \refref{Alfonsi:2019ggg}.

The local symmetries of a metric and two-form are now given by diffeomorphisms, generated by vectors $\xi^\mu$, plus gauge transformations of the two-form, generated by one-forms, $\lambda_\mu$:
\be
\delta g_{\mu\nu}  = L_\xi g_{\mu\nu} \,,\quad
\delta B_{\mu\nu}  = L_\xi B_{\mu\nu} + 2 \partial_{[\mu} \lambda_{\nu]} \,.
\ee
A new feature of these one-form gauge symmetries is that transformations $\lambda_\mu = \partial_\mu \varphi$ are trivial: they do not generate any change of the $B$-field. Thus there are gauge symmetries of the gauge symmetries.
The gauge invariant strength of the B-field is:
\be
H_{\mu\nu\rho} = 3 \partial_{[\mu} B_{\nu\rho]}\,.
\ee 
For later use, we note again that combined diffeomorphism and gauge symmetry can be written as
\be
\delta B_{\mu\nu} = \xi^\rho H_{\rho \mu\nu} + 2 \partial_{[\mu} \lambda_{\nu]}  
\label{deltaBH}
\ee
after shifting $\lambda_\mu \rightarrow \lambda_\mu -  \xi^\nu B_{\nu\mu}$.

\subsubsection{Generalised diffeomorphisms}

We want to again combine the parameters generating the local symmetries into one geometrical quantity.
Here, the combination of $d$-dimensional vectors and $d$-dimensional one-forms gives what we call a \emph{generalised} vector field:
\be
\Lambda^M = ( \xi^\mu,\lambda_\mu  ) \,,
\label{genvecinit}
\ee
which is a $2d$-dimensional object. 
For such generalised vectors, one might try to generate $\mathrm{GL}(2d)$ diffeomorphisms on introducing (as in Kaluza-Klein theory) $d$ extra coordinates, $\tilde x_\mu$. 
However, if you do this, it will soon become clear that a conventional Kaluza-Klein approach based on such a theory cannot reproduce the symmetries, field content or action of the theory with the two-form. For instance, a $d$-dimensional metric and $B$-field have a combined total of $d^2$ independent components, whereas a $2d$-dimensional metric $g_{MN}$ would have $d(2d+1)$ components. 

We should therefore think more critically about what sort of geometry is going to appear. 
Geometrically, denoting the spacetime manifold by $M$, a generalised vector \refeqn{genvecinit} is a section of the sum $TM \oplus T^*M$ of the tangent and cotangent bundles. 
This is known as the generalised tangent bundle, and is the starting point of the {\it{generalised geometry}} developed by Hitchin and Gualtieri \cite{Gualtieri:2003dx,Hitchin:2004ut}. 

Given this setup, we can already define a natural \emph{indefinite} metric, namely a split signature bilinear form
\be
\eta_{MN} = \begin{pmatrix} 0 & \delta_\mu^\nu \\ \delta^\mu_\nu & 0\end{pmatrix} \,,
\label{etafirstlook}
\ee
corresponding to the natural pairing between vectors and one-forms, expressed here as:
\be
\eta_{MN} \Lambda^M \Lambda^{\prime N} = \xi^\mu \lambda^\prime_\mu + \xi^{\prime \mu} \lambda_\mu \,.
\ee
The bilinear form \refeqn{etafirstlook} is preserved by the group $O(d,d)$. For this reason we refer to it as the $O(d,d)$ structure or $O(d,d)$ metric.

We therefore seek a notion of \emph{generalised} diffeomorphism that preserves this pairing i.e. which preserves $O(d,d)$.
We define this notion using the \emph{generalised Lie derivative}.
We denote the generalised Lie derivative with respect to some generalised vector $U$ by $\mathcal{L}_U$, and we require it to be compatible with the $O(d,d)$ structure \refeqn{etafirstlook}, meaning that:
\be
\mathcal{L}_U \eta_{MN} = 0 \,.
\label{Leta}
\ee
To explicitly define the generalised Lie derivative, we first assume it acts on a scalar quantity $S$ in the usual fashion, $\delta_U S = U^M \partial_M S$.
Then we will specify its action on generalised vectors, and require that it obey the Leibniz property in order to define its action on generalised covectors, $V_M$, and hence generalised tensors. For instance, $\eta_{MN}$ is a generalised tensor of rank 2. 
We also require that generalised Lie derivatives form a closed algebra in a sense that we will make precise shortly.
The (unique) definition that results is:
\be
\mathcal{L}_U V^M = U^N \partial_N V^M - V^N \partial_N  U^M + \eta^{MN} \eta_{PQ} \partial_N U^P V^Q \,.
\label{ODDGLD}
\ee
Here we have explicitly written partial derivatives with respect to a set of doubled coordinates $X^M = (x^\mu, \tilde x_\mu)$.
The theory itself will soon tell us how we should think of the $d$ extra coordinates we have included here. 

This generalised Lie derivative can be rewritten educationally in various ways. For instance, we can write:
\be
\mathcal{L}_U V^M =L_U V^M + Y^{MN}{}_{PQ} \partial_N U^P V^Q \,,\quad Y^{MN}{}_{PQ} = \eta^{MN} \eta_{PQ} \,,
\label{ODDGLDY}
\ee
in order to show that it consists of the ordinary Lie derivative plus a modification, the so-called Y-tensor, which is itself $O(d,d)$ invariant.
Alternatively, we can write:
\be
\mathcal{L}_U V^M =U^N \partial_N V^M - 2 \mathbb{P}^M{}_N{}^P{}_Q \partial_P U^Q V^N\,,\quad
\mathbb{P}^M{}_N{}^P{}_Q = \tfrac{1}{2} (\delta^M_Q \delta^P_N - \eta^{MP} \eta_{NQ}) \,,
\label{ODDGLDproj}
\ee
where $\mathbb{P}^M{}_N{}^P{}_Q$ projects the $\mathrm{GL}(2d)$ matrix $\partial_P U^Q$ into the adjoint of $O(d,d)$. This makes the $O(d,d)$ compatibility condition \refeqn{Leta} immediate, and displays the generalised Lie derivative as a transport term plus an $O(d,d)$ transformation.

Unlike the usual Lie derivative, the generalised Lie derivative is not antisymmetric in $U \leftrightarrow V$.
The skew-symmetrisation of \refeqn{ODDGLD} defines an antisymmetric bracket
\be
[U,V]_C^M \equiv \frac{1}{2} ( \mathcal{L}_U V - \mathcal{L}_V U )^M\,,
\ee
which we refer to as the C-bracket. (The ``C'' signifies that for $\tilde\partial^\mu = 0$ this bracket matches what is known as the Courant bracket of generalised geometry.)

The ordinary Lie derivative forms a closed algebra under the Lie bracket, $[L_U, L_V] = L_{[U,V]}$.
We require the algebra of generalised Lie derivatives to behave similarly:
\be
[\mathcal{L}_{U}, \mathcal{L}_V ] = \mathcal{L}_{[U,V]_C} \,.
\ee
However, if we compute this requirement on a generalised vector $W^M$ we get: 
\be
\begin{split}
[\mathcal{L}_U,\mathcal{L}_V]W^M & = \mathcal{L}_{[ U,V ]_C} W^M  \\ & + \eta^{NP}
\eta_{QR}  \Big(  -(\partial_PV^Q)(\partial_NU^M)W^R 
+ (\partial_PU^Q)(\partial_NV^M)W^R  \\ & 
 \qquad - \frac{1}{2}(\partial_N W^M)(\partial_P U^Q) V^R
 + \frac{1}{2}(\partial_N W^M)(\partial_P V^Q) U^R \Big) \,.
\end{split}
\label{eq:LieClosure}
\ee
At first, this appears to be a failure: we do not have a closed algebra for arbitrary $U^M$ and
$V^M$. For closure we must impose some consistency conditions (note that this calculation also fixes the coefficients of the various terms in the definition of the generalised Lie derivative).
The standard way to do so is to restrict the coordinate dependence of all quantities, so that the following condition is obeyed: 
\be
\eta^{MN} \partial_M \Psi \partial_N \Psi^\prime = 0 
\ee
where $\Psi,\Psi^\prime$ stand for any fields or gauge parameters of the theory. This is known as the \emph{strong
constraint}.
There is also a \emph{weak constraint}, which is: 
\be
\label{eq:weak}
\eta^{MN} \partial_M \partial_N \Psi = 0 \,.
\ee
This has an origin in string theory as arising from the level-matching constraint acting on the massless fields.
Collectively, these two
constraints give the \emph{section condition} or \emph{section constraint} of the theory, which we write as:
\be
\eta^{MN} \partial_M \otimes \partial_N = 0\,.
\label{sectionfirstlook}
\ee
Using the definition \refeqn{etafirstlook}, this is saying that $\partial_\mu \otimes \tilde \partial^\mu + \tilde \partial^\mu \otimes \partial_\mu = 0$.
The obvious solution to this constraint is to say that in fact nothing depends on the extra $d$ coordinates $\tilde x_\mu$, i.e. to set $\tilde\partial^\mu = 0$ acting on anything. 
In that case, the extra coordinates $\tilde x_\mu$ are a (useful) fiction which appear in this reformulation, but appear to not have physical meaning. 
However, in string theory, we can interpret these extra coordinates as being the coordinates conjugate to winding modes, in backgrounds with compact cycles.
They are then T-dual to the coordinates $x^\mu$ in the compact directions.
For this reason, we will refer to the $\tilde x_\mu$ as dual coordinates.

It may seem at this point that the constraint $\tilde \partial^\mu = 0$ appears to be of the same nature as the Kaluza-Klein truncation condition \refeqn{KKconstraint}.
However the theory that we are constructing is rather different in nature to a genuine Kaluza-Klein uplift.
The section condition solution $\partial_\mu \neq 0$, $\tilde \partial^\mu =0$ is more than a truncation to the zero mode sector; the strong constraint $\eta^{MN} \partial_M \Psi \partial_N \Psi^\prime = 0$ forbids the existence of the tower of Kaluza-Klein modes in the $\tilde x_\mu$ directions that one would normally have.
In addition, picking a solution to \refeqn{sectionfirstlook} breaks the $O(d,d)$ symmetry that would otherwise be present.
We will return later to the interpretation of these doubled coordinates, and the role of $O(d,d)$, in string theory.

Assuming that the generalised vectors do not depend on the $\tilde x_\mu$, i.e. if we set $\tilde\partial^\mu = 0$, then letting $U^M = (u^\mu, \lambda_\mu)$, $V^M = (v^\mu, \eta_\mu)$, the definition \refeqn{ODDGLD} is equivalent to:
\be
\mathcal{L}_U V= ( L_u v , L_u \eta - \iota_v d\lambda ) \,,
\label{ODDGLDcpts}
\ee
using the ordinary Lie derivative, $L_u$, the exterior derivative $d$ and the interior product $\iota$ defined such that $(\iota_v d\lambda)_\mu= 2 v^\nu \partial_{[\nu} \lambda_{\mu]}$.
In particular, the vector part is just the ordinary Lie derivative. 
We see from \refeqn{ODDGLDcpts} that generalised vectors of the form $U^M = ( 0, \partial_\mu\varphi)$, for $\varphi$ an arbitrary scalar, do not generate any transformation.
These correspond to the trivial, or reducible, gauge transformations of the $B$-field. 

Using the section condition, generalised vectors corresponding to trivial transformations can be written as follows: 
\be
U^M = \eta^{MN} \partial_N \varphi \,,
\label{trivialgenvec}
\ee
and it follows using the section condition that generalised vectors of this form give $\mathcal{L}_U = 0$ acting on anything.
The symmetrisation of the generalised Lie derivative is of this form:
\be
\{ U, V\}^M \equiv \frac{1}{2} ( \mathcal{L}_U V+\mathcal{L}_V U)^M = \frac{1}{2} \eta^{MN} \partial_N ( \eta_{PQ} U^P V^Q ) \,.
\ee
The presence of such trivial transformations is another important difference between the generalised and ordinary Lie derivatives.
One place such transformations appear is when one checks whether the C-bracket obeys the Jacobi identity.
Unlike the Lie bracket, the C-bracket does not obey Jacobi, even on imposing the section condition, but instead corresponds to a generalised vector generating a trivial transformation:
\be
[ [ U,V]_C , W]_C + \text{cyclic} = \frac{1}{3} \{ [U,V]_C, W \} + \text{cyclic}\,.
\ee
Hence the Jacobi identity holds acting on fields.

\subsubsection{Generalised metric and dilaton}

We now define fields with definite tensorial properties under generalised diffeomorphisms.
We define the \emph{generalised metric} as a rank two symmetric generalised tensor, transforming as 
\be
\delta_\Lambda \cH_{MN} = \mathcal{L}_{\Lambda} \cH_{MN} 
 = \Lambda^P\partial_P \cH_{MN} + 2 \partial_{(M} \Lambda^P \cH_{N)P} - 2\eta^{PQ} \eta_{K(M|} \partial_P \Lambda^K \cH_{|N)Q} \,,
\label{gldfirstlook}
\ee
and itself defining an element of $O(d,d)$ via the compatibility requirement:
\be
\cH_{MK} \eta^{KL} \cH_{LN} = \eta_{MN} \,,
\ee
This requirement plus imposing that $\cH_{MN}$ is symmetric implies that it parametrises the coset $\tfrac{O(d,d)}{O(1,d-1) \times O(1,d-1)}$, which has $d^2$ components.
We can compare this to the ordinary $d$-dimensional metric parametrising $\tfrac{ \mathrm{GL}(d) }{\mathrm{SO}(1,d-1)}$, where the denominator subgroup is the local Lorentz symmetry.

The $d^2$ components of the metric $g$ and $B$-field $B$ now appear as components of the generalised metric, which can be parametrised as:
\be
\cH_{MN} = \begin{pmatrix}
g_{\mu\nu} - B_{\mu\rho}g^{\rho\sigma} B_{\sigma\nu} & B_{\mu\rho}g^{\rho \nu} \\
- g^{\mu\rho} B_{\rho \nu} & g^{\mu\nu}
\end{pmatrix} \,,
\label{Hfirstlook}
\ee
If we solve the section condition \refeqn{sectionfirstlook} by imposing $\tilde \partial^\mu = 0$ then it is a straightforward calculation to check that the components of \refeqn{gldfirstlook} evaluated on the parametrisation \refeqn{Hfirstlook} of the generalised metric reproduce the standard diffeomorphism and gauge transformations of $g$ and $B$.

We now want to construct an action from which the dynamics will derive. 
For consistency, we have to introduce an additional degree of freedom, analogous to the Kaluza-Klein scalar.
The generalised metric \refeqn{Hfirstlook} has unit determinant, which precludes us from writing down an invariant measure.
We therefore introduce a scalar $\gendil$ such that $e^{-2\gendil}$ has weight one; it transforms under generalised diffeomorphisms as
\be
\delta_\Lambda e^{-2\gendil} = \partial_M (\Lambda^M e^{-2\gendil}) \,.
\ee
We can therefore identify it with
\be
e^{-2\gendil} = e^{-2\Phi} \sqrt{|g|} \,,
\label{gendilfirstlook}
\ee
where $\Phi$ is an ordinary scalar field. We will refer to $\gendil$ as the generalised, or doubled, dilaton.

As our theory is not invariant under $\mathrm{GL}(2d)$ diffeomorphisms, its dynamics do not follow from the usual Einstein-Hilbert action.
Although one can construct notions of generalised connections and curvatures, it is also possible to proceed by brute force: writing down a combination of all fully contracted terms quadratic in derivatives of $\cH_{MN}$ and $\gendil$ and fixing their coefficients by the requirement that this be a scalar under generalised diffeomorphisms (up to terms which vanish by the section condition).
Remarkably, there is a unique solution:
\be
S_{\text{DFT}}=
\int \dd X \, e^{-2\gendil} \mathcal{R}_{\text{DFT}}(\cH, \gendil) 
\label{SDFTfirstlook}
\ee
with
\be
\begin{split}
\mathcal{R}_{\text{DFT}}  = \,&
\frac{1}{8} \cH^{M N} \partial_M  \cH^{PQ} \partial_N  \cH_{PQ}
- \frac{1}{2}  \cH^{MN} \partial_M  \cH^{PQ} \partial_P  \cH_{QN}\\
& + 4 \partial_M  \cH^{MN} \partial_N d 
- 4  \cH^{MN} \partial_M d \partial_N d\\
& - \partial_M \partial_N  \cH^{MN} + 4  \cH^{MN} \partial_M \partial_N d
 \,.
\end{split}
\label{RDFTfirstlook}
\ee
This is the action of double field theory. It has been developed in \refref{Siegel:1993th, Siegel:1993xq,Hull:2009mi,Hull:2009zb, 
Hohm:2010jy,Hohm:2010pp}.

On inserting the parametrisations \refeqn{Hfirstlook} and \refeqn{gendilfirstlook} of the generalised metric and scalar, and setting $\tilde \partial^\mu = 0$, this recovers the following action:
\be
S=\int \dd^{d} x \sqrt{-g} e^{-2\Phi}\left(R(g) -\frac{1}{12} H_{\mu\nu\rho} H^{\mu\nu\rho}  + 4 \nabla_\mu \nabla^\mu  \Phi  -4  \nabla_\mu \Phi \nabla^\mu \Phi  \right)   \, .  
\label{NSsugrafirstlook}
\ee
Observe that just as in Kaluza-Klein reduction, we arrive not just at an arbitrary theory of a metric and two-form, but one in which the couplings to the scalar are of a very particular form. 

This action is well-known in string theory.
For $d=26$, it describes the low energy effective action for the massless fields of the bosonic string.
For $d=10$, it describes the low energy effective action for the massless fields in the NSNS sector of the type II superstring.
In this case, we refer to the metric as being in string frame, meaning that there is a factor of $e^{-2\Phi}$ multiplying the Einstein-Hilbert term.
The scalar $\Phi$ is the string dilaton.

The action \refeqn{NSsugrafirstlook} exhibits no obvious signs of the $O(d,d)$ symmetry. 
However, if we Kaluza-Klein reduce this action on $D$ dimensions, an $O(D,D)$ duality symmetry is present in the reduced theory. 
This is the T-duality symmetry of string theory on a $D$-dimensional torus.
Here we have found that we can much more generally use a reformulation based around $O(d,d)$ to describe the full theory in any background! 

Indeed, it was searching for ``duality symmetric'' reformulations of string theory and supergravity that first led to the development of double field theory, as was historically pioneered in \refref{Duff:1989tf, Tseytlin:1990nb, Tseytlin:1990va, Siegel:1993th, Siegel:1993xq}.
In string theory, these coordinates $\tilde x_\mu$ appear as the coordinates conjugate to string winding modes on a torus as in the derivation of \refref{Hull:2009mi}. The full double field theory with the section constraint applied is not restricted to such backgrounds, however in this case the dual coordinate dependence is restricted by the strong section constraint.

Finally, let us say something about the field equations for double field theory.
That for the generalised dilaton, $\gendil$, can be directly found to be:
\be
\mathcal{R}_{\text{DFT}}= 0 \,.
\ee
The field equation for the generalised metric must be obtained more carefully.
This is because any variation of the field $\cH_{MN}$ must respect the fact it is $O(d,d)$ valued.
Varying the $O(d,d)$ compatibility condition $\cH \eta^{-1} \cH = \eta$ implies that the variation of the generalised metric is projected,
\be
\delta \cH_{MN} =  P_{MN}{}^{KL} \delta \cH_{KL} 
\,,\quad
P_{MN}{}^{KL} 
= \frac{1}{2}(\delta_{M}^{(K} \delta_N^{L)}- \cH_{MP} \eta^{P(K}\eta_{NQ}\cH^{L)Q})\,.
\label{projdeltaH}
\ee
(Note that $\cH_{MP} \eta^{PN} = \eta_{MP} \cH^{PN}$.)
Hence formally the equations of motion are:
\begin{align}
\mathcal{R}_{MN} \equiv 
P_{MN}{}^{KL} \frac{\delta S}{\delta {\cH}^{KL}} =0 \,,
\label{Reom}
\end{align}
where $\mathcal{R}_{MN}$ geometrically defines a generalised Ricci tensor.

\subsubsection{Solving the section condition}

We will re-encounter DFT later on in this review, when we consider its (standard and exotic) brane solutions in sections \ref{branes} and \ref{nongeoexotic}, as well as some interesting non-Riemannian solutions to its field equations in section \ref{nonrie}. 
Here we encourage the reader to view it as a simpler prototype of the ideas of ExFT that we will now develop in more depth.
For further reading on DFT, we refer to the reviews \refref{Aldazabal:2013sca,Berman:2013eva, Hohm:2013bwa}.
We will now just make some comments about the nature of the section condition.
These comments will also apply to exceptional field theory.

The DFT description does not assume any Kaluza-Klein reduction, and instead extends the geometry by introducing the generalised Lie derivative.
We can conveniently describe the extended geometry in terms of the extra coordinates $X^M$, whose physical nature is restricted by the imposition of the section condition. 
In a genuinely toroidal background, where we have $\partial_M = 0$ and a true $O(d,d)$, the extra coordinates should be thought of as being conjugate to string winding modes. 

In general, how should we interpret the section condition?
Let's consider DFT for simplicity.
We saw above that \refeqn{sectionfirstlook} may be solved by imposing $\tilde \partial^\mu = 0$ acting on all fields and gauge parameters.
This resembles the form of the Kaluza-Klein constraint, but the strong constraint means that this is more than a truncation to the zero mode sector: we are really killing off all coordinate dependence on the $\tilde x_\mu$. 

However, if we start with the doubled description, we can equally well solve the section condition by taking:
\begin{align}
\partial_\mu =0  \, . \label{altsec}
\end{align}
Then we view the $\tilde x_\mu$ as the physical coordinates.

How do we interpret these different choices in satisfying the constraint \refeqn{sectionfirstlook}?
There are in fact two distinct issues. 
We must be able to solve simultaneously the equations of motion and the section constraint, and we should identify what we call {\it{spacetime}}.\footnote{A frequently asked question is whether it is possible to impose the section condition on-shell. This would be preferable to imposing it by hand on all field configurations, however so far such an approach has not been realised.} Spacetime should be a half-dimensional submanifold of the doubled space. The natural thing to do is to use the solution of the section condition to determine our choice of how we identify spacetime inside the doubled space. (This is not required by double field theory but doing so allows a supergravity interpretation of the solution.) 

If we have a solution to the equations of motion of DFT that does not depend on a coordinate or its dual i.e. all fields are independent of some particular pair of coordinates $x$ and ${\tilde{x}}$ then there is an ambiguity in identifying spacetime inside the doubled space. One can take $x$ as being the spacetime coordinate and the space having an isometry in the $x$ direction or ${\tilde{x}}$ as being the spacetime coordinate and the space having an isometry in the ${\tilde{x}}$ direction.

These alternative perspectives in the presence of isometries of spacetime lead to the conventional notion of T-duality. From the double field theory point of view T-duality is just an ambiguity in identifying the spacetime submanifold inside the double space. 
Supposing that $x$ and $\tilde x$ parametrise a doubled torus, with $R$ the radius of the $x$ circle, the $O(1,1)$ generalised metric is $\cH_{MN} = \diag( R^2 , 1/ R^2)$. 
When we declare that $x$ is the physical coordinate we interpret this as $\cH_{MN} = \diag( g_{xx} , g^{xx} )$, however if we take $\tilde x$ to be the physical coordinate then 
we would view $\cH_{MN} = \diag ( \tilde g^{\tilde x \tilde x}, \tilde g_{\tilde x \tilde x})$ and the physical metric is $\tilde g_{\tilde x\tilde x} = 1/R^2$
(here we kept the generalised metric fixed but changed which components of generalised vectors correspond to spacetime vectors, we can also take the opposite point of view).
In this way we see the standard radius inversion T-duality from the point of view of DFT.
For $d$ isometries the duality group then becomes $O(d,d)$ as a change in choice of identifying the spacetime coordinates becomes an $O(d,d)$ transformation on $\cH_{MN}$.

\subsection{Exceptional field theory}
\label{genKK_EFT}

The next stage is to repeat with a three-form theory what we just did with one- and two-forms.
This is a more tricky problem. The gauge transformations of a three-form $C_{ijk}$ are:
\be
\delta C_{ijk} = 3 \partial_{[i} \lambda_{jk]} \,.
\ee
(In this section, we shift our notation somewhat, and denote $d$-dimensional spacetime indices by $i,j,\dots$.)
We want to combine the two-form gauge parameters, $\lambda_{ij}$, with vectors $v^i$, to again obtain generalised vectors $\Lambda^M = (v^i, \lambda_{ij})$.
The number of components of a generalised vector is now $d + \frac{1}{2} d(d-1)$.
The counting suggests that these could arise from an \emph{antisymmetric} representation of the group $\mathrm{GL}(d+1)$, rather than a vector representation as in Kaluza-Klein theory (and also as for $O(d,d)$ in double field theory).

Meanwhile, the metric and three-form have $\frac{1}{2} d(d+1) + \frac{1}{6}d(d-1)(d-2)$ components.
In general, this does not naturally fit into a $\mathrm{GL}(d+1)$ metric.
We are forced to pick out $d=4$ (which happens to the smallest value of $d$ for which we can define the field strength $F_{ijkl} = 4 \partial_{[i}C_{jkl]}$). 
In this case, the metric and three-form together comprise $14$ independent components.
This matches the number of components of a symmetric unit determinant five-by-five matrix and hence of the coset $\tfrac{\mathrm{SL}(5)}{\mathrm{SO}(5)}$  (for convenience, we assume implicitly here that our metric has Euclidean signature; the theory we will obtain below will be viewed as a truncation from a larger theory of Lorentzian signature).
If we further include, again, an extra scalar, this can be described by a $\mathrm{GL}(5)$ metric. We will for now ignore this scalar and introduce it separately below, in order to more closely follow our description of DFT.

\subsubsection{Generalised diffeomorphisms}

We will now describe the $\Gfour$ theory. 
Let $\fM,\fN,\dots = 1,\dots,5$ denote five-dimensional indices. 
Generalised vectors are in fact ten-dimensional (with four vector and six two-form components). Geometrically, these are sections of an extended tangent bundle, $TM \oplus \Lambda^2 T^*M$ \cite{Hull:2007zu}.
We define a generalised vector $\Lambda^{\fM \fN} = - \Lambda^{\fN \fM}$ with components
\be
\Lambda^{i 5} = v^i \,,\quad 
\Lambda^{ij} = \frac{1}{2} \epsilon^{ijkl} \lambda_{kl} \,,
\ee
where $\epsilon^{ijkl}$ with $\epsilon^{1234} = 1$ is the alternating symbol.
We then further extend our theory, again along Kaluza-Klein lines, by introducing dual coordinates $\tilde x_{ij}$ with a pair of antisymmetric indices; we group these with the original coordinates into an antisymmetric set of coordinates $X^{\fM \fN}=-X^{\fN \fM}$ with components $X^{i5} = x^i$, $X^{ij} = \frac{1}{2} \epsilon^{ijkl} \tilde x_{kl}$.
We define derivatives $\partial_{\fM \fN}$, with components $\partial_{i5} = \partial_i$ and $\partial_{ij} = \frac{1}{2} \epsilon_{ijkl} \tilde\partial^{kl}$.\footnote{This dualisation using the alternating symbol is convenient when using the five-dimensional indices to construct the antisymmetric representation (we could instead write directly $\Lambda^M = (v^i, \lambda_{ij})$ with mixed upper and lower indices).}

We could define the ordinary Lie derivative involving two ten-dimensional generalised vectors, but this would give a $\mathrm{GL}(10)$ Lie derivative and not capture the symmetries we want.
Instead, let's think about the group $\mathrm{SL}(5)$. 
This has the totally antisymmetric invariants $\epsilon_{\fM \fN \fP \fQ \fK}$ and $\epsilon^{\fM \fN \fP \fQ \fK}$.
A generalised Lie derivative which preserves these invariants is defined by \cite{Berman:2011cg, Berman:2012vc}:
\be
\mathcal{L}_\Lambda W^{\fM} = \frac{1}{2} \Lambda^{\fP \fQ} \partial_{\fP \fQ} W^{\fM} - W^{\fP} \partial_{\fP \fQ} \Lambda^{\fM \fQ} + \frac{1}{5} \partial_{\fP \fQ} \Lambda^{\fP\fQ} W^{\fM}\,,
\ee
acting on a field $W^{\fM}$ carrying a single five-dimensional index. The factor of $1/2$ in the first term is inserted to prevent overcounting.
Using the Leibniz rule, this implies on a second generalised vector $V^{\fM \fN}$ we have:
\be
\begin{split}
\mathcal{L}_\Lambda V^{\fM\fN}& =  \frac{1}{2} \Lambda^{\fP \fQ} \partial_{\fP \fQ} V^{\fM \fN} -  \frac{1}{2} V^{\fP \fQ} \partial_{\fP \fQ} \Lambda^{\fM \fN} \\
& \quad + \frac{1}{8} \epsilon^{\fM \fN \fP \fQ \fT} \epsilon_{\fK \fL \fR \fS \fT} \partial_{\fP \fQ} \Lambda^{\fK \fL} V^{\fR \fS} 
- \frac{1}{5} \frac{1}{2} \partial_{\fP \fQ} \Lambda^{\fP\fQ} V^{\fM \fN} 
\,,
\end{split}
\ee
or in terms of a single 10-dimensional index $M \equiv [\fM \fN]$, letting $V^M \equiv V^{\fM \fN}$, $\Lambda^M \equiv \Lambda^{\fM \fN}$, we can write
\be
\mathcal{L}_\Lambda V^M = \Lambda^N \partial_N V^M - V^N \partial_N \Lambda^M + Y^{MN}{}_{PQ} \partial_N \Lambda^P V^Q - \frac{1}{5} \partial_N \Lambda^N V^M
\label{gldYfirstlook} \,.
\ee
letting $Y^{MN}{}_{PQ} \equiv \epsilon^{MN \mathcal{K}} \epsilon_{PQ \mathcal{K}}$.
We can compare \refeqn{gldYfirstlook} to \refeqn{gldfirstlook}.
The final term with the $\tfrac{1}{5}$ coefficient is a consequence of choosing to define an $\Gfour$ rather than $\mathrm{GL}(5)$ Lie derivative.
In practice it is convenient to eliminate this from many expressions by declaring all generalised vectors to have weight $+\frac{1}{5}$ (note this means $\partial_M$ has weight $-\frac{1}{5}$).
We will discuss weights in more detail in section \ref{exft}.

The consistency of the theory, in particular closure of the algebra generated by generalised Lie derivatives, again requires a section condition, which this time takes the form:
\be
\epsilon^{\fM \fN \fP \fQ \fK} \partial_{\fM \fN} \partial_{\fP \fQ} \Psi = 0 \,,\quad
\epsilon^{\fM \fN \fP \fQ \fK} \partial_{\fM \fN} \Psi \partial_{\fP \fQ} \Psi^\prime = 0 \,,
\label{section5firstlook}
\ee
where again $\Psi,\Psi^\prime$ stand for any quantities in the theory.

The solution of this constraint which returns us to the $d$-dimensional theory is $\tilde \partial^{ij} = 0$.
Again, though the requirement $\tilde \partial^{ij} = 0$ appears to be of the same nature as the Kaluza-Klein truncation condition \refeqn{KKconstraint}, this is really a more stringent condition.

In this case, for two generalised vectors $U^M = ( u, \lambda_{(2)})$, $V^M = (v,\eta_{(2)})$ viewed as pairs of vectors $u,v$ and two-forms $\lambda_{(2)}$, $\eta_{(2)}$, the generalised Lie derivative is 
\be
\mathcal{L}_U  V = ( L_u v, L_u \eta_{(2)} - \iota_v d \lambda_{(2)} ) \,,
\ee
which is the direct generalisation of the $O(d,d)$ expression \refeqn{ODDGLDcpts}.

We will discuss the further properties of generalised Lie derivatives of this form in section \ref{exft}, and move on now to describing the metric and three-form in this new $\Gfour$ extended geometry.

\subsubsection{Generalised metric}

We define a $\Gfour$-valued generalised metric $m_{\fM \fN}$ which is symmetric and has unit determinant.
This means we should take it to have weight zero under the above $\Gfour$ generalised diffeomorphisms.
One way to parametrise this generalised metric is\cite{Duff:1990hn}
\be
m_{\cal{M} \cal{N}}
=\begin{pmatrix} 
g^{-2/5} g_{ij} & - g^{-2/5} g_{ik} C^k \\
- g^{-2/5} g_{jk} C^k & g^{3/5} + g^{-2/5} g_{kl} C^k C^l 
\end{pmatrix} \,,\quad
C^i \equiv \frac{1}{3!} \epsilon^{ijkl} C_{jkl} \,.
\label{5gmfirstlook}
\ee
Now acting with the generalised Lie derivative on $m_{\fM \fN}$, we find that for $\tilde \partial^{ij} = 0$ and $\Lambda^{\fM \fN}= (v^i, \frac{1}{2}\epsilon^{ijkl} \lambda_{kl})$ we recover the expected symmetry transformations of the metric and three-form.

In order to write an action, we introduce a scalar $\Delta$ such that $e^{-2\Delta}$ has weight $\tfrac{7}{5}$ under generalised diffeomorphisms. (This funny choice of weight is a consequence of again working with an $\Gfour$ compatible generalised Lie derivative.)
This can be expressed in terms of a scalar field $\varphi$ and the determinant of the metric as:
\be
e^{-2\Delta} = e^{-2\varphi} |g|^{7/10} \,.
\ee
We can write the action most compactly by introducing a ten-by-ten representation of the generalised metric:
\be
\gM_{MN} \equiv \gM_{\fM \fM^\prime, \,\fN \fN^\prime} = 2 m_{\fM [ \fN} m_{\fN^\prime]\fM^\prime} \,.
\ee 
The parametrisation of $\gM_{MN}$ resulting from \refeqn{5gmfirstlook} is:
\be
\gM_{MN} = |g|^{\tfrac{1}{5}}\begin{pmatrix}
g_{ik} + \frac{1}{2} C_{i mn } C_{k}{}^{mn} & \frac{1}{2} C_{i}{}^{mn} \epsilon_{ kl mn}   \\
\frac{1}{2} C_{k}{}^{mn} \epsilon_{ijmn} & 2 |g|^{-1} g_{i[k} g_{l] j} 
\end{pmatrix} \,,
\ee
where indices on the three-form are raised using $g^{ij}$.
This is the direct generalisation of the $O(d,d)$ generalised metric, compare with \refeqn{Hfirstlook}.
Using $\gM_{MN}$ and $\Delta$, we can then search for a quantity quadratic in derivatives which is a scalar under generalised diffeomorphisms (up to terms vanishing by the section condition). The result leads to the action:
\be
\begin{split}
S   =\int \dd^{10} X\, e^{-2\Delta} \,\Big(&
\frac{1}{12} \gM^{MN} \partial_M \gM^{KL} \partial_N \gM_{KL} - \frac{1}{2} \gM^{MN} \partial_M \gM^{KL} \partial_K \gM_{LN} 
\\ & \qquad
+ \frac{24}{7} \left(
\partial_M \gM^{MN} \partial_N \Delta - \gM^{MN}\partial_M \Delta \partial_N \Delta + \gM^{MN} \partial_M \partial_N \Delta 
\right)
\\ & \qquad
-\partial_M \partial_N \gM^{MN} 
\Big)\,.
\end{split}
\label{babyexft}
\ee
Truncating the coordinate dependence via $\tilde \partial^{ij} = 0$ we obtain 
\be
S = \int \dd^4x \sqrt{|g|} e^{-2\varphi} \Big(
R(g) - \frac{1}{48} F_{ijkl} F^{ijkl} + \frac{24}{7} g^{ij} \partial_i \varphi \partial_j \varphi
\Big) \,.
\label{s4trunc}
\ee
Now we should ask if this result is familiar from string theory or -- given that a three-form couples not to strings but to membranes -- M-theory.
The place we should look to find a theory with a metric and a three-form is 11-dimensional supergravity \cite{Cremmer:1978km}. The bosonic part of the action for this theory is
\be
\begin{split}
S & = \int \dd^{11}x \sqrt{|\hat g|} \left( R(\hat g) - \frac{1}{48} \hat F_{\hmu\hnu\hrho\hsigma} \hat F^{\hmu \hnu \hrho \hsigma} \right)
\\ & \qquad + \int\dd^{11} x \frac{1}{144^2} \epsilon^{\hmu_1 \dots \hmu_{11}} \hat F_{\hmu_1 \dots \hmu_4} \hat F_{\hmu_5 \dots \hmu_8} \hat C_{\hmu_9 \dots \hmu_{11}} 
\,,
\end{split}
\ee
where $x^{\hmu}$ denote the 11-dimensional coordinates.
It can be checked that the action \refeqn{s4trunc} is a truncation of this theory to four-dimensions, assuming that the full 11-dimensional metric $\hat g_{\hmu\hnu}$ and three-form $\hat C_{\hmu\hnu\hrho}$ are of the form
\be
\begin{split}
d\hat s_{11}^2 & = e^{-\tfrac{4}{7} \varphi} \eta_{\mu\nu} dx^\mu dx^\nu + g_{ij} dx^i dx^j \,,\\
\hat C_3 & = \frac{1}{6} C_{ijk} dx^i \wedge dx^j \wedge dx^k \,,
\end{split}
\label{11trunc}
\ee
and that no fields depend on the seven-dimensional coordinates $x^\mu$. (Here $\eta_{\mu\nu}$ denotes a seven-dimensional Minkowskian constant metric.) 
The action \refeqn{s4trunc} was first obtained (modulo the absence of the extra scalar) in \refref{Berman:2010is}, while the truncation \refeqn{11trunc} is as in \refref{Coimbra:2011ky,Coimbra:2012af}.

Why does $\Gfour$ appear? 
Suppose we made the opposite truncation of 11-dimensional supergravity: assuming there was no coordinate dependence on the directions $x^i$ and Kaluza-Klein reducing.
Then we would obtain a seven-dimensional theory with an $\Gfour$ U-duality symmetry.
Here we have managed to do the opposite and still found that the structure of the theory is controlled by $\Gfour$!

In general, if we reduce eleven-dimensional supergravity on a $d$-dimensional torus, we obtain an $n$-dimensional theory ($n=11-d$) with a symmetry of the group $\Edd$.
These are the U-duality symmetry groups which we listed in table \ref{tab:eddinfo}.
For $d=6,7,8$ we obtain the exceptional Lie groups of Cartan's classification, while for $d \leq 5$ we have by definition $E_{5(5)} = \mathrm{SO}(5,5)$, $E_{4(4)} = \Gfour$, $E_{3(3)} = \mathrm{SL}(3) \times \mathrm{SL}(2)$. 
For $d=9$, there is an infinite dimensional symmetry, based on the affine Lie algebra $E_9$. 
For $d > 9$, it is conjectured that there is a symmetry based on the further infinite-dimensional Lie algebra extensions defined through the generalisations of the $\Edd$ Dynkin diagram, the maximal case being $E_{11}$ 	\cite{West:2001as}. 

For this exceptional sequence of groups, actions of the form \refeqn{babyexft} can be obtained, based on a notion of $\Edd$ generalised diffeomorphisms, and with generalised metrics which are elements of $\tfrac{\Edd}{H_d}$, where $H_d$ is the maximal compact subgroup listed in table \ref{tab:eddinfo}.
These theories, first developed for instance in \refref{Hillmann:2009ci,Berman:2010is, Berman:2011pe,Berman:2011jh,Coimbra:2011ky,Coimbra:2012af,Godazgar:2013rja}, are equivalent to truncations of eleven-dimensional supergravity.
For $d > 4$, these truncations involve not only the three-form, but components of its electromagnetic dual six-form, which is required for the representation theory to work.

In order to go beyond truncations, the full exceptional field theory includes also the $n$-dimensional spacetime that we ignored in making the truncation \refeqn{11trunc}. 
We replace the scalar $\Delta$ with the extra degrees of freedom needed to describe the full eleven-dimensional metric and three-form.
These extra degrees of freedom are the $n$-dimensional external metric and a ``tensor hierarchy'' of fields carrying both $n$-dimensional and $\Edd$ indices, denoted by $\Aa_\mu{}^M$, $\Ab_{\mu\nu}{}^{(MN)}$, $\dots$.
These can be thought of as generalisations of $p$-forms.

The full exceptional field theories were first developed in \refref{Hohm:2013pua,Hohm:2013vpa,Hohm:2013uia,Hohm:2014fxa} for the cases of $\Gsix, \Gseven$ and $\Geight$, and then extended to the other lower rank groups in \refref{Hohm:2015xna,Abzalov:2015ega,Musaev:2015ces,Berman:2015rcc} (and more recently have been partially extended to the infinite-dimensional cases in \refref{Bossard:2017aae,Bossard:2018utw,Bossard:2019ksx}).

These theories are all restricted by versions of the section condition. 
It turns out that there are two distinct ways to solve this condition \refref{Hillmann:2009ci, Hohm:2013vpa, Blair:2013gqa}.
One of these takes $d$ of the extended coordinates to be physical, and reduces exceptional field theory to 11-dimensional supergravity (which can be reduced to 10-dimensional type IIA supergravity).
The other takes $d-1$ of the extended coordinates to be physical, and reduces exceptional field theory to 10-dimensional type IIB supergravity.
These two solutions cannot be related into each other by an $\Edd$ transformation.

Hence exceptional field theory provides, via the introduction of an extended or exceptional geometry based on $\Edd$ generalised diffeomorphisms, a unification of the 11- and 10-dimensional maximal supergravities, as well as a higher-dimensional origin of the $\Edd$ symmetries that appear as the U-dualities on dimensional reduction.
The next part of this review will develop exceptional field theory in more detail.

%% file: sec3SUGRAAndDuality.tex

\section{Supergravity decompositions and $\Edd$} 
\label{SUGRAduality}

In this section, we will take a closer look at 11-dimensional supergravity.
We will make an $(n+d)$-dimensional split of the coordinates, and rewrite the fields, symmetries and action in a form adapted to Kaluza-Klein reduction, without actually carrying out any truncation of the coordinate dependence or making any assumptions about the backgrounds we are describing. 
This will introduce many features which are important for understanding the structure of exceptional field theory, and will also be necessary for precisely matching the latter to supergravity. Here we follow loosely the presentation of \refref{Hohm:2013vpa} and also draw on the Kaluza-Klein ``gauging'' perspective stressed in \refref{Hohm:2019wql}. 
In addition, by making this split we will see the first signs of a natural reorganisation of the field content into $\Edd$ multiplets.

\subsection{Revisiting Kaluza-Klein} 
\label{KKsplit}

\subsubsection{The decomposition of the metric}

First, we will revisit Kaluza-Klein reductions of pure gravity.
We start with a metric $\hat g_{\hmu \hnu}$ in $(n+d)$ dimensions, whose dynamics is governed by the Einstein-Hilbert term.
We are going to rewrite this theory in an unfamiliar fashion: as if we are going to perform a Kaluza-Klein reduction, but without actually truncating the coordinate dependence.
We split the coordinates into $n$ ``external'' coordinates and $d$ ``internal'' coordinates and mimic the decomposition we would do if we were going to reduce on the $d$ ``internal'' directions. However, \emph{we assume throughout that all fields depend on all $(n+d)$ coordinates.}

Accordingly, we split the $(n+d)$-dimensional index as $\hmu = (\mu,i)$, where $\mu$ is an $n$-dimensional index and $i$ a $d$-dimensional index.
A Kaluza-Klein-esque decomposition of the metric is
\be
\hat g_{\mu\nu} = \begin{pmatrix}
\phi^{\omega}  g_{\mu\nu} + A_\mu{}^k \phi_{kl} A_\nu{}^l & 
\phi_{j k} A_\mu{}^k \\
\phi_{ik} A_\nu{}^k & \phi_{ij} 
\end{pmatrix} \,,
\label{metricsplit}
\ee
with $g_{\mu\nu}$ the external metric, $\phi_{ij}$ the internal metric, and $A_\mu{}^i$ the Kaluza-Klein vector.
A factor of the determinant $\phi \equiv \det \phi$ of the internal metric appears in the decomposition of the external metric components, raised to a power $\omega$.
This is convenient for controlling whether the action viewed in $n$-dimensional terms is in Einstein frame or not. 
If we take
\be
\omega = -\frac{1}{n-2} 
\ee
then we will obtain the $n$-dimensional Ricci scalar with no powers of the determinant $\phi$ multiplying it.\footnote{In order to go from 10-dimensional string frame (where the dilaton appears multiplying the Ricci scalar as $e^{-2\Phi} R$) to $n$-dimensional Einstein frame, we should replace $\phi^{\omega}  g_{\mu\nu} \rightarrow \phi^{\omega} e^{-4\Phi \omega} g_{\mu\nu}$. This is important because U-dualities are manifest in lower-dimensional Einstein frame, while T-dualities are manifest in lower-dimensional string frame, for which we would accordingly use the same decomposition but with $\omega=0$.}

The price we pay for splitting the coordinates in this fashion, and writing the metric as in \refeqn{metricsplit}, is that we fix the local Lorentz symmetry, $\mathrm{SO}(1,n+d-1) \rightarrow \mathrm{SO}(1,n-1)\times \mathrm{SO}(d)$.
The Lorentz gauge fixing  underlying the metric parametrisation of \refeqn{metricsplit} is made explicit by writing the vielbein and its inverse
\be
\hat e^{\hat a}{}_{\hmu} = \begin{pmatrix}
\phi^{\tfrac{\omega}{2}}  e^a{}_\mu &  0 \\
 A_\mu{}^j \phi^{\bar j}{}_j &  \phi^{\bar j}{}_j
\end{pmatrix} 
\,,\quad
\hat e^{\hmu}{}_{\hat a}{} = \begin{pmatrix}
\phi^{-\tfrac{\omega}{2}}  e^\mu{}_a &  0 \\
- \phi^{-\tfrac{\omega}{2}}  e^\nu{}_a A_\nu{}^i  & \phi^i{}_{\bar i}
\end{pmatrix} \,,
\label{vielbeinKK}
\ee
where $e^a{}_\mu e^b{}_\nu \eta_{ab} = g_{\mu\nu}$ and $\phi^{\bar i}{}_i \phi^{\bar j}{}_j \delta_{\bar i \bar j} = \phi_{ij}$, i.e. $a,b,\dots$ denote flat $\mathrm{SO}(1,n-1)$ indices and $\bar i, \bar j,\dots$ denote flat $\mathrm{SO}(d)$ indices.

We require that the lower triangular form of $\hat e^{\hat a}{}_{\hmu}$ be preserved under symmetry transformations.
Diffeomorphisms act on the vielbein as
\be
\delta_{\hat\Lambda} \hat e^{\hat a}{}_{\hmu}  = \hat\Lambda^{\hnu} \partial_{\hnu} \hat e^{\hat a}{}_{\hmu}  + \partial_{\hmu} \hat\Lambda^{\hnu} \hat e^{\hat a}{}_{\hnu} \,.
\label{deltahate}
\ee
We stress again we make no assumptions about the coordinate dependence.
Then, under $d$-dimensional diffeomorphisms, with $\hat\Lambda^{\hmu} = (0,\Lambda^i)$, the form \refeqn{vielbeinKK} of the vielbein is preserved, and we find
\be
\begin{split}
\delta_\Lambda e^a{}_\mu & = L_\Lambda e^a{}_\mu = \Lambda^k \partial_k e^a{}_\mu -\omega \partial_k \Lambda^k e^a{}_\mu \,,\\
\delta_\Lambda A_\mu{}^i & = D_\mu \Lambda^i \,,\\
\delta_\Lambda \phi^{\bar i}{}_{i} &= L_\Lambda \phi^{\bar i}{}_{i} = \Lambda^j \partial_j \phi^{\bar i}{}_i + \partial_i \Lambda^j \phi^{\bar i}{}_j \,.
\end{split}
\label{deltaKKsplit}
\ee
We define $L_\Lambda$ to be the internal Lie derivative acting as usual on fields carrying internal indices, where the internal vielbein $\phi^{\bar i}{}_{i}$ has weight 0, but the external vielbein $e^a{}_\mu$ has weight $-\omega$.
In the transformation of the Kaluza-Klein vector $A_\mu{}^i$ we introduce 
\be
D_\mu \equiv \partial_\mu - L_{A_\mu} \,,
\ee
which is an $n$-dimensional derivative made covariant under $d$-dimensional diffeomorphisms, with $A_\mu{}^i$ the connection.
Covariance means that if $T$ is some quantity which transforms tensorially under $d$-dimensional diffeomorphisms, $\delta_\Lambda T = L_\Lambda T$, then so does $D_\mu T$, i.e. $\delta_\Lambda D_\mu T = L_\Lambda D_\mu T$.
We can define a field strength
\be
F_{\mu\nu}{}^i = 2 \partial_{[\mu} A_{\nu]}{}^i - 2 A_{[\mu|}{}^j \partial_j A_{|\nu]}{}^i \,,
\label{kkf}
\ee
which appears in the commutator $[D_\mu, D_\nu] = - L_{{F}_{\mu\nu}}$.

If we were to impose the Kaluza-Klein reduction condition $\partial_i = 0$, then the transformation \refeqn{deltaKKsplit} of the fields reduces to a standard $U(1)^d$ transformation of $A_\mu{}^i$ under $\Lambda^i(x)$, and the field strengths \refeqn{kkf} take the usual abelian form.
The second term in \refeqn{kkf} can be thought of as a non-Abelian modification in the presence of the infinite-dimensional gauge symmetry corresponding to diffeomorphisms $\Lambda^i(x,y)$ depending on both the $n$- and $d$-dimensional coordinates.

Under $n$-dimensional diffeomorphisms, generated by $\hat\Lambda^{\hmu} = (\xi^\mu,0)$, a vielbein component $\hat e^{a}{}_i \neq 0$ is generated.
This does not preserve our gauge fixing.
We must therefore remove this with a compensating (infinitesimal) Lorentz transformation of the form
\be
\delta_{\hat \lambda} \hat e^{\hat a}{}_{\hmu} = \lambda^{\hat a}{}_{\hat b} \hat e^{\hat b}{}_{\hmu}
\,,\quad
\lambda^a{}_{\bar i} = - \phi^i{}_{\bar i} \partial_i \xi^\nu \phi^{ \tfrac{\omega}{2}} e^a{}_\nu = - \delta_{\bar i \bar j} \eta^{a b} \lambda^{\bar j}{}_{b}\,.
\ee
The resulting expressions are not manifestly covariant under internal diffeomorphisms.
This can be improved by applying also a field-dependent internal diffeomorphism $\Lambda^i = - \xi^\nu A_\nu{}^i$.
The result of external diffeomorphisms plus compensating Lorentz and field-dependent internal diffeomorphisms is then
\be
\begin{split}
\delta_\xi e^a{}_\mu & = \xi^\nu D_\nu e^a{}_\mu + D_\mu \xi^\nu e^a{}_\nu\,,\\
\delta_\xi A_\mu{}^i & = \xi^\nu F_{\nu \mu}{}^i + \phi^{\omega}\phi^{ij}g_{\mu\nu} \partial_j \xi^\nu \,,\\
\delta_\xi \phi^{\bar i}{}_{i} & = \xi^\nu D_\nu \phi^{\bar i}{}_i \,. 
\end{split}
\label{KKndiff}
\ee
When $\partial_i = 0$, the total symmetry transformation of $A_\mu{}^i$ coming from \refeqn{deltaKKsplit} and \refeqn{KKndiff} has the form of \refeqn{deltaAF}.

\subsubsection{Including $p$-forms}

Let's discuss the general treatment of form-fields in line with the above decomposition.
As well as simply splitting the indices of the form fields, we will also redefine them systematically in order to obtain fields with more natural symmetry transformations.
The procedure is to first flatten the indices with the vielbein, such that for instance for a one-form $\hat \Omega_{\hmu}$ we define $\hat \Omega_a$ and $\hat \Omega_{\bar i}$ by
\be
\hat \Omega_a \equiv \hat e_a{}^{\hmu} \hat \Omega_{\hmu} = \phi^{- \tfrac{\omega}{2}} e_a{}^\mu ( \hat \Omega_\mu - A_\mu{}^i \hat \Omega_i ) \,,\quad
\hat \Omega_{\bar i} \equiv \hat e_{\bar i}{}^{\hmu} \hat \Omega_{\hmu} = \phi_{\bar i}{}^{i} \hat \Omega_i \,,
\ee
and then unflatten using $\hat e_\mu{}^a$ (\emph{not} $\hat e_\mu{}^{\hat a}$) and $\hat e_i{}^{\bar i}$, in order to obtain the following redefined form fields
\be
\Omega_\mu \equiv \hat \Omega_{\mu} - A_\mu{}^i \hat \Omega_i\,,
\quad
\Omega_i \equiv \hat \Omega_i \,.
\ee
This ensures that the components carrying external indices transform nicely under internal diffeomorphisms, $\delta_\Lambda \Omega_i = L_\Lambda \Omega_i$ and $\delta_\Lambda \Omega_\mu = L_\Lambda \Omega_\mu$ where $\Omega_i$ is a covector of zero weight and $\Omega_\mu$ a scalar of zero weight with respect to the internal Lie derivative.
The field strength and gauge transformation parameters of the form-field can be similarly redefined, as we will soon see.

Let's apply this to the case of a three-form, $\Celeven_{\hmu\hnu\hrho}$, which is relevant for 11-dimensional SUGRA.
We make the following redefinition of its components
\be
\begin{split}
\Cdef_{ijk} & = \Celeven_{ijk} \,,\\
\Cdef_{\mu ij} & = \Celeven_{\mu ij} - A_\mu{}^k \Celeven_{k ij} \,,\\ 
\Cdef_{\mu \nu i} & = \Celeven_{\mu \nu i} - 2A_{[\mu}{}^k \Celeven_{\nu] i k} 
+ A_\mu{}^k A_\nu{}^l \Celeven_{i k l }\,,\\ 
\Cdef_{\mu\nu\rho} & = \Celeven_{\mu\nu\rho} - 3 A_{[\mu}{}^k \Celeven_{\nu\rho]k} 
+ 3 A_{[\mu}{}^k A_\nu{}^l \Celeven_{\rho]kl} 
- A_\mu{}^k A_\nu{}^l A_\rho{}^m \Celeven_{klm}\,,
\end{split} 
\label{redefC3}
\ee
and do the same for the field strengths, leading to the expressions
\be
\begin{array}{cccl}
&\Fdef_{mnpq} & =& 4 \partial_{[m} \Cdef_{npq]}\,, \\
&\Fdef_{\mu mnp} & =& D_\mu \Cdef_{mnp} -3 \partial_{[m} \Cdef_{|\mu|np]}\,,\\
&\Fdef_{\mu\nu m n} & = &2 D_{[\mu} \Cdef_{\nu] mn} + F_{\mu\nu}{}^p \Cdef_{pmn} +2\partial_{[m} \Cdef_{|\mu\nu|n]}\,,\\
&\Fdef_{\mu\nu\rho m} & = &
3 D_{[\mu} \Cdef_{\nu \rho]m} + 3 F_{[\mu\nu}{}^n \Cdef_{\rho]mn}
- \partial_m \Cdef_{\mu\nu\rho} \,,\\
&\Fdef_{\mu \nu \rho \sigma} & =&4 D_{[\mu} \Cdef_{\nu\rho \sigma]} + 6 F_{[\mu\nu}{}^m \Cdef_{\rho \sigma ] m} \,,
\end{array}
\label{F=dA}
\ee
where on the right-hand side we see the Kaluza-Klein field strength $F_{\mu\nu}{}^m$ defined in \refeqn{kkf}.

A consequence of these redefinitions is the resulting field strengths obey modified Bianchi identities
\be
\begin{array}{cccl}
&D_\mu \Fdef_{mnpq} & =& 4 \partial_{[m} \Fdef_{|\mu|npq]}\,, \\
&2 D_{[\mu} \Fdef_{\nu] mnp} & =& - 3 \partial_{[m|} \Fdef_{\mu\nu|np]} - F_{\mu\nu}{}^q \Fdef_{qmnp} \,,\\
& 3 D_{[\mu} \Fdef_{\nu\rho ] m n} & = & 2 \partial_{[m|} \Fdef_{\mu\nu\rho|n]} + 3 F_{[\mu\nu}{}^p \Fdef_{\rho]pmn}\,,\\
&4 D_{[\mu} \Fdef_{\nu\rho\sigma] m} & = & - \partial_m \Fdef_{\mu\nu\rho\sigma} + 6 F_{[\mu\nu}{}^p \Fdef_{\rho\sigma]mp} \,,\\
&5D_{[\mu} \Fdef_{\nu \rho \sigma \lambda]} & =& 10 F_{[\mu\nu}{}^m \Fdef_{\rho\sigma \lambda] m} \,.
\end{array}
\label{FBI_mod}
\ee
Excluding the transformation under external diffeomorphisms, which is more complicated and not immediately relevant for us, the gauge transformations of the above fields are as follows.
\begin{itemize}
\item 
Under internal diffeomorphisms, the redefined components of the three-form, $\Cdef$, transform according to their index structure via the internal Lie derivative, i.e. for example $\delta_\Lambda \Cdef_{\mu \nu i} = \Lambda^k \partial_k \Cdef_{\mu\nu i} + \partial_i \Lambda^k \Cdef_{\mu\nu k}$.
(The redefinition \refeqn{redefC3} guarantees that no terms involving $\partial_\mu \Lambda^i$ appear.)
Under these transformations the field strengths are covariant, also transforming under internal Lie derivatives (and hence are invariant when setting $\partial_i=0$).

\item The original three-form transformed under abelian gauge transformations as $\delta_{\hat\lambda} \Celeven_{\hmu\hnu\hrho} = 3 \partial_{[\hmu} \hat\lambda_{\hnu\hrho]}$.
Redefining $\hat\lambda$ in the same way,
\be
\begin{split}
\lambdadef_{ij} & = \hat\lambda_{ij} \,,\\
\lambdadef_{\mu i} & = \hat \lambda_{\mu i} - A_\mu{}^k \lambda_{ki} \,,\\
\lambdadef_{\mu\nu} & = \hat \lambda_{\mu\nu} + 2 A_{[\mu}{}^k \hat \lambda_{\nu] k} + A_{\mu}{}^k A_\mu{}^l \hat \lambda_{kl}\,,
\end{split}
\label{redeflambda}
\ee
we have
\be
\begin{split}
\delta_\lambda \Cdef_{ijk} & = 3 \partial_{[i} \lambdadef_{jk]} \,,\\
\delta_\lambda \Cdef_{\mu i j} & = D_\mu \lambdadef_{ij} - 2 \partial_{[i} \lambdadef_{|\mu|j]} \,,\\
\delta_\lambda \Cdef_{\mu\nu i} & = 2 D_{[\mu} \lambdadef_{\nu] i} - \lambdadef_{ij} F_{\mu\nu}{}^j + \partial_i \lambdadef_{\mu\nu} \,,\\
\delta_\lambda \Cdef_{\mu\nu\rho} & = 3 D_{[\mu} \lambdadef_{\nu\rho]}  - 3 F_{[\mu \nu}{}^k \lambdadef_{\rho]k} \,. 
\end{split} 
\label{redefCgauge}
\ee
Under these transformations the field strengths are invariant (by definition).
\end{itemize}

Note that by using the vielbein in this fashion, various calculations become simpler by first going to flat indices before carrying out the split of the indices.
For instance,
\be
\begin{split} 
\hat \Omega_{\hmu} \hat \Omega_{\hnu} \hat g^{\hmu \hnu} 
& = \hat \Omega_a \hat \Omega_b \eta^{ab} + \hat \Omega_{\bar i} \hat \Omega_{\bar j} \delta^{\bar i \bar j} \\
& = \hat \Omega_a \hat \Omega_b g^{\mu\nu} e_\mu{}^a e_\nu{}^b + \hat \Omega_{\bar i} \hat \Omega_{\bar j} \phi_i{}^{\bar i} \phi_j{}^{\bar j} \phi^{ij} \\
& = \phi^{-\omega} \Omega_\mu \Omega_\nu g^{\mu\nu}  + \phi^{ij} \Omega_i \Omega_j \,.
\end{split} 
\ee
Most relevantly, for a $p$-form field strength $\hat F_{\hmu_1 \dots \hmu_p}$ redefined as above we will get
\be
\frac{1}{p!} \hat F_{\hmu_1 \dots \hmu_p} \hat F^{\hmu_1 \dots \hmu_p}
= \sum_{q=0}^p \frac{1}{q! (p-q)!} \phi^{-q \omega} \Fdef_{\mu_1 \dots \mu_q i_{1} \dots i_{p-q} } \Fdef^{\mu_1 \dots \mu_q i_{1} \dots i_{p-q} }
\,, 
\label{magicF}
\ee
using $g^{\mu\nu}$ and $\phi^{ij}$ to raise indices.

\subsection{Decomposition of 11-dimensional supergravity}
\label{11red}

Now we want to illustrate the outcome of the above procedure for the bosonic part of the action of 11-dimensional supergravity. 
This action is
\be
\begin{split}
S = \int \dd^{11} x  &\left( \sqrt{|\hat g|}\left[ R( \hat g ) - \frac{1}{48} \Feleven^{\hmu\hnu\rho\sigma} \Feleven_{\hmu\hnu\hrho\hsigma} \right] + \frac{1}{144^2} \epsilon^{\hmu_1\dots\hmu_{11}} \Feleven_{\hmu_1\dots\hmu_4} \Feleven_{\hmu_5\dots\hmu_8} \Celeven_{\hmu_9\hmu_{10} \hmu_{11} }\right)\,,
\end{split} 
\label{Seleven}
\ee
where $\Feleven_{\hmu\hnu\rho\sigma} = 4 \partial_{[\hmu} \Celeven_{\hnu\hrho\hsigma]}$.

There are several ways to calculate the decomposition of the action.
We will not give details (see \refref{Hohm:2013vpa} for one way to do it) but merely state that the result, entirely equivalent to \refeqn{Seleven}, can be organised into
\be
\begin{split}
S & = \int \dd^{n} x \,\dd^d Y 
\sqrt{|g|}\left( 
 R_{\text{ext}}(g) + \mathcal{L}_{\text{kin}} 
 + \mathcal{L}_{1} + \mathcal{L}_2 + \mathcal{L}_3 
 + \mathcal{L}_{\text{int}}
+ \sqrt{|g|}^{-1} \mathcal{L}_{\text{CS}}
\right) \,,
\end{split}
\label{Splitnd}
\ee
where we group the terms as follows.
First, the (improved) external Ricci scalar (ignoring total derivatives) is
\be
\begin{split}
R_{\text{ext}}(g)  = &
\frac{1}{4} g^{\mu \nu} D_\mu g_{\rho \sigma} D_\nu g^{\rho \sigma} 
- \frac{1}{2} g^{\mu \nu} D_\mu g^{\rho \sigma} D_\rho g_{\nu \sigma} \\
&\qquad\qquad 
+ \frac{1}{4} g^{\mu \nu} D_\mu \ln g D_\nu \ln g 
+ \frac{1}{2} D_\mu \ln g D_\nu g^{\mu \nu} \,,
\end{split}
\label{11Rext}
\ee
which includes contributions involving the Kaluza-Klein vector $A_\mu{}^i$ via the derivatives $D_\mu = \partial_\mu - L_{A_\mu}$.
Secondly, we have ``kinetic terms'' for the purely internal fields
\be
\mathcal{L}_{\text{kin}} 
= + \frac{1}{4} D_\mu \phi^{ij} D^\mu \phi_{ij} - \frac{1}{4(n-2)} D_\mu \ln \phi D^\mu \ln \phi - \frac{1}{12}  \Fdef^{\mu ijk} \Fdef_{\mu ijk} \,.
\label{11kin_int}
\ee
Then we have kinetic terms for 1-forms, 2-forms and 3-forms
\be
\mathcal{L}_1 = - \frac{1}{4} \phi^{\tfrac{1}{n-2}}  {F}^{\mu\nu i} {F}_{\mu\nu}{}_i- \frac{1}{8} \phi^{\tfrac{1}{n-2}} \Fdef^{\mu\nu ij} \Fdef_{\mu\nu ij}\,,
\label{11kin_1}
\ee
\be
\mathcal{L}_2  = - \frac{1}{12} \phi^{\tfrac{2}{n-2}} \Fdef^{\mu\nu\rho i} \Fdef_{\mu\nu\rho i}\,,
\label{11kin_2}
\ee
\be
\mathcal{L}_3 = - \frac{1}{48} \phi^{\frac{3}{n-2}} \Fdef^{\mu\nu\rho\sigma} \Fdef_{\mu\nu\rho\sigma} \,,
\label{11kin_3}
\ee
We use the internal metric $\phi_{ij}$ and its inverse to raise and lower internal indices, and the external metric $g_{\mu\nu}$ and its inverse to raise and lower external indices.
Finally, we have terms involving solely internal derivatives
\be
\begin{split}
\mathcal{L}_{\text{int}}
= \phi^{-\tfrac{1}{n-2}} 
\bigg( &
\frac{1}{4} \phi^{ij} \left(  \partial_i g^{\mu\nu} \partial_j g_{\mu\nu} +  \partial_i \ln g \partial_j \ln g \right)
+ \frac{1}{2} \phi^{\tfrac{1}{n-2}} \partial_i ( \phi^{ij}\phi^{-\tfrac{1}{n-2}} ) \partial_j \ln g
\\ & \quad
+\frac{1}{4} \phi^{i j} \partial_i \phi_{k l} \partial_j \phi^{k l} 
- \frac{1}{2} \phi^{i j} \partial_j \phi^{k l} \partial_{k} \phi_{i l}
\\ & \qquad
 + \frac{1}{4} \frac{4-n}{(n-2)^2} \phi^{i j} \partial_i \ln \phi \, \partial_j \ln \phi
- \frac{1}{n-2} \partial_i \ln \phi \, \partial_j \phi^{i j} \,.
\\ & \qquad
 - \frac{1}{48} 
\Fdef^{ijkl} \Fdef_{ijkl}
\bigg)\,.
\end{split}
\label{11int}
\ee
The details of the Chern-Simons term are best presented on a case-by-case basis.

How do we interpret this result? 

One immediate application is that we have rewritten the theory in a way which makes Kaluza-Klein reduction immediate.
Imposing the truncation $\partial_i=0$, we find an $n$-dimensional theory consisting of a metric, various $p$-forms, and scalars.
This theory inherits a $\mathrm{GL}(d)$ global symmetry acting on the $d$-dimensional indices $i,j$ in the obvious way.
This is a global ``duality'' symmetry that the $n$-dimensional theory has inherited from its higher-dimensional origin.
Indeed, if we let $n=4$ and $d=1$ in \refeqn{Splitnd}, and drop the three-form, we immediately recover our earlier reduction \refeqn{4dem} for $\partial_i = 0$.

Now let's return from $n$ to $n+d$ dimensions.
We reinstate the dependence on the $d$-dimensional coordinates in all fields and symmetry parameters.
The global $\mathrm{GL}(d)$ is augmented to the local $\mathrm{GL}(d)$ of $d$-dimensional diffeomorphisms, which are generated by the $d$-dimensional vectors $\Lambda^i(x,y)$.
Looking at the action \refeqn{Splitnd} and the symmetry transformations \refeqn{KKndiff} and \refeqn{deltaKKsplit} we see that the effect of turning back on the dependence on the $d$-dimensional coordinates is to realise this ``gauging'' by the replacement $\partial_\mu \rightarrow D_\mu = \partial_\mu - L_{A_\mu}$, modifying the standard Kaluza-Klein field strength $F_{\mu\nu}{}^i =2 \partial_{[\mu} A_{\nu]}{}^i$ to the definition \refeqn{kkf}, and generating the additional terms $\mathcal{L}_{\text{int}}$, such that the whole action is invariant under both $d$- and $n$-dimensional diffeomorphisms.

This pure gravity theory in the $(n+d)$ split could be called the $\mathrm{GL}(d)$ exceptional field theory.
Starting with an $n$-dimensional theory with a global $\mathrm{GL}(d)$ symmetry, and $U(1)$ gauge fields transforming under this $\mathrm{GL}(d)$, it realises the local diffeomorphism and gauge symmetries in terms of $\mathrm{GL}(d)$ diffeomorphisms of the higher-dimensional theory.
Furthermore, it exhibits the origin of the lower-dimensional $\mathrm{GL}(d)$ symmetry manifestly, without reducing.

Of course, this is an entirely convoluted way to think about Kaluza-Klein reductions and oxidations. 
That is because we understand very well the geometry of theories with metrics, and also understand very well the geometry of one-form gauge fields.

In a generic theory, that would be the end of the story.
However, in Kaluza-Klein reductions of 11-dimensional supergravity, the metric and 3-form components conspire to transform under global duality symmetries that enhance the naive $\mathrm{GL}(d)$ to the $\Edd$  U-duality symmetry. 
In particular, the Kaluza-Klein one-forms $A_\mu{}^i$ combine with the form field components to give a set of one-forms $\Aa_\mu{}^M = ( A_{\mu}{}^i, C_{\mu i j }, \dots )$ which transform in a representation of the duality group.
One way to view exceptional field theory, as emphasised particularly in \refref{Hohm:2019wql}, is as the result of ``gauging'' this global symmetry, using the whole set of one-forms $\Aa_\mu{}^M$, replacing $\partial_\mu \rightarrow \D_\mu = \partial_\mu - \mathcal{L}_{\Aa_\mu}$, where $\mathcal{L}$ is the generalised Lie derivative.

From the point of view that we introduced in section \ref{genKK}, there is then first of all the generalised Kaluza-Klein interpretation of the $d$-dimensional ``internal'' fields, here denoted $\phi_{ij}$ and $\Cdef_{ijk}$, which unify into the generalised metric $\gM_{MN}$ and transform under generalised diffeomorphisms.
To enlarge our theory to include the $n$-dimensional external space, we have to introduce the gauge fields $\Aa_\mu{}^M$ (in order to covariantise $n$-dimensional partial derivatives) in which case the full theory resembles a Kaluza-Klein decomposition of an even larger theory, with external metric, $g_{\mu\nu}$, Kaluza-Klein vector $\Aa_{\mu}{}^M$ and internal metric $\gM_{MN}$.
The irreducible complexity coming from the fact our original supergravity contained $p$-form fields for $p>1$ will now enter in the need to introduce a tensor hierarchy of additional $n$-dimensional forms, which will be necessary for constructing a field strength for $\Aa_\mu{}^M$. 
All this we will explain fully in section \ref{exft}. 

\begin{figure}[ht]
\centering
	\begin{tikzpicture}
	\tikzset{boxes/.style={inner sep=5pt,draw,align=center, rounded corners, fill=gray!20,text width=2.5cm}};

	\draw (-4.25,0) node (higher) [boxes] {\small 10/11-d SUGRA \\$x^{\hmu}$};

	\draw (0,0) node (split) [boxes] {\small 10/11-d SUGRA \\$x^{\hmu}=(x^\mu, Y^i)$};

	\draw (4.25,3) node (ExFT) [boxes] {\small DFT/ExFT \\$(x^\mu, Y^M)$};
	
	\draw (4.25,-3) node (reduced) [boxes] {\small $n$-dim SUGRA \\ $x^\mu$};
	
	\draw [thick,->] (split) to [bend left] node [below left=-0.2cm and -0.75cm,inner sep=10pt, align=center, text width=2.5cm] {\footnotesize KK reduction \\ \small $\partial_i =0$}	 (reduced);
	
	\draw [dashed,blue,thick,->] (split) to [bend right] node [ above left=0cm and -0.8cm, text width=3cm,align=center, ] {\footnotesize ``Generalised oxidation'' \\ \scriptsize $L_{A_\mu} \rightarrow \mathcal{L}_{\mathcal{A}_\mu}$}  (ExFT) ;

	\draw [thick,dotted,<->] (higher) -- (split) node (gfpos) [midway] {}; 
	\draw (-4,1.2) node (gaugefix) [] {\scriptsize Split coords, gauge fix Lorentz};
	\draw [->] (gaugefix) to [out=0,in=90] (gfpos);
	\end{tikzpicture}
\caption{The procedure described in this subsection is represented by the dotted line in this diagram, and is the starting point for either a Kaluza-Klein reduction, or for a ``generalised oxidation'' leading to double or exceptional field theory.
}
\label{fig:split}
\end{figure}
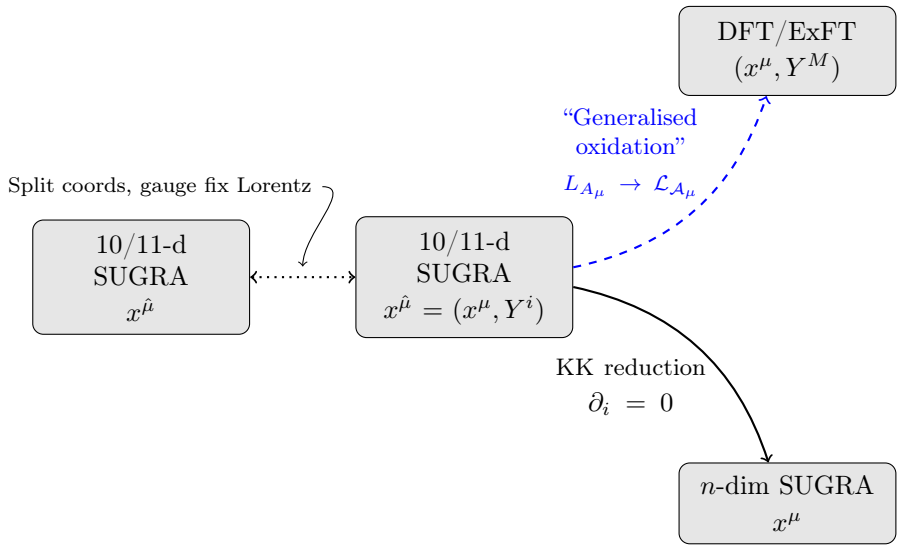

Remarkably, the resulting theory is equivalent not only to the Kaluza-Klein reduction of supergravity, on setting $\partial_M = 0$, but is in fact equivalent to the original 10- and 11-dimensional theories, on choosing appropriate solutions of the section condition.
More precisely, ExFT is most easily seen to be equivalent to supergravity in terms of the $(n+d)$-dimensional partially Lorentz gauge-fixed form described in section \ref{11red} above.
This gives the following picture, shown schematically in figure \ref{fig:split}.
On making such an $(n+d)$-dimensional split of supergravity, we can either:
\begin{itemize}
\item \emph{Reduce:} we immediately obtain the Kaluza-Klein reduction on setting $\partial_i\rightarrow 0$, $D_\mu \rightarrow \partial_\mu$, in which internal diffeomorphisms join with the gauge transformations of the form fields present and become gauge transformations of the fields of the reduced theory. With the field redefinitions as above it is simple to easily verify the U-duality symmetry of the reduced theory.
\item \emph{Extend:} using exceptional field theory, we can extend $\partial_i \rightarrow \partial_M$, with internal diffeomorphisms (generated by the Lie derivative) combining with gauge transformations into generalised diffeomorphisms (generated by a generalised Lie derivative). In this framework, the Kaluza-Klein vectors $A_\mu{}^i$ are packaged with external one-forms arising from the gauge fields into a generalised Kaluza-Klein vector $\mathcal{A}_\mu{}^M$. This field $\mathcal{A}_\mu{}^M$ and the partial derivatives $\partial_M$ are taken to formally transform in conjugate representations of $O(d,d)$ or $\Edd$. This gives a theory on an extended geometry, involving $O(d,d)$ or $\Edd$, and equivalent to the full 10- or 11-dimensional theory on solving the section condition on the $\partial_M$.
\end{itemize}

We will discuss the full ExFT construction in section \ref{exft}. 
First, we will set the scene by explaining how to identify $\Edd$ multiplets in the same $(n+d)$-dimensional split of the supergravity fields.

\subsection{$\Edd$ multiplets}
\label{counting}

We have described an $(n+d)$-dimensional split of 11-dimensional supergravity.
The $\Edd$-covariant multiplets that appear in exceptional field theory, and in $n$-dimensional supergravity on Kaluza-Klein reduction, are hiding amidst the details of this decomposition.
Let us describe how these multiplets appear, for the cases of $d=4$ to $d=8$.

\subsubsection{$\Gfour$}

Let's take $d=4$, $n=7$.
The claim is that we can rearrange all field components into representations of the group $\mathrm{SL}(5)$.
We organise this counting according to how the field components appear from the $n$-dimensional point of view.

We have 14 scalars, $g_{ij}$ and $\Cdef_{ijk}$. These comprise the coset $\tfrac{\mathrm{SL}(5)}{\mathrm{SO}(5)}$, as we saw in section \ref{genKK_EFT}.

We have 10 one-forms, four from the metric $A_\mu{}^i$ and six from the three-form $\Cdef_{\mu ij}$.
This must correspond to the 10-dimensional antisymmetric representation of $\Gfour$.
So far, so good.

However, we have only 4 two-forms, $\Cdef_{\mu \nu i}$, and a single three-form, $\Cdef_{\mu\nu\rho}$.
A singlet we could live with, but there is no non-trivial four-dimensional representation of $\mathrm{SL}(5)$.
In order to obtain a genuine $\Gfour$ representation, we have to use the fact that in seven-dimensions, two- and three-forms are (electromagnetically) dual. 
Hence if we dualise our lone three-form into a two-form we obtain five two-forms, which we can represent in the five-dimensional representation (specifically, the antifundamental).
Equivalently, we could dualise the four two-forms into three-forms, to obtain five of those.

The dualisation of form fields reflects the underlying non-perturbative nature of U-duality in the full M-theory. 
(For U-duality, such dualisations were studied extensively in the seminal work of \refref{Cremmer:1997ct, Cremmer:1998px}.)
U-duality transformations relate branes with their electromagnetic duals. 
In the case at hand, the three-form couples electrically to the M2 brane, and magnetically to the M5 brane.
The latter can also be viewed as being electrically charged (in eleven dimensions) under a dual six-form.
Reducing on a four-torus, this dual six-form gives a single extra two-form.
We can therefore explicitly relate the dual two-form we obtain from dualising the components $\Cdef_{\mu\nu\rho}$  of the three-form to the components $\Cdef_{\mu\nu ijkl}$ of the six-form we would obtain starting in eleven-dimensions with the latter. 

The five-dimensional representation we are constructing corresponds in brane terms to M2 branes wrapped once on the four-torus (giving four possible strings in seven-dimensions) plus the M5 brane wrapped completely on the four-torus (giving one possible string in seven dimensions).
Conversely, if we carried out the opposite dualisation (turning the two-forms into three-forms), we would find the fields coupling to the brane multiplet consisting of an M2 brane not wrapping the four-torus (giving one membrane in seven-dimensions) and the M5 brane wrapping three directions of the torus (giving four possible membranes in seven-dimensions).

We summarise the field content of 11-dimensional SUGRA in terms of $\Gfour$ multiplets in table \ref{sl5count}.
To include all the original three-form components, we include the 5-dimensional representation of three-forms, which is dual to the 5-dimensional representation of two-forms. 
In fact, we do not need to stop there.
The one-form $\mathcal{A}_\mu{}^M$ in the $\mathbf{10}$ would be dual in seven-dimensions to a four-form, which we would denote by $\mathcal{D}_{\mu\nu\rho\sigma M}$.
The degrees of freedom of this four-form would be the six components $\Cdef_{\mu\nu\rho\sigma ij}$ dual to $\Cdef_{\mu ij}$ plus four degrees of freedom $\tilde A_{\mu \nu\rho\sigma i}$ dual to the Kaluza-Klein vector $A_\mu{}^i$.
This brings into play the \emph{dual graviton} in 11-dimensions.

\renewcommand{\arraystretch}{1.4}
\begin{table}[ht]
\tbl{The bosonic field content of 11-dimensional SUGRA in $\Gfour$ multiplets. The two- and three-forms are electromagnetically dual, as are the one- and four-forms.}	
{
\centering
\begin{tabular}{cccc}
Metric & $g_{\mu\nu}$ & & \\
Scalars & $\gM_{MN} \in \frac{\Gfour}{\Hfour}$ & $g_{ij}$ (10) & $\Cdef_{ijk}$ (4) \\
One-forms & $\mathcal{A}_\mu \in \mathbf{10}$ & $A_\mu{}^i$ (4) & $\Cdef_{\mu ij}$ (6) \\
Two-forms & $\mathcal{B}_{\mu\nu} \in \mathbf{\bar{5}}$ & $\Cdef_{\mu \nu i}$ (4) & $\Cdef_{\mu\nu ijkl}$ (1)\\
Three-forms & $\mathcal{C}_{\mu\nu\rho} \in \mathbf{5}$ & $\Cdef_{\mu\nu\rho}$ (1) & $\Cdef_{\mu\nu\rho ijk}$ (4)\\
Four-forms & $\mathcal{D}_{\mu\nu\rho\sigma}\in \mathbf{\overline{10}}$ & $\Cdef_{\mu\nu\rho\sigma ij}$ (6) & $\tilde A_{\mu\nu\rho\sigma i}$ (4) 
\end{tabular}
}
\label{sl5count}
\end{table}

One of the remarkable features of exceptional field theory is that it describes simultaneously all the possible duality relations that are needed to realise $\Edd$ multiplets at the level of the full $(n+d)$-dimensional SUGRA, and allows one to deal efficiently with the description of exotic dual fields. 
This will be explained in section \ref{compensator}.

\subsubsection{$\Gfive$}

Now we take $d=5$, $n=6$. 
The representations of $\Gfive$ that appear are listed in table \ref{so55count}.
In this case, we need to dualise the single three-form $\Cdef_{\mu\nu\rho}$ into an extra one-form, to realise the 16-dimensional Majorana-Weyl spinor representation of $\Gfive$.
Alternatively, we can describe this three-form as part of a multiplet of three-forms $\mathcal{C}_{\mu\nu\rho}$ in the Majorana-Weyl spinor representation of opposite chirality, again requiring dualisations of the metric. 

\renewcommand{\arraystretch}{1.4}
\begin{table}[ht]
\tbl{The bosonic field content of 11-dimensional SUGRA in $\Gfive$ multiplets. The two-form is electromagnetically self-dual, and the one-form is dual to three-form.}	
{
\centering
\begin{tabular}{ccccc}
Metric & $g_{\mu\nu}$ & & & \\
Scalars & $\gM_{MN} \in \frac{\Gfive}{\Hfive}$ & $g_{ij}$ (15) & $\Cdef_{ijk}$ (10) & \\
One-forms & $\mathcal{A}_\mu \in \mathbf{16}$ & $A_\mu{}^i$ (5) & $\Cdef_{\mu ij}$ (10) & $\Cdef_{\mu ijklm}$ (1)\\
Two-forms & $\mathcal{B}_{\mu\nu} \in \mathbf{{10}}$ & $\Cdef_{\mu \nu i}$ (5) & $\Cdef_{\mu\nu ijkl}$ (5)\\
Three-forms & $\mathcal{C}_{\mu\nu\rho} \in \mathbf{\overline{16}}$ & $\Cdef_{\mu\nu\rho}$ (1) & $\Cdef_{\mu\nu\rho ijk}$ (10) & $\tilde A_{\mu \nu \rho i}$ (5)
\end{tabular}
}
\label{so55count}
\end{table}

\subsubsection{$\Gsix$}

Now we take $d=6$, $n=5$. 
The representations of $\Gsix$ that appear are listed in table \ref{e6count}.
In this case, we need to dualise the single three-form $\Cdef_{\mu\nu\rho}$ into an extra scalar.
We further include a two-form $\Ab_{\mu\nu} \in \mathbf{\overline{27}}$ which would be dual to the one-form $\Aa_\mu$ in the fundamental, and include dual metric components.

\renewcommand{\arraystretch}{1.4}
\begin{table}[ht]
\tbl{The bosonic field content of 11-dimensional SUGRA in $\Gsix$ multiplets. The two-form is dual to the one-form.}	
{
\centering
\begin{tabular}{ccccc}
Metric & $g_{\mu\nu}$ & &  &\\
Scalars & $\gM_{MN} \in \frac{\Gsix}{\Hsix}$ & $g_{ij}$ (21) & $\Cdef_{ijk}$ (20) &  $\Cdef_{ijklmn}$ (1)\\
One-forms & $\mathcal{A}_\mu \in \mathbf{27}$ & $A_\mu{}^i$ (6) & $\Cdef_{\mu ij}$ (15) & $\Cdef_{\mu ijklm}$ (6)\\
Two-forms & $\mathcal{B}_{\mu\nu} \in \mathbf{\overline{27}}$ & $\Cdef_{\mu \nu i}$ (6) & $\Cdef_{\mu\nu ijkl}$ (15) & $\tilde A_{\mu \nu i}$ (6)\\
\end{tabular}}
\label{e6count}
\end{table}

\subsubsection{$\Gseven$}

Now we take $d=7$, $n=4$.
The representations of $\Gseven$ that appear are listed in table \ref{e7count}.
In this case, the one-form $\Aa_\mu$ in the 56-dimensional fundamental representation contains dual components of the three- and six-form, and necessarily also includes dual metric components in order to obtain the full $\Gseven$ representation.

\renewcommand{\arraystretch}{1.4}
\begin{table}[ht]
\tbl{The bosonic field content of 11-dimensional SUGRA in $\Gseven$ multiplets. The one-form is electromagnetically self-dual.}	
{
\centering
\begin{tabular}{cccccc}
Metric & $g_{\mu\nu}$ & & & & \\
Scalars & $\gM_{MN} \in \frac{\Gseven}{\Hseven}$ & $g_{ij}$ (28) & $\Cdef_{ijk}$ (35) &  $\Cdef_{ijklmn}$ (7)\\
One-forms & $\mathcal{A}_\mu \in \mathbf{56}$ & $A_\mu{}^i$ (7) & $\Cdef_{\mu ij}$ (21) & $\Cdef_{\mu ijklm}$ (21) & $\tilde A_{\mu i}$ (7)\\
\end{tabular}}
\label{e7count}
\end{table}

\subsubsection{$\Geight$}

Finally we come to $d=8$, $n=3$.
The representations of $\Geight$ that appear are listed in table \ref{e8count}.
Here we have to dualise the Kaluza-Klein vector to obtain a further eight scalars, in order to fill up the expected coset $\frac{\Geight}{\Height}$ of dimension 128.
The one-form $\mathcal{A}_\mu$ in the 248-dimensional adjoint involves many further ``exotic'' dualisations of the standard form fields. 

We motivated the inclusion of dual components of fields in these $\Edd$ multiplets using the fact that U-duality relates branes and their electromagnetic duals.
The branes that couple to these form field components are exotic branes, which we will discuss in section \ref{nongeoexotic}.

\renewcommand{\arraystretch}{1.4}
\begin{table}[ht]
\tbl{The bosonic field content of 11-dimensional SUGRA in $\Geight$ multiplets. The one-form contains more degrees of freedom consisting of exotic dualisations of the supergravity fields, which couple electrically to \emph{exotic branes}.}{
\centering
\begin{tabular}{ccccccc}
Metric & $g_{\mu\nu}$ & & & &\\
Scalars & $\gM_{MN} \in \frac{\Geight}{\Height}$ & $g_{ij}$ (36) & $\Cdef_{ijk}$ (56) &  $\Cdef_{ijklmn}$ (28) & $\tilde A_{i}$ (8)\\
One-forms & $\mathcal{A}_\mu \in \mathbf{248}$ & $A_\mu{}^i$ (8) & $\Cdef_{\mu ij}$ (28) & $\Cdef_{\mu ijklm}$ (56) & $\tilde A_{\mu i}$ (8) & $\dots$\\
\end{tabular}
}
\label{e8count}
\end{table}

%% file: sec4and5ExFTIntroduction.tex
\section{Exceptional field theory}
\label{exft}

In this section, we will describe the building blocks of exceptional field theory in a systematic fashion.
We will begin in section \ref{exgld} with the generalised Lie derivative, which defines $\Edd$ compatible diffeomorphisms.
This provides the local symmetry of the extended geometry introduced in the exceptional field theory approach. 
We will then construct the generalised gauge fields -- the tensor hierarchy -- of ExFT in section \ref{thier}, including a detailed breakdown for $\Gfour$ in section \ref{exsl5th}, and discuss the treatment of dual graviton degrees of freedom within the theory in section \ref{compensator}.
Finally, we will explain how to put everything together to formulate the action principle of ExFT, describing the common features of the action in section \ref{action}, and giving more explicit details for $\Gfour$ in section \ref{action5}, $\Gseven$ in section \ref{actionE7} and $\Geight$ in section \ref{action8}.

\subsection{The exceptional generalised Lie derivative}
\label{exgld}

\subsubsection{Generalised vectors and the representation $R_1$}

The generalised Lie derivative captures the local symmetries of the ``internal'' sector of 11-dimensional supergravity. 
We therefore (temporarily) restrict to a $d$-dimensional submanifold $M_d$ of the full 11-dimensional geometry.
On $M_d$ we have a metric, $g_{ij}$, a three-form, $C_{ijk}$, and (for sufficiently large $d$) a six-form, $C_{ijklmn}$, which in 11-dimensions is defined as the dual of the three-form by
\be
dC_6 = \star F_4 + \frac{1}{2} C_3 \wedge F_4 \,. 
\label{defC6}
\ee 
The local symmetries are $d$-dimensional diffeomorphisms, generated by vector fields $\Lambda^i$, and gauge transformations of the form fields, defined in terms of two- and five-form gauge parameters. Altogether we have
\be
\begin{split}
\delta g_{ij} & = L_\Lambda g_{ij} \,,\\
\delta C_{ijk} & = L_\Lambda C_{ijk} + 3 \partial_{[i} \lambda_{jk]} \,,\\
\delta C_{ijklmn} & = L_\Lambda C_{ijklmn} + 6 \partial_{[i} \lambda_{jklmn]} + 
30 C_{[ijk} \partial_{l} \lambda_{mn]} \,.
\end{split}
\label{11dlocal}
\ee
We combine these parameters of local symmetries into a generalised vector
\be
\Lambda^M = (\Lambda^i, \lambda_{ij} , \lambda_{ijklm} 
)\,,
\label{excgenvec}
\ee
which is a section of the exceptional tangent bundle
\be
TM_d \oplus \Lambda^2 T^*M_d \oplus \Lambda^5 T^*M_d 
 \,.
\label{exctang}
\ee
For $O(d,d)$, generalised vectors formed the $2d$-dimensional vector representation of the group $O(d,d)$, and the generalised tangent bundle was equipped with a natural bilinear form $\eta_{MN}$ preserved by $O(d,d)$ transformations.
Exceptional generalised vectors as in \refeqn{excgenvec} form representations of the group $\Edd$, and on the exceptional tangent bundle \refeqn{exctang} we can then define in a natural way $\Edd$-invariant tensors. We will see such tensors shortly: these will not be rank two symmetric tensors so cannot be interpreted as a quadratic form, as in the $O(d,d)$ case.

It is common in the literature to denote the representation of $\Edd$ in which the generalised vector appears by $R_1$.
The conjugate representation -- corresponding to generalised covectors -- is then $\bar R_1$.
We will soon encounter further representations. The most relevant of these are listed in table \ref{GHR}.

For $d \geq 7$, generalised vectors become more complicated, taking the form
\be
\Lambda^M = (\Lambda^i, \lambda_{ij} , \lambda_{ijklm}, \tilde \Lambda_{i_1\dots i_7; i}, \dots
)\,,
\label{excgenvec1}
\ee
which is a section of
\be
TM_d \oplus \Lambda^2 T^*M_d \oplus \Lambda^5 T^*M_d \oplus ( T^*M_d \otimes \Lambda^7 T^*M_d ) \oplus \dots
 \,.
\label{exct7plus}
\ee
The dots indicate additional terms which in fact present only for $d \geq 8$. 
The extra mixed symmetry tensors that appear can be viewed as gauge symmetries of exotic dual fields, starting with the dualisation of the metric.

Why does this happen for $d=7$? Recall the arguments of section \ref{counting} and consider the one-forms obtained by reducing the metric, three-form and dual six-form.
These are not sufficient to fill out a representation of $\Gseven$.
In $n=4$ one-forms are dual to one-forms, and it follows directly from the duality relation \refeqn{defC6} that the 21 one-forms $C_{\mu ij}$ coming from the 11-dimensional three-form are dual to the 21 one-forms $C_{\mu ijklm}$ coming from the six-form. 
If we introduce a further seven one-forms $\tilde A_{\mu i}$ which are dual to the Kaluza-Klein vector $A_\mu{}^i$, then we obtain a total of 56 one-forms, which fill out the fundamental representation of $\Gseven$: $\mathcal{A}_\mu{}^M = ( A_\mu{}^i , C_{\mu ij}, C_{\mu ijklm}, \tilde A_{\mu i})$. 
We can then impose a $\Gseven$-compatible self-duality constraint on the field strength of $\mathcal{A}_\mu{}^M$, as we will see in more detail below.

If we lift this back to eleven-dimensions, then the $\tilde A_{\mu i}$ have an interpretation as components of the ``dual graviton''.
This can be defined in linearised gravity by treating one index of the linearised metric as a form index and dualising on it.
This leads to the dual graviton $h_{\hmu_1 \dots \hmu_8; \hnu}$, where the first 8 indices are antisymmetric, and gauge symmetries $\Lambda_{\hmu_1 \dots \hmu_7; \hnu}$. The components $h_{i_1 \dots i_7 \mu; i}$ then correspond for $d=7$ to the extra dual one-forms, and the gauge transformation parameters $\Lambda_{i_1 \dots i_7; j}$ to their $U(1)$ symmetry transformations.

Although for $d=7$ these appear at the level of the generalised diffeomorphism symmetry of ExFT, for $d<7$ such transformations are also present as gauge transformations of the tensor hierarchy gauge fields of the theory. 
We will discuss how this is treated in exceptional field theory in section \ref{compensator}.

\subsubsection{The generalised Lie derivative}

\renewcommand{\arraystretch}{1.5}
\begin{table}
\tbl{Groups and (selected) representations appearing in ordinary geometry, DFT and ExFT. Generalised vectors are valued in $R_1$, and the section condition in $R_2$. The special weight is given by $\omega = -\tfrac{1}{n-2}$ in ExFT and $\omega = 0$ in DFT. Generally, $R_{8-d} = \bar{R}_1$ and $R_{9-d} = \mathbf{adj}$.}
{
\centering
\begin{tabular}{cccccccccc} 
$G$ &
 $\alpha$ & $-\omega$ & $R_1$ & $R_2$  & $R_3$ & $R_4$ & $R_5$ & $R_6$ \\\hline 
$\mathrm{GL}(d)$ &
 1 & 0 & $\mathbf{d}$ &  &  &  &  \\
$\mathrm{O}(d,d)$ & 
  $2$ & $0$ & $\mathbf{2d}$ & $\mathbf{1}$ & & &  \\
 $\Gthree$ & 
 n/a & $\frac{1}{6}$ & $\mathbf{(3,2)}$ & $\mathbf{(\bar{3},1)}$ & $\mathbf{(1,2)}$ & $\mathbf{(3,1)}$ & $\mathbf{(\bar{3}, \bar{2})}$  &  $\mathbf{(8,1)\oplus(1,3)}$	\\
$\Gfour$ & 
 $3$ & $\frac{1}{5}$ & $\mathbf{10}$ & $\mathbf{\bar{5}}$ & $\mathbf{5}$ & $\mathbf{\overline{10}}$  & $\mathbf{24}$\\
$\Gfive$ & 
$4$ & $\frac{1}{4}$ & $\mathbf{16}$ & $\mathbf{10}$ & $\mathbf{\overline{16}}$ & $\mathbf{45}$\\
$\Gsix$ & 
$6$ & $\frac{1}{3}$ & $\mathbf{27}$ & $\mathbf{\overline{27}}$ & $\mathbf{78}$\\
$\Gseven$ & 
$12$  &$\frac{1}{2}$ &$\mathbf{56}$ & $\mathbf{133}$\\
$\Geight$ & 
$60$ & $1$ & $\mathbf{248}$ & $\mathbf{1 \oplus 3875}$\\
\end{tabular}}
\label{GHR} 
\end{table} 
\renewcommand{\arraystretch}{1}

We will now describe the $\Edd$ generalised Lie derivative \cite{Coimbra:2011ky, Berman:2012vc, Hohm:2013pua}.
We introduce derivatives $\partial_M$ with respect to an extended set of coordinates, $Y^M$, which lie in the $R_1$ representation.
The extended coordinates $Y^M$ can therefore be decomposed in the same manner as generalised vectors, \refeqn{excgenvec1}, with $Y^M = ( y^i, \tilde y_{ij}, \tilde y_{ijklm}, \dots )$ and so contain a host of dual coordinates with multiple indices. A physical viewpoint on these coordinates is that they can be thought of as conjugate to wrapping modes of the M-theory branes (M2, M5, KKM and ultimately more exotic objects that we will meet in section \ref{nongeoexotic}), which we will discuss further in section \ref{windingcentral}. (Although we refer here, and throughout this section, to the M-theory interpretation, as explained in section \ref{ornotIIB} there is also a dual IIB description.)

Given a generalised vector $U^M$, its derivative $\partial_N U^M$ is then valued in $R_1 \otimes \bar{R}_1 = \mathbf{1} \oplus \mathbf{adj} \oplus \dots$. 
We introduce a projector $\mathbb{P}_{}{}^M{}_N{}^P{}_Q$ which projects from $R_1 \otimes \bar{R}_1$ onto the adjoint representation.
The generalised Lie derivative of a generalised vector $V^M$ of weight $\lambda_V$ can then be defined as
\be
\mathcal{L}_U V^M = U^N \partial_N V^M - \alpha \mathbb{P}^M{}_N{}^P{}_Q \partial_P U^Q  V^N + \lambda_V \partial_N U^N V^M \,,
\label{gldproj}
\ee
where $\alpha$ is a numerical constant which depends on the group under consideration (see table \ref{GHR}).
An equivalent form is
\be
\mathcal{L}_U V^M = U^N \partial_N V^M -  V^N \partial_N U^M   + Y^{MN}{}_{PQ} \partial_N U^P V^Q + ( \lambda_V + \omega)  \partial_N U^N V^M \,,
\label{gldY}
\ee
where the $Y$-tensor is by definition related to the adjoint projector by
\be
Y^{MN}{}_{PQ} = - \alpha \mathbb{P}_{}{}^{M}{}_{Q}{}^N{}_{P} + \delta^M{}_P \delta^N{}_Q  - \omega \delta^M{}_Q\delta^N{}_P 
\,.
\label{yfrompadj}
\ee
In practice, the adjoint projector and hence the $Y$-tensor are defined in terms of invariant tensors of the group $\Edd$.  

The numerical constant $\omega$ is the same that we saw in the decomposition \refeqn{metricsplit} of the 11-dimensional metric, and is given by
\be
\omega = - \frac{1}{n-2} \,.
\ee
We need to pick a convention for the weights of all objects.
We will follow \refref{Hohm:2013pua} and take the generalised metric to be a true element of $\Edd$, with determinant one, and this to be an unweighted tensor.
We then have to require that generalised vectors generating generalised diffeomorphisms should themselves have weight $- \omega$ (which means the final term in \refeqn{gldY} vanishes when considering the generalised Lie derivative of such generalised vectors).
Alternatively, we could omit the term involving $\omega$ from the definition \refeqn{gldY} which in effect defines an $\Edd \times \mathbb{R}^+$ generalised Lie derivative, as in \refref{Coimbra:2011ky}.

Given the generalised Lie derivative as defined either via \refeqn{gldproj} or \refeqn{gldY} we fix the coefficient $\alpha$, or equivalently the form of the Y-tensor, by requiring closure
\be
[ \mathcal{L}_U , \mathcal{L}_V ] W = \mathcal{L}_{[U,V]_E} W \,, \quad [U,V]_E \equiv \frac{1}{2} ( \mathcal{L}_U V - \mathcal{L}_V U )\,,
\label{Eclosure}
\ee
where in order to use \refeqn{gldY} we assume $U,V$ and $W$ are generalised vectors of weight $-\omega$.
The condition \refeqn{Eclosure} holds if \cite{Berman:2012vc}
\be
\begin{split}
0=
&
\frac{1}{2}Y^{N P}{}_{K L} U^{L} \partial_{P}{V^{K}} \partial_{N}{W^{M}} +Y^{Q K}{}_{N P} W^{P} \partial_{Q}{V^{M}} \partial_{K}{U^{N}}
\\ &
+\frac{1}{2}\left( Y^{M K}{}_{N P}\delta_Q^R- Y^{M K}{}_{L Q} Y^{L R}{}_{N P}  \right)  V^{P} W^{Q}\partial_{K}\partial_{R}{U^{N}}
\\&
+ W^{Q} 
\Big(Y^{M K}{}_{P Q}  \partial_{K}{V^{N}} \partial_{N}{U^{P}} 
+Y^{M K}{}_{P L} Y^{L R}{}_{N Q}  \partial_{R}{V^{N}} \partial_{K}{U^{P}}
\\ &\qquad\qquad
+\frac{1}{2}Y^{M K}{}_{N P}  \partial_{K}{V^{N}} \partial_{Q}{U^{P}}
+\frac{1}{2}Y^{M K}{}_{L Q} Y^{L R}{}_{N P}  \partial_{R}{V^{N}} \partial_{K}{U^{P}}
\Big)
\\ & - (U \leftrightarrow V) \,.
\end{split}
\ee
For the $\Edd$ generalised Lie derivatives defined below, this will be true for the particular definitions of $Y$-tensor following from \refeqn{yfrompadj}, assuming
\be
Y^{MN}{}_{PQ} \partial_M \Psi \partial_N \Psi^\prime = 0 \,,\quad
Y^{MN}{}_{PQ} \partial_M \partial_N \Psi= 0 \,,
\label{sectionY}
\ee
where again $\Psi,\Psi^\prime$ are any fields or gauge parameters in the theory.
This is the section condition of ExFT.

\subsubsection{The section condition and the representation $R_2$}

The section condition can be interpreted as the requirement that the tensor product $\partial_M \otimes \partial_N$ vanishes when projected onto a particular representation. 
In double field theory, the section condition was $\eta^{MN}\partial_M \otimes \partial_N = 0$, and this representation was the trivial one. 
In exceptional field theory, the structure of the representation theory involved is much richer.
Some notation is therefore needed. 
We denote by $R_2$ the representation found in the symmetric tensor product of $R_1 \otimes R_1$ listed in table \ref{GHR}, and $\bar R_2$ its conjugate.
The section condition is equivalent to
\be
\partial_M \otimes \partial_N \big|_{\bar{R}_2 } = 0 \,.
\ee
For $d \leq 6$, the Y-tensor will be proportional to the projector onto this representation:
\be
Y^{MN}{}_{PQ}  = 2(d-1)( \mathbb{P}_{R_2} )^{MN}{}_{PQ}\,,
\ee
and in particular is symmetric on both upper and lower indices.

For $d>6$, the general story receives some modifications.

For $d=7$, the representation $R_1$ is the fundamental $\mathbf{56}$, and the antisymmetric part of the tensor product $R_1 \otimes R_1$ includes the trivial representation.
We then in fact require $\partial_M \otimes \partial_N \big|_{\bar{R}_2  \oplus \mathbf{1}} =0$. In this case it turns out that
\be
Y^{MN}{}_{PQ}\big|_{\Gseven} = 12 ( \mathbb{P}_{R_2} )^{MN}{}_{PQ} +  28 ( \mathbb{P}_{\mathbf{1}} )^{MN}{}_{PQ} \,,
\ee
which no longer has fixed symmetry.

For $d=8$, the representation is the fundamental $\mathbf{248}$ which is also the adjoint.
We require $\partial_M \otimes \partial_N \big|_{\bar{R}_2 \oplus \bar{R}_1} =0$.
Due to complications resulting from the presence of dual graviton degrees of freedom appearing already in the generalised metric, the definition of the generalised Lie derivative in terms of the Y-tensor is not sufficient for a closed algebra, and has to be modified further, as will be explained in section \ref{action8}.

\subsubsection{The $\Gthree$ generalised Lie derivative}
\label{gld3}

This is the case $d=3$. A generalised vector combines a vector $v^i$ with a two-form $\lambda_{ij}$ for a total of six components. 
Dualising the antisymmetric pair of indices, $\lambda^i \equiv \frac{1}{2} \epsilon^{ijk} \lambda_{jk}$, we can see this as an $\mathrm{SL}(2)$ doublet of $\mathrm{SL}(3)$ vectors, hence we have representation $R_1 = \mathbf{(3,2)}$. 
In a convention where $V^M \equiv V^{i \alpha} = ( v^i, \lambda^i)$, with $\alpha$ an index declared to transform in the $\mathbf{2}$ of $\mathrm{SL}(2)$, the Y-tensor is
\be
Y^{i\alpha,j\beta}{}_{k\gamma, l\delta} = \epsilon^{ijm} \epsilon_{klm} \epsilon^{\alpha \beta} \epsilon_{\gamma \delta} \,.
\ee
In this case, as the group is a product, the adjoint representation is $(\mathbf{8},\mathbf{1}) \oplus (\mathbf{1}, \mathbf{3})$, and in terms of the separate adjoint projectors on each factor we have
\be
\begin{split}
\mathcal{L}_U V^M & = U^N \partial_N V^M - 2 \mathbb{P}_{(\mathbf{8},\mathbf{1})}{}^M{}_N{}^P{}_Q \partial_P U^Q  V^N 
\\ & \qquad - 3 \mathbb{P}_{(\mathbf{1},\mathbf{3})}{}^M{}_N{}^P{}_Q \partial_P U^Q  V^N 
+ \lambda_V \partial_N U^N V^M \,,
\end{split}
\label{gldproj32}
\ee
with projectors as defined in \refref{Hohm:2015xna}.

The representation $R_2$ is $(\mathbf{\bar{3}},\mathbf{1})$ and the section condition is then
\be
\epsilon^{ijk} \epsilon^{\alpha \beta} \partial_{j\alpha} \otimes \partial_{k\beta} = 0 \,,
\ee
and the solution that recovers the $d=3$ physical spacetime is $\partial_{i1}\neq0$, $\partial_{i2} = 0$.

\subsubsection{The $\Gfour$ generalised Lie derivative}
\label{gld4}

This is the case $d=4$. A generalised vector combines a vector $v^i$ with a two-form $\lambda_{ij}$ for a total of ten components.
This arranges into the antisymmetric representation of $\Gfour$, so $R_1 = \mathbf{10}$.
As in section \ref{genKK_EFT}, we introduce a five-dimensional fundamental index $\fM = 1,\dots, 5$, and we write a generalised vector as carrying a pair of antisymmetric indices, hence $V^{\fM \fN} = ( v^i, \lambda^{ij})$, i.e. $V^{i5} \equiv v^i$, $V^{ij} \equiv \lambda^{ij}$, after dualising the antisymmetric pair of indices, $\lambda^{ij} \equiv \frac{1}{2} \epsilon^{ijkl} \lambda_{kl}$.
We adopt a convention where we include a factor of $1/2$ when summing over antisymmetric indices. 
Hence for instance $V^M \partial_M \equiv \frac{1}{2} V^{\fM \fN} \partial_{\fM \fN} = V^{i5} \partial_{i5} + \frac{1}{2} V^{ij}\partial_{ij}$.
Then the Y-tensor is
\be
Y^{\fM \fM^\prime , \fN \fN^\prime}{}_{\fP\fP^\prime,\fQ\fQ^\prime} = 
\epsilon^{\fM\fM^\prime\fN\fN^\prime \fK} \epsilon_{\fP\fP^\prime \fQ\fQ^\prime \fK} \,.
\ee
The representation $R_2$ is the $\mathbf{\bar 5}$ and the section condition is then
\be
\partial_{[\fM \fN}\otimes \partial_{\fP \fQ]} = 0 \,,
\label{sl5sec}
\ee
and the solution that recovers the $d=4$ physical spacetime is $\partial_{i5} \neq 0$, $\partial_{ij} = 0$.

\subsubsection{The $\Gfive$ generalised Lie derivative}
\label{gld5}

This is the case $d=5$. A generalised vector combines a vector $v^i$ with a two-form $\lambda_i$ and a five-form $\lambda_{ijklm}$ for a total of sixteen components.
We treat the five-form as a scalar, $\tilde \lambda \equiv \frac{1}{5!} \epsilon^{ijklm} \lambda_{ijklm}$.
Altogether we can see this as a Majorana-Weyl spinor representation of $\Gfive$, which means $R_1 = \mathbf{16}$.
We denote the spinor index by $M$ and the vector index of the ten-dimensional fundamental representation by $I$.
The gamma matrices of $\Gfive$ are 32 $\times$ 32 matrices and can be decomposed into off-diagonal 16 $\times$ 16 blocks, $\gamma_I{}^{MN}$, $\gamma^I{}_{MN}$, which are symmetric on the $MN$ indices, and obey
\be
\gamma^I{}_{MN} \gamma^{JNP} + \gamma^J{}_{MN} \gamma^{INP} = 2 \eta^{IJ}\delta^P_M\,,
\ee
where $\eta^{IJ}$ is the defining $\Gfive$ structure. Note that $\gamma^I{}_{MN}\gamma_J{}^{MN} = 16 \delta^I_J$.
The Y-tensor is then
\be
Y^{MN}{}_{PQ} = \frac{1}{2} \gamma_I{}^{MN} \gamma^I{}_{PQ} \,.
\ee
The representation $R_2$ is the $\mathbf{10}$ and the section condition is
\be
\gamma_I{}^{MN} \partial_M\otimes \partial_N = 0 \,.
\label{so55sec}
\ee
Decomposing a generalised vector as $V^M = ( V^i, V_{ij}, V^{z})$, where the index $M=z$ denotes the single five-form component, we can parametrise these gamma matrices as 
\be
\begin{array}{ccccc}
(\gamma_I)^{MN} & \rightarrow & ( \gamma^i)^j{}_{kl} = 2 \delta^{ij}_{kl}\,, & (\gamma_i)^j{}_z = \sqrt{2} \delta_i^j\,, & (\gamma_i)_{jklm} = \frac{1}{\sqrt{2}} \epsilon_{ijklm}\,, \\
(\gamma^I)_{MN} & \rightarrow & ( \gamma_i)_j{}^{kl} = 2 \delta_{ij}^{kl}\,, & (\gamma^i)_j{}^z = \sqrt{2} \delta^i_j\,, & (\gamma^i)^{jklm} = \frac{1}{\sqrt{2}} \epsilon^{ijklm} \,.
\end{array} 
\ee
The solution of the section condition \refeqn{so55sec} giving a $d=5$ physical spacetime is $\partial_i \neq 0$, $\partial^{ij} = 0$, $\partial_z = 0$.

\subsubsection{The $\Gsix$ generalised Lie derivative}
\label{gld6} 

Finally, for $d=6$ we reach a genuinely exceptional group.
A generalised vector combines a vector $v^i$ with a two-form $\lambda_{ij}$ and a five-form $\lambda_{ijklm}$ for a total of 27 components.
We dualise the five-form into a second vector, $\tilde \lambda^{\bar i} \equiv \frac{1}{6!} \epsilon^{ijklmn} \lambda_{jklmn}$ (the $\bar i$ is used simply to distinguish from the index on the actual vector).
Altogether this gives the fundamental representation of $\Gsix$, so $R_1 = \mathbf{27}$.
There is a conjugate representation $\mathbf{\overline{27}}$, and two totally symmetric cubic invariant tensors, which are denoted by $d^{MNP}$ and $d_{MNP}$.
These are normalised such that 
\be
d^{MPQ} d_{NPQ} = \delta^M_N \,,
\ee
and obey a cubic identity
\be
d_{K(MN} d_{PQ)L} d^{KL S} = \frac{2}{15} \delta^{S}_{(M} d_{NPQ)} \,,\quad
d^{K(MN} d^{PQ)L} d_{KL S} = \frac{2}{15} \delta_{S}^{(M} d^{NPQ)} \,,\quad
\ee
The Y-tensor is then
\be
Y^{MN}{}_{PQ} = 10 d^{MNK} d_{PQK}\,.
\ee
The representation $R_2$ is the $\mathbf{\overline{27}}$ and the section condition is
\be
d^{MNK} \partial_M \otimes \partial_N = 0 \,.
\label{e6sec}
\ee
The components of the cubic invariant are (in the normalisation of \refref{Hohm:2013vpa}, where we do not include an explicit factor of $1/2$ in contractions)
\be
\begin{array}{cccc}
d_{MNP} & \rightarrow & d_{i \bar j}{}^{kl} = \frac{1}{\sqrt{5}} \delta^{kl}_{[ij]} \,,
& d^{ij,kl,mn} = \frac{1}{4\sqrt{5}} \epsilon^{ijklmn}\,, \\
d^{MNP} & \rightarrow & d^{i \bar j}{}_{kl} = \frac{1}{\sqrt{5}} \delta^{ij}_{[kl]} \,,
& 
d_{ij,kl,mn} =\frac{1}{4 \sqrt{5}} \epsilon_{ijklmn} \,.
\end{array} 
\label{dm}
\ee
The solution of the section condition \refeqn{e6sec} giving $d=6$ physical directions is then $\partial_i \neq 0$, $\partial^{ij} = 0 = \partial_{\bar i}$.

\subsubsection{The $\Gseven$ generalised Lie derivative}

This is the case $d=7$, and exhibits different properties to the preceding examples.
A generalised vector consists of a vector $v^i$, two-form $\lambda_{ij}$, five-form $\lambda_{ijklmn}$ and now the additional one-form $\tilde \Lambda_i$ associated to gauge transformations of the dual graviton. 
We dualise the five-form into a bivector $\tilde\lambda^{ij} = \frac{1}{5!} \epsilon^{ij klm np} \lambda_{klmnp}$.
These form the fundamental representation of $\Gseven$, $R_1 = \mathbf{56}$.

The group $\Gseven$ is characterised by an invariant antisymmetric rank two tensor, $\Omega_{MN}$, such for two generalised vectors $V^M = (V^i, V_{ij}, \tilde V^{ij}, \tilde V_i)$ and $W^M = (W^i, W_{ij}, \tilde W^{ij} , \tilde W_i )$ we have  
\be
\Omega_{MN} V^M W^N = V^i \tilde W_i - \tilde V_i W^i + \frac{1}{2} \tilde V^{ij} W_{ij} - \frac{1}{2} V_{ij} \tilde W^{ij}\,.
\ee
This means that the fundamental representation is self-conjugate; we can raise and lower indices using $\Omega_{MN}$ and its inverse, which is defined such that $\Omega^{MP} \Omega_{NP} = \delta^M_N$. We raise as $V^M = \Omega^{MN} V_N$ and lower by $V_M = V^N \Omega_{NM}$. 
There is further a quartic symmetric invariant $c_{MNPQ}$, however it is more convenient to introduce the generators $(t_\alpha)_M{}^N$ in the adjoint, in terms of which the Y-tensor is defined by
\be
Y^{MN}{}_{PQ} = - 12 (t_\alpha)_{PQ} (t^\alpha)^{MN} + \frac{1}{2} \Omega^{MN} \Omega_{PQ} \,,
\label{Y7}
\ee
where the Cartan-Killing form $\kappa_{\alpha\beta} = (t_\alpha)_M{}^N (t_\beta)_N{}^M$ is used to contract adjoint indices.
The generators with fundamental indices fully raised or lowered are symmetric on these indices.

The representation $R_2$ is the $\mathbf{133}$ (the adjoint representation), however the section condition requires both
\be
t^{\alpha M N} \partial_M \otimes \partial_N = 0 \,,\quad
\Omega^{MN} \partial_M \otimes \partial_N = 0 \,.
\label{e7sec}
\ee
The solution of the section condition \refeqn{e6sec} giving $d=7$ physical directions is then $\partial_i \neq 0$, $\partial^{ij} = \tilde \partial_{ij} = \tilde \partial^i$.

\subsubsection{The $\Geight$ generalised Lie derivative}

This is the case $d=8$.
The vector, two-form, five-form and dual graviton symmetry transformation parameters do not fill out an $\Geight$ representation.
We have to add additional mixed symmetry objects.
However, even after doing so, the construction described above fails to give a consistent algebra \cite{Berman:2012vc}.
In order to obtain a generalised Lie derivative for this group, one needs to add an additional term. This additional term can be viewed as an additional (constrained) symmetry transformation as in \cite{Hohm:2014fxa}, as will be explained in section \ref{action8}, or alternatively it can be written in terms of the Weitzenb\"ock connection as in \refref{Rosabal:2014rga,Cederwall:2015ica}, in which case no additional symmetry parameters are introduced.

\subsubsection{Further features of generalised Lie derivatives}

Let us discuss some properties of the generalised Lie derivative that we will need to further develop exceptional field theory. 

Generalised vectors of the form 
\be
\chi^M = Y^{MN}{}_{PQ} \partial_{N} \chi^{PQ}
\label{trivial}
\ee
for $\chi^{PQ} = \chi^{QP}$ symmetric, generate trivial transformations
\be
\begin{split}
\mathcal{L}_{\chi} V^M = 0 \,,
\end{split}
\ee
using the section condition.
For $\Gseven$ there are additional trivial transformations, consisting of generalised vectors of the form $\chi^M = \Omega^{MN} B_N$ where $B_N$ is constrained to obey the same constraint as the derivatives $\partial_M$.
We will return to this in more detail below.

The symmetrisation is
\be
\begin{split}
2 \{ U , V \}^M & \equiv (\mathcal{L}_{U} V + \mathcal{L}_V U)^M \\&
= Y^{MN}{}_{(PQ)} \partial_N ( U^P V^Q) 
+ Y^{MN}{}_{[PQ]} ( \partial_N U^P V^Q - U^P \partial_N V^Q ) \,.
\end{split}
\label{symmgld}
\ee
For $d=2,\dots,6$ the Y-tensor is symmetric on upper and lower indices.
Then $\{U,V\}$ is exactly of the form of a trivial transformation \refeqn{trivial}.
For $d=7$, the antisymmetric part of the Y-tensor is $Y^{MN}{}_{[PQ]} = \frac{1}{2} \Omega^{MN} \Omega_{PQ}$. 
The second term in \refeqn{symmgld} is thus also a trivial transformation in this case.
Hence for $d=2,\dots,7$, $\{U,V\}$ constitutes a generalised vector whose generalised Lie derivative is zero acting on all fields, using the section condition.

The Jacobiator of the E-bracket is
\be
[ [ U, V ]_E , W ]_E + \text{cyclic} = \frac{1}{3} \{ [U,V]_E, W \} + \text{cyclic} \,,
\label{jacobiatorE}
\ee
and so is non-zero but trivial acting on fields.

The condition for closure of the generalised Lie derivative, in terms of generalised vectors $U,V,W$, is:
\be
\mathcal{L}_U \mathcal{L}_V W - \mathcal{L}_V \mathcal{L}_U W = \mathcal{L}_{[U,V]_E} W 
\ee
and this can be rewritten as 
\be
\mathcal{L}_U ( \mathcal{L}_V W ) = \mathcal{L}_V ( \mathcal{L}_U W) + \mathcal{L}_{\mathcal{L}_U V} W \,,
\label{leibniz}
\ee
using the fact that
\be
\mathcal{L}_U V = [ U,V]_E  + \{ U, V \}
\ee
and the symmetric part $\{U,V\}$ always generates a trivial transformation, $\mathcal{L}_{\{U,V\}} = 0$.
The condition \refeqn{leibniz} is the Leibniz property for the binary operation
\be
U \circ V \equiv \mathcal{L}_U V\,,\quad
U \circ ( V \circ W ) = V \circ ( U \circ W) + (U \circ V) \circ W \,,
\ee
expressing the fact that the algebra of generalised Lie derivatives is not a Lie algebra but a  Leibniz algebra i.e. the generalisation of a Lie algebra to the case where the fundamental bracket, is not antisymmetric, $U \circ V \neq - V \circ U$.
This provides an interesting mathematical perspective on the definition of generalised diffeomorphisms, discussed in \refref{Hohm:2018ybo,Hohm:2019wql}, which we will not pursue in this review.

\subsection{The tensor hierarchy}
\label{thier}

The generalised Lie derivative defines generalised diffeomorphisms of the extended geometry of exceptional field theory.
In particular, the generalised metric will transform tensorially under such transformations.
The full exceptional field theory contains also generalised form-fields, which are $n$-dimensional $p$-forms arranged into representations of $\Edd$ which we denote by $R_p$.
These transform under generalised diffeomorphisms as gauge fields. 
This sequence of generalised form fields is called the ``tensor hierarchy''\footnote{This name was originally used to describe the $p$-form fields of gauged supergravity \cite{deWit:2005hv,deWit:2008ta}, in which the same sequence of representations and similar gauge structures appear. This is no coincidence: the tensor hierarchy of gauged SUGRA is automatically generated by generalised Scherk-Schwarz reductions of ExFT, see section \ref{genSS}.}.
We will now explain how it is treated in ExFT, following the systematic approach worked out in \refref{Cederwall:2013naa,Aldazabal:2013via,Wang:2015hca}.

\subsubsection{A need of hierarchies}
\label{needhier}

Let's first recall the situation in the Kaluza-Klein-esque decomposition of a theory of an $(n+d)$-dimensional metric, which we studied in section \ref{KKsplit}.
There, without imposing the Kaluza-Klein reduction ansatz, we saw that the Kaluza-Klein vector $A_\mu{}^i$ appeared as a gauge field for internal diffeomorphisms, being used to define covariant $n$-dimensional partial derivatives $D_\mu = \partial_\mu - L_{A_\mu}$.

In exceptional field theory, we introduce a one-form $\mathcal{A}_\mu{}^M$, valued in the same $R_1$ representation as generalised vectors, which plays the same role with respect to \emph{generalised diffeomorphisms}. 
To that end, we define it to transform as (compare \refeqn{deltaKKsplit})
\be
\delta_\Lambda \Aa_\mu{}^M = \mathcal{D}_\mu \Lambda^M \equiv \partial_\mu \Lambda^M - \mathcal{L}_{\Aa_\mu} \Lambda^M \,,
\label{deltaA}
\ee
which also defines the covariant partial derivative $\mathcal{D}_\mu = \partial_\mu -\mathcal{L}_{\Aa_\mu}$. 
This means we assume $\Aa_\mu{}^M$ has weight $-\omega=\tfrac{1}{n-2}$ in order that we can take the generalised Lie derivative with respect to it.
Then if $\mathcal{T}$ is some generalised tensor, a short calculation using the properties of the generalised Lie derivative shows that
\be
\delta_\Lambda \mathcal{D}_\mu \mathcal{T} = \mathcal{L}_\Lambda \mathcal{D}_\mu \mathcal{T} \,.
\ee
Of course, this one-form corresponds to the one-form multiplets described in section \ref{counting}, where we discussed the decomposition and rearrangement into $\Edd$ multiplets of the bosonic field contents of SUGRA (in the 11-dimensional case).

We now seek a field strength for $\Aa_\mu{}^M$.
Our definition will be valid for $d \leq 6$, i.e. up to and including $\Gsix$.
For $\Gseven$, modifications are necessary, which we will describe in section \ref{compensator}.
In particular we assume below that the Y-tensor is symmetric in both its upper and lower pairs of indices: for $\Gseven$ this is not the case.

The naive field strength appearing in the commutator $[\mathcal{D}_\mu, \mathcal{D}_\nu] = - \mathcal{L}_{{F}_{\mu\nu}}$ is
\be
F_{\mu\nu} = 2 \partial_{[\mu} \Aa_{\nu]} - [ \Aa_\mu , \Aa_\nu ]_E  \,,
\ee
for which a short calculation gives
\be
\delta F_{\mu\nu}{}^M = 2 \mathcal{D}_{[\mu} \delta \Aa_{\nu]}{}^M + Y^{MN}{}_{PQ} \partial_N ( \Aa_{[\mu}{}^P \delta \Aa_{\nu]}{}^Q ) 
\label{deltanaive}
\ee
and hence
\be
\delta_\Lambda F_{\mu\nu}{}^M = \mathcal{L}_{\Lambda} F_{\mu\nu}{}^M + Y^{MN}{}_{PQ} \partial_N ( - \Lambda^P F_{\mu\nu}{}^Q +  \Aa_{[\mu}{}^P \delta \Aa_{\nu]}{}^Q ) \,.
\ee
Hence this does not transform tensorially under generalised diffeomorphisms.
To cure this, we introduce a two-form transforming in the $R_2$ representation.
This naturally introduces the $\Edd$ multiplet of two-forms of section \ref{counting}, containing (perhaps dual) physical degrees of freedom of the bosonic fields of SUGRA.
We here write this two-form as $\Ab_{\mu\nu}{}^{MN}$, carrying a symmetric pair of $R_1$ indices such that 
\be
\Ab_{\mu\nu}{}^{MN} = \frac{1}{2(d-1)} Y^{MN}{}_{PQ} \Ab_{\mu\nu}{}^{PQ} \,,
\ee
recalling that for $d \leq 6$ the $Y$-tensor projects onto the $R_2$ representation, with 
\be
Y^{MN}{}_{PQ} Y^{PQ}{}_{KL} = 2 (d-1) Y^{MN}{}_{KL}\,.
\ee
This new two-form enters the fully covariant field strength defined by
\be
\Fa_{\mu\nu}{}^M =  2 \partial_{[\mu} \Aa_{\nu]}{}^M - [ \Aa_\mu , \Aa_\nu ]_E^M + Y^{MN}{}_{PQ} \partial_N \Ab_{\mu\nu}{}^{PQ} \,,
\label{fullcovF}
\ee
where it is convenient in what follows to explicitly include the Y-tensor acting as the $R_2$ projector.
It then follows that
\be
\delta \Fa_{\mu\nu}{}^M = 2 \mathcal{D}_{[\mu} \delta \Aa_{\nu]}{}^M + Y^{MN}{}_{PQ} \partial_N \Delta \Ab_{\mu\nu}{}^{PQ} \,,
\ee
having let
\be
\Delta \Ab_{\mu\nu}{}^{MN} \equiv \delta \Ab_{\mu\nu}{}^{MN} + \frac{1}{2(d-1)} Y^{MN}{}_{PQ} \Aa_{[\mu}{}^P \delta \Aa_{\nu]}{}^Q \,,
\ee
As a result, the field strength \refeqn{fullcovF} transforms as a generalised vector of weight $-\omega$ under
\be
\begin{split}
\delta_\Lambda \Aa_\mu{}^M & = \mathcal{D}_\mu \Lambda^M \,,\\
\Delta_\Lambda \Ab_{\mu\nu}{}^{MN} & = \frac{1}{2(d-1)} Y^{MN}{}_{PQ} \Lambda^P \Fa_{\mu\nu}{}^Q \,.
\end{split} 
\ee
The weight of $\Ab_{\mu\nu}{}^{MN}$ must be taken to be $-2\omega= \tfrac{2}{n-2}$.
The field strength \refeqn{fullcovF} is further invariant under gauge transformations
\be
\begin{split}
\delta_\lambda \Aa_\mu{}^M & = - Y^{MN}{}_{PQ} \partial_N \lambda_\mu{}^{PQ} \,,\\
\Delta_\lambda \Ab_{\mu\nu}{}^{MN} & = 2 D_{[\mu} \lambda_{\nu]}{}^{MN}  \,.
\end{split} 
\ee
Next we must define a field strength for the two-form.
Our definition will be valid for $d \leq 5$, i.e. up to and including $\Gfive$.
For $\Gsix$, modifications are necessary, which will be the subject of section \ref{compensator}. 

We can (partially) discover the field strength for the two-form by considering the Bianchi identity for the field strength \refeqn{fullcovF}:
\be
\begin{split}
3 \mathcal{D}_{[\mu} \Fa_{\nu\rho]}{}^M &
= - 6 \mathcal{L}_{\Aa_{[\mu}} \partial_{\nu} \Aa_{\rho]}{}^M 
- 3 \mathcal{L}_{\partial_{[\mu} \Aa_{\nu}} \Aa_{\rho]}{}^M - 3 \mathcal{L}_{\Aa_{[\nu}} \partial_\mu \Aa_{\rho]}{}^M
\\ & \quad
+ 3 \mathcal{L}_{\Aa_{[\mu}} \mathcal{L}_{\Aa_{\nu}} \Aa_{\rho]}{}^M
- 3 \mathcal{D}_{[\mu} ( Y^{MN}{}_{PQ} \partial_N \Ab_{\nu\rho]}{}^{PQ})\,.
\end{split}
\label{bianchiF1}
\ee
This simplifies on rewriting the Jacobi identity \refeqn{jacobiatorE} as
\be
3 \mathcal{L}_{\Aa_{[\mu}} \mathcal{L}_{\Aa_{\nu}} \Aa_{\rho]}{}^M = 2 \{ [\Aa_{[\mu} , \Aa_\nu]_E, A_{\rho]} \}^M
 = Y^{MN}{}_{PQ} \partial_N\left([\Aa_{[\mu} , \Aa_\nu]_E^P, A_{\rho]}^Q\right)\,.
 \label{jacobiatorEnicer}
\ee
After some further manipulation of \refeqn{bianchiF1} we obtain
\be
3 \mathcal{D}_{[\mu} \Fa_{\nu\rho]}{}^M = Y^{MN}{}_{PQ} \partial_N H_{\mu\nu\rho}{}^{PQ} \,,
\label{bianchiFf}
\ee
in which the following naive field strength appears:
\be
H_{\mu\nu\rho}{}^{MN} = 3 \mathcal{D}_{[\mu} \Ab_{\nu\rho]}{}^{MN} + \frac{1}{2(d-1)} Y^{MN}{}_{PQ} \left( - 3 \Aa_{[\mu}{}^P \partial_\nu \Aa_{\rho]}{}^Q 
 + \Aa_{[\mu}{}^P [ \Aa_\nu, \Aa_{\rho]}]_E^Q \right) \,.
\label{Hnaive}
\ee
This can be checked to again not be automatically covariant under generalised diffeomorphisms (although it is up to terms which vanish when projected with $Y^{MN}{}_{PQ} \partial_N$, as in \refeqn{bianchiFf}, so that the Bianchi identity of $\Fa$ is covariant).
The fix is to then introduce a three-form $\Ac_{\mu\nu\rho}{}^{MN,P}$, transforming in the representation $R_3 \subset R_1 \otimes R_2$, and of weight $-3\omega=  \tfrac{3}{n-2}$.
To understand how this field should appear, we can compute the variation of \refeqn{Hnaive}:
\be
\begin{split}
\delta H_{\mu\nu\rho}{}^{MN} & = 3 \mathcal{D}_{[\mu} \Delta \Ab_{\nu\rho]}{}^{MN} - 3 \frac{1}{2(d-1)} Y^{MN}{}_{PQ} \Fa_{[\mu\nu}{}^P \delta \Aa_{\rho]}{}^Q 
 \\ & 
\qquad
+ \frac{1}{2(d-1)}\left( Y^{M N}{}_{R S}  \delta_L^K- Y^{MN}{}_{PL}Y^{PK}{}_{RS} \right)
\times
\\ & \qquad \qquad\qquad\qquad
\times\partial_K 
\left( 
- 3 \delta \Aa_{[ \mu}{}^L \Ab_{\nu\rho]}{}^{RS}  +   \delta A_{[ \mu}{}^L \Aa_{\nu}{}^R \Aa_{\rho]}{}^S 
\right) 
\end{split}
\ee
using the identity
\be
\delta_P{}^{(K} Y^{MN)}{}_{RS} - Y^{(MN}{}_{QP} Y^{K)Q}{}_{RS} = 0 \,,
\label{yidentity5}
\ee
which is true for $d \leq 5$.
We then define 
\be
\begin{split}
\Fb_{\mu\nu\rho}{}^{MN} = &3 \mathcal{D}_{[\mu} \Ab_{\nu\rho]}{}^{MN} + \frac{1}{2(d-1)} Y^{MN}{}_{PQ} \left( - 3 \Aa_{[\mu}{}^P \partial_\nu \Aa_{\rho]}{}^Q 
 + \Aa_{[\mu}{}^P [ \Aa_\nu, \Aa_{\rho]}]_E^Q \right) 
 \\&\qquad
 + \frac{1}{2(d-1)}\left( Y^{M N}{}_{R S}  \delta_L^K- Y^{MN}{}_{PL}Y^{PK}{}_{RS} \right)\partial_K  \Ac_{\mu\nu\rho}{}^{RS,L} 
 \,,
\end{split}
\label{Hfull}
\ee
such that
\be
\begin{split}
\delta H_{\mu\nu\rho}{}^{MN} & = 3 \mathcal{D}_{[\mu} \Delta \Ab_{\nu\rho]}{}^{MN} - 3 \frac{1}{2(d-1)} Y^{MN}{}_{PQ} \Fa_{[\mu\nu}{}^P \delta \Aa_{\rho]}{}^Q 
 \\ & 
\qquad
+ \frac{1}{2(d-1)}\left( Y^{M N}{}_{R S}  \delta_L^K- Y^{MN}{}_{PL}Y^{PK}{}_{RS} \right)\partial_K \Delta \Ac_{\mu\nu\rho}{}^{RS,L}
\end{split}
\ee
with
\be
\begin{split}
\Delta \Ac_{\mu\nu\rho}{}^{MN,K} & = \delta \Ac_{\mu\nu\rho}{}^{MN,K}
\\ &\qquad
+ \frac{1}{2(d-1)}\left( Y^{M N}{}_{R S}  \delta_L^K- Y^{MN}{}_{PL}Y^{PK}{}_{RS} \right)\times
\\ &\qquad\qquad\qquad\qquad\times
\left( 
- 3 \delta \Aa_{[ \mu}{}^L \Ab_{\nu\rho]}{}^{RS}  +   \delta A_{[ \mu}{}^L \Aa_{\nu}{}^R \Aa_{\rho]}{}^S 
\right)\,.
\end{split}
\ee
Then we can begin to specify the gauge transformations of $\Ac_{\mu\nu\rho}{}^{MN,P}$ such that this field strength is fully covariant or invariant.

Although we can then again describe this solely in terms of the Y-tensor, it is convenient to take a step back from the coalface and think about the structures involved in these calculations.
The form fields $\Aa_\mu, \Ab_{\mu\nu}, \Ac_{\mu\nu\rho}$ generalise the notions of $p$-forms, and what secretly appears in the above constructions are operations generalising the wedge product and exterior derivative.
To describe the full tensor hierarchy, it is convenient to introduce some notation with which to express this.

\subsubsection{Systematic approach to the tensor hierarchy}

Standard $p$-forms, viewed as sections of the bundles $\Lambda^p T^*M$, can be manipulated using the exterior derivative and wedge product:
\be
d : \Lambda^p T^*M \rightarrow \Lambda^{p+1} T^*M \,,\quad d^2 = 0 \,,
\ee
\be
\wedge: \Lambda^p T^*M \otimes \Lambda^q T^*M \rightarrow \Lambda^{p+q} T^*M\,,\quad \omega_p \wedge \eta_q = (-1)^{pq} \eta_q \wedge \omega_p \,.
\ee
In addition, given a vector field $v$ there is the interior product 
\be
\iota_v : \Lambda^p T^*M \rightarrow \Lambda^{p-1} T^*M 
\ee
with $\iota_v \omega_p$ a $p-1$ form defined by contraction on the first slot.
The Lie derivative of a $p$-form can be expressed using Cartan's magic formula as
\be
L_v \omega_p = \iota_v d \omega_p + d ( \iota_v \omega_p) \,.
\label{cartanmagic}
\ee
These operations can be generalised to the tensor hierarchy $p$-forms of exceptional field theory \cite{Cederwall:2013naa,Wang:2015hca}. 
As seen above, these are $p$-forms from the point of view of the ``external'' space -- i.e. they transform in the totally antisymmetric rank $p$ representation of $\mathrm{GL}(11-d)$ -- and are assigned to specific representations $R_p$ of $\Edd$, with weight $-p\omega$ under generalised diffeomorphisms.

Let $A \in \RR_1$, $B \in \RR_2$, $C \in \RR_3$, etc., denote fields in these representations with these weights.
Further let $\sing$ denote the trivial representation of $\Edd$ with weight $1$ under generalised diffeomorphisms.

\vspace{1em}\noindent
\emph{ExFT exterior derivative $\hat \partial$} 

\noindent A generalisation of the exterior derivative involving only the derivatives $\partial_M$ can be defined, denoted by $\hat\partial$, with the properties that
\be
\hat \partial : \RR_p \rightarrow \RR_{p-1} \,,\quad \hat \partial^2 = 0 \,,
\ee
where the nilpotency holds using the section condition. Note that this acts to take ExFT $p$-forms to ExFT $(p-1)$-forms, i.e. in the opposite direction to the usual exterior derivative.
We want $\hat \partial$ to be well-defined in the sense that it takes generalised tensors to generalised tensors.
Thus given an ExFT $p$-form $F_p \in R_p$, transforming covariantly under the generalised Lie derivative, then $\hat \partial F_p \in R_{p-1}$ transforms covariantly under the generalised Lie derivative. It is only possible to define such a covariant derivative operation for $2 \leq p \leq 8-d$.
This gives a sequence
\be
\RR_1 
 \stackrel{\hat\partial}{\longleftarrow}
\RR_2 
 \stackrel{\hat\partial}{\longleftarrow}
\RR_3 
 \stackrel{\hat\partial}{\longleftarrow}
\dots \stackrel{\hat\partial}{\longleftarrow} \RR_{7-d} 
 \stackrel{\hat\partial}{\longleftarrow}\RR_{8-d} 
 \,.
\ee
For example, the map $\hat \partial: R_2 \rightarrow R_1$ is always defined as
\be
(\hat \partial B)^M = Y^{M N}{}_{P Q} \partial_N B^{P Q} \,.
\ee
We do not extend the action of $\hat \partial$ to act on $R_1$. However, we can see that $\hat \partial B$ is of the form of a generalised vector that generates a trivial generalised Lie derivative, $\mathcal{L}_{\hat \partial B} = 0$.

The map $\hat \partial: R_3 \rightarrow R_2$ is defined by
\be
(\hat \partial C)^{MN} =  \frac{1}{2(d-1)}\left( Y^{M N}{}_{R S}  \delta_L^K- Y^{MN}{}_{PL}Y^{PK}{}_{RS} \right)\partial_K  C^{RS,L} \,.
\ee
Then
\be
\begin{split}
(\hat \partial^2 C)^M
&  =\left( Y^{M N}{}_{R S}  \delta_L^K- Y^{M N}{}_{TL}Y^{TK}{}_{RS} \right) \partial_N \partial_K  C^{RS,L}
\\ 
&  = \frac{3}{2} \left( Y^{(M N}{}_{R S}  \delta_L^{K)}- Y^{(M N}{}_{TL}Y^{|T|K)}{}_{RS} \right) \partial_N \partial_K  C^{RS,L}
\\ & \qquad
- \frac{1}{2} \left( Y^{NK}{}_{R S}  \delta_L^{M}- Y^{NK}{}_{TL}Y^{TM}{}_{RS} \right) \partial_N \partial_K  C^{RS,L}
\\ 
 & = 0
\,.
\end{split}
\ee
using both the algebraic identity \refeqn{yidentity5}, which holds for $d \leq 5$, and the section condition.

\vspace{1em}\noindent
\emph{ExFT exterior product $\p$} 

\noindent A generalisation of the exterior product can be defined, namely:
\be
\p : \RR_p \otimes \RR_{q} \rightarrow \RR_{p+q} \,,
\ee
for $p+q \leq 8-d$, and 
\be
\p : \RR_{p} \otimes \RR_{9-d-p} \rightarrow \sing \,,
\ee
for $p=1, \dots, 9-d$.
The precise expressions for $\p$ just express the corresponding projections onto the appropriate $\Edd$ representations. 
This is possible because in general $R_{p+1} \subset R_1 \otimes R_{p}$.
For $X \in \RR_p$ and $X^\prime \in \RR_q$, we take $X \p X^\prime = X^\prime \p X$ for $p \neq q$; for $p=q$ the symmetry properties follow from the representation theory.

For example, the product $\p: \RR_1 \otimes \RR_1 \rightarrow \RR_2$ can be defined in general for $A,A^\prime \in \RR_1$ as
\be
( A\otimes A^\prime )^{MN} = \frac{1}{2(d-1)} Y^{(MN)}{}_{PQ} A^PA^{\prime Q} \,,
\ee
which is always symmetric.

\vspace{1em}\noindent
\emph{ExFT magic formula} 

\noindent For $\Lambda \in \RR_1$ and $X \in \RR_p$, with $1 < p < 8-d$, the generalised Lie derivative can be re-expressed as
\be
\mathcal{L}_\Lambda X = \Lambda \p \hat\partial X + \hat\partial ( \Lambda \p X) \,,
\ee
generalising the magic formula \refeqn{cartanmagic}.
For $\Lambda_1, \Lambda_2 \in \RR_1$, we can also write the generalised Lie derivative in terms of its antisymmetric and symmetric parts as:
\be
\mathcal{L}_{\Lambda_1} \Lambda_2 = [ \Lambda_1, \Lambda_2 ]_E + \hat \partial ( \Lambda_1 \p \Lambda_2 ) \,,
\ee
while the failure of the Jacobi identity \refeqn{jacobiatorE} or \refeqn{jacobiatorEnicer} becomes
\be
3 \mathcal{L}_{A_{[1}} \mathcal{L}_{A_2} A_{3]} = \hat \partial ( [ A_{[1} , A_{2} ]_E \p A_{3]}) \,.
\ee

\vspace{1em}\noindent
\emph{Tensor hierarchy in general} 

\noindent We can write down the gauge transformations and field strengths associated to the first few fields of the tensor hierarchy. 
Let us concentrate on the fields $\Aa_\mu \in \RR_1$, $\Ab_{\mu\nu} \in \RR_2$, $\Ac_{\mu\nu\rho} \in \RR_3$, $\Ad_{\mu\nu\rho\sigma} \in \RR_4$ and the field strengths $\Fa_{\mu\nu} \in \RR_1$, $\Fb_{\mu\nu\rho} \in \RR_2$ and $\Fc_{\mu\nu\rho\sigma} \in \RR_3$.
This will suffice to describe the tensor hierarchy for $\Gfour$ for which $d=4$ and $8-d=4$. In this case, the field strength for $\Ad_{\mu\nu\rho\sigma}$ would require some additional modifications of the type to be described in section \ref{compensator}. However as we will see in section \ref{action5} this does not affect the construction of the action. 
The extension of the tensor hierarchy for the lower rank groups $\Gthree$ and $\Gtwo$ can be found in \refref{Hohm:2015xna,Wang:2015hca,Berman:2015rcc}.

The field strengths are: 
\be
\begin{split} 
\Fa_{\mu\nu} &= 2\partial_{[\mu} \Aa_{\nu]} - [\Aa_\mu,\Aa_\nu]_E + \hat\partial\Ab_{\mu\nu} \,, \\
\Fb_{\mu\nu\rho} &= 3\D_{[\mu}\Ab_{\nu\rho]} - 3\partial_{[\mu}\Aa_{\nu}\p\Aa_{\rho]} + \Aa_{[\mu}\p[\Aa_\nu,\Aa_{\rho]}]_E + \hat\partial\Ac_{\mu\nu\rho} \,, \\
\Fc_{\mu\nu\rho\sigma} & = 4\D_{[\mu}\Ac_{\nu\rho\sigma]} + 3\hat\partial \Ab_{[\mu\nu}\p\Ab_{\rho\sigma]} - 6\Fa_{[\mu\nu}\p\Ab_{\rho\sigma]} + 4\Aa_{[\mu}\p(\Aa_\nu\p\partial_\rho\Aa_{\sigma]}) \\
 & \quad - \Aa_{[\mu}\p(\Aa_\nu\p[\Aa_\rho,\Aa_{\sigma]}]_E) + \hat\partial\Ad_{\mu\nu\rho\sigma} \,, 
\end{split} 
\label{FABC}
\ee
and their variations are given by:
\be
\begin{split}
 \delta\Fa_{\mu\nu} &= 2\D_{[\mu}\delta\Aa_{\nu]} + \hat\partial\Delta\Ab_{\mu\nu} \,, \\
 \delta\Fb_{\mu\nu\rho} &= 3\D_{[\mu}\Delta\Ab_{\nu\rho]} - 3\delta\Aa_{[\mu}\p\Fa_{\nu\rho]} + \hat\partial\Delta\Ac_{\mu\nu\rho} \,,\\
 \delta \Fc_{\mu\nu\rho\sigma} & = 4\D_{[\mu} \Delta \Ac_{\nu\rho\sigma]} - 4\delta\Aa_{[\mu}\p\Fb_{\nu\rho\sigma]} - 6\Fa_{[\mu\nu}\p\Delta\Ab_{\rho\sigma]} + \hat\partial\Delta\Ad_{\mu\nu\rho\sigma}\,,
\end{split}
\ee
where
\be
\begin{split}
\Delta\Ab_{\mu\nu} & = \delta\Ab_{\mu\nu} + \Aa_{[\mu}\p\delta\Aa_{\nu]} \,, \\
\Delta\Ac_{\mu\nu\rho} & = \delta\Ac_{\mu\nu\rho} - 3\delta\Aa_{[\mu}\p\Ab_{\nu\rho]} + \Aa_{[\mu}\p(\Aa_\nu\p\delta\Aa_{\rho]}) \,,\\
\Delta \Ad_{\mu\nu\rho\sigma} & = \delta \Ad_{\mu\nu\rho\sigma} - 4\delta\Aa_{[\mu}\p\Ac_{\nu\rho\sigma]}
 + 3\Ab_{[\mu\nu}\p\left(\delta\Ab_{\rho\sigma]}+2\Aa_{\rho}\p\delta\Aa_{\sigma]}\right) 
 \\ & \qquad+ \Aa_{[\mu}\!\p(\Aa_\nu\p(\Aa_\rho\p\delta\Aa_{\sigma]})) \,.
\end{split}
\label{DeltaGauge}
\ee
The symmetry transformations under generalised diffeomorphisms parametrised by $\Lambda \in R_1$ and gauge transformations $\lambda_\mu \in R_2$, $\Theta_{\mu\nu} \in R_3$, and $\Omega_{\mu\nu\rho} \in R_4$ are:
\be
\begin{split} 
\delta \Aa_{\mu} &= \D_{\mu} \Lambda - \hat{\partial}\, \lambda_\mu \,, \\
\Delta \Ab_{\mu\nu} &= \Lambda\, \p \Fa_{\mu\nu} + 2 \D_{[\mu} \lambda_{\nu]} - \hat{\partial} \Theta_{\mu\nu} \,, \\
\Delta \Ac_{\mu\nu\rho} &= \Lambda\, \p \Fb_{\mu\nu\rho} + 3 \Fa_{[\mu\nu} \p \lambda_{\rho]} + 3 \D_{[\mu} \Theta_{\nu\rho]} - \hat{\partial} \Omega _{\mu\nu\rho} \,, \\
\Delta \Ad_{\mu\nu\rho\sigma} & = \Lambda\, \p \Fc_{\mu\nu\rho\sigma} - 4 \Fb_{[\mu\nu\rho} \p \lambda_{\sigma]} + 6 \Fa_{[\mu\nu} \p \Theta_{\rho\sigma]} + 4 \D_{[\mu} \Omega_{\nu\rho\sigma]} 
\,,
\end{split} 
\label{deltaGauge}
\ee
under which the field strengths transform as generalised tensors in the appropriate representations.

The Bianchi identities obeyed by the field strengths are:
\begin{equation}
\begin{split}
3\D_{[\mu}\Fa_{\nu\rho]} &= \hat\partial\Fb_{\mu\ldots\rho} \,, \\
4\D_{[\mu}\Fb_{\nu\rho\sigma]} + 3\Fa_{[\mu\nu}\p\Fa_{\rho\sigma]} &= \hat\partial\Fc_{\mu\nu\rho\sigma} \,, \\
5\D_{[\mu}\Fc_{\nu\rho\sigma\lambda]} + 10\Fa_{[\mu\nu}\p\Fb_{\rho\sigma\lambda]} &= \hat\partial\Fd_{\mu\nu\rho\sigma\lambda} \,, \\
\end{split}
\label{eq:Bianchi}
\end{equation}
where $\Fd_{\mu\nu\rho\sigma\lambda}$ denotes the field strength for $\Ad_{\mu\nu\rho\sigma}$ which would be constructed using the modifications presented in section \ref{compensator}.

\subsection{Example: tensor hierarchy for $\Gfour$} 
\label{exsl5th}

To make the ideas of the ExFT tensor hierarchy more explicit, let's go into more detail on the example of the $\Gfour$ ExFT.
The tensor hierarchy representations are:
\be
R_1 = \mathbf{10} \,,\quad
R_2 = \mathbf{\bar{5}} \,,\quad
R_3 = \mathbf{5} \,,\quad
R_4 = \mathbf{\overline{10}} \,,
\ee
and with $\fM,\fN = 1,\dots, 5$ we will denote objects transforming in these representations by
\be
A^{\fM \fN} \in R_1 \,,\quad
B_{\fM} \in R_2 \,,\quad
C^{\fM} \in R_3 \,,\quad
D_{\fM\fN} \in R_4 \,,
\ee
where $A^{\fM\fN}$ and $D_{\fM\fN}$ are antisymmetric.
We take $A,B,C,D$ to have weights $1/5, 2/5,3/5,4/5$ respectively under generalised diffeomorphisms.
We use $\epsilon_{\fM\fN\fP\fQ\fK}$ to represent the alternating symbol with $\epsilon_{12345} = \epsilon^{12345}=1$. 

The nilpotent exterior derivative $\hat\partial$ is:
\be
\begin{split}
( \hat \partial B)^{\fM\fN} & = \frac{1}{2} \epsilon^{\fM\fN\fP\fQ\fK} \partial_{\fP\fQ} B_{\fK} \,,\\
( \hat \partial C)_{\fM} & = \partial_{\fN\fM} C^{\fN} \,,\\
(\hat \partial D)^{\fM} & = \frac{1}{2} \epsilon^{\fM\fN\fP\fQ\fK} \partial_{\fN\fP} D_{\fQ\fK} \,.
\end{split} 
\label{sl5d}
\ee
Nilpotency is easily confirmed using the section condition $ \epsilon^{\fM\fN\fP\fQ\fK} \partial_{\fN \fP} \partial_{\fQ \fK} =0$.

The products $\p$ are defined as:
\begin{align}
  \left( A_1 \p A_2 \right)_{\fM} &= \frac14 A_1{}^{\fN\fP} A_2{}^{\fQ\fK} \epsilon_{\fM\fN\fP\fQ\fK} \,, &  \left( A \p B \right)^{\fM} &= A^{\fM\fN} B_{\fN} \,,
     \label{sl5wedge1}
   \\
  \left( A \p C \right)_{\fM\fN} &= \frac14 \epsilon_{\fM\fN\fP\fQ\fK} A^{\fP\fQ} C^{\fK} \,, &
  A \p D &= \frac{1}{2} A^{\fM \fN} D_{\fM \fN} \,, 
   \label{sl5wedge2}\\
  \left( B_1 \p B_2 \right)_{\fM\fN} &= B_{2[\fM} B_{|1|\fN]} \,, &
  B \p C &= B_{\fM} C^{\fM} \,. 
 \label{sl5wedge3}
\end{align}
With the Y-tensor $Y^{\fM\fN\fP\fQ}{}_{\fM^\prime\fN^\prime\fP^\prime\fQ^\prime} = \epsilon^{\fM\fN\fP\fQ\fK} \epsilon_{\fM^\prime\fN^\prime \fP^\prime \fQ^\prime\fK}$ the  generalised Lie derivative acts as \cite{Berman:2011cg, Wang:2015hca}:
\be
\begin{split} 
\mathcal{L}_\Lambda A^{\fM\fN} & = 
\frac{1}{2} \Lambda^{\fP\fQ} \partial_{\fP\fQ} A^{\fM\fN}
+ \frac{1}{2} \partial_{\fP\fQ} \Lambda^{\fP\fQ} A^{\fM\fN} 
+  2\partial_{\fP\fQ} \Lambda^{\fP[\fM} A^{\fN] \fQ} \,,
\end{split} 
\label{gldA5}
\ee
\be
\mathcal{L}_\Lambda B_{\fM} = \frac{1}{2} \Lambda^{\fP\fQ} \partial_{\fP\fQ} B_{\fM} + B_{\fP} \partial_{\fM\fQ} \Lambda^{\fP\fQ} \,,
\label{gldB5}
\ee
\be
\mathcal{L}_\Lambda C^{\fM} = \frac{1}{2} \Lambda^{\fP\fQ} \partial_{\fP\fQ} C^{\fM} - \partial_{\fP\fQ} \Lambda^{\fP\fM} C^{\fQ} + \frac{1}{2} \partial_{\fP\fQ} \Lambda^{\fP\fQ} C^{\fM} \,,
\label{gldC5}
\ee
\be
\mathcal{L}_\Lambda D_{\fM\fN} = \frac{1}{2} \Lambda^{\fP\fQ} \partial_{\fP\fQ} D_{\fM\fN}
- 2 \partial_{[\fM| \fP } \Lambda^{\fP \fQ} D_{\fQ|\fN]}\,.
\label{gldD5}
\ee
Now we let $\Aa_\mu \in R_1$, $\Ab_{\mu\nu} \in R_2$, $\Ac_{\mu\nu\rho} \in R_3$ and $\Ad_{\mu\nu\rho\sigma} \in R_4$ denote the fields in the tensor hierarchy.
The explicit expressions for their field strengths follow from the above formulae and from the general results \refeqn{FABC}:
\be
\Fa_{\mu\nu}{}^{\fM\fN} = 2 \partial_{[\mu} \Aa_{\nu]}{}^{\fM\fN} - [\Aa_\mu,\Aa_\nu]_E{}^{\fM\fN} 
+ \frac{1}{2} \epsilon^{\fM\fN\fP\fQ\fK} \partial_{\fP\fQ} \Ab_{\mu \nu \fK} \,,
\label{5Fa}
\ee
\be
\begin{split}
\Fb_{\mu\nu\rho \fM} & =  3 \D_{[\mu } \Ab_{\nu \rho ]\fM}
- \frac{3}{4} \epsilon_{\fM\fN\fP\fQ\fK} \partial_{[\mu} \Aa_\nu{}^{\fN\fP} \Aa_{\rho]}{}^{\fQ\fK} 
\\ & \qquad
+\frac{1}{4} \epsilon_{\fM\fN\fP\fQ\fK} \Aa_{[\mu}{}^{\fN\fP} [ \Aa_\nu, \Aa_{\rho]}]_E{}^{\fQ\fK} 
+ \partial_{\fN\fM} \Ac_{\mu\nu\rho}{}^{\fN} \,,
\end{split}
\label{5Fb}
\ee
\be
\begin{split}
\Fc_{\mu\nu\rho \sigma}{}^{\fM} & =  4 \D_{[\mu } \Ac_{\nu \rho\sigma ]}{}{^\fM}
+ 3 \frac{1}{2} \epsilon^{\fM \fN \fP \fQ \fK} \partial_{\fP \fQ} \Ab_{[\mu\nu |\fK|}\Ab_{\rho\sigma] \fN} 
\\ & \qquad- 6\Fa_{[\mu\nu}{}^{\fM \fN} \Ab_{\rho\sigma] \fN} +   \epsilon_{\fN \fP \fQ \fK \fL}\Aa_{[\mu}^{\fM \fN} \Aa_\nu{}^{\fP\fQ} \partial_\rho\Aa_{\sigma]}{}^{\fK \fL} \\
 & \qquad -  \frac{1}{4}  \epsilon_{\fN \fP \fQ \fK \fL}\Aa_{[\mu}^{\fM \fN}\Aa_\nu{}^{\fP \fQ} [\Aa_\rho,\Aa_{\sigma]}]_E{}^{\fK \fL} + \frac{1}{2} \epsilon^{\fM \fN \fP \fQ \fK} \partial_{\fN \fP} \Ad_{\mu\nu\rho\sigma \fQ \fK} \,, \\
\end{split}
\label{5Fc}
\ee
These expressions can seem rather baroque, and it may not be clear exactly what the underlying significance of the operations $\hat\partial$ and $\p$ really is. 
To demystify the situation, let's solve the section condition such that $\fM = (i,5)$ with $\partial_{i5} \equiv \partial_i \neq 0$. 
We can determine the four-dimensional geometric nature of objects in $R_1, R_2, R_3, R_4$ by seeing how they transform under generalised diffeomorphisms with $\Lambda^{i5} \equiv v^i$ and $\Lambda^{ij} = 0$, interpreting the vector $v^i$ as the generator of four-dimensional spacetime diffeomorphisms.

Using \refeqn{gldA5} we see that $A^{i5}$ is a vector and $A^{ij}$ is a bivector of weight one. 
In four dimensions, the latter is equivalent to a two-form (of weight zero), with $A^{ij} = \frac{1}{2} \epsilon^{ijkl} A_{kl}$.
Using \refeqn{gldB5} we find that $B_{i}$ is a one-form and that $B_5$ is a scalar of weight one, which is equivalent to a four-form, $B_5 = \frac{1}{4!} \epsilon^{ijkl} B_{ijkl}$.
Then using \refeqn{gldC5} we find that $C^i$ is a vector of weight one, and so equivalent to a three-form $C^i = \frac{1}{3!} \epsilon^{ijkl} C_{jkl}$, while $C^5$ is a scalar of weight zero.
Finally using \refeqn{gldD5} we find that $D_{ij}$ is a two-form and $D_{i5} = D_i$ is a one-form.

What we have described are sections (in the usual sense) of the following generalised tangent bundles $\mathcal{R}_p$ with fibres $R_p$:
\be
\mathcal{R}_1 \cong TM \oplus \Lambda^2 T^*M \,,
\ee
\be
\mathcal{R}_2 \cong T^*M \oplus \Lambda^4 T^*M \,,
\ee
\be
\mathcal{R}_3 \cong \mathbb{R} \oplus \Lambda^3 T^*M\,,
\ee
\be
\mathcal{R}_4 \cong \Lambda^2 T^*M \oplus T^*M \,,
\ee
where $M$ denotes a four-dimensional manifold.
In particular, sections of these bundles correspond to the vector, two- and five-form gauge transformation parameters of 11-dimensional supergravity organised according to how they act on the fields in the $(7+4)$-dimensional split of the coordinates.
Thus sections of $\mathcal{R}_1$ describe four-dimensional diffeomorphisms and gauge transformations $(v^i,\lambda_{ij})$, sections of $\mathcal{R}_2$ describe gauge transformation parameters carrying one $n$-dimensional index, i.e. of the form $(\lambda_{\mu i} , \lambda_{\mu ijkl})$, while sections of $\mathcal{R}_3$ are gauge transformation parameters of the form $(\lambda_{\mu\nu}, \lambda_{\mu\nu ijk})$ with two external legs, and finally sections of $\mathcal{R}_4$ describe gauge transformation parameters with three external legs $(\lambda_{\mu\nu\rho ij}, \Lambda_{\mu\nu\rho ijkl;m})$ in which to obtain a $\Gfour$ representation we have to include gauge transformations associated to the dual graviton. These additional gauge transformations show up exactly at the point beyond which the naive tensor hierarchy construction does not apply.

The derivative $\hat\partial$ given in \refeqn{sl5d} then corresponds to the usual exterior derivative:
\be
(\hat \partial B)^{i5} = 0 \,,\quad
(\hat \partial B)^{ij} = \frac{1}{2} \epsilon^{ijkl} \partial_k B_{l} \,,
\ee
\be
(\hat \partial C)_i = -\partial_i C^5 \,,\quad
(\hat \partial C)_5 = \frac{1}{6} \epsilon^{ijkl} \partial_i C_{jkl}\,,
\ee
\be
(\hat \partial D)^{i}  = \epsilon^{ijkl} \partial_j D_{kl} \,,\quad
(\hat\partial D)^5 = 0 \,.
\ee
The products $\p$ given in \refeqn{sl5wedge1} to \refeqn{sl5wedge3} can then be seen to correspond to all possible wedge and interior products of the components of these generalised tensors:
\be
( A_1 \p A_2)_i = - A_1{}^j A_{2 ji}- A_2{}^j A_{1 ji}
\,,\quad
(A_1 \p A_2)_5 = \frac{1}{2} \epsilon^{ijkl} A_{1 ij } A_{2 kl}  \,,
\ee
\be
(A\p B)^i = \frac{1}{2} \epsilon^{ijkl} A_{jk} B_l \,,\quad
(A\p B)^5 = - A^j B_j \,,
\ee
\be
(A\p C)_{ij} = \frac{1}{2} A_{ij} C^5 + \frac{1}{2} A^k C_{kij} \,,\quad
(A\p C)_{i5} = - \frac{1}{6} \epsilon^{jklm} A_{ij} C_{klm}\,,
\ee
\be
(A\p D) = A^i D_i + \frac{1}{4} \epsilon^{ijkl} A_{ij} D_{kl} \,,
\ee
\be
(B_1 \p B_2)_{i5} = \frac{1}{4!} \epsilon^{jklm}( B_{2 i } B_{1 jklm } - B_{1i} B_{2jklm} )\,,\quad
(B_1 \p B_2)_{ij} = B_{2[i|} B_{1 |j]}\,,
\ee
\be
(B\p C) = \frac{1}{6} \epsilon^{ijkl} B_i C_{jkl} + \frac{1}{4!} \epsilon^{ijkl} B_{ijkl} C^5 \,.
\ee
We can use these expressions to identify the components of the tensor hierarchy fields with the degrees of freedom of 11-dimensional supergravity.
Here we refer to section \ref{KKsplit} and in particular the redefined three-form components of \refeqn{redefC3} and the field strengths \refeqn{F=dA}. 

Using the above formulae, we find that the components of the ExFT field strength $\Fa_{\mu\nu}$ are:
\be
\begin{split} 
\Fa_{\mu\nu}{}^i & =  2 \partial_{[\mu} \Aa_{\nu]}{}^i - 2 \Aa_{[\mu}{}^j\partial_j \Aa_{\nu]}{}^i \,,\\
\Fa_{\mu\nu}{}^{ij} & = 2 \partial_{[\mu} \Aa_{\nu]}{}^{ij} 
-2 \Aa_{[\mu|}{}^k \partial_k \Aa_{|\nu]}{}^{ij} 
- 3 \partial_k \Aa_{[\mu}{}^{[ij} \Aa_{\nu]}{}^{k]}
- 3 \partial_k \Aa_{[\mu}{}^{[i} \Aa_{\nu]}{}^{jk]}
- \epsilon^{ijkl} \partial_k \Ab_{\mu\nu l} 
\end{split}
\ee
First of all, it is clear that $\Fa_{\mu\nu}{}^i$ coincides with the field strength \refeqn{kkf} of the Kaluza-Klein vector $A_\mu{}^i$ appearing in the metric decomposition \refeqn{metricsplit}, on making the obvious identification $\Aa_\mu{}^i = A_\mu{}^i$.
Then if we dualise $\Aa_{\mu ij} = \frac{1}{2} \epsilon_{ijkl} \Aa_\mu{}^{kl}$, and define $\Fa_{\mu\nu ij} = \frac{1}{2} \epsilon_{ijkl} \Fa_{\mu\nu}{}^{kl}$, a short calculation reveals that
\be
\Fa_{\mu\nu ij} = 2 D_{[\mu} \Aa_{\nu]ij} + 2  \partial_{[i|} ( -\Ab_{\mu \nu|j]} - \Aa_{[\mu}{}^k \Aa_{\nu ]|j] k }) 
\ee
Here we meet again the covariant derivative $D_\mu = \partial_\mu - L_{A_\mu}$ involving the conventional Lie derivative with respect to $d$-dimensional diffeomorphisms.
We are led to use a further field redefinition in order to identify the appearance of the redefined three-form components $\Cdef_{\mu ij}$ and $\Cdef_{\mu\nu i}$
\be
\Aa_{\mu ij} \equiv \Cdef_{\mu ij} 
\,,
\quad
-\Ab_{\mu \nu i} - \Aa_{[\mu}{}^k \Aa_{\nu] i k } \equiv \Cdef_{\mu \nu i}\,,
\ee
such that referring to the expressions \refeqn{F=dA} for the redefined field strengths we can identify
\be
\Fa_{\mu \nu ij} \equiv \Fdef_{\mu \nu ij} - F_{\mu \nu}{}^k \Cdef_{ijk} \,.
\ee
The ``twisting'' by the internal components of the three-form is a standard feature of quantities which transform tensorially under generalised diffeomorphisms (and reflects the patching used to precisely define how sections of the bundles $\mathcal{R}_p$ transform on overlaps \cite{Coimbra:2011ky}, a detail we have glossed over in the above presentation).

We move on to the field strength $\Fb_{\mu\nu\rho}$.
If we define
\be
\Ab_{\mu \nu 5} \equiv \frac{1}{4!} \epsilon^{ijkl} \Ab_{\mu\nu ijkl} 
\,,\quad
\Ac_{\mu\nu \rho}{}^i \equiv \frac{1}{3!} \epsilon^{ijkl} \Ac_{\mu\nu\rho jkl} \,,
\ee
then
\be
\begin{split}
\Fb_{\mu\nu\rho i} & = 3 D_{[\mu} \Ab_{\nu \rho ]i}
- 2 \Aa_{[\mu|}{}^i \partial_{[i|} \Ab_{\nu \rho] j]}
\\ & \qquad
- ( \partial_{[\mu} \Aa_{\nu}{}^j + \Fa_{[\mu\nu}{}^j ) \Aa_{\rho]ij}
- ( \partial_{[\mu} \Aa_{\nu|ij|} + \Fa_{[\mu\nu |ij|} ) \Aa_{\rho]}{}^j 
- \partial_i \Ac_{\mu\nu\rho}{}^5 \,,
\\ 
\Fb_{\mu\nu \rho 5} & =
\frac{1}{4!} \epsilon^{ijkl} 
\Big( 
3 D_{[\mu} \Ab_{\nu \rho ] ijkl} 
+ \frac{3 \cdot{4!} }{2}  \partial_i  \Aa_{[\mu | kl} \Ab_{\nu \rho ] j} 
\\ & \qquad\qquad \qquad\qquad+ \frac{4!}{2} \left(  
- \frac{1}{2} (\partial_{[\mu} \Aa_{\nu |ij|} + \Fa_{[\mu \nu |ij} )
\Aa_{\rho] kl} - \Aa_{[\mu|kl} \partial_i \Ab_{\nu\rho] j}\right)
\\ & \qquad\qquad\qquad\qquad
+ \frac{4!}{3!} \partial_i \Ac_{\mu\nu\rho jkl}
\Big)\,,
\end{split} 
\ee
and substituting in the previous identifications leads to
\be
\Fb_{\mu\nu\rho i} = - 3D_{[\mu} \Cdef_{\nu \rho]i} - 3 F_{\mu\nu}{}^j \Cdef_{\rho i j} + \partial_i ( - C_{\mu\nu\rho}{}^5 + A_\mu{}^j A_\nu{}^k \Cdef_{\rho j k} )\,,
\ee
so that if
\be
C_{\mu\nu\rho}{}^5 = - \Cdef_{\mu\nu\rho} + A_{[\mu}{}^j A_\nu{}^k \Cdef_{\rho] jk}\,,
\ee
we find again by comparison with \refeqn{F=dA}
\be
\Fb_{\mu\nu\rho i} = - \Fdef_{\mu\nu\rho i}\,.
\ee
This, along with the confirmation that $\Fc_{\mu\nu\rho\sigma}{}^5 = - \Fdef_{\mu\nu\rho \sigma}$, which proceeds similarly, completes the dictionary relating the components of the three-form to the components of the ExFT fields. (We have already seen back in section \ref{genKK_EFT} how the generalised metric contains the internal components $\Cdef_{ijk}$, see \refeqn{5gmfirstlook}.)
The remaining degrees of freedom, for instance $\Ab_{\mu\nu 5}$ and $\Ac_{\mu\nu\rho}{}^i$, contain no further new physical degrees of freedom, and can be identified with components of the dual 11-dimensional six-form. 
We will see later how the dynamics of the $\Gfour$ ExFT enforce the duality relation between the three-form and its dual in terms of a duality relationship between the ExFT gauge fields.

Finally, let's note that to match the gauge symmetries directly requires similar identifications.
For instance, the gauge transformation of the one-form 
\be
\delta \Aa_\mu{}^{\fM\fN} = \D_\mu \Lambda^{\fM\fN} - \frac{1}{2} \epsilon^{\fM\fN\fP\fQ\fK} \partial_{\fP\fQ} \lambda_{\mu \fK} \,,
\ee
implies for the components $\Aa_\mu{}^i = A_\mu{}^i$ and $\Aa_{\mu ij} = \Cdef_{\mu ij}$ that
\be
\begin{split} 
\delta A_\mu{}^i & = D_\mu \Lambda^i \\
\delta \Cdef_{\mu ij} & = D_\mu \Lambda_{ij} + 2 \partial_{[i} ( \lambda_{|\mu|j]} + \Lambda^k \Cdef_{|\mu| j] k} )
+ L_\Lambda \Cdef_{\mu ij} \,,
\end{split} 
\ee
so on comparing with \refeqn{redefCgauge} we identify
\be
\Lambda_{ij} = \lambdadef_{ij} \quad,\quad
\lambda_{\mu i} = - \lambdadef_{\mu i} - \Lambda^j \Cdef_{\mu ij}\,,
\ee
where $\lambdadef_{ij}$ and $\lambdadef_{\mu i}$ denote here the redefined gauge symmetry parameters of \refeqn{redeflambda}.

\subsection{Dual graviton gauge transformations and constrained compensator fields} 
\label{compensator}

The tensor hierarchy construction that we have described has to be modified when we reach the representation $R_p$ which coincides with $\bar R_1$.
This happens for $p=8-d$ (we assume $d < 8$ in this subsection; for $d=8$ similar modifications must be made at the level of generalised diffeomorphisms themselves and will be described in section \ref{action8}).
To describe these modifications, we follow the paper \refref{Hohm:2013uia} discussing the $\Gseven$ case as well as \refref{Hohm:2018qhd} which is a detailed review of the dual graviton and its appearance in exceptional field theory. (Explicit details for $\Gsix$ can be found in \refref{Musaev:2014lna, Arvanitakis:2018hfn}.)

The representation after $R_{8-d} = \bar R_1$ is the adjoint, $R_{9-d} = \textbf{adj}$.
Let $\alpha = 1,\dots,\text{dim}\,\Edd$ be an adjoint index, and $(t^\alpha)_M{}^N$ denote the adjoint generators valued in the $R_1$ representation.
Given $V_\alpha \in R_{9-d}$ of weight $-(9-d) \omega = 1$ under generalised diffeomorphisms, the map $\hat \partial: R_{9-d} \rightarrow R_{8-d}$ defined by
\be
(\hat \partial V)_M = (t^\alpha)_M{}^N \partial_N V_\alpha \,,
\ee
fails to be covariant.
Rewriting the adjoint projector in the definition \refeqn{gldproj} of the generalised Lie derivative as $\mathbb{P}^M{}_N{}^P{}_Q = (t_\alpha)_N{}^M (t^\alpha)_Q{}^P$, and defining the structure constants by $[t_\alpha, t_\beta] = f_{\alpha \beta}{}^\gamma t_\gamma$, we can obtain the generalised Lie derivative acting on $V_\alpha$:
\be
\delta_{\Lambda} V_\alpha = \Lambda^K \partial_K V_\alpha - \alpha f_{\beta \alpha}{}^\gamma (t^\beta)_{N}{}^M \partial_M \Lambda^N V_\gamma + \partial_N \Lambda^N V_\alpha 
\ee
by requiring that $(t^\alpha)_M{}^N$ themselves be invariant.
It follows that
\be
\delta_\Lambda 
(\hat \partial V)_M = 
\mathcal{L}_{\Lambda} (\hat \partial V)_M 
+ (t^\beta)_N{}^P \partial_M \partial_P \Lambda^N V_\beta\,.
\ee
The key feature of the non-covariant second term is that it involves the derivative $\partial_M$ which is constrained to obey the section condition.
To compensate for this, we can introduce fields $\Xi_M \in \bar R_1$ which are themselves constrained, and taken to be of weight $-(8-d) \omega = 1+\omega$.
Such a ``constrained compensator'' field must obey
\be
Y^{MN}{}_{PQ} \Xi_M \Xi_N = 0 = Y^{MN}{}_{PQ} \Xi_M \partial_N \,.
\label{ccconstraints}
\ee
We declare it to have the non-covariant transformation 
\be
\delta_\Lambda \Xi_M = \mathcal{L}_\Lambda \Xi_M - (t^\beta)_N{}^P \partial_M \partial_P \Lambda^N V_\beta\,.
\ee
such that the combination
\be
(\hat \partial V)_M +\Xi_M
\ee
will be a generalised tensor.

In practice, one makes use of these as follows. 
Denote the tensor hierarchy fields in $R_{8-d}$ and $R_{9-d}$ as follows:
\be
\Ac_{\mu_1 \dots \mu_{8-d} M} \in R_{8-d} = \bar R_1\,,\quad
\Ac_{\mu_1 \dots \mu_{9-d} \alpha} \in R_{9-d} = \textbf{adj}\,,
\ee
and introduce $\Ac_{\mu_1 \dots \mu_{9-d} M}$ as a constrained compensator field obeying \refeqn{ccconstraints}.
Then the field strength $\Fa_{\mu_1 \dots \mu_{9-d} M }$  of $\Ac_{\mu_1 \dots \mu_{8-d} M}$ which is obtained from the general tensor hierarchy construction described above is modified to
\be
\Fa_{\mu_1 \dots \mu_{9-d} M} = (9-d) \D_{[\mu_1} \Ac_{\mu_2 \dots \mu_{9-d}] M } + \dots + ( \hat \partial  \Ac_{\mu_1 \dots \mu_{9-d}} )_M 
+  \Ac_{\mu_1 \dots \mu_{9-d} M }
\ee
where the ellipsis denotes terms involving lower rank $p$-forms determined using the general tensor hierarchy methods and formulae that we previously described.
With both the adjoint-valued $(9-d)$-form and the constrained $(9-d)$-form given appropriate gauge transformations, this field strength is then invariant.
In particular, the gauge transformations of the constrained $(9-d)$-form will act as shift symmetries on the $(8-d)$-form, and these can be used to gauge away the dual graviton degrees of freedom appearing in the latter.

How this works will be clearest if we do a particular example.
Let's now pick the case $d=7$, where we encounter these features in the definition of the field strength $\Fa_{\mu\nu}{}^M$ of the one-form $\Aa_{\mu}{}^M$. 
The representation $R_1$, which is the fundamental $\mathbf{56}$ of $\Gseven$, is self-conjugate.
We can use the antisymmetric invariant $\Omega_{MN}$ and its inverse $\Omega^{MN}$ to raise and lower indices, with $\Omega^{MP} \Omega_{NP} = \delta^M_N$, $V^M = \Omega^{MN} V_N$ and $V_M = V^N \Omega_{NM}$. 
The adjoint generator with indices fully raised or lowered, $t_{\alpha MN}$ or $t_\alpha{}^{MN}$, is symmetric in $MN$.
The adjoint projector is
\be
\begin{split}
\mathbb{P}^M{}_N{}^P{}_Q  & = (t_\alpha)_N{}^M (t^\alpha)_Q{}^P
\\ & 
= \frac{1}{24} \delta^M_N \delta_Q^P  + \frac{1}{12} \delta^M_Q \delta_N^P + (t_\alpha)_{NQ} (t^\alpha)^{MP}  - \frac{1}{24} \Omega^{MP} \Omega_{NQ} \,,
\end{split}
\ee
and $\mathbb{P}^M{}_N{}^N{}_M = 133$ implying $(t_\alpha)_{MN} (t^\alpha)^{MN} = -133$.
The generalised Lie derivative on generalised vectors is
\be
\mathcal{L}_\Lambda V^M = \Lambda^N \partial_N V^M - 12 \mathbb{P}^M{}_N{}^P{}_Q \partial_P \Lambda^Q V^N + \lambda_V \partial_N \Lambda^N V^M \,.
\ee
For the special case of $\lambda_V = -\omega= \frac{1}{2}$ this becomes
\be
\mathcal{L}_\Lambda V^M = \Lambda^N \partial_N V^M -V^N \partial_N V^M -  12 (t_\alpha)_{NQ} (t^\alpha)^{MP} \partial_P \Lambda^Q V^N 
+ \frac{1}{2} \Omega^{MP} \Omega_{NQ} \partial_P \Lambda^Q V^N \,,
\ee
consistent with the definition of the Y-tensor as in \refeqn{Y7}.
If we now define as before the naive field strength, $F_{\mu\nu} = 2 \partial_{[\mu} \Aa_{\nu]} - [ \Aa_\mu , \Aa_\nu ]_E$, its variation is
\be
\begin{split}
\delta F_{\mu\nu}{}^M 
& 
= 2 \D_{[\mu} \Aa_{\nu]}{}^M - 12 (t^\alpha)^{MN} (t_\alpha)_{PQ}  \partial_N ( \Aa_{[\mu}{}^P \delta \Aa_{\nu]}{}^Q)
\\ & \qquad
- \frac{1}{2} \Omega^{MN} \Omega_{PQ} ( \Aa_{[\mu}{}^Q  \partial_N \delta \Aa_{\nu]}{}^P - \partial_N \Aa_{[\mu}{}^Q \delta \Aa_{\nu]}{}^P)
\,.
\end{split}
\ee	
This as expected does not lead to a covariant transformation under generalised diffeomorphisms but now we have two types of terms to worry about.
Thus we introduce a two-form $\Ab_{\mu\nu\alpha}$ in the adjoint and a further constrained two-form $\Ab_{\mu\nu M}$, obeying
\be
(t_\alpha)^{MN}  \Ab_M  \Ab_N = 0 = (t_\alpha)^{MN}  \Ab_M \partial_N \,,\quad
\Omega^{MN}  \Ab_M  \Ab_N = 0 = \Omega^{MN}  \Ab_M \partial_N \,,
\ee
where we suppressed the four-dimensional indices.
The full field strength is defined as:
\be
\Fa_{\mu\nu}{}^M = F_{\mu\nu}{}^M - 12 (t^\alpha)^{MN} \partial_N \Ab_{\mu\nu \alpha} -\frac{1}{2} \Omega^{MN} \Ab_{\mu\nu N} \,,
\label{e7fullF} 
\ee
and obeys
\be
\delta \Fa_{\mu\nu}{}^M = 2 \D_{[\mu} \delta \Aa_{\nu]}{}^M - 12 (t^\alpha)^{MN} \partial_N \Delta \Ab_{\mu\nu \alpha} -\frac{1}{2} \Omega^{MN} \Delta \Ab_{\mu\nu N} \,,
\label{e7fullFdelta}
\ee
with
\be
\begin{split}
\Delta \Ab_{\mu\nu \alpha } & = \delta \Ab_{\mu\nu\alpha}  + (t_\alpha)_{MN} A_{[\mu}{}^M \delta A_{\nu]}{}^N\,,\\
\Delta  \Ab_{\mu\nu M} & = \delta  \Ab_{\mu\nu M }  + \Omega_{PQ} (\Aa_{[\mu}{}^Q  \partial_M \delta \Aa_{\nu]}{}^P - \partial_M \Aa_{[\mu}{}^Q \delta \Aa_{\nu]}{}^P)\,.
\end{split} 
\ee
For the adjoint-valued field, this conforms to the general expressions for the tensor hierarchy on defining 
$\Ab_{\mu\nu}{}^{MN} = - (t^\alpha)^{MN} \Ab_{\mu\nu \alpha}$.
The field strength \refeqn{e7fullFdelta} is then covariant under generalised diffeomorphisms, and invariant under gauge transformations altogether acting as:
\be
\begin{split}
\delta \Aa_\mu{}^M & = \D_\mu \Lambda^M + 12 (t^\alpha)^{MN} \partial_N \lambda_{\mu \alpha} + \frac{1}{2} \Omega^{MN}\Xi_{\mu N} \,,\\
\Delta \Ab_{\mu\nu \alpha} & = (t_\alpha)_{MN} \Lambda^M \Fa_{\mu\nu}{}^N + 2 \D_{[\mu} \lambda_{\nu]\alpha} \,,\\
\Delta  \Ab_{\mu\nu M} & =-\Omega_{PQ}( \Fa_{\mu\nu}{}^P \partial_M \Lambda^Q - \Lambda^Q\partial_M \Fa_{\mu\nu}{}^P ) 
 \\ & \qquad + 2 \D_{[\mu} \Xi_{\nu] M} + 4 8(t^\alpha)_Q{}^P (\partial_P\partial_M \Aa_{[\mu}{}^Q ) \Xi_{\nu] \alpha}\,.
\end{split}
\ee
Now we see the new gauge symmetry associated to the constrained field, generated by the parameter $\Xi_{\mu M}$, which must also be covariantly constrained.
Importantly, this acts as a shift symmetry of the one-form $\Aa_\mu{}^M$.
We can therefore use it -- after solving the section condition constraints on the derivatives and the constrained fields -- to gauge away the components of $\Aa_{\mu}{}^M$ that would correspond to dual graviton degrees of freedom.

One can further obtain field strengths from the Bianchi identities:
\be
3 \D_{[\mu} \Fa_{\nu\rho]}{}^M = - 12 (t^\alpha)^{MN} \partial_N \Fb_{\mu\nu\rho \alpha} - \frac{1}{2} \Omega^{MN}\Fb_{\mu\nu\rho  N} \,.
\ee
However, these are not necessary for the formulation of the dynamics of the theory.

\subsection{The action of exceptional field theory}
\label{action}

Now let's gather together everything that we need to write down the bosonic dynamics of exceptional field theory. 
We organise the field content of ExFT according to their transformation properties under generalised diffeomorphisms.
We have:
\begin{itemize}
\item the $n$-dimensional metric, 
\be
g_{\mu\nu} \in \mathrm{GL}(n) / \mathrm{SO}(1,n-1) \,. 
\ee
This is a scalar of weight $-2\omega = \tfrac{2}{n-2}$ under generalised diffeomorphisms, $\delta_\Lambda g_{\mu\nu} = \Lambda^N \partial_N g_{\mu\nu} + \frac{2}{n-2} \partial_N \Lambda^N g_{\mu\nu}$. 
\item the generalised metric
\be
\gM_{MN} \in \Edd / H_d \,, 
\ee
where $H_d$ denotes the maximal compact subgroup of $\Edd$.
The generalised metric is a tensor of weight zero under generalised diffeomorphisms, $\delta_\Lambda \gM_{MN} = \mathcal{L}_\Lambda \gM_{MN}$.
\item the tensor hierarchy gauge fields:
\be
\begin{array}{l}
\Aa_\mu \in R_1 \,,\\
\Ab_{\mu\nu} \in R_2 \,,\\
\vdots \\
\Ac_{\mu_1 \dots \mu_{8-d}} \in R_{8-d} = \bar R_1 \,,\\
\Ac_{\mu_1 \dots \mu_{9-d}} \in R_{9-d} = \textbf{adj}\,,
\end{array}
\ee
plus the constrained compensator field $\tilde \Ac_{\mu_1 \dots \mu_{9-d}} \in \bar R_1$.
Each $p$-form has weight $- p\omega = \tfrac{p}{n-2}$.
The one-form $\Aa_\mu$ transforms as a connection under generalised diffemorphisms, $\delta_\Lambda \Aa_\mu = \D_\mu \Lambda$, and the additional gauge transformations and field strengths of the tensor hierarchy have been described in sections \ref{thier} and \ref{compensator} above.
\end{itemize}

Using the transformation properties of the fields under generalised diffeomorphisms, the building blocks of the ExFT action can be constructed in terms of quantities which are scalars under generalised diffeomorphisms. 
The full action can be organised in terms of such blocks as:
\be
\begin{split}
S_{\text{ExFT}} = \int \dd^n x \, \dd Y \, \sqrt{|g|} \Big(
R_{\text{ext}}(g) 
+ \mathcal{L}_{\text{kin}} 
+ \sum_p \mathcal{L}_p 
+ \mathcal{L}_{\text{int}} (\mathcal{M},g)
+ \frac{1}{\sqrt{|g|}} \mathcal{L}_{\text{top}} 
\Big)\,.
\end{split}
\label{sexft}
\ee
Here we have:
\begin{itemize}
\item the $n$-dimensional Ricci scalar, 
\be
\begin{split}
R_{\text{ext}}(g)  = &
\frac{1}{4} g^{\mu \nu} \D_\mu g_{\rho \sigma} \D_\nu g^{\rho \sigma} 
- \frac{1}{2} g^{\mu \nu} \D_\mu g^{\rho \sigma} \D_\rho g_{\nu \sigma} \\
&\qquad\qquad 
+ \frac{1}{4} g^{\mu \nu} \D_\mu \ln g \D_\nu \ln g 
+ \frac{1}{2} \D_\mu \ln g \D_\nu g^{\mu \nu} \,,
\end{split}
\label{Rext}
\ee
where $\D_\mu = \partial_\mu - \mathcal{L}_{\Aa_\mu}$.
\item a term involving $n$-dimensional derivatives of the generalised metric, which for $4 \leq d \leq 8$ is:
\be
 \mathcal{L}_{\text{kin}} = \frac{1}{4\alpha} \D_\mu \gM^{MN} \D^\mu \gM_{MN} \,,
\ee
For $d < 4$, the details depend on the group, as the generalised metric does not solely appear in its $R_1$ representation.
\item kinetic terms for the field strengths, where if we let $\Sigma, \Pi$ denote indices for the $R_p$ representation and $\gM_{\Sigma\Pi}$ the corresponding representation of the generalised metric 
\be
\mathcal{L}_{p} = - \frac{1}{2\cdot p!} \gM_{\Sigma \Pi} \Fa_{\mu_1 \dots \mu_p}{}^\Sigma \Fa^{\mu_1 \dots \mu_p}{}^\Pi 
\ee
In practice, kinetic terms for only a subset of the tensor hierarchy gauge fields are included. 
Furthermore when $n$ is even, the ExFT action is only a pseudo-action, and one must also impose a twisted self-duality constraint on the $\tfrac{n}{2}$-form field strength, at the level of the equations of motion, in which case if its kinetic term is included in the action it comes multiplied by a further factor of 1/2.
\item terms involving extended derivatives of the generalised metric and external metric, which for $4 \leq d \leq 7$ take the universal form:
\be
\begin{split}
\mathcal{L}_{\text{int}} (\gM, g) & =  
\frac{1}{4 \alpha} \gM^{MN} \partial_M \gM^{KL} \partial_N \gM_{KL} - \frac{1}{2} \gM^{MN} \partial_M \gM^{KL} \partial_K \gM_{LN} 
\\ & \qquad + \frac{1}{2} \partial_M \gM^{MN} \partial_N \ln |g|
\\
& \qquad + \frac{1}{4} \gM^{MN} \left( \partial_M g^{\mu\nu} \partial_N g_{\mu\nu} + \partial_M \ln |g| \partial_N \ln |g| \right)
\end{split}
\label{Luniversal}
\ee
This can be called the ``internal Lagrangian'' and is often referred to as the ExFT ``potential'' $V(\gM,g) = - \mathcal{L}_{\text{int}} (\gM, g)$ (as on dimensional reduction to $n$-dimensions it is this term which generates a potential for the components of $\gM_{MN}$, which become the scalar fields of the reduced theory).
For $d =2,3$ the generalised metric is reducible and does not appear solely in its $R_1$-valued representation. 
For $d=8$, there are additional terms involving the structure constants of the algebra, which we will see in section \ref{action8}.
\item the topological term, which is a Chern-Simons-like term constructed using only the gauge fields of the tensor hierarchy and their field strengths, schematically let us indicate this as:
\be
\mathcal{L}_{\text{top}} \sim \Aa \p \Fa \p \Fb + \dots
\ee
but the precise details depend on the dimension under consideration.
\end{itemize}

Each of the components of \refeqn{sexft} is separately invariant under generalised diffeomorphisms.
The full (pseudo-)action is fixed uniquely by further requiring invariance under \emph{external} diffeomorphisms $\xi^\mu(x,Y)$.
These are specified by requiring the metric-like degrees of freedom to transform as tensors, and for the gauge fields mirroring the transformations \refeqn{deltaAF}, \refeqn{deltaBH} and \refeqn{KKndiff}.
The resulting set of transformations are, in general:
\be
\begin{split}
\delta_\xi g_{\mu\nu} & = L_\xi g_{\mu\nu} \,, \qquad \delta_\xi \gM_{MN}  = L_\xi \gM_{MN} \,,\\
\delta_\xi \Aa_\mu{}^M & = \xi^\nu \Fa_{\nu\mu}{}^M  + \gM^{MN}  g_{\mu\nu} \partial_N \xi^\nu \,,\\
\Delta_\xi \Ab_{\mu\nu} & = \xi^\rho \Fb_{\rho \mu\nu} \,,\\
\Delta_\xi \Ac_{\mu\nu\rho} & = \xi^\sigma \Fc_{\sigma\mu\nu\rho} \,,\\
\end{split}
\ee
and so on.
Invariance under such transformations fixes all relative coefficients in the action: what is crucial here is the $Y$-dependence of the symmetry parameters $\xi^\mu(x,Y)$ which is precisely what ``stitches together'' the different terms in the action, and is what makes the difference between the ExFT action and the $n$-dimensional Kaluza-Klein reduced action one would obtain by truncating the dependence on these coordinates. For the full calculation details for the different cases, see the original papers \refref{Hohm:2013pua, Hohm:2013vpa, Hohm:2013uia,Hohm:2014fxa,Abzalov:2015ega,Musaev:2015ces,Hohm:2015xna,Berman:2015rcc}. We will also present an edited version of the calculation for $\Gfour$ in \sayappendix \ref{fixingsl5}.

The $\Edd$ exceptional field theories for $d=2$ to $d=8$, from smallest to largest, are:

\begin{itemize}
\item $\Gtwo$: The simplest exceptional group is just $\mathrm{SL}(2)$. This case is interesting as the ExFT is $(9+3)$-dimensional, and from the IIB point of view corresponds to adding two extra coordinates: this gives a natural correspondence with (and an action for) F-theory, where one introduces an auxiliary two-torus whose complex structure from the ExFT point of view is (part of) the generalised metric. 
The details of this theory were worked out in \refref{Berman:2015rcc} following the partial description of the tensor hierarchy in \refref{Wang:2015hca}.
\item $\Gthree$: The complete $\Gthree$ ExFT was constructed in \refref{Hohm:2015xna} (an earlier truncation of the theory had been given in \refref{Malek:2012pw}). 
\item $\Gfour$: The $\Gfour$ ExFT was pioneered in \refref{Berman:2010is,Berman:2011cg} in a form closely related to that which we used to introduce ExFT in section \ref{genKK_EFT}.
Subsequently the complete ExFT was written down in \refref{Musaev:2015ces}. 
\item $\Gfive$: The complete ExFT description was given in \refref{Abzalov:2015ega} (earlier work on the internal sector can be found in \refref{Berman:2011pe}).
\item $\Gsix$: This was formulated in \refref{Hohm:2013pua, Hohm:2013vpa}.
\item $\Gseven$: An early version of what we would now call ExFT was in fact introduced for the $\Gseven$ case in \refref{Hillmann:2009ci}.
A truncated ExFT version appear in \refref{Berman:2011jh} and the full ExFT description was finally given in \refref{Hohm:2013uia}.
\item $\Geight$: The full ExFT including the explanation of how generalised diffeomorphisms work appeared in \refref{Hohm:2014fxa} (an earlier truncation of the theory had been given in \refref{Godazgar:2013rja}).
\end{itemize}

There has also been remarkable progress in extending the construction to the infinite dimensional Lie algebras which appear for $d>8$:
\begin{itemize}
\item $\Gnine$: The exceptional field theory based on the affine Lie algebra $E_9$ has been constructed at the level of generalised diffeomorphisms and the potential involving soley internal derivatives in \refref{Bossard:2017aae, Bossard:2018utw}. 
\item $\Geleven$: The role of $E_{11}$ as an underlying symmetry of M-theory has been studied for many years by West and collaborators \cite{West:2001as,West:2003fc,Kleinschmidt:2003jf,West:2004st,West:2004kb,Riccioni:2007ni,West:2010ev,Tumanov:2016abm}. Here the (infinitely many) extended coordinates can be associated to a ``vector'' representation of $E_{11}$, and the dynamics follows from self-duality equation imposed in the description of the non-linear realisation of $E_{11}$ acting on the coordinate representation, with the latter containing the coordinates, and truncating to finite subalgebras reproduces features of  DFT and ExFT \cite{West:2010ev,Berman:2011jh,West:2011mm,West:2012qz,Tumanov:2015iea}. More recent work has shown to how include the (generalisation of the) full tensor hierarchy of ExFT, including constrained compensator fields, see \refref{Bossard:2017wxl, Bossard:2019ksx}.
\end{itemize}

One minor comment is that in the above we assumed that the timelike direction was part of the $n$-dimensional unextended spacetime. 
It is also possible to choose the $n$-dimensional space to be Euclidean, in which case the generalised metric will parametrise a coset $\Edd/H_d^*$ involving now a non-compact subgroup $H_d^*$ of different signature \cite{Hull:1998br}.
This has been discussed in \refref{Malek:2013sp, Blair:2013gqa,Berman:2019izh,Hohm:2019bba}.

The equations of motion that follow from the action, together with the self-duality constraint if required, then describe the dynamics of ExFT.
When obtaining the equation of motion of the generalised metric, the fact that it is valued in a coset must be taken into account. This means one cannot just consider arbitrary variations of the generalised metric. The variations must obey the coset structure and so we should consider only $\delta \gM_{MN}$ that obey \cite{Berkeley:2014nza, Berman:2019izh}
\be
\delta \gM_{MN} = P_{MN}{}^{KL} \delta \gM_{KL}
\ee
with the following projector:
\be
\begin{split}
P_{MN}{}^{KL} & = \gM_{MQ} \mathbb{P}^Q{}_N{}^{(K}{}_R \gM^{L) R} 
\\ &
= \frac{1}{\alpha} \left(
\delta^{(K}_M \delta^{L)}_N - \omega \gM_{MN} \gM^{KL} - \gM_{MQ} Y^{Q(K}{}_{RN} \gM^{L)R}
\right)\,.
\end{split} 
\ee
The field equation of the generalised metric is then 
\be
P_{MN}{}^{KL} \frac{\delta S}{\delta \gM^{KL}} = 0\,.
\ee
To conclude our development of the core ideas of ExFT, we will now describe in more detail the structure of the $\Gfour$, $\Gseven$ and $\Geight$ ExFTs.
For $\Gfour$, building on what we have seen in sections \ref{genKK_EFT} and \ref{exsl5th}, we will describe explicitly how to relate the theory to 11-dimensional supergravity.
For the other theories, we refer to the literature for the complete details.

\subsection{Example: the $\Gfour$ exceptional field theory}
\label{action5}

\subsubsection{The action} 

This exceptional field theory was pioneered in \refref{Berman:2010is, Berman:2011cg}. The details below follow the full construction from \refref{Musaev:2015ces} (and also \refref{Bosque:2016fpi}). We have already seen a truncated form of this theory in section \ref{genKK_EFT}, and described the tensor hierarchy fields in detail in section \ref{exsl5th}.
Recall we use $\fM, \fN =1,\dots 5$ to denote five-dimensional fundamental indices, while the $R_1$ representation of generalised vectors is the 10-dimensional antisymmetric representation, for which we write a 10-dimensional index $M = [\fM \fN]$ as an antisymmetric pair of five-dimensional indices.

The field content of the $\Gfour$ exceptional field theory is
\begin{equation}
\left\{ g_{\mu\nu}, \MM_{\fM \fN, \fP \fQ}, {\AAA_\mu}^{\fM \fN}, \Ab_{\mu\nu \fM}, \Ac_{\mu\nu \rho}{}^{\fM},\dots\right\} \, .
\end{equation}
Here we have the $7$-dimensional metric $g_{\mu\nu}$, the generalised metric $\gM_{\fM \fN, \fP\fQ}$ parametrizing the coset $\Gfour/\Hfour$, plus the tensor hierarchy fields: the one-form $\Aa_{\mu}{}^{\fM \fN}$, two-form $\Ab_{\mu\nu \fM}$, and $\Ac_{\mu\nu\rho}{}^{\fM}$.
The corresponding field strengths of the tensor hierarchy fields are $\Fa_{\mu\nu}{}^{\fM\fN}$, $\Fb_{\mu\nu\rho \fM}$ and $\Fc_{\mu\nu\rho\sigma}{}^{\fM}$, defined explicitly in \refeqn{5Fa}, \refeqn{5Fb} and \refeqn{5Fc}. The four-form $\Ad_{\mu\nu\rho\sigma \fM \fN}$ appears in the definition of $\Fc_{\mu\nu\rho\sigma}{}^{\fM}$, but drops out of the field equations. Hence this does not describe additional physical degrees of freedom.

All these fields are taken to depend on the 7-dimensional coordinates, $x^\mu$, and the 10-dimensional extended coordinates, $Y^M$.
The coordinate dependence of the fields on the latter is subject to the physical section condition which picks a subspace of the exceptional extended space. This section condition can be formulated in terms of the $\Gfour$ invariant $\epsilon^{\fM\fN\fP\fQ\fK}$
\be
\epsilon^{\fM \fN \fP \fQ \fK} \partial_{\fM \fN} \partial_{\fP \fQ} \Phi = 0 \, , \quad
\epsilon^{\fM \fN \fP \fQ \fK} \partial_{\fM \fN}\Phi \partial_{\fP \fQ} \Psi = 0 \, , 
\label{eq:section5}
\ee
where $\Phi$ and $\Psi$ denote any field or gauge parameter.

It is convenient to decompose the generalised metric as \cite{Duff:1990hn}
\be
\gM_{\fM\fN, \fP \fQ} = m_{\fM \fP} m_{\fQ \fN} - m_{\fM \fQ} m_{\fP \fN} \,,
\ee
where $m_{\fM \fN}$ is symmetric and has unit determinant. We denote its inverse by $m^{\fM \fN}$.

Then we can write the action \refeqn{sexft} specialised to $\Gfour$ as
\be
\begin{split}
S_{\Gfour} = \int \dd^7 x \, \dd Y \, \sqrt{|g|} \bigg(&
R_{\text{ext}}(g) 
+ \frac{1}{4} \D_\mu m^{\fM \fN} \D^\mu m_{\fM \fN}
\\ & 
- \frac{1}{8} m_{\fM \fP} m_{\fN \fQ} \Fa_{\mu\nu}{}^{\fM \fN} \Fa^{\mu\nu \fP \fQ} 
-  \frac{1}{12} m^{\fM \fN} \Fb_{\mu\nu\rho \fM} \Fb^{\mu\nu\rho}{}_{\fN}
\\ & + \mathcal{L}_{\text{int}} (m,g)
+ \sqrt{|g|}^{-1} \mathcal{L}_{\text{top}} 
\bigg)\,.
\end{split}
\label{sexft5}
\ee
In this case, the internal Lagrangian or potential \refeqn{Luniversal} can be expressed as
\be
\begin{split}
\mathcal{L}_{\text{int}}(m,g) &=
 \frac{1}{8} m^{\fM \fP} m^{\fN \fQ} \partial_{\fM \fN} m_{\fK \fL} \partial_{\fP \fQ} m^{\fK \fL} 
 +\frac{1}{2} m^{\fM \fP} m^{\fN \fQ} \partial_{\fM \fN} m^{\fK \fL} \partial_{\fK \fP} m_{\fQ \fL}
 \\ &  \qquad
 + \frac{1}{2} \partial_{\fM \fN} m^{\fM \fP} \partial_{\fP \fQ} m^{\fN \fQ} 
 + \frac{1}{2} m^{\fM \fP} \partial_{\fM \fN} m^{\fN \fQ} \partial_{\fP \fQ} \ln |g|
 \\ & \qquad
 + \frac{1}{8} m^{\fM \fP} m^{\fN \fQ} ( \partial_{\fM \fN} g^{\mu\nu} \partial_{\fP \fQ} g_{\mu\nu} 
 + \partial_{\fM \fN} \ln |g| \partial_{\fP \fQ} \ln |g| )\,.
\end{split}
\label{Lint5}
\ee
It can be explicitly checked that this is a scalar under generalised diffeomorphisms. (This is the direct generalisation of the miniature $\Gfour$ ExFT we wrote down in \refeqn{babyexft}, with the scalar $\Delta$ there replaced by the full 7-dimensional metric $g_{\mu\nu}$ here.) 

The topological term is best represented by writing it in terms of an integral over an auxiliary 8-dimensional spacetime:
\be
S_{\text{top}} 
= \kappa \int \dd^8 x\,\dd Y \epsilon^{\mu_1\dots\mu_8}\left(
\frac{1}{4} \hat\partial \Fc_{\mu_1 \dots \mu_4} \p \Fc_{\mu_5 \dots \mu_8}
- 4 \Fa_{\mu_1 \mu_2} \p ( \Fb_{\mu_3\mu_4\mu_5} \p \Fb_{\mu_6\mu_7\mu_8 })
\right)\,,
\ee
where the coefficients have been chosen so that its variation is a total derivative
\be
\begin{split}
\delta S_{\text{top}} = 2 \kappa
\int \dd^8 x\,\dd Y \epsilon^{\mu_1\dots\mu_8}\ \D_{\mu_1} \Big(&
- 4 \delta \Aa_{\mu_2} \p ( \Fb_{\mu_3\mu_4\mu_5} \p \Fb_{\mu_6\mu_7\mu_8}) 
\\ &
- 12 \Fa_{\mu_2 \mu_3} \p ( \Delta \Ab_{\mu_4\mu_5} \p \Fb_{\mu_6\mu_7\mu_8})
\\ & \qquad + ( \hat \partial \Delta \Ac_{\mu_2 \mu_3 \mu_4} ) \p \Fc_{\mu_5 \dots \mu_8}
\Big) \,,
\end{split}
\label{deltastopfive}
\ee
and coefficient $\kappa$ is determined to be
\be
\kappa = \frac{1}{12\cdot 4!} \,.
\ee
In \sayappendix \ref{fixingsl5}, we demonstrate this by requiring invariance under $7$-dimensional diffeomorphisms.

Kinetic terms are included for the generalised metric and the gauge fields $\Aa_\mu$ and $\Ab_{\mu\nu}$.
On the other hand, the field strength $\Fc_{\mu\nu\rho\sigma}$ of the gauge field $\Ac_{\mu\nu\rho}$ only appears in the topological term.
This gauge field also appears in the field strength $\Fb_{\mu\nu\rho}$.
We can find its equation of motion:
\be
\partial_{\fN \fM} \left(
 \sqrt{|g|} m^{\fM \fP} \Fb^{\mu\nu\rho}{}_{\fP} - 12 \kappa \epsilon^{\mu\nu\rho \sigma_1\dots \sigma_4} \Fc_{\sigma_1\dots\sigma_4}{}^{\fM} 
\right)=0 \,.
\label{5duality}
\ee
This implies a duality relation between the gauge field $\Ac_{\mu\nu\rho}$ and the gauge field $\Ab_{\mu\nu}$.

Finally, note that although the four-form $\Ad_{\mu\nu\rho\sigma \fM \fN}$ appears in the definition of $\Fc_{\mu\nu\rho\sigma}{}^{\fM}$, it does so in the form $\hat \partial \Ad_{\mu\nu\rho\sigma}$ and consequently drops out of the variation \refeqn{deltastopfive}, as $\Fc_{\mu\nu\rho\sigma}{}^{\fM}$ only appears accompanied by $\hat \partial$ (specifically, we can integrate by parts and use nilpotency of $\hat \partial$ to see this).

\subsubsection{Relationship to 11-dimensional supergravity}

Let's explain how to match the $\Gfour$ ExFT action \refeqn{sexft5} to that of 11-dimensional SUGRA.
We rely on the decomposition of the latter described in section \ref{11red}. 

The general idea (for any ExFT) is to identify the $d$-dimensional coordinates $y^i \subset Y^M$ which give a solution of the section condition.
This also allows one to identify the components $\Lambda^i$ corresponding to $d$-dimensional diffeomorphisms, which we can then use to start identifying the normal tensorial properties of the components of the ExFT fields. 

We then check the transformation of the $n$-dimensional metric, $\delta_\Lambda g_{\mu\nu} = \mathcal{L}_\Lambda g_{\mu\nu}$ with weight $-2\omega$. Being a scalar under generalised diffeomorphisms, the generalised Lie derivative here reduces directly to the ordinary Lie derivative on solving the section condition.
This allows us to identify the ExFT $d$-dimensional metric $g_{\mu\nu}$ directly with the metric denoted using the same symbol in the decomposition \refeqn{metricsplit}.

The generalised metric transforms as $\delta_\Lambda \gM_{MN} = \mathcal{L}_{\Lambda} \gM_{MN}$ with weight 0. 
By checking the transformations of its components, one can write down a parametrisation in terms of the $d$-dimensional components of the eleven-dimensional metric,  $\phi_{ij} \equiv \hat g_{ij}$, and the $d$-dimensional components of the three-form, $\Cdef_{ijk} \equiv \hat C_{ijk}$, along with components of the dual six-form.

For $\Gfour$ in particular, the physical coordinates are $y^i \equiv Y^{i5}$.
Then we identify the components of the generalised diffeomorphism parameters as $\Lambda^i \equiv \Lambda^{i5}$, generating four-dimensional diffeomorphisms, and $\Lambda^{ij} = \frac{1}{2} \epsilon^{ijkl} \lambda_{kl}$, generating gauge transformations of the three-form.
Using this identification, it can be checked that one can parametrise the metric exactly as in \refeqn{5gmfirstlook}, that is as
\be
m_{\cal{M} \cal{N}}
=\begin{pmatrix} 
\phi^{-2/5} \phi_{ij} & - \phi^{-2/5} \phi_{ik} C^k \\
- \phi^{-2/5} \phi_{jk} C^k & \phi^{3/5} + \phi^{-2/5} \phi_{kl} C^k C^l 
\end{pmatrix} \,,\quad
C^i \equiv \frac{1}{3!} \epsilon^{ijkl} \Cdef_{jkl} \,.
\label{5gm_again}
\ee
In order to check the below results, we also write for the readers' convenience the inverse metric:
\be
m^{\cal{M} \cal{N}}
=\begin{pmatrix} 
\phi^{2/5} \phi^{ij} + \phi^{-3/5} C^i C^j &  \phi^{-3/5} C^i \\
\phi^{-3/5}C^j & \phi^{-3/5}  
\end{pmatrix} \,.
\ee
For the tensor hierarchy gauge fields, we use the results of section \ref{exsl5th}:
\be
\Fa_{\mu\nu}{}^{\fM \fN}
= \begin{pmatrix}
{F}_{\mu\nu}{}^i \\
\frac{1}{2} \epsilon^{ijkl} ( \Fdef_{\mu\nu kl} - {F}_{\mu\nu}{}^p \Cdef_{klp} )
\end{pmatrix} \,,\quad
\mathcal{H}_{\mu\nu\rho i}= - \Fdef_{\mu\nu\rho i} \,,\quad
\mathcal{J}_{\mu\nu\rho\sigma}{}^5 = - \Fdef_{\mu\nu\rho\sigma} \,. 
\ee
We have not explicitly determined $\mathcal{H}_{\mu\nu\rho 5}$ or $\mathcal{J}_{\mu\nu\rho\sigma}{}^i$, but we will see that we can define these through the duality relation \refeqn{5duality}.

Now we want to insert these results into the $\Gfour$ ExFT action, \refeqn{sexft}, and compare with the decomposition \refeqn{Splitnd} of the original 11-dimensional action \refeqn{Seleven}.
Inserting \refeqn{5gm_again} into the $\Gfour$ ExFT action, a straightforward calculation shows that $\mathcal{L}_{\text{int}}$ agrees with the corresponding term \refeqn{11int} (with $n=7$) appearing in the decomposition of the eleven-dimensional supergravity action.
The terms $\mathcal{L}_{\text{kin}}$ of \refeqn{11kin_int} are reproduced from the analogous expression $\frac{1}{4} \D_\mu m_{\fM \fN} \D^\mu m^{\fM \fN}$ in the action \refeqn{sexft5}. 
As $\D_\mu g_{\nu \rho} = D_\mu g_{\nu\rho}$ on solving the section condition, the $n$-dimensional Ricci scalar $R_{\text{ext}}(g)$ of \refeqn{sexft5} immediately matches that appearing in \refeqn{Splitnd}.

Next we consider the kinetic terms for the field strength.
The $\mathcal{F}^2$ term is easily seen to match the corresponding expression \refeqn{11kin_1} appearing in \refeqn{Splitnd}.
The $\mathcal{H}^2$ term is:
\be
\begin{split}
-  &\frac{1}{12} m^{\fM \fN} \Fb_{\mu\nu\rho \fM} \Fb^{\mu\nu\rho}{}_{\fN}
= 
\\ & \quad - \frac{1}{12} \phi^{2/5} \phi^{ij} \Fdef_{\mu\nu\rho i} \Fdef^{\mu\nu\rho}{}_j 
 - \frac{1}{48} \phi^{-3/5} ( \mathcal{H}_{\mu\nu\rho 5} - C^i \Fdef_{\mu\nu\rho i} )(\mathcal{H}^{\mu\nu\rho}{}_5 - C^j \Fdef^{\mu\nu\rho}{}_j )
 \,.
 \label{5kin2immediate}
\end{split}
\ee
To deal with $\mathcal{H}_{\mu\nu\rho 5}$, we consult the duality relationship $m^{\fM \fN} \mathcal{H}_{\fN} \sim \star \Fc^{\fM}$ appearing under the derivative in \refeqn{5duality}.
Taking the $\fM = 5$ component, we have explicitly that:
\be
\phi^{-3/5} ( \mathcal{H}_{\mu\nu\rho 5} - C^i \Fdef_{\mu\nu\rho i} ) = -\frac{1}{4!} \frac{1}{\sqrt{|g|}} g_{\mu \mu^\prime} g_{\nu\nu^\prime} g_{\rho \rho^\prime} \epsilon^{\mu^\prime \nu^\prime\rho^\prime \sigma_1 \dots \sigma_4 } \Fdef_{\sigma_1 \dots \sigma_4} \,,
\label{detailed5duality}
\ee
which implies that on-shell
\be
 - \frac{1}{12} \phi^{-3/5} ( \mathcal{H}_{\mu\nu\rho 5} - C^i \Fdef_{\mu\nu\rho i} )(\mathcal{H}^{\mu\nu\rho}{}_5 - C^j \Fdef^{\mu\nu\rho}{}_j )
= -\frac{1}{12} \phi^{3/5} \Fdef_{\mu\nu\rho\sigma} \Fdef^{\mu\nu\rho\sigma}\,.
\ee
which is \refeqn{11kin_3}.
To be a bit more accurate, if we start with the decomposition of 11-dimensional supergravity described in \ref{11red} then we can dualise the three-form component $\Cdef_{\mu\nu\rho}$ by introducing a field $\tilde \Cdef_{\mu\nu}$ whose equation of motion is the Bianchi identity for $\Cdef_{\mu\nu\rho}$, i.e. add to the Lagrangian a term
\be
- 
\frac{1}{5!\,2!} \epsilon^{\mu_1 \dots \mu_7} \tilde C_{\mu_1 \mu_2} ( 5 D_{\mu_3} \Fdef_{\mu_4 \dots \mu_7} - 10 F_{\mu_3\mu_4}{}^i \Fdef_{\mu_5 \dots \mu_7i} ) \,.
\ee
Integrating by parts allows us to write all terms involving $\Fdef_{\mu\nu\rho\sigma}$ as:
\be
\begin{split}
- \frac{1}{48}\sqrt{|g|} \phi^{\tfrac{3}{5}} \left(
\Fdef^{\mu\nu\rho\sigma} \Fdef_{\mu\nu\rho\sigma}
-  \frac{1}{3}\sqrt{|g|}^{-1} \phi^{-\tfrac{3}{5}} \epsilon^{\mu_1 \dots \mu_7} \Fdef_{\mu_1 \dots \mu_4}
\tilde \Fdef_{\mu_5\mu_6\mu_7}
\right)
\end{split}
\label{dualisingF4}
\ee
with a field strength for the two-form:
\be
\tilde \Fdef_{\mu\nu\rho} = 3 D_{[\mu} \tilde \Cdef_{\nu \rho]} + \dots  \,,
\ee
where the dots denote extra terms arising from the Chern-Simons term.
Now we treat $\Fdef_4$ as an independent field.
Its equation of motion is algebraic:
\be
\Fdef_{\mu_1 \dots \mu_4} - \frac{1}{6} \sqrt{|g|}^{-1} \phi^{-\tfrac{3}{5}} \epsilon_{\mu_1\dots \mu_4}{}^{\mu_5 \dots \mu_7} \tilde \Fdef_{\mu\nu\rho}=0\,,
\label{5dualityorig}
\ee
and we can backsubstitute to eliminate $\Fdef_4$ from the action, giving the dual description in which $\tilde \Cdef_{\mu\nu}$ is dynamical and $\Cdef_{\mu\nu\rho}$ has been integrated out.
We can then identify $\mathcal{H}_{\mu\nu\rho 5} - C^i \Fdef_{\mu\nu\rho i} =  \tilde\Fdef_{\mu\nu\rho}$, in which case the duality relation \refeqn{5dualityorig} matches \refeqn{detailed5duality}, and we 
generate a kinetic term
\be
-\frac{1}{12} \sqrt{|g|} \phi^{-\tfrac{3}{5}} \tilde \Fdef_{\mu\nu\rho}\tilde \Fdef^{\mu\nu\rho}\,,
\ee
which corresponds directly to that coming from ExFT, \refeqn{5kin2immediate}.
Carrying out the full calculation in detail, as for instance for $\Gsix$ in \refref{Hohm:2013vpa}, shows that the ExFT topological term then matches that of supergravity. 

\subsection{Example: the $\Gseven$ exceptional field theory}
\label{actionE7}

Now we describe the $\Gseven$ exceptional field theory, written using four-dimensional coordinates $x^\mu$, and 56-dimensional extended coordinates $Y^M$ in the fundamental representation of $\Gseven$.
The 56-dimensional exceptional extended geometry was originally studied in \refref{Hull:2007zu, Pacheco:2008ps, Hillmann:2009ci, Coimbra:2011ky, Coimbra:2012af, Berman:2011jh}, and the full exceptional field theory that we describe below is based on \refref{Hohm:2013uia}.

The field content of the $\Gseven$ exceptional field theory is
\begin{equation}
\left\{ g_{\mu\nu}, \MM_{MN}, {\AAA_\mu}^M, \Ab_{\mu\nu \alpha}, \Ab_{\mu\nu M}\right\} \, .
\end{equation}
Here we have the usual $4$-dimensional metric $g_{\mu\nu}$, the metric $\gM_{MN}$ parametrizing the coset $\Gseven/\Hseven$, plus the tensor hierarchy fields: the one-form $\Aa_{\mu}{}^M$ and a pair of two-forms $\Ab_{\mu\nu \alpha}$ and $\Ab_{\mu\nu M}$. Here $\alpha= 1,\dots,133$ is an adjoint index.

The section condition is expressed in terms of the $\Gseven$ generators $(t_\alpha)^{MN}$ and the invariant symplectic form $\Omega_{MN}$ of $Sp(56)\supset \Gseven$ as
\begin{align}
(t_\alpha)^{MN}\partial_M\partial_N \Phi &= 0 \, , &
(t_\alpha)^{MN}\partial_M\Phi\partial_N \Psi &= 0 \, , &
\Omega^{MN}\partial_M\Phi\partial_N \Psi &= 0\,
\label{eq:section}
\end{align}
where $\Phi,\Psi$ stand for any field and gauge parameter. 
We further require, as explained in section \ref{compensator}, that the two-form $\Ab_{\mu\nu M}$ be similarly constrained, so (suppressing four-dimensional indices)
\begin{align}
(t_\alpha)^{MN}\Ab_M \Ab_N &= 0 \, , &
(t_\alpha)^{MN}\Ab_M \partial_N \Phi &= 0 \, , &
\Omega^{MN} \Ab_M \Ab_N  &= 0\,,&
\Omega^{MN} \Ab_M \partial_N \Phi  &= 0\,
\label{eq:sectionB}
\end{align}
The equations of motion describing the dynamics of the fields can be derived from the following pseudo-action
\begin{equation}
\begin{aligned}
S = \int \dd^4x \,\dd^{56}Y \sqrt{|g|}&\left(R_{\text{ext}}(g)
		+ \frac{1}{48}g^{\mu\nu}\DD_\mu\MM^{MN}\DD_\nu\MM_{MN} \right. \\
		&\quad \left.	- \frac{1}{8}\MM_{MN}\FF^{\mu\nu M}{\FF_{\mu\nu}}^N
		+ \mathcal{L}_{\text{int}} (\MM_{MN},g_{\mu\nu}) + \sqrt{|g|}^{-1}\LL_{\mathrm{top}}\right) \, ,
\end{aligned}
\label{eq:action}
\end{equation}
together with the twisted self-duality constraint for the 56 ExFT gauge vectors ${\AAA_\mu}^M$ 
\begin{equation}
{\FF_{\mu\nu}}^M =  \frac{1}{2}\sqrt{|g|} \epsilon_{\mu\nu\rho\sigma} \Omega^{MN} \MM_{NK} \FF^{\rho\sigma K} 
\label{eq:selfduality}
\end{equation}
which relates the 28 ``electric'' vectors to the 28 ``magnetic'' ones, with the field strength defined in \refeqn{e7fullF}.

The first term in \refeqn{eq:action} is the covariantized Einstein-Hilbert term, \refeqn{Rext}.
The second term is the kinetic term for the generalized metric $\MM_{MN}$.
The third term is a Yang-Mills-type kinetic term for the gauge vectors ${\AAA_\mu}^M$.
Note the coefficient of the Yang-Mills term is $1/8$ rather than $1/4$, reflecting the fact that we also impose a self-duality condition on the field strength.
The two-forms here do not have kinetic terms, and so represent dual degrees of freedom.
We then have the internal Lagrangian, or potential, which is as in \refeqn{Luniversal} with $\alpha= 12$.
The final term in the action \refeqn{actionE8} is the topological term, again most easily written as an action in one dimension higher on an auxiliary five-manifold with the physical space given by the boundary of the auxiliary manifold:
\begin{equation}
S_{\text{top}}= -\frac{1}{24} \int \dd^5x\,\dd^{56}Y \epsilon^{\mu \nu \rho \sigma \lambda} \FF_{\mu\nu}{}^M \DD_\rho \FF_{\sigma \lambda M} \, .
\end{equation}
This again varies into a total derivative, and the variation can be simply expressed in four-dimensions as 
\be
\begin{split}
\delta S_{\text{top}} = - \frac{1}{4} \int \dd^4x\,\dd^{56}Y \epsilon^{\mu \nu \rho \sigma} 
\Big(&
\delta \Aa_{\mu}{}^M \D_\nu \Fa_{\nu\rho\sigma M} 
\\ & \qquad
+
6 (t^\alpha)^{MN} \partial_N \Delta \Ab_{\rho\sigma \alpha} \Fa_{\mu\nu M} 
\\ & \qquad
+ \frac{1}{4} \Omega^{MN}\Delta\tilde\Ab_{\rho\sigma N} \Fa_{\mu\nu M} 
\Big)
 \, .
\end{split}
\ee
Observe that, making use of the variation \refeqn{e7fullFdelta}, it is easy to see the equations of motion of the two-forms merely give back (components of) the self-duality equation.

\subsection{Example: the $\Geight$ exceptional Field Theory}
\label{action8}

Finally, let's describe the exceptional field theory with $n=3$.
Here, the representation $R_1 = \mathbf{248}$ is the adjoint. 
It is necessary to modify the definition of generalised diffeomorphisms in order to achieve closure and to take into account the presence of would-be dual graviton degrees of freedom
entering the generalised metric. For all lower rank cases, these degrees of freedom appeared solely in the tensor hierarchy, where the mechanisms of section \ref{compensator} were necessary to deal with them.
In particular, there an extra constrained compensator gauge field was introduced, whose accompanying gauge symmetry served to shift away the dual graviton degrees of freedom.
Here, for $\Geight$, we have to introduce an extra shift symmetry acting alongside generalised diffeomorphisms themselves.
(An alternative approach in which the parameter of this extra shift symmetry is instead replaced by a connection term is described in \refref{Rosabal:2014rga,Cederwall:2015ica}.)

Following \refref{Hohm:2014fxa}, introduce the generators $T^M$ of $\Geight$ obeying  $[T^M, T^N] = - f^{MN}{}_K T^K$, and normalised such that the Killing form is
\begin{align}\label{eq:Killing}
\kappa^{MN} \equiv \frac{1}{60} \operatorname{Tr} (T^M T^N) = \frac{1}{60} f^{MP}{}_Q f^{NQ}{}_P\,.
\end{align}
This can be used to raise and lower adjoint indices.

The generalised Lie derivative acting on a generalised vector of weight $\lambda$ is
\begin{align}
\mathcal{L}_\Lambda V^M = \Lambda^K \partial_K V^M - 60 {\left( \mathbb{P}_{\mathbf{248}} \right)}^M{}_K{}^N{}_L \partial_N \Lambda^LV^K + \lambda(V) \partial_N \Lambda^N V^M\,,
\label{E8gld}
\end{align}
with the adjoint projector
\begin{align}
{\left( \mathbb{P}_{\mathbf{248}} \right)}^M{}_K{}^N{}_L = \frac{1}{60} f^M{}_{KP} f^{PN}{}_L\,.
\end{align}
This corresponds to the $Y$-tensor:
\be
Y^{MN}{}_{KL} = - f^M{}_{LP} f^{PN}{}_K + 2 \delta^{(M}_K \delta^{N)}_L \,.
\ee
However, this generalised Lie derivative does not give rise to a closed algebra, even with the section condition.
To compensate for the lack of closure, an extra gauge symmetry is introduced, under which generalised vectors transform as:
\be
\delta_\Sigma V^M = - \Sigma_L f^{LM}{}_N V^N \,,
\label{constrainedgaugetransformation}
\ee
where the gauge parameter $\Sigma_M$ is not an arbitrary covector but is constrained as part of the section condition of the $\Geight$ ExFT.
This section condition applies to any two quantities $F_M$, $F^\prime_M$ which are covariantly constrained meaning that their tensor product projected into the representations $\mathbf{1 \oplus 248 \oplus3875} \subset \mathbf{248 \otimes 248}$ vanishes, i.e.
\be
\kappa^{MN} F_M \otimes F^\prime_N  = 0\,,\quad
f^{MNK} F_N \otimes F^\prime_K  = 0\,,\quad
{(\mathbb{P}_{\mathbf{3875}})}^{KL}{}_{MN} F_K \otimes F^\prime_L = 0\,.
\label{E8section}
\ee
The covariantly constrained quantities of the $\Geight$ ExFT include the partial derivatives $\partial_M$, as usual, as well as the gauge parameters $\Sigma_M$. 

This section condition then ensures closure of the algebra of the combined action of generalised diffeomorphisms and constrained $\Sigma_M$ transformations, which we denote by
\be
\mathbb{L}_{( \Lambda, \Sigma )} \equiv \mathcal{L}_\Lambda + \delta_\Sigma\,.
\ee 
The covariant $3$-dimensional partial derivative is then defined as:
\be
\D_\mu \equiv \partial_\mu - \mathbb{L}_{( \Aa_\mu, \Ab_\mu)} \,,
\ee
introducing alongside the usual one-form $\Aa_\mu{}^M$ an additional one-form $\Ab_{\mu M}$ which is covariantly constrained by \refeqn{E8section}.

The field content of the $\Geight$ ExFT is then
\be
\{ g_{\mu\nu} , \gM_{MN}, \Aa_{\mu}{}^M , \Ab_{\mu M} \} \,.
\ee
To construct fully gauge invariant field strengths for the one-forms, it is as usual necessary to continue the tensor hierarchy, and introduce two-forms, namely $\Ac_{\mu\nu}$ in the trivial representation, $\Ac_{\mu\nu}{}^{MN}$ taken to lie in the $\mathbf{3875}$, and $\tilde\Ac_{\mu\nu M}{}^N$ which is covariantly constrained on its lower index.
The expressions for the field strengths are determined to be
\be
\begin{split}
\mathcal{F}_{\mu\nu}{}^M & = 2 \partial_{[\mu} \Aa_{\nu]}{}^M 
 - 2 \Aa_{[\mu}{}^N \partial_N \Aa_{\nu]}{}^M
 + 14 (\mathbb{P}_{3875})^{MN}{}_{KL} \Aa_{[\mu}{}^K \partial_N \Aa_{\nu]}{}^L
 \\ & \quad
 + \frac{1}{4} \Aa_{[\mu}{}^N \partial^M \Aa_{\nu]N} - \frac{1}{2} f^{MN}{}_P f^{P}{}_{KL} \Aa_{[\mu}{}^K \partial_N \Aa_{\nu]}{}^L  
 \\ & \quad + 14 (\mathbb{P}_{3875})^{MN}{}_{KL} \partial_N \Ac_{\mu\nu}{}^{KL} + \frac{1}{4} \partial^M \Ac_{\mu\nu} + 2 f^{MN}{}_K\tilde\Ac_{\mu\nu N}{}^K \,,\\
\mathcal{G}_{\mu\nu M} & = 2 \D_{[\mu} \Ab_{\nu] M} - f^N{}_{KL} \Aa_{[\mu}{}^K \partial_M \partial_N \Aa_{\nu]}{}^L 
 \\ & \quad + 2 \partial_N \tilde\Ac_{\mu \nu M}{}^N + 2 \partial_M \tilde\Ac_{\mu\nu N}{}^N \,.
\end{split} 
\ee
The action for the $\Geight$ ExFT constructed in \refref{Hohm:2014fxa} is
\be
\begin{split}
S & = \int \textrm{d}^{3}x \textrm{d}^{248} Y   \sqrt{|g|} \,\Bigg(
R(g) +  \frac{1}{240}  g^{\mu \nu} \D_\mu \mathcal{M}_{MN} \D_\nu \mathcal{M}^{MN}
+ \mathcal{L}_{\text{int}}( \mathcal{M},g ) + \frac{1}{\sqrt{|g|}} \mathcal{L}_{\text{top}}
\Bigg) \,.
\end{split}
\label{actionE8}
\ee
The internal Lagrangian, or potential, is now
\begin{align}
\begin{aligned}
	\mathcal{L}_{\text{int}} (\mathcal{M}, g) & = 
	+ \frac{1}{240} \mathcal{M}^{MN} \partial_M \mathcal{M}^{KL} \partial_N \mathcal{M}_{KL}
	 - \frac{1}{2} \mathcal{M}^{MN} \partial_M \mathcal{M}^{KL} \partial_L \mathcal{M}_{NK}\\
	& \qquad - \frac{1}{7200} f^{NQ}{}_P f^{MS}{}_R \mathcal{M}^{PK} \partial_M \mathcal{M}_{QK} \mathcal{M}^{RL} \partial_N \mathcal{M}_{SL}\\
	& \qquad + \frac{1}{2} \partial_M \ln |g| \partial_N \mathcal{M}^{MN} 
	+ \frac{1}{4} \mathcal{M}^{MN} \left(  \partial_M \ln |g| \partial_N \ln |g| + \partial_M g^{\mu \nu} \partial_N g_{\mu \nu} \right)\,,
\label{V8}
\end{aligned}
\end{align}
in which we see the explicit appearance of the structure constants of $\Geight$. 
Finally, the Chern-Simons term is:
\begin{align}
S_{\text{top}} = \frac{1}{4} \int \textrm{d}^4x \textrm{d}^{248} Y \left( \mathcal{F}^M \wedge \mathcal{G}_M - \frac{1}{2} f_{MN}{}^K \mathcal{F}^M \wedge \partial_K \mathcal{G}^{\mathcal{N}} \right)\,,
\end{align}
written again as an integral over an auxiliary four-dimensional spacetime, where $\wedge$ here denotes the standard four-dimensional wedge product.
One finds that its variation is
\be
\delta S_{\text{top}} = \frac{1}{2} \int \dd^3 x \,\dd^{248} Y \epsilon^{\mu\nu\rho} \left( \Fa_{\mu\nu}{}^M \delta \Ab_{\rho M} + ( \mathcal{G}_{\mu\nu M} - f_{MN}{}^K \partial_K \Fa_{\mu\nu}{}^N ) \delta A_{\rho}{}^M \right) \,.
\ee
Observe that the one-forms have no kinetic terms: their field equations imply that they are on-shell dual to the scalar degrees of freedom contained in the generalised metric. 
For instance, the equation of motion of $\Ab_{\mu M}$ implies that
\be
\delta \Ab_{\mu M} \left(  \frac{1}{2} \epsilon^{\mu\nu\rho} \Fa_{\nu\rho}{}^M - \frac{1}{60} |g|^{1/2}f^{MP}{}_N  \D^\mu \gM^{KN} \gM_{KP} \right) = 0 \,,
\ee
setting certain components of $\Fa_{\mu\nu}{}^M$ dual to components of the scalar currents $\D^\mu \gM^{KN} \gM_{KP}$.
Solving the section condition with $\partial_M =( \partial_i, 0 ,\dots ,0)$, with $i=1,\dots,8$ we will similarly have only $\Ab_{\mu i} \neq 0$.
Then the equations of motion of $\Ab_{\mu i}$ provide the duality relation between the Kaluza-Klein vectors $A_\mu{}^i$ and scalar dual graviton degrees of freedom.
However, the shift symmetry generated by $\Sigma_M$ allows us to gauge away these degrees of freedom; and they do not appear in the explicit decomposition of the action after imposing the section condition.

\clearpage
\section{Features of Exceptional Field Theory}
\label{features}

We have now described the basic elements of exceptional field theory.
In this section, we will discuss some further properties and applications.

\subsection{M-theory and type IIB solutions of the section constraint}
\label{ornotIIB}

Although we have motivated and in part constructed ExFT using 11-dimensional supergravity, this was simply for pedagogical reasons.
The theory really is a unifying formulation of both the 10- and 11-dimensional maximal supergravities.
In particular, one can obtain the ExFT description of type IIB theory by choosing a different solution of the section condition (while type IIA follows directly from the 11-dimensional supergravity, as usual).

The general statement is that solutions of the section condition other than $\partial_M = 0$ break $\Edd$.
Solutions that break $\Edd$ to some $\mathrm{SL}(D)$ subgroup lead to $D$-dimensional diffeomorphisms and hence to a standard spacetime picture.

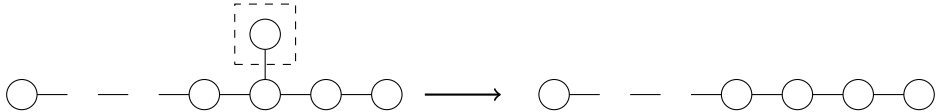
\begin{figure}[ht]
\centering
	\begin{tikzpicture}[scale=1]
	\begin{scope}[xshift=-1cm,scale=2]
	\foreach \x in {0,1,2,3,6}
	{
	\draw (-\x*0.4,0) circle (0.1cm);
	}
	\foreach \x in {0,...,5}
	\draw (-0.1-0.4*\x,0) -- (-0.3 -0.4*\x,0);

	\draw (-0.8,0.4) circle (0.1cm);
	\draw (-0.8,0.3) -- (-0.8,0.1); 
	\draw [dashed] (-1,0.2) rectangle (-0.6,0.6);
	\end{scope}
	
	\draw[thick,->] (-0.5,0) -- (0.5,0);
	
	\begin{scope}[xshift=6cm,scale=2]
	\foreach \x in {0,1,2,3,6}
	{
	\draw (-\x*0.4,0) circle (0.1cm);
	}
	\foreach \x in {0,...,5}
	\draw (-0.1-0.4*\x,0) -- (-0.3 -0.4*\x,0);

	\end{scope}
	\end{tikzpicture}
	\caption{Deleting the indicated node gives an $\mathrm{SL}(d)$ subalgebra of $\Edd$, corresponding to an M-theory solution of the section condition.}
	\label{Mthsection}
\end{figure}

One way to search for subalgebras is to start with the Dynkin diagram and delete nodes. This gives the Dynkin diagrams of subalgebras. 
For the Dynkin diagram of $\Edd$, there are two separate ways to obtain an $\mathrm{SL}(D)$ subalgebra.\footnote{The construction of generalised geometries by starting with $\mathrm{SL}(D)$ and adding nodes to its Dynkin diagram was studied in \refref{Strickland-Constable:2013xta} and a general $\mathrm{SL}(N)$ ExFT was constructed in \refref{Park:2014una}.}
The first is shown in figure \ref{Mthsection}. This gives an $\mathrm{SL}(d)$ subalgebra. This corresponds to the $d$-dimensional solution of the section condition that reduces exceptional field theory to 11-dimensional supergravity.
 
\begin{figure}[ht]
\centering
	\begin{tikzpicture}[scale=1]
	\begin{scope}[xshift=-1cm,scale=2]
	\foreach \x in {0,1,2,3,6}
	{
	\draw (-\x*0.4,0) circle (0.1cm);
	}
	\foreach \x in {0,...,5}
	\draw (-0.1-0.4*\x,0) -- (-0.3 -0.4*\x,0);

	\draw (-0.8,0.4) circle (0.1cm);
	\draw (-0.8,0.3) -- (-0.8,0.1); 
	
	\draw [dashed] (-0.6,-0.2) rectangle (-0.2,0.2);
	\end{scope}
	
	\draw[thick,->] (-0.5,0) -- (0.5,0);

	\begin{scope}[xshift=6cm,scale=2]
	\foreach \x in {0,2,3,6}
	{
	\draw (-\x*0.4,0) circle (0.1cm);
	}
	\foreach \x in {2,3,4,5}
	\draw (-0.1-0.4*\x,0) -- (-0.3 -0.4*\x,0);
	\draw (-0.8,0.4) circle (0.1cm);
	\draw (-0.8,0.3) -- (-0.8,0.1); 
	\end{scope}
	\end{tikzpicture}
	\caption{Deleting the indicated node gives an $\mathrm{SL}(d-1)\times \mathrm{SL}(2)$ subalgebra of $\Edd$, corresponding to a IIB solution of the section condition.}
	\label{IIBsection}
\end{figure}
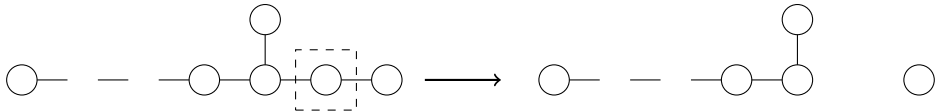

The second is shown in figure \ref{IIBsection}. This gives an $\mathrm{SL}(d-1) \times \mathrm{SL}(2)$ subalgebra. This corresponds to a $(d-1)$-dimensional solution of the section condition, that must therefore give a ten-dimensional supergravity.
The unbroken $\mathrm{SL}(2)$ means\cite{Hillmann:2009ci, Hohm:2013pua} that this must be the type IIB supergravity, with its $\mathrm{SL}(2)$ S-duality symmetry.
This IIB solution of the section condition is independent of the M-theory one, in that they cannot be rotated into each other by an $\Edd$ transformation.

In order to make explicit contact with type IIB supergravity, one decomposes all fields and gauge parameters under $\mathrm{SL}(d-1) \times \mathrm{SL}(2)$. 
Let's show this in more detail by focusing on the example of $\Gfour$, following \refref{Blair:2013gqa}.
The section condition is
\be
\partial_{[\fM \fN} \otimes \partial_{\fP \fQ]} = 0 \,,
\ee
where $\partial_{\fM\fN} = - \partial_{\fN \fM}$ and $\fM$ is a five-dimensional index.
The M-theory solution takes $\fM = (i,5)$, with $i=1,\dots,4$, and sets
\be
\partial_i \equiv \partial_{i5} \neq 0 \,,\quad
\partial_{ij} = 0 \,.
\label{5M}
\ee
This reduces to a type IIA solution of the section condition when there is a further isometry, $\partial_{45} =0$, for instance.

The IIB solutions takes $\fM=(i,\alpha)$, where now $i=1,2,3$ and $\alpha=4,5$ is an $\mathrm{SL}(2)$ doublet index.
The section condition is satisfied if
\be
\partial_{ij} \neq 0 \,,\quad
\partial_{i\alpha} = 0 = \partial_{\alpha \beta} \,.
\label{5IIB}
\ee
The three physical coordinates are therefore chosen to be $Y^{ij}$. It can be convenient to define $y_i = \frac{1}{2} \epsilon_{ijk} Y^{jk}$ so that the coordinates carry one index, albeit a lower one.

We therefore decompose a generalised diffeomorphism parameter as $\Lambda^{\fM \fN} = ( \epsilon^{ijk} \Lambda_k, \Lambda^{i\alpha}, \Lambda^{\alpha \beta})$.
The vector $\Lambda_i$ generates diffeomorphisms and $\Lambda^{i\alpha}$ is a pair of gauge transformations for two-forms. The final component $\Lambda^{\alpha \beta}$ is an $\mathrm{SL}(2)$ singlet and can be identified as $\Lambda^{\alpha \beta} = \frac{1}{6} \epsilon^{\alpha \beta} \epsilon_{ijk} \lambda^{ijk}$ where $\lambda^{ijk}$ is a gauge transformation parameter of the four-form. In fact the latter drops out of the generalised Lie derivative completely. 
Evaluating the transformation of the generalised metric for the choice \refeqn{5IIB}, one finds that it can be parametrised in terms of a three-dimensional metric, $g^{ij}$,\footnote{These are the three-dimensional components of the ten-dimensional Einstein frame metric. Note that in \refeqn{littlemCB} that $g \equiv |\det (g^{ij})|$.} an $\mathrm{SL}(2)$ doublet of two-forms, $C^{ij \alpha}$ and the axio-dilaton, represented as an $\mathrm{SL}(2)/\mathrm{SO}(2)$ valued matrix $\cH_{\alpha \beta}$, as:
\be
m_{\fM \fN} = 
g^{1/10} 
\begin{pmatrix}
g^{1/2}
g_{ij}  + \frac{1}{4} g^{-1/2} \cH_{\alpha \beta} \epsilon_{ikl} \epsilon_{jmn} C^{kl ,\alpha} C^{mn,\beta}  &
\frac{1}{2} g^{-1/2}\cH_{\beta \gamma}  \epsilon_{ikl} C^{kl,\gamma} \\
 \frac{1}{2} g^{-1/2}\cH_{\alpha \gamma} \epsilon_{jkl} C^{kl,\gamma} & g^{-1/2} \cH_{\alpha \beta} 
\end{pmatrix} \,,
\label{littlemCB}
\ee
with
\be
C^{ij,\alpha} = ( C^{ij} , B^{ij} ) \,,\quad
\cH_{\alpha \beta} = e^\Phi \begin{pmatrix}
1 & C_0 \\
C_0 & C_0^2 + e^{-2\Phi} 
\end{pmatrix}\,.
\ee
One can go to identify the components of the tensor hierarchy fields with the appropriate components of the type IIB fields.

Recall that type IIB contains a self-dual four-form field, and so does not have a true 10-dimensional action.
To make contact with ExFT we break 10-dimensional Lorentz invariance. This allows us to explicitly construct both ExFT actions and pseudo-actions describing the components of the four-form, with the duality relations arising either directly as equations of motion within ExFT, or being imposed by hand in the cases when $n$ is even, reproducing the self-duality constraint of type IIB.

In general, if we are working with the extended space of $\Edd$ exceptional field theory, we must choose a $d$ or $d-1$ dimensional submanifold of this space which we will call spacetime. The natural choice will be determined by the dependence of the fields on the coordinates (obeying the section condition). However if there are two isometries (note that two isometries are required as opposed previously one isometry for T-duality in DFT) then there is an ambiguity and one can choose whether you wish to be in M-theory frame or IIB frame. This reflects the statement that M-theory on a $T^2$ is dual to IIB on a circle. The dimension jumping is a result of the fact that a wrapped membrane mode requires two dimensions, and this is swapped with a single momentum mode in IIB.

We can also note here that one can further choose a partial solution of the section condition, which breaks $\Edd$ to its $SO(d-1,d-1)$ subgroup (by deleting the rightmost node in the Dynkin diagram). This allows one to directly reduce from ExFT to DFT, as in \refref{Thompson:2011uw}.

\subsection{Generalised Scherk-Schwarz and gauged supergravities} 
\label{genSS}

We learn from the above discussion that exceptional field theory contains the 11-dimensional maximal supergravity and both the type II 10-dimensional maximal supergravities under one extended umbrella.
These theories are obtained on solving the section condition by keeping as physical only $d$ or $d-1$ coordinates.

Exceptional field theory therefore succeeds at a Kaluza-Klein-esque unification of the 11- and 10-dimensional maximal supergravities.
This unification achieves the further goal of providing a direct link with the $\Edd$ symmetries that appear on dimensional reduction.
This answers the long-standing question as to what extent one these symmetries are present in the original higher-dimensional theories, and the exceptional geometry hints at the structure of string and M-theory geometries beyond supergravity.

More practically, it is also a useful toolkit.
In particular, the reorganisation of the supergravity fields based around $\Edd$ is naturally adapted for dimensional reduction in general, not just the truncation to the $n$-dimensional maximal supergravity with global $\Edd$ duality symmetry obtained by setting $\partial_M = 0$ in the ExFT action. 
This means that one can consider direct reductions from ExFT to $n$-dimensions, and then later re-interpret these in terms of reductions from 10- or 11-dimensions.
This gives a very useful way to study dimensional reductions and consistent truncations involving gauged supergravity. 

The dimensional reduction procedure that is used is that of generalised Scherk-Schwarz reduction.
A Scherk-Schwarz reduction \cite{Scherk:1979zr} generalises the Kaluza-Klein procedure to allow us to reduce higher-dimensional theories on more complicated internal manifolds than flat tori, retaining information about the internal coordinate dependence via a factorisation ansatz on all fields.
The generalised Scherk-Schwarz of ExFT not only contains and generalises this procedure, but allows us to relax the section condition in a controlled manner, by retaining information about \emph{dual} coordinate dependence via a similar factorisation ansatz.\footnote{Generalised Scherk-Schwarz reductions and their applications to non-geometry were pioneered first for DFT \cite{Geissbuhler:2011mx, Aldazabal:2011nj,Grana:2012rr,Dibitetto:2012rk}.}

In this section, we will describe how these ideas work in ExFT.

\subsubsection{Scherk-Schwarz reductions}
\label{ssred}

In a theory with a metric and a $p$-form field, the Scherk-Schwarz ansatz for the $d$-dimensional ``internal'' components is:
\be
\begin{split}
g_{ij}(x,y) &= u^a{}_i (y) u^b{}_j (y) \bar g_{ab}(x) \,,\\
C_{i_1 \dots i_p}(x,y) & = u^{a_1}{}_{i_1}(y) \dots u^{a_p}{}_{i_p}  (y)  \bar C_{a_1\dots a_p }(x) +  c_{i_1 \dots i_p}(y) \,,
\end{split}
\label{ssg}
\ee
where the fields $\bar g_{ab}$ and $\bar C_{a_1\dots a_p}$ depend only on the coordinates $x^\mu$ of the $n$-dimensional ``external'' spacetime, and the dependence on the $d$-dimensional coordinates $y^i$ enters through the ``twists'' $u^a{}_i(y)$ and $c_{i_1\dots i_p}(y)$.
Here $i=1,\dots,d$ are the usual spacetime indices, and we use the separate indices $a=1,\dots,d$ for quantities in the effective theory resulting from the reduction. Mathematically, the twist $u^a{}_i$ can be viewed as a parallelisable frame for the internal manifold.

In carrying out the dimensional reduction, derivatives with respect to the internal coordinates now do not truncate immediately out of the action.
Instead, they hit these twists giving rise to ``fluxes'' carrying geometric information about the background on which we are reducing (this background is obtained by setting $\bar{g}_{ab} = \delta_{ab}, \bar C_{a_1 \dots a_p} = 0$).
In order that the reduced theory be independent of $y$, these fluxes need to be constant. 
The usual Kaluza-Klein reduction is the special case $u^a{}_i (y) = \delta^a{}_i$, $c_{i_1 \dots i_p}(y) = 0$.

An obvious example of a background to which we can apply this procedure is a $(p+1)$-dimensional torus with $u^a{}_i = \delta^a_i$ and constant flux $h_{a_1 \dots a_{p+1}} \equiv \delta_{a_1}^{i_1} \dots \delta_{a_{p+1}}^{i_{p+1}}(p+1) \partial_{[i_1} c_{i_2 \dots i_{p+1}]}$.

A class of backgrounds with a non-trivial ``geometric'' twist $u^a{}_i(y)$ are group manifolds (e.g. the three-sphere which is $\mathrm{SU}(2)$), where the twists $u^a{}_i$ can be taken to be the left-invariant one-forms obeying $du^a = \tfrac{1}{2} f_{bc}{}^a u^b \wedge u^c$, with the $f_{bc}{}^a$ corresponding to the structure constants of the group.
These left-invariant one-forms give a global frame, or parallelisation, describing the group manifold (and we see that the indices $a,b,\dots$ are exactly tangent space i.e. Lie algebra indices).
In this case the dual vector fields $u^i{}_a$ obey 
\be
L_{u_a} u_b = - f_{ab}{}^c u_c\,,
\label{groupmanifold}
\ee
where $L_{u_a}$ denotes the usual Lie derivative with respect to $u_a$.
The structure constants $f_{ab}{}^c$ may be viewed as a sort of ``geometric flux''.

The $n$-dimensional theory resulting from this reduction procedure can be interpreted as a deformation of the ordinary Kaluza-Klein reduction.
The fluxes $f_{ab}{}^c$ and $h_{a_1 \dots i_{a+1}}$ have the effect of ``gauging'' the symmetry transformations of the $n$-dimensional fields.
For instance, the ansatz for the Kaluza-Klein vector 
\be
A_{\mu}{}^{i}(x,y) = u^i{}_a(y)\bar A_{\mu}{}^a (x)\,,
\ee
means that the field $\bar A_\mu{}^a$ becomes a non-Abelian gauge field: factorising $d$-dimensional diffeomorphisms as $\Lambda^i(x,y) = u^i{}_a (y) \bar \Lambda^a(x)$ leads to
\be
\delta \bar A_\mu{}^a = \partial_\mu \Lambda^a + f_{bc}{}^a \bar A_\mu{}^b\bar \Lambda^c \,.
\ee
In addition, a scalar potential is generated for the scalar moduli fields $\bar g_{ab}$ and $\bar c_{a_1 \dots a_p}$
\be
V = - \frac{1}{4} \bar g^{ab} \bar g^{cd} \bar g_{ef} f_{ac}{}^e f_{bd}{}^f - \frac{1}{2} \bar g^{ab} f_{ca}{}^d f_{db}{}^c + \dots \,.
\ee
Phenomenologically, this is a desirable feature as otherwise we will have many massless scalars.

\subsubsection{Gauged ExFT}
\label{gaugedExFT}

The generalised Scherk-Schwarz ansatz in ExFT assumes a factorisation of the fields generalising \refeqn{ssg} in an $\Edd$ covariant way \cite{Berman:2012uy,Aldazabal:2013mya,Musaev:2013rq,Hohm:2014qga,Ciceri:2016dmd}.
This means we introduce an $\Edd$ valued twist matrix $U^A{}_M(Y)$, of unit determinant, its inverse $U^M{}_A(Y)$, and also a function $\rho(Y)$ of weight $\omega$ under generalised diffeomorphisms. 
These are functions of the extended coordinates only. 
Then the factorisation ansatz for the ExFT fields, taking into account their $\Edd$ representations and generalised diffeomorphism weights, is:
\be
\begin{split} 
\gM_{MN}(x,Y) & = U^A{}_M(Y) U^B{}_N(Y) \bar\gM_{AB}(x,Y) \,,\\
g_{\mu\nu}(x,Y) & = \rho^{-2} (Y) \bar{g}_{\mu\nu} (x,Y)\,,\\
\Aa_\mu{}^M(x,Y) & = \rho^{-1} (Y) U^M{}_A(Y) \bar{\Aa}_{\mu}{}^A (x,Y) \,,\\
\end{split}
\label{SSExFT}
\ee
plus the obvious extensions to the higher rank tensor hierarchy fields.\footnote{Observe that we do not introduce $p$-form twists for the latter, with a view to preserving Lorentz invariance in reductions to an $n$-dimensional theory: in principle we can incorporate such twists by moving to a bigger ExFT and sticking with the ansatz \refeqn{SSExFT}.}
Now $M,N,\dots$ label the usual $R_1$ representation of ExFT while $A,B,\dots$ label the $R_1$ representation of the descendant ``gauged'' ExFT which will result from imposing the generalised Scherk-Schwarz conditions below.  

The symmetry parameters are factorised similarly, in particular for generalised diffeomorphisms we have
\be
\Lambda^M (x,Y)  = \rho^{-1}(Y) U^M{}_A (Y) \bar\Lambda^A(x,Y)\,.
\ee
The structure of ExFT implies that $U^A{}_M$ and $\rho^{-1}$ only appear in the symmetry transformations and dynamics in specific combinations.
We can call these combinations  ``generalised fluxes''.
They can be defined geometrically using the generalised Lie derivative.
Let $\tilde U^M{}_A \equiv \rho^{-1} U^M{}_A$, which has weight $-\omega$ and so can be used to take generalised Lie derivatives.
We define
\be
\mathcal{L}_{\tilde U_A} \rho^{-1} \equiv \theta_A \rho^{-1} \,,
\ee
which gives the so-called ``trombone'' gauging 
\be
\theta_A \equiv\frac{1}{n-2} \rho^{-1} \left( (n-1) U^N{}_A \partial_N \ln \rho^{-1} + \partial_N U^N{}_A  \right) \,.
\label{fluxtrombone} 
\ee
The twist matrices themselves form an algebra under generalised Lie derivatives:
\be
\mathcal{L}_{ \tilde U_A } \tilde U^M{}_B \equiv - \tau_{AB}{}^C \tilde U^M{}_C \,,
\label{genpar}
\ee
with
\be
\tau_{AB}{}^C = \Theta_{AB}{}^C + \frac{n-2}{n-1} \left( 2 \delta^C_{[A} \theta_{B]} - Y^{CD}{}_{AB} \theta_D \right) \,,
\ee
in which we define $\Theta_{AB}{}^C$ given by
\be
\begin{split} 
\Theta_{AB}{}^C= 
\rho^{-1} \Big( &
 2 U^C{}_M U^N{}_{[B} \partial_{|N|} U^M{}_{A]} - Y^{MN}{}_{PQ} U^C{}_M \partial_N U^P{}_A U^Q{}_B 
\\ & - \frac{1}{n-1} \left(
 2 \delta^C_{[A} \partial_{|N|} U^N{}_{B]} - Y^{CD}{}_{AB} \partial_N U^N{}_D 
 \right) 
\Big)\,.
\end{split} 
\label{fluxembedding} 
\ee
For an arbitrary background, $\theta_A$ and $\Theta_{AB}{}^C$ will be non-constant.
If however we can find twists $U^A{}_M$ and $\rho$ such that the generalised fluxes are constant, then we can use the ansatz \refeqn{SSExFT} to formulate a variant of ExFT in which the constant fluxes $\theta_A$ and $\Theta_{AB}{}^C$ deform the structure of the theory into a \emph{gauged} ExFT.
This amounts to performing a generalised Scherk-Schwarz reduction of exceptional field theory.

In this case, the condition \refeqn{genpar} shows the twists form a genuine algebra.
In general, this will not be a Lie algebra, but a Leibniz algebra. 

Geometrically, the $\tilde U^A{}_M$ can be thought of as a choice of generalised frame fields (for the background obtained from \refeqn{SSExFT} on setting $\gM_{AB} \rightarrow \delta_{AB}$). If they obey \refeqn{genpar} with constant generalised structure constants $\tau_{AB}{}^C$, and are everywhere linearly independent and non-vanishing, then they provide a \emph{generalised parallelisation} of the underlying generalised geometry, which is a useful way of thinking about this construction \cite{Grana:2008yw} (for instance, all spheres can be made generalised parallelisable using an appropriate generalised geometry \cite{Lee:2014mla}).

In a background admitting the factorisation \refeqn{SSExFT} with constant generalised fluxes, the algebra of generalised Lie derivatives can be closed without using the section condition.
This is because the twists $U^A{}_M$ and $\rho^{-1}$ may depend on dual coordinates as long as the generalised fluxes \refeqn{fluxtrombone} and \refeqn{fluxembedding} do not.
Closure of the algebra of generalised diffeomorphisms now implies a weaker set of constraints.

To describe these, we should firstly think of the gauged ExFT obtained after factorising out the twists as being described in terms of $\Edd$-valued quantities carrying the indices $A,B,\dots$.
We define partial derivatives $\partial_A \equiv \delta_A^M \partial_M$, such that for $\bar\Psi(x,Y)$ any field or symmetry parameter obtained after this factorisation, we require
\be
\rho^{-1} U^N{}_{A} \partial_N\bar\Psi(x,Y) =  \partial_{A} \bar\Psi (x,Y) \,,
\label{twistonpartial}
\ee
i.e. the twist is trivial in the directions on which $\bar\Psi$ depends (the presence of the $\rho^{-1}$ here is due to the fact that the derivatives $\partial_M$ must be taken to have non-trivial weight under generalised diffeomorphisms). 
We can use \refeqn{twistonpartial} and the definition \refeqn{genpar} to show that generalised diffeomorphisms of a generalised vector $V^M = \tilde U^M{}_A \bar V^A$ of weight $-\omega$ can be written as
\be
\delta_\Lambda V^M = \tilde U^M{}_A ( \bar{\mathcal{L}}_{\bar{\Lambda}} \bar V^A - \tau_{BC}{}^A \bar\Lambda^B \bar V^C )\,.
\ee 
leading to the definition
\be
\delta_{\bar \Lambda} \bar{V}^A =  \bar{\mathcal{L}}_{\bar{\Lambda}} \bar V^A - \tau_{BC}{}^A \bar\Lambda^B \bar V^C \,,
\ee
of gauged generalised diffeomorphisms in the gauged ExFT.
Here $\bar{\mathcal{L}}_{\bar{\Lambda}}$ denotes the usual generalised Lie derivative expressed in terms of the barred quantities and the indices $A,B,\dots$.
Closure now follows from the conditions
\be
\tau_{AB}{}^C \partial_C  \bar\Psi = 0 \,,\quad
Y^{MN}{}_{PQ} \partial_M U^Q{}_{A} \partial_N \bar \Psi = 0 \,,
\label{gaugedconstraints}
\ee
plus the ``quadratic constraint''
\be
2 \tau_{[C|B}{}^E \tau_{|D]E}{}^{A} + \tau_{CD}{}^E\tau_{EB}{}^A = 0 \,,
\label{QC}
\ee
and the requirement that the section condition hold in terms of the coordinate dependence of the barred quantities, i.e. $Y^{AB}{}_{CD} \partial_{A}\otimes \partial_{B} = 0$.

The generalised metric $\bar{\gM}_{AB}$ transforms under generalised diffeomorphisms of the factorised theory as:
\be
\delta_{\bar\Lambda} \bar{\gM}_{AB}
= \bar{\mathcal{L}}_{\bar{\Lambda}} \bar{\gM}_{AB} + 2 \bar \Lambda^C \Theta_{C(A}{}^D \bar{\gM}_{B) D}
+ 2 \alpha \frac{(n-2)}{n-1} \bar{\gM}_{D(A} \mathbb{P}^D{}_{B)}{}^{E}{}_F\bar \Lambda^E \theta_F \,.
\label{gaugeddeltaM}
\ee
The one-form $\bar{\Aa}_{\mu}{}^A$ transforms as
\be
\delta_{\bar \Lambda} \bar{\Aa}_{\mu}{}^A = \bar{\D}_\mu \bar\Lambda^A + \tau_{BC}{}^A \bar\Aa_{\mu}{}^B \bar\Lambda^C\,,
\label{gaugeddeltaA}
\ee
and its field strength is similarly modified
\be
\bar{\Fa}_{\mu\nu}{}^{\bar A} = 2 \bar{D}_{[\mu} \bar \Aa_{\nu]}{}^A + \tau_{[BC]}{}^A \bar \Aa_{\mu}{}^B \bar \Aa_{\nu}{}^C + \dots \,.
\label{gaugedF}
\ee
(We omit the terms involving the $B$-field.)

The dynamics of the gauged ExFT now follow from those of ExFT on making the above replacements.
When $\theta_A \neq 0$, these dynamics can solely be expressed in terms of the field equations, and there is no underlying action principle.
This is a general feature of having a non-zero trombone gauging (what is being gauged here is an overall scaling of the action, which is a symmetry of the equations of motion but not the action itself).
When $\theta_A = 0$ on the other hand we can work with the action directly.
In this case, we find that alongside the modification of $\D_\mu$ and the tensor hierarchy field strengths, we also have
\be
\mathcal{L}_{\text{int}}(\gM , g) \rightarrow
\mathcal{L}_{\text{int}}(\bar{\gM}, \bar g) + \mathcal{L}_{\text{cross}}(\bar{\gM}, \Theta) + \mathcal{L}_{\text{pot}}(\bar{\gM}, \Theta) 
\ee
with
\be
 \mathcal{L}_{\text{cross}}(\bar{\gM}, \Theta)  = - \frac{1}{\alpha}\bar{\gM}^{AB} \bar{\gM}^{CD} \Theta_{AC}{}^E \partial_{B} \bar{\gM}_{DE} \,,
\label{SScross}
\ee
\be
\mathcal{L}_{\text{pot}}(\bar{\gM}, \Theta ) = -\frac{1}{2\alpha} \left( a \bar{\gM}^{AB} \bar{\gM}^{CD} \bar{\gM}_{EF} \Theta_{AC}{}^E \Theta_{BD}{}^F +\bar{\gM}^{AB} \Theta_{AC}{}^D \Theta_{BD}{}^C \right) \,,
\label{SSV}
\ee
where the coefficients $(\alpha,a)$ are given by $(12, \tfrac{1}{7})$ for $\Gseven$, $(6,\tfrac{1}{5})$ for $\Gsix$, $(4,\frac{1}{4})$ for $\Gfive$. 
Here the coefficient $\alpha$ appears in the relationship between the Y-tensor and the adjoint projector, \refeqn{yfrompadj}, while the coefficient $a$ is related to the representation theory of $\Theta$ -- it is the $a_1$ of appendix $C$ of \refref{Aldazabal:2013mya}.
For $\Gfour$, the embedding tensor $\Theta$ contains two irreducible representations rather than one, so the potential has a slightly different form.  
In general, the $\Gfour$ embedding tensor $\Theta_{\fA \fB , \fC \fD}{}^{\fE \fF}$ can be written as
\be
\Theta_{\fA \fB , \fC \fD}{}^{\fE \fF} 
= 4 \Theta_{\fA \fB [\fC}{}^{[\fE} \delta_{\fD]}^{\fF]} \,,
\ee
and one has a further decomposition into irreps 
\be
\Theta_{\fA\fB\fC}{}^{\fD} = Z_{\fA\fB\fC}{}^{\fD} + \tfrac12 \delta_{[\fA}^\fD Y_{\fB]\fC} \,,
\ee
where $Z_{\fA\fB\fC}{}^{\fD} = Z_{[\fA\fB\fC]}{}^{\fD}$ (in the $\mathbf{40}$ of $\Gfour$) and $Y_{\fA\fB} = Y_{(\fA\fB)}$ (in the $\mathbf{15}$ of $\Gfour$).
In terms of these quantities, and the five-by-five generalised metric $\bar m_{\fA\fB}$, one has
\be
\begin{split} 
-\mathcal{L}_{\text{pot}}(\bar{m}, \Theta ) & =
\tfrac{1}{32} \left(
 2\bar m^{\fA\fB}\bar m^{\fC\fD} Y_{\fA\fC} Y_{\fB\fD} - ( \bar m^{\fA\fB} Y_{\fA\fB})^2 
\right)
\\ & \qquad
+ \tfrac{1}{12} Z_{\fA\fB\fC}{}^{\fD}Z_{\fA'\fB'\fC'}{}^{\fD'}
\bar m^{\fA\fA'} \bar m^{\fB\fB'}\bar m^{\fC\fC'}\bar m_{\fD\fD'} 
\\ & \qquad
+ \tfrac14 Z_{\fA\fB\fC}{}^{\fD} Z_{\fA'\fB'\fD}{}^{\fC}\bar m^{\fA\fA'}\bar m^{\fB\fB'}
 \,.
\end{split}
\label{SSVfive}
\ee
The above construction extends the utility of exceptional field theory and provides the natural way to describe deformations and gaugings of supergravity in $\Edd$ covariant fashion.
We now describe some applications and examples.

\subsubsection{Gauged supergravities and consistent truncations}

We can truncate to $n$-dimensional supergravity by using the ansatz \refeqn{SSExFT} with the fields $\bar\gM_{AB}$, $\bar{g}_{\mu\nu}, \dots$ and gauge parameters $\bar\Lambda^A,\dots$ independent of the extended coordinates.
Then, setting $\partial_A =0$ and $\bar{\D}_{\mu} \rightarrow \partial_\mu$ in our gauged ExFT, we obtain an $n$-dimensional \emph{gauged supergravity}.
In an $n$-dimensional gauged supergravity, a subgroup of the global $\Edd$ duality group is promoted to a local (hence gauged) symmetry. 
We can see how this appears from ExFT from the expression \refeqn{gaugeddeltaA} for the gauge transformations of the one-form, which is clearly of a non-abelian form; furthermore the potential term \refeqn{SSV} coincides with the scalar potential of maximal gauged supergravity. 
We then identify $\Theta_{AB}{}^C$ with the \emph{embedding tensor} of gauged supergravity, and $\theta_A$ with the so-called trombone gauging.
The embedding tensor describes the embedding of the gauged subgroup into the adjoint of $\Edd$, i.e. we can write $\Theta_{AB}{}^C = \Theta_{A}{}^\alpha (t_\alpha)_B{}^C$ in terms of the adjoint generators $(t_\alpha)_B{}^C$.
Closure of the gauge algebra requires that the embedding tensor must obey the quadratic constraint \refeqn{QC}, while supersymmetry requires that it also obey a linear constraint, namely that $\Theta_{AB}{}^C$ is valued in certain representations in the product $\mathbf{\overline{R}_1} \otimes \mathbf{adj}$.
The structure of the generalised Lie derivative of ExFT turns out to guarantee the latter constraint, always producing the embedding tensor in the correct representations.
This is another signal that the bosonic ExFT secretly knows about supersymmetry.

This approach allows one to very efficiently describe reductions to gauged supergravities in exceptional field theory.
This turns out to have many advantages.
As we re-order the supergravity fields in a manner directly adapted to dimensional reductions, it is often much simpler to express results on reductions and uplifts in terms of ExFT.
For instance, one can use ExFT to prove that truncations about certain backgrounds are \emph{consistent} (i.e. that any solution of the lower-dimensional theory can be uplifted to a solution of the higher-dimensional one) by making use of the consistency conditions of the generalised Scherk-Schwarz procedure rather than working with much more cumbersome formulae in supergravity.
A sampling of such work, in the context of exceptional generalised geometry or ExFT, includes \refref{Lee:2014mla,Baron:2014yua, Hohm:2014qga,Baguet:2015sma,Baguet:2015iou, Baron:2014bya,Inverso:2017lrz,Malek:2015hma, Malek:2016bpu, Malek:2017njj, Malek:2018zcz, Malek:2019ucd,Cassani:2019vcl}.
As examples of the power of ExFT in this context we can mention the first proof of the consistency of the truncation of type IIB on AdS${}_5 \times S^5$ \cite{Lee:2014mla, Baguet:2015sma}, and recent progress in obtaining a general method to compute Kaluza-Klein mass spectra \cite{Malek:2019eaz}. 

\subsubsection{Massive IIA}

As well as gauged supergravities in dimensions lower than ten, we can also use the generalised Scherk-Schwarz procedure to engineer deformations of the ten-dimensional supergravities themselves.
The surprising feature here is that in order to do so, it is necessary to make explicit use of dual coordinate dependence in the twist matrices. 
In to explicitly demonstrate this feature, we will describe two examples.
The first is the massive IIA theory, also known as the Romans supergravity \cite{Romans:1985tz}, which was studied in DFT in \refref{Hohm:2011cp} and in ExFT in \refref{Ciceri:2016dmd,Cassani:2016ncu}.

The massive IIA theory is a (supersymmetry preserving) deformation of the ten-dimensional type IIA supergravity via a deformation parameter $m$ known as the Romans mass, which can be interpreted as the Hodge dual of a 10-form field strength (and so appears when D8 branes, which couple electrically to an RR 9-form, are present \cite{Polchinski:1995mt}).
The Romans deformation modifies the gauge transformations, such that under a gauge transformation of the NSNS two-form we have
\be
\delta B_2 = d \lambda_1 \,,\quad \delta C_1 = -m \lambda_1 \,,\quad \delta C_3 = - m B_2 \wedge \lambda_1 \,,
\label{massivegauge}
\ee
leading to modified field strengths
\be
F_2 = d C_1 + m B_2 \,,\quad  
F_4 = d C_3 - H_3 \wedge C_1 + \frac{m}{2} B_2 \wedge B_2 \,.
\label{massiveF}
\ee
The Romans mass also appears in the action as a cosmological constant-like term
\be
S_{\text{Romans}} \supset - \frac{1}{2} \int \dd^{10} x \, \sqrt{|\hat g|} m^2 \,,
\label{massiveS}
\ee
where here $\hat g$ denotes the 10-dimensional string frame metric.
This cannot be obtained from a dimensional reduction of the 11-dimensional maximal supergravity (though it can be obtained from a non-covariant modification of the latter involving a Killing vector \cite{Bergshoeff:1997ak}).

In both double and exceptional field theory, massive IIA can be incorporated by choosing a solution of the section condition in which the fields depend on the coordinates of a IIA section and on dual coordinates associated to a type IIB section.
This dual coordinate dependence must be of a particular form such that the theory still makes sense: this form is exactly that of a generalised Scherk-Schwarz factorisation.

Let us see how this works explicitly, using our favourite example of the $\Gfour$ ExFT.
We want to discuss now both IIB and IIA solutions of the section condition.
We therefore write the five-dimensional index $\fM = (i,4,5)$ with $i$ a three-dimensional coordinate.
The extended coordinates used in the $\Gfour$ ExFT are $Y^{\fM \fN} = ( y^{i5}, y^{ij}, y^{i4}, y^{i5} , y^{45})$.
Here we can identify both IIB physical coordinates, $y^{ij}$, and IIA physical coordinates, $y^{i5}$ (these are T-dual sets of coordinates - we can make other choices of IIB and IIA coordinates but it is convenient to use this symmetric pairing).

The generalised metric written as a five-by-five matrix $m_{\fM \fN}$ can be factorised using a generalised vielbein $E^{\fA}{}_{\fM}$ such that
\be
m_{\fM \fN} = E^{\fA}{}_{\fM} E^{\fB}{}_{\fN} \delta_{\fA \fB} \,,
\ee
where $\fA,\fB$ are five-dimensional flat indices.
We can use a frame adapted to the IIB solution of the section condition (recall that we dualise the physical indices $y^{ij} \rightarrow y_i$ hence the three-dimensional indices appear upside down):
\be
E^{\fA}{}_{\fM} = e^{1/10} \begin{pmatrix}
e^{1/2} e_i{}^a & 0 \\
\frac{1}{2} e^{-1/2} h^{\bar\alpha}{}_{\alpha} \epsilon_{ikl} C^{kl,\alpha} & e^{-1/2} h^{\bar \alpha}{}_{\alpha}
\end{pmatrix}\,,
\label{littleECB}
\ee
where $e_i{}^a$ is a (inverse) vielbein for the three-dimensional (Einstein frame) metric, $C^{ij,\alpha}$ are the RR and NSNS two-forms, and $h^{\bar \alpha}{}_{\alpha}$ is a vielbein for the $\mathrm{SL}(2)/\mathrm{SO}(2)$ coset element containing the dilaton and RR zero-form.
Alternatively, we can use a frame adapted to the IIA solution of the section condition:
\be
E^{\fA}{}_{\fM} = 
e^{4\Phi/5} e^{1/10} 
\begin{pmatrix}
 e^{-1/2} e^{-\Phi} e^a{}_i & 0 & - e^{-1/2} e^{-\Phi} e^a{}_i B^i \\
e^{-1/2} C_i& e^{-1/2} & e^{-1/2} (C-C_k B^k)\\
 0 & 0 & e^{-\Phi} e^{1/2} 
\end{pmatrix} 
 \,.
\label{littleECA}
\ee
Here $e^a{}_i$ is a vielbein for the three-dimensional (string frame) metric, $B^i \equiv \frac{1}{2} \epsilon^{ijk} B_{jk}$, $C \equiv \frac{1}{6} \epsilon^{ijk} C_{ijk}$, and $\Phi$ is the dilaton.

Now suppose we take the following $\Gfour$ element to define a generalised Scherk-Schwarz twist of the $\Gfour$ ExFT:
\be
U^{\fA}{}_{\fM} = \begin{pmatrix} \delta^a_i & 0 & 0 \\ -\frac{1}{6} m\epsilon_{ijk} y^{jk} & 1 & 0 \\ 0 & 0 & 1 \end{pmatrix} 
\label{Umassive}\,,
\ee
where we now take the indices $\fA,\fB,\dots$ to correspond to the indices of the gauged exceptional field theory (with in this case the ten-dimensional index $A \equiv [ \fA \fB]$).
Comparing \refeqn{Umassive} with \refeqn{littleECB} and \refeqn{littleECA}, we see that in the type IIB theory this corresponds to expanding about a background in which the RR two-form has	 a linear dependence on the physical coordinates, $C^{ij} \sim m y^{ij}$, and in the IIA theory to the RR one-form having a linear dependence on \emph{dual} coordinates, $C_i \sim m \frac{1}{2} \epsilon_{ijk} y^{jk}$.
In effect, this set up corresponds to starting with a IIB theory with a constant RR three-form flux, $F_3 \sim m$, and then T-dualising on the three-directions in which this flux is present in order to obtain the Romans IIA theory with RR 0-form flux, $F_0 \sim m$ (of course, this is not the only possible route to massive IIA\cite{Ciceri:2016dmd}).

Let us define $\tilde y_i = \frac{1}{2} \epsilon_{ijk} y^{jk}$ and take the following factorisation ansatz:
\be
\begin{split}
g_{\mu\nu}(x,Y) & = g_{\mu\nu}(x,\bar Y) \,,\\
m_{\fM \fN} (x,Y)& = U^{\fA}{}_{\fM}(\tilde y) U^{\fB}{}_{\fN}(\tilde y)\bar m_{\fA \fB}(x, \bar Y) \,,\\
\Aa_{\mu}{}^{\fM \fN} (x,Y) & = U^{\fM}{}_{\fA}(\tilde y)U^{\fN}{}_{\fB}(\tilde y) \bar{\Aa}_{\mu}{}^{\fA \fB} (x,\bar Y) \,,
\end{split}
\label{massiveSS}
\ee
(and so on for the tensor hierarchy fields) where $\bar Y^A = (y^{a5}, y^{ab}, y^{a4}, y^{45})$ denote the coordinates of the gauged ExFT, in which the physical dependence of the fields is constrained according to the constraints above.

For the specific twist matrix \refeqn{Umassive}, the trombone \refeqn{fluxtrombone} vanishes, $\theta_{\fA \fB}= 0$, and $\Theta_{\fA \fB \fC}{}^{\fD}$ has components $\Theta_{abc}{}^{4} = Z_{abc}{}^4= - m \epsilon_{abc}$, hence
\be
\tau_{ab,c5}{}^{45} = \Theta_{ab,c5}{}^{45} = - m \epsilon_{abc} \,,\quad
\tau_{ab,cd}{}^{4e} = \Theta_{ab,c5}{}^{4e} = -2 m \epsilon_{ab[c} \delta_{d]}^e \,.
\ee
In addition to the section condition $\partial_{[AB}\otimes \partial_{CD]} = 0$ in the gauged ExFT, we require from the first constraint of \refeqn{gaugedconstraints} that $\partial_{45} = \partial_{a4} = 0$, and from the second we must have $\partial_{ab} =0$. This selects as expected the coordinates $y^{a5} = \delta^a{}_i y^{i5}$ as being physical, as we would expect. Additionally, one can check that the quadratic constraint \refeqn{QC} is obeyed.

The generalised metric describing the three-dimensional ``internal'' components of the massive IIA fields is:
\be
\small
\bar m_{\fA \fB} = 
e^{8 \Phi/5} \begin{pmatrix}
 \phi^{-2/5} ( e^{-2\Phi} \phi_{ab} + C_a C_b ) & \phi^{-2/5} C_a & \phi^{-2/5} (- e^{-2\Phi} \phi_{ac} B^c + C_a ( C - C_c B^c ) ) \\
  & \phi^{-2/5} & \phi^{-2/5} (C-C_c B^c ) \\
   & & \phi^{3/5} ( e^{-2\Phi} (1+\phi_{cd} B^cB^d) + (C-C_cB^c)^2)
\end{pmatrix}
\label{barm}
\ee
where $B^a \equiv \frac{1}{2} \epsilon^{abc} B_{bc}$, $C \equiv \frac{1}{6} \epsilon^{abc} C_{abc}$, and $\Phi$ the dilaton.
With $\theta_{\fA\fB}=0$, generalised diffeomorphisms induce massive gauge transformations that follow from:
\be
\delta_{\bar \Lambda} m_{\fA \fB} \supset  \Lambda^{\fC \fD} \Theta_{\fC \fD ( \fA}{}^{\fE} m_{\fB) \fE}
= - \Lambda^{cd} m \epsilon_{cde} \delta^e_{(\fA} m_{\fB)4}\,,
\ee
which defining $\lambda_a = \frac{1}{2} \epsilon_{abc} \Lambda^{bc}$ implies
\be
\delta C_a =- m\lambda_a \,,\quad
\delta C_{abc} =- m  3\lambda_{[a} B_{bc]} \,,
\ee
reproducing the modified gauge transformations \refeqn{massivegauge}.
From this it follows using the generalised diffeomorphism invariance of ExFT that the complete massive IIA theory is reproduced, as can be explicitly checked in detail using the complete ExFT to IIA dictionary.
In particular, the modification of the ExFT action \refeqn{SSVfive} reproduces the Romans cosmological term \refeqn{massiveS}.
(To verify this it is needed to note that in passing from the 10-dimensional string frame metric to the $n$-dimensional Einstein frame metric used in ExFT, there is an additonal conformal redefinition involving the dilaton, such that $\hat g_{\mu\nu} = |\phi|^{-1/(n-2)} e^{4\Phi/(n-2)} g_{\mu\nu}^{\text{ExFT}}+ \dots$.)

\subsubsection{Generalised IIB}

A second case where a deformation of one of the ten-dimensional theories can be realised in ExFT using dual coordinate dependence is the so-called generalised IIB theory.
This theory arose when studying remarkable deformations of the $AdS_5 \times S^5$ superstring which preserve integrability.
Certain such integrable deformations produce string models for which the target space does not satisfy the usual supergravity field equations, but rather obeys the equations of motion of a \emph{generalised IIB theory}. 
These equations of motion turn out to be implied by the conditions of $\kappa$-symmetry for the type II superstring \cite{Arutyunov:2015mqj, Wulff:2016tju} and are \emph{weaker} than the usual Weyl invariance conditions. 

The generalised IIB equations involve the usual metric and form fields, but in place of the usual dilaton $\Phi$ we introduce a Killing vector field $K$ and a one-form $Z$ such that $dZ = \iota_K H_3$, where $H_3$ is the field strength of the NSNS two-form.
When the Killing vector vanishes, we can introduce the dilaton $\Phi$ by writing $Z = d \Phi$.
If the Killing vector $K$ is chosen to be adapted to a given direction $y^*$, i.e. $K^i = \delta^i_{\star}$, then we can locally write $Z_i = \partial_i \Phi - B_{i \star}$. This resembles a ``gauging'' of the dilaton derivative, $\partial_i \Phi \rightarrow \partial_i \Phi - B_{i \star}$.

This can be accommodated within both DFT and ExFT by introducing a generalised Scherk-Schwarz ansatz which amounts to requiring the IIB dilaton depend linearly on a coordinate associated to a dual IIA section \cite{Sakatani:2016fvh,Baguet:2016prz}. Note that this amounts to gauging the trombone, and so the corresponding gauged ExFT is defined solely in terms of its equations of motion, exactly as is the case for the generalised IIB theory which does not have a known action.
(A recent review of integrable deformations and generalised IIB with more details of the applications of DFT/ExFT to this interesting area of research is \refref{Orlando:2019his}.)

\subsection{Supersymmetry}
\label{susy}

Everything we have presented has been entirely bosonic, but did in a sense know about supersymmetry.
This was apparent in the fact that the bosonic symmetries of ExFT fixed all coefficients of the bosonic action, whereas conventionally one would use supersymmetry to fix the normalisation of the Chern-Simons term. 
Furthermore, the structure of the generalised Lie derivative gave rise to the precise representations of the embedding tensor of gauged maximal supergravity, which are also normally fixed by supersymmetry.

The explicit supersymmetrisation of the bosonic ExFT action can be constructed.
The fermions transform as spinors of $\mathrm{SO}(1,n-1)$ and in representations of the denominator subgroup $H_d$ of the $\Edd/H_d$ coset.
The details for $\Gseven$ were worked out in \refref{Godazgar:2014nqa} (with a superspace formulation provided in \refref{Butter:2018bkl}), those for $\Gsix$ in \refref{Musaev:2014lna}, and $\Geight$ in \refref{Baguet:2016jph}.

We can also use ExFT to describe (reductions on) backgrounds that break some amount of supersymmetry.
In fact, this has long been a powerful motivation for \cite{Hitchin:2004ut} and application of generalised geometry \cite{Grana:2004bg, Grana:2005sn} (so below we somewhat freely are referring on the same footing both to exceptional generalised geometry and ExFT with the section condition solved).
The main idea captured by generalised diffeomorphisms is the unification of metric and form-field symmetries and degrees of freedom.
This means that generalised geometric quantities can naturally be used to describe features of flux compactifications, including the Killing spinor equations and their solutions in the presence of fluxes.

One useful feature here is that one can characterise the existence of a particular number of Killing spinors, equivalent to the presence of a particular amount of supersymmetry, in terms of bosonic \emph{generalised} tensors which must be nowhere vanishing on spacetime, and which will generically obey certain compatibility and integrability conditions. These objects generalise the holomorphic three-form and symplectic two-form of Calabi-Yau manifolds.
See for example \refref{Ashmore:2015joa,Ashmore:2016qvs, Malek:2016bpu, Malek:2017njj} 

In general, supergravities with less than maximal supersymmetry have different bosonic spectra and ultimately different U-duality groups and thus theories with lower supersymmetry will have different associated generalised geometries. So far this has been relatively unexplored with the majority of work being done for the maximal case. 
One approach has been to consider ExFT on K3 where by introducing additonal vector fields the duality between M-theory on K3 and the heterotic theory on $T^3$ can be described \cite{Malek:2016vsh}.
Another (not unrelated approach) to enlarge the ``duality web'' of supergravities that can be described in ExFT is to take quotients by discrete subgroups of $\Edd$ which eliminate degrees of freedom while preserving some amount of supersymmetry.
We will discuss how this can be done for half-maximal theories next.

\subsection{O-folds and half-maximal theories} 
\label{O}
The idea here is to introduce generalised notions of orbifolds and orientifolds within ExFT \cite{Blair:2018lbh}.
Given $Z^M{}_N \in G_{\textrm{discrete}} \subset E_{d(d)}$, we impose a quotient via an identification on the extended coordinates 
\be
(X^\mu, Y^M ) \sim (X^\mu , Y^{\prime M} ) = ( X^\mu , Z^M{}_N Y^N ) \,,
\ee
and the ExFT fields:
\be\begin{split} 
g_{\mu\nu} ( X, Y) &\sim g_{\mu \nu} ( X, Y^\prime ) \,,\\
\gM_{MN} ( X,Y) &\sim (Z^{-1})^P{}_M (Z^{-1})^Q{}_N \gM_{PQ} (X, Y^\prime ) \,,\\
\mathcal{A}_\mu{}^M(X,Y) &\sim Z^M{}_N \mathcal{A}_\mu{}^N(X,Y^\prime) \,,
\end{split}
\ee
plus similar transformations of the other tensor hierarchy fields according to their representations.
In general, the identification $Y^M \sim Z^M{}_N Y^N$ may be entirely geometric, in that physical coordinates are identified with physical coordinates, or non-geometric, in that physical coordinates are identified with dual coordinates.
In the former case, the quotient will describe orbifolds or orientifolds, while the latter case corresponds to asymmetric orbifolds and non-perturbative generalisations.
We can refer to such constructions as generalised orientifolds \cite{Dasgupta:1996ij} or, by analogy with T- and U-folds, ``O-folds'' (indeed in general T/U-fold compactifications such O-fold quotients appear at fixed points in moduli space \cite{Dabholkar:2002sy}).

As we mentioned above, the preservation of some given amount of supersymmetry by some ExFT configuration can be described in terms of an appropriate set of generalised tensors, whose (global, non-vanishing) existence is equivalent to that of the requisite amount of Killing spinors of the background. 
Given such a ``half-maximal structure'' for a given ExFT \cite{Malek:2017njj}, quotients which preserve the existence of the half-maximal structure can be classified, as was done for $\Gfour$ in \refref{Blair:2018lbh}.\footnote{Some earlier occurences of orientifold constructions in doubled or exceptional geometry include \refref	{Grana:2012zn,Baraglia:2013xqa,Ciceri:2016hup}. Notably in the doubled case the $\mathbb{Z}_2$ orientifold quotient is not an element of $O(d,d)$ but rather sends the $O(d,d)$ metric $\eta$ to minus itself\cite{Baraglia:2013xqa}.}

The simplest such quotient uses a $\mathbb{Z}_2 \subset \Gfour$ which is generated by
\be
Z^{\fM}{}_{\fN} = \mathrm{diag}(-1,-1,-1,-1,+1) \,,
\label{Z}
\ee
acting on the fundamental representation.
Quotienting by this action defines a $\mathbb{Z}_2$ O-fold without referring to a specific solution of the section condition.
The transformation \refeqn{Z} acts on the extended coordinates as $Y^{\fM\fN} \rightarrow Z^{\fM}{}_{\fP}Z^{\fN}{}_{\fQ} Y^{\fP\fQ}$, so that there are six even and four odd coordinates. 
Out of these, we have to choose 4 or 3 physical coordinates such that the section condition is obeyed. These physical coordinates will then either be even or odd according to the precise match between the ExFT coordinates and the physical ones.

In an M-theory solution of the section condition, the single $\mathbb{Z}_2$ \refeqn{Z} describes two distinct quotients of 11-dimensional supergravity.
In the first, the physical coordinates $y^i \equiv Y^{i5}$ have parity $(-+++)$, and this together with  the action induced on the spacetime fields ($\hat g, \hat C_3$) $\rightarrow (\hat g, - \hat C_3$) matches the Ho\v{r}ava-Witten quotient of M-theory on an interval. 
In the second, the physical coordinates $y^i \equiv Y^{i5}$ have parity $(----)$, and the set-up describes a geometric orbifold of M-theory on $\mathbb{R}^4/\mathbb{Z}_2$ or $T^4/\mathbb{Z}_2$.

Conversely, in a IIB solution, the $\mathbb{Z}_2$ \refeqn{Z} describes three separate case.
In two of these, the physical coordinates $y_i \equiv \frac{1}{2}\epsilon_{ijk} Y^{jk}$ have parity $(+++)$. The quotients then differ according to whether they act as $\diag(+1,-1)$ or $(-1,+1)$ on the $\mathrm{SL}(2)$ doublet of 2-forms. 
One possibility gives a ten-dimensional theory in which $C_0, B_2,C_4$ are projected out, while the other gives a ten-dimensional theory in which $C_0, C_2, C_4$ are projected out.
We identify the former with the orientifold of type IIB giving the type I theory, and the latter with the S-dual heterotic $\mathrm{SO}(32)$ theory.
The third possibility has physical coordinates of parity $(--+)$ and can be identified with type IIB in the presence of O7 orientifold planes.
(The T-duals of these theories, namely heterotic $E_8 \times E_8$ and type IIA in the presence of O8 and O6 orientifold planes, can be found in the corresponding type IIA solutions of the section condition).\footnote{It is possible to include the extra vector multiplets in the ExFT construction \cite{Blair:2018lbh} however the precise identification of the gauge group as $\mathrm{E}_8 \times \mathrm{E}_8$ and $\mathrm{SO}(32)$ is done by hand, in the absence of an ExFT version of anomaly cancellation arguments.}

An extended geometric interpretation is as follows. Four of the coordinates $Y^M$ being odd, we have $2^4$ fixed points of the $\mathbb{Z}_2$ quotient.
Each of these fixed points gives a $7+6$ dimensional ``O-fold plane''
These can intersect with the $7+4$ or $7+3$ dimensional physical spacetime in a variety of ways, giving conventional orientifold planes as well as other fixed point hyperplanes.
When the odd coordinates are all dual, the fixed points do not occur in the physical directions and so the O-fold planes fill the entire physical spacetime.
Then, we obtain true 10-dimensional theories corresponding to type IIB in the presence of O9 planes (i.e. type I), or the heterotic theories (which can perhaps be associated to certain NS9 planes \cite{Bergshoeff:1998re}). 
Alternatively, there can be genuine fixed points in some of the physical directions $Y^i$, so that the O-fold plane is not spacetime filling but rather becomes for example an ordinary orientifold plane (perhaps automatically accompanied by D-branes) in spacetime, or some even more non-perturbative generalisation (in the M-theory geometric orbifold, the fixed point planes can be thought of as the strong coupling limits of O6 plus D6 brane configurations).

\subsection{Higher derivative corrections} 
\label{corrections}

Supergravity is the low energy effective action of string and M-theory, and therefore receives corrections at high energies.
In string theory, these corrections can be organised into an expansion in $\alpha^\prime$, the string length, and also in $g_s$, the string coupling. 
Given that duality is believed to be a symmetry of the full quantum string theory, it is expected that the structure of these expansions should be compatible with $O(d,d)$ or $\Edd$ -- for the former case evidence for the $O(d,d)$ symmetry in the first order corrections to the cosmological truncation (keeping time dependence only) of the NSNS sector was shown long ago in \refref{Meissner:1996sa}. Without making such truncations, one might hope that the new approaches of DFT and ExFT may provide a useful starting point for computing the form of such corrections directly in ten- or eleven-dimensions.
And indeed, in the ten-dimensional case, there has been much progress on working out the form of $\alpha^\prime$ corrections of the bosonic and heterotic supergravities using extensions of doubled geometry \cite{Hohm:2013jaa,Godazgar:2013bja, Bedoya:2014pma,Hohm:2014xsa,Coimbra:2014qaa, Hohm:2014eba,Hohm:2015mka, Marques:2015vua,Hohm:2016yvc,Baron:2017dvb, Baron:2018lve, Eloy:2020dko}.
 
Meanwhile, a remarkable application of exceptional field theory is to compute the higher derivative corrections to maximal supergravity in $n$-dimensions in an $\Edd$ covariant fashion. 
This can be done by computing amplitudes in ExFT \cite{Bossard:2015foa, Bossard:2017kfv, Bossard:2020xod}.
In this set-up, one first uses the result that the one loop amplitude in maximal supergravity can be reduced to the scalar box diagram. 
The extended space is taken to be toroidal and the scalar field expanded in a Fourier expansion involving the generalised momenta $p_M$ associated to the directions $Y^M$. These generalised momenta then are allowed to run in loops, as long as the generalised momentum $p_M$ obeys the section condition in momenta space. Thus one arrives at a sum over a lattice of momenta subject to the algebraic constraint corresponding to the section condition. These sums may be carried out and give generalised Eisenstein series associated to $\Edd$ symmetry. This is the exceptional field theory extension of the work done in \refref{Green:1997as} where the one-loop box in eleven dimensions with a toroidal background allowed the calculation of $\mathrm{SL}(2)$ invariant amplitudes for the IIB string which could be used to determine the 1/2 BPS protected higher derivative corrections.
Even more impressive results in \refref{Bossard:2020xod} going to higher loops and lower supersymmetry show the utility of the ExFT approach in encoding non-trivial properties of string theory and as a calculation tool for the quantum theory.

%% file: sec6BranesInExFT.tex
\subsection{Waves and monopoles in Kaluza-Klein theory}
\label{kksolns}

We began, in section \ref{genKK} by describing Kaluza-Klein theory. 
We will now discuss the Kaluza-Klein perspective on solutions of the lower-dimensional theory which are electrically and magnetically charged under the Kaluza-Klein vector. 
These become purely geometric solutions in the higher-dimensional theory.
This will point the way to interpreting brane solutions in the extended geometry of both double and exceptional field theory. 

We first need a brief discussion on the nature of charges in general relativity. 
There are various ways to define conserved charges associated to solutions, and here we will use the ADM definition.\cite{Arnowitt:1960es} 
Given a Killing vector $\xi$, the associated ADM conserved charge is defined as an integral over a surface at spatial infinity:
\be
Q [ \xi ]  = 
\frac{1}{16\pi G}
\int_{\infty} d \Sigma_{\mu\nu} \sqrt{|g|} 
\left( -2 \nabla^{[\mu}  \xi^{\nu]} - 2 \xi^{[\mu} B^{\nu]} \right)  
\label{ADMcharge}
\ee
where $\nabla_\mu$ is the Levi-Civita covariant derivative and $B^\mu = \partial_\nu g^{\mu\nu} + g^{\mu\nu}\partial_\nu \ln g$ is the boundary vector. 
A time-like Killing symmetry implies a conserved energy, which is the ADM mass, and a space-like translational symmetry implies a conserved momentum. 

Now let's consider a Kaluza-Klein reduction from $d+1$ to $d$ dimensions, and denote the extra dimension by $z$.
In the Kaluza-Klein reduced theory, electric charge corresponds to momentum in the Kaluza-Klein direction, which is defined using \refeqn{ADMcharge}, $P_z \equiv Q[\frac{\partial}{\partial z} ]$.
The mass of the solution follows from $P_0 \equiv Q[\frac{\partial}{\partial t}]$. 
Of particular relevance for the generalisation to branes are BPS-type solutions where the charge determines the energy, which here implies $P_0 = |P_z|$. A solution with such a property is a null wave. Exact solutions of this type are known in relativity and are called pp-waves:
\begin{align}
ds^2=-H^{-1} dt^2 + H(dz + (H^{-1}-1)dt)^2 + \delta_{ij} dx^i dx^j \label{wave}
\end{align}
The function $H$ is a harmonic function of the transverse directions $x^i$, and one solution is $H(r)=1+ p/r^{d-3}$ with $r=\sqrt{x^i x^j \delta_{ij}}$.
Evaluating the conserved charges according to \refeqn{ADMcharge}, the parameter $p$ in the harmonic function corresponds (with appropriate normalisation) to the momentum.
As the solution \refeqn{wave} has a Killing symmetry along $z$, it is independent of the $z$ coordinate and obeys the Kaluza-Klein constraint $\partial_{z} = 0$.
The Kaluza-Klein vector $A_\mu$ has one non-zero component, $A_t = H^{-1} - 1$, and is the gauge potential for an a electric field.

We can further construct magnetic charges, by interpreting the Kaluza-Klein spacetime as only locally a product spacetime, with the Kaluza-Klein circle fibred over the lower-dimensional spacetime.
An example of such a solution is the Kaluza-Klein monopole first described (for $d=4$) by Sorkin \refref{Sorkin:1983ns} and Gross and Perry \cite{Gross:1983hb}: 
\begin{align}
ds^2=-dt^2 + H^{-1}(dz +A_i dx^i)^2 +H \delta_{ij} dx^i dx^j \, .
\label{KKM}
\end{align}
Here the field $A_i$, which is the Kaluza-Klein vector and corresponds to a Dirac monopole potential in the reduced theory, is related to the harmonic function $H$ as follows:
\begin{align}
\partial_{[i} A_{j]}= \frac{1}{2} \epsilon_{ij}{}^k \partial_k H \,  \, , \quad H= 1+ \frac{g}{r} \, .
\label{Dirac}
\end{align}
The four-dimensional spacetime can be viewed as $M_4 = \mathbb{R}_t \times \mathbb{R}^+ \times S^2$, where $r \in \mathbb{R}^+$ is the radial direction, and the Kaluza-Klein circle $S^1$ is fibred over the $S^2$ using the Hopf fibration of $S^3$, so that $M_5 = \mathbb{R}_t \times \mathbb{R^+} \times S^3$.

The mass of the KKM monopole will be given by the ADM formula, \refeqn{ADMcharge}.
The magnetic charge, corresponding to the parameter $g$ in the harmonic function in \refeqn{Dirac}, is however measured as the first Chern class of the $S^1$ fibration.
That is, 
\begin{equation}
g=\frac{1}{2 \pi} \int_{S^2} F  \, ,
\end{equation}
where locally $F=dA$ as defined above in equation \refeqn{Dirac}.

\subsection{Branes are waves and monopoles in double field theory}
\label{branesDFT}

Now, again following the intuition developed in section \ref{genKK}, we will extend the above discussion to brane solutions in double field theory. 

First of all, we can define a doubled version of the ADM charge \refeqn{ADMcharge}. 
We define a generalised Killing vector to be a generalised vector field $\Lambda^M$ which preserves the generalised metric and dilaton, $\mathcal{L}_\Lambda \cH_{MN} = 0$ and $\mathcal{L}_\Lambda \gendil = 0$.\footnote{Generalised vectors which generate trivial transformations always lead to a vanishing charge.}
Given such a generalised Killing vector, the corresponding generalisation of the ADM charge \refeqn{ADMcharge} is: \cite{Blair:2015eba,Park:2015bza,Naseer:2015fba}
\be
\mathcal{Q}[\Lambda] \sim \int_\infty d \Sigma_{MN} e^{-2\gendil} ( - 4 ( \bar P^{Q[M} P_P{}^{N]}  - P^{Q[M} \bar P_P{}^{N]}  )\nabla_Q \Lambda^P  - 2 B^{[M} \Lambda^{N]} ) \,,
\label{ADMDFT} 
\ee 
where $\nabla_M$ denotes the DFT covariant derivative \cite{Siegel:1993th, Siegel:1993xq, Jeon:2010rw, Jeon:2011cn, Hohm:2011si} which annihilates the generalised metric, $O(d,d)$ structure, and the generalised dilaton, $P^M{}_N = \frac{1}{2} ( \delta^M_N - \mathcal{H}^M{}_N)$, $\bar P^M{}_N = \frac{1}{2} ( \delta^M_N + \cH^M{}_N)$ are orthogonal projectors and  the DFT boundary vector \cite{Berman:2011kg} is $B^M = -\partial_N \cH^{MN} + 4 \cH^{MN} \partial_N \gendil$. (We use $\eta_{MN}$ and its inverse to raise and lower indices.)

To describe solutions in DFT language, we will write
\be
ds^2|_{\text{DFT}}  = \cH_{MN} dX^M dX^N \,,
\label{dftline}
\ee
which must be interpreted carefully as this is not a true line element.
One can think of it purely as a useful shorthand for solutions, or in the context of brane worldvolume actions one can be more precise. 
Let us take a paragraph to expand on this: note though this is somewhat of a technical digression. 

The issue with \refeqn{dftline} is that it does not transform covariantly under generalised diffeomorphisms. 
If one writes string or brane actions with a doubled target space, one would like to pull back \refeqn{dftline} to the worldvolume.
This can be done by ``gauging'' the non-physical dual directions, as originally discussed in the doubled sigma model of \refref{Hull:2004in, Hull:2006va} and later in the context of DFT in \refref{Lee:2013hma}. 
This gauging replaces $dX^M \rightarrow dX^M + V^M$, where $V^M$ is an auxiliary worldvolume one-form.
This one-form is constrained to obey $V^M \partial_M = 0$, and is assigned a transformation under generalised diffeomorphisms acting as $\delta_\Lambda X^M = \Lambda^M$ such that the gauged line element $\Delta s^2 \equiv \cH_{MN} (dX^M+V^M) (dX^N+V^N)$ transforms covariantly, $\delta_\Lambda (\Delta s^2) = \mathcal{L}_\Lambda \cH_{MN} ( dX^M +V^M)(dX^N+V^N)$.
This ensures that when $\Lambda$ is a generalised Killing vector, we obtain a symmetry on the worldvolume.
On choosing a solution of the section condition, we can write $V^M = (0, \tilde V_\mu)$.
Then the dual coordinates only appear in the combination $d\tilde x_\mu + \tilde V_\mu$ and we can integrate these out using the algebraic equation of motion for $\tilde V_\mu$. 
As no fields depend on the $\tilde x_\mu$, there is a shift symmetry $\tilde x_\mu \rightarrow \tilde x_\mu +\varphi_\mu$, which is what we gauge to eliminate these coordinates, with $\tilde V_\mu \rightarrow \tilde V_\mu -d \varphi_\mu$.
With this caveat in mind, we will continue to use \refeqn{dftline} as a shorthand for expressing solutions in terms of the generalised metric.

Returning to the description of brane solutions in DFT, we start with the DFT wave. 
The form of this solution is most easily found by writing the pp-wave solution \refeqn{wave} in terms of the generalised metric. 
Let the doubled coordinates be $X^M = (t , z, x^i, \tilde t, \tilde z, \tilde x_i)$.
Here $i=1,\dots, d-2$ labels the doubled transverse coordinates. 
The wave solution in DFT is: \cite{Berkeley:2014nza}
\be
\begin{split} 
ds^2|_{\text{DFT}} &=  (H-2)(dt^2-dz^2)-H(d \tilde{t}{}^2- d \tilde{z}{}^2)  
\\ & \quad + 2(H-1)(dt d\tilde{z} +d \tilde{t} dz)  
\\ & \quad+ \delta_{ij} dx^i dx^j + \delta^{ij} d \tilde x_i d \tilde x_j  \,,\\
e^{2\gendil} & = \text{constant} \,.
\end{split}
\label{DFTwave}
\ee
Analysing the DFT equations of motion shows that the function $H$ must be harmonic in the coordinates $(x^i, \tilde x_i )$, obeying the equation
\begin{align}
(\delta^{ij} \partial_i \partial_j + \delta_{ij} \tilde \partial^i \tilde \partial^j) H = \delta
\label{harmonicH}
\end{align}
and furthermore $H$ must be subject to weak constraint, $\partial_i \tilde \partial^i H = 0$. (The $\delta$ function in \refeqn{harmonicH} is such that $H$ is actually harmonic away from the position of the wave itself. For a detailed discussion where the wave is in fact sourced by the doubled string, see \refref{Blair:2016xnn}.)
It is simplest to set $\tilde \partial^i = 0$, and identify the $x^i$ with the physical spacetime coordinates, such that one solution is $H = 1 + \frac{p}{r^{d-4}}$.

The DFT wave \refeqn{DFTwave} is independent of the coordinates $(t,z,\tilde t, \tilde z)$.
In order to use the generalised ADM expression \refeqn{ADMDFT}, we commit to taking $t$ as the physical timelike direction (as the integration in \refeqn{ADMDFT} is over a constant time hypersurface). 
Then the generalised Killing vectors $\frac{\partial}{\partial t}$ and $\frac{\partial}{\partial \tilde z}$ turn out to give non-zero charges:
\be
\mathcal{Q}[\partial/\partial t] = \pm  \mathcal{Q}[\partial/\partial \tilde z]  \sim p \,.
\ee
This makes precise the interpretation of \refeqn{DFTwave} as a null wave carrying momentum in the $\tilde z$ direction.
(The sign allows for the wave with the opposite orientation, $\tilde z \rightarrow - \tilde z$.)

If we then interpret $(t,\tilde z, x^i)$ as the coordinates of the physical spacetime, we find exactly the $d$-dimensional pp-wave solution of the form \refeqn{wave}.

On the other hand, if we interpret $(t,z,x^i)$ as the coordinates of the physical spacetime, we find a different spacetime solution:
\be
\begin{split}
ds^2 & = H^{-1} ( - dt^2 + dz^2) + \delta_{ij} dx^i dx^j \,,\\
B & = (H^{-1} - 1 ) dt \wedge dz \,,\\
e^{-2\Phi} & = H\,,
\end{split}
\label{F1soln} 
\ee
which is exactly that sourced by a fundamental string.
From the DFT perspective, this carries momentum in the \emph{dual} direction, $\tilde z$, and the DFT momentum $\mathcal{Q}[\partial/\partial \tilde z]$ corresponds to the electric charge of the $B$-field.

Thus, wave-like solutions of DFT correspond to either pp-wave or fundamental string solutions of supergravity, depending on the direction of orientation of the wave.
The electric charge of the $B$-field is geometrised in doubled space into dual momentum.

We have done the electric charges, and so the next step is to find the magnetic charges, now following the form of the Gross-Perry-Sorkin solution.
We write our doubled coordinates as $X^M = (x^a,z, y^i, \tilde x_a, \tilde z, \tilde y_i)$, where $i=1,2,3$ and $a=0,\dots,5$.
We take the coordinate $\tilde z$ to correspond to the fibre of the $S^3$, and write the DFT monopole as \cite{Berman:2014jsa}:
\be
\begin{split}
ds^2|_{\text{DFT}} &=  H(1+H^{-2}\delta^{ij}A_i A_j)dz^2 + H^{-1} d \tilde{z}^2 \\
&\quad + 2H^{-1} A_i(dy^i d\tilde{z}- \delta^{ij} d\tilde{y}_j dz)  \\
&\quad+ H(\delta_{ij} +H^{-2} A_i A_j)dy^i dy^j +H^{-1} \delta^{ij} d \tilde{y}_i d \tilde{y}_j  \\
&\quad+ \eta_{ab} dx^a dx^b + \eta^{ab} d\tilde{x}_a d \tilde{x}_b \,\\
e^{-2\gendil} &= H e^{-2 \phi_0} \, ,
\end{split}
\ee
with $A_i$ and $H$ are related as for the Gross-Perry-Sorkin monopole, and are chosen to depend on the transverse coordinates $y^i$.
Choosing the physical coordinates to be $(x^a, \tilde z, y^i)$ this exactly corresponds to the string theory Kaluza-Klein monopole.
On the other hand, choosing the physical coordinates to be $(x^a, z, y^i)$, we find a solution which has magnetic charge under the $B$-field and corresponds to the NS5 brane solution.
In this case, the dual coordinate $\tilde z$ is fibred over the physical space in a non-trivial manner, and we can interpret the magnetic charge of the $B$-field as measuring the {\it{twisting}} of the novel extended space over the usual spacetime.
One can use the (non-constant) generalised fluxes $f_{AB}{}^C$ of DFT (defined in terms of a choice of doubled vielbein, $E^M{}_A$ with $\mathcal{L}_{E_A} E_B = - f_{AB}{}^C E_C$) to write down an $O(d,d)$ covariant expression unifiying the magnetic charge of the $B$-field with the charge associated to the first Chern class of the monopole fibration - see the discussion in \refref{Blair:2015eba}.

\subsection{Branes in exceptional field theory}
\label{branesinexft}

In exceptional field theory, we similarly have that charged brane solutions corresponds to solutions either carrying momentum in the extended space or exhibiting a monopole-type fibration in the extended space\cite{Berkeley:2014nza,Berman:2014jsa,Berman:2014hna}.

We can denote our extended coordinates (in an M-theory solution of the section condition) as $Y^M = (y^i, y_{ij}, y_{ijklm}, \dots)$.
Then, while null waves travelling in the physical $y^i$ directions give the usual pp-waves, a null wave in one of the dual $y_{ij}$ directions gives an M2 solution (with the brane wrapping the $ij$ directions) and momentum in one of the $y_{ijklm}$ directions corresponds to an M5 (wrapping the $ijklm$ directions).

At the same time, we can consider monopole-type solutions, and take one of the coordinates $Y^M$ to be the monopole fibre coordinate.
For example, a Hopf fibration of $y_{ij}$ over some $S^2$ in the physical spacetime will create a monopole-like solution that corresponds to an M5 brane.  

This means that exceptional field theory has a further novelty.
Once we have sufficiently large $\Edd$ that the coordinates $y_{ijklm}$ appear, then momentum in these directions will produce M5 brane solutions and Hopf fibrations of these coordinates will produce M2 branes. 
This reflects the fact that U-duality transforms membranes and fivebranes into each other.
In addition, for $d=7$ we have coordinates $y_{i_1\dots i_7;i}$ which are to be thought of as conjugate to the winding of the Kaluza-Klein monopole itself (including the fibre directions).
This points to the existence of {\it{self-dual}} solutions in the full exceptional field theory that are a combination of the wave and monopole-type solutions. 
We will explictly describe such a self-dual solution for the $\Gseven$ ExFT below.

A further complication in ExFT is the split into an $n$-dimensional ``external'' space and the extended space.
There are thus solutions of wave or monopole type which are purely external, in that they only exhibit non-trivial charges in the $n$-dimensional space, as well as solutions which are purely internal (for the wave-type solutions, this involves including the timelike direction in the extended space). In addition, there are mixed solutions in which the solutions are charged under the tensor hierarchy gauge fields and hence are not simply geometrical in the sense of being described solely by the external or generalised metric. 
Ultimately, ExFT should be extended all the way to $d=11$ in which case all these distinctions will vanish, and one would expect all possible single charge ExFT solutions to be subsumed into a single solution, described by generalised self-duality constraints \cite{West:2001as,Tumanov:2015yjd,Bossard:2017wxl,Bossard:2019ksx}.

In section \ref{salgebras} below we will revisit the characterisation of purely internal 1/2 BPS brane solutions using the superalgebra of ExFT.

For now, we will specialise to the self-dual solution of the $\Gseven$ ExFT in which we take the external four-dimensional spacetime to have coordinates $x^\mu = (t,x^i)$, $i=1,2,3$, and consider solutions which are charged under the one-form $\Aa_{\mu}{}^M$. 
We can think of this field as a generalisation of the Kaluza-Klein vector, in which case solutions which are charged under this can naturally be thought of as waves or monopoles, as we will see. Put differently, these are solutions whose charges are associated to invariance under (global) generalised diffeomorphisms $\Lambda^M$, just as is the case for the DFT wave. We can also think of solutions charged under $\Aa_\mu{}^M$ as appearing as point particles in the external spacetime. 

The self-dual solution described in \refref{Berman:2014hna} is as follows.
The four-dimensional external metric of the $\Gseven$ ExFT is
\begin{equation}
g_{\mu\nu} = \diag [-H^{-1/2},H^{1/2}\delta_{ij}]\, , \qquad H(r) = 1 + \frac{h}{r}\,,
\label{eq:exmetric}
\end{equation}
where $r^2=\delta_{ij}x^ix^j$ and $h$ is some constant which will parameterise the solution and correspond to its electric or magnetic charge.

The vector potential ${\AAA_\mu}^M$ has ``electric'' and ``magnetic'' components that are given by
\begin{equation}
{\AAA_t}^M 	= \frac{H-1}{H} a^M \,,\quad
{\AAA_i}^M	= A_i \ta^M \, ,
\label{eq:vecpot}
\end{equation}
where $A_i$ is magnetic potential obeying the familiar Kaluza-Klein monopole equation:
\begin{equation}
\partial_{[i}A_{j]} = \frac{1}{2}{\epsilon_{ij}}^k\partial_k H \, .
\label{eq:AH}
\end{equation}
The (constant) generalised vector $a^M$ in the extended space points in any one of the 56 extended directions. This is the direction of the momentum mode. The dual vector $\ta^M$ denotes the dual direction corresponding to the monopole fibration, and is determined by demanding $a^M \sim \Omega^{MN}\MM_{NK}\ta^K$, using the self-duality equation \refeqn{eq:selfduality}:
\begin{align}
{\FF_{\mu\nu}}^M &= 
 \frac{1}{2}\sqrt{|g|} \epsilon_{\mu\nu\rho\sigma}  g^{\rho\lambda}g^{\sigma\tau}\Omega^{MN} \MM_{NK}{\FF_{\lambda\tau}}^K\,.
\end{align}
Finally, we have the generalised metric $\gM_{MN}$ which is diagonal, with non-zero components equal to $\{ H^{3/2},H^{1/2},H^{-1/2},H^{-3/2}\}$. The powers of $H^{\pm 3/2}$ appear once each, and correspond to the special wave and monopole fibration directions, and the powers of $H^{\pm 1/2}$ appear 27 times each. The precise order of the 56 entries of course depends on a coordinate choice, but once this is fixed it characterizes the solution together with the choice of direction for $a^M$.

For $a^M = \delta^M_z$ and $\tilde a^M = \delta^M_{\tilde z}$, given \refeqn{eq:exmetric}, \refeqn{eq:vecpot} and using \refeqn{eq:AH} and the fact the generalised metric is diagonal the self-duality condition on the field strength requires 
\be
\Omega^{z \tilde z} = 
-1 
\,,\quad \gM_{zz} = H^{3/2} \,,\quad \gM_{\tilde z \tilde z} = H^{-3/2} \,,
\ee
i.e. the generalised vectors $a^M$ and $\tilde a^M$ point in directions which are mutually dual with respect to the symplectic pairing encoded by $\Omega_{MN}$, i.e. $\Omega_{MN} a^M \tilde a^N = 
 1$ and the direction $z$ corresponds always to the generalised metric component $\gM_{zz}$.
As there is no coordinate dependence on any of the extended coordinates, these are isometries and we are free to change our choice of spacetime section and in doing so generate the whole orbit of brane solutions. From this point of view, we keep the vectors $a^M$, $\tilde a^M$ fixed and change what we mean by the physical directions.

In terms of the internal components $\phi_{ij} = \hat g_{ij}$ of the 11-dimensional metric, the $\Gseven$ generalised metric with $C_{ijk} = C_{ijklmn} = 0$ is always parametrised as:
\be
\gM_{MN} = \diag (
\phi^{1/2} \phi_{ij}  
,
2\phi^{1/2} \phi^{i[k} \phi^{l]j} ,
2\phi^{-1/2} \phi_{i[k} \phi_{l]j} ,
\phi^{-1/2} \phi_{ij} 
) \,,
\ee
while the ExFT external metric is related to the external components of the eleven-dimensional metric by $g_{\mu\nu} = \phi^{1/2} ( \hat g_{\mu\nu} - A_{\mu}{}^k A_{\nu}{}^l \phi_{kl})$, where $\mathcal{A}_\mu{}^i = A_\mu{}^i$ is the Kaluza-Klein vector.

Let's explain how to organise this solution to describe waves, monopoles and branes.

If we take the direction $z$ to be one of the physical directions, the solution corresponds in spacetime to a pp-wave, with:
\be
\begin{split}
\gM_{MN} = \diag (
H^{3/2}, H^{1/2} 1_6, &
H^{-1/2} 1_6, H^{1/2} 1_{15} ,
\\ &
H^{1/2} 1_6, H^{-1/2} 1_{15},
H^{-3/2}, H^{-1/2} 1_6
) \,.
\end{split}
\ee
This also describes the KKM on swapping $z \leftrightarrow \tilde z$ (which interchanges the first and last seven-by-seven blocks, and the middle 21$\times$21 blocks, in this generalised metric).

If we take the direction $z$ to be on the M2 winding directions, the solution corresponds in spacetime to the M2 brane solution, with:
\be
\begin{split}
\gM_{MN} = 
\diag(
H^{-1/2} , H^{1/2} 1_6,&
H^{3/2}, H^{1/2} 1_{10}, H^{-1/2} 1_{10}, 
\\ & 
H^{-3/2}, H^{-1/2} 1_{10}, H^{1/2} 1_{10},
H^{1/2} , H^{-1/2} 1_6
) \,.
\end{split}
\ee
This then describes an M5 solution on swapping $z\leftrightarrow \tilde z$.

We can define electric and magnetic charges: 
\be
\mathcal{Q}_M = \int_{S^2_\infty} \dd^2\Sigma \,\dd^{56}Y\, \gM_{MN} \star \Fa^{N} \,,\quad
\tilde{\mathcal{Q}}_M = \int_{S^2_\infty} \dd^2\Sigma \,\dd^{56}Y\, \Omega_{MN} \Fa^N 
\ee
using the four-dimensional Hodge star and integrating over a sphere at infinity in the four-dimensional spacetime.
These will obey a similar self-duality constraint. 

The table \ref{tab:solutions} below illustrates the plethora of brane and geometric solutions captured by this single ExFT solution. The orientation of the wave part of the solution, or equivalently the electric part of the $\Gseven$ one-form potential, determines the supergravity solution in eleven dimensions. We also list the IIA and IIB solutions associated with the different choices of section condition. 

Interestingly, one can easily generate brane bound states by by orienting the wave in a superposition of directions. For example the M2/M5 bound state requires a superposition of a membrane direction $y_{mn}$ and a fivebrane direction $y^{mn}$. 
These more general solutions can be generated by acting on the vector $a^M$ with solution generating transformations not necessarily in $\Gseven$.

\begin{table}[h]
\tbl{This table shows all supergravity solutions in ten and eleven dimensions as coming from the single $\Gseven$ self-dual solution. Here $m,n=1,\dots,7$ and $\bar m, \bar n =1,\dots,6$ label the internal directions in 11- and 10-dimensions, and $\alpha$ is an $\mathrm{SL}(2)$ doublet index.}
{
\centering
\begin{tabular}{|c|c|clcl|cc|}
\hline
theory & solution & \multicolumn{2}{c}{orientation}
 & \multicolumn{2}{c|}{\begin{tabular}[c]{@{}c@{}}ExFT\\vector\end{tabular}} 
& ${\AAA_t}^M$ & ${\AAA_i}^M$ \\ \hline
\multirow{5}{*}{$D=11$} 
& WM 	&\ \ \ & $y^m$ 		&& ${\AAA_\mu}^m$ 	& KK-vector 		& dual graviton 	\\
 & M2 	&& $y_{mn}$ 		&& $\AAA_{\mu mn}$ 		& $C_3$ 			& $C_6$ 			\\
 & M2/M5 && 	\ *	&&		\ *	&  $C_3\oplus C_6$ 	&		$C_6\oplus C_3$		\\
 & M5 	&& $y^{mn}$ 		&& ${\AAA_\mu}^{mn}$ 	& $C_6$ 			& $C_3$ 			\\
 & KK7 	&& $y_m$ 		&& $\AAA_{\mu m}$ 		& dual graviton 	& KK-vector 		\\ \hline
\multirow{8}{*}{\begin{tabular}[c]{@{}c@{}}$D=10$\\Type IIA\end{tabular}} 
 & WA 	&& $y^{\barm}$ 			&& ${\AAA_\mu}^{\barm}$			& KK-vector & dual graviton 	\\
 & D0 	&& $y^\theta$ 		&& ${\AAA_\mu}^\theta$		& $C_1$ 			& $C_7$ 		\\
 & D2 	&& $y_{\barm\bn}$ 		&& $\AAA_{\mu \barm\bn}$		& $C_3$ 			& $C_5$ 		\\
 & F1 	&& $y_{\barm\theta}$ 	&& $\AAA_{\mu \barm\theta}$	& $B_2$ 			& $B_6$ 		\\
 & KK6A && $y_{\barm}$ 			&& $\AAA_{\mu \barm}$			& dual graviton 	& KK-vector 	\\ 
 & D6 	&& $y_\theta$ 		&& $\AAA_{\mu\ \theta}$		& $C_7$ 			& $C_1$ 		\\
 & D4 	&& $y^{\barm\bn}$		&& ${\AAA_\mu}^{\barm\bn}$		& $C_5$ 			& $C_3$ 		\\
 & NS5 	&& $y^{\barm\btheta}$	&& ${\AAA_\mu}^{\barm\theta}$	& $B_6$ 			& $B_2$ 		\\\hline
\multirow{5}{*}{\begin{tabular}[c]{@{}c@{}}$D=10$\\ Type IIB\end{tabular}} 
 & WB 		&& $y^{\barm}$ 			&& ${\AAA_\mu}^{\barm}$ 			& KK-vector 		& dual graviton \\
 & F1 / D1 	&& $y_{\barm \alpha}$ 		&& $\AAA_{\mu \barm  \alpha}$ 		& $B_2$ / $C_2$ 	& $B_6$ / $C_6$ 	\\
 & D3 		&& $y_{\barm\bn\bk}$ 	&& $\AAA_{\mu \barm\bn\bk}$ 	& $C_4$ 			& $C_4$ 			\\
 & NS5 / D5 	&& $y^{\barm \alpha}$ 		&& ${\AAA_\mu}^{\barm \alpha}$		& $B_6$ / $C_6$ 	& $B_2$ / $C_2$ 	\\
 & KK6B 		&& $y_{\barm}$			&& $\AAA_{\mu \barm}$			& dual graviton 	& KK-vector 		\\ 
\hline
\end{tabular}}
\label{tab:solutions}
\end{table}


As a comment, one might think of this $\Gseven$ self-dual solution as a gravitational analogue of the self-dual string in the (0,2) world volume theory of the fivebrane. The reduction of the self-dual string to 4-d produces the spectrum of N=4 Yang-Mills and the $\mathrm{SL}(2)$ duality group is then a product of the reduction or how the string winds the Kaluza-Klein torus. This is similar to then how the reduction on the full extended space leads to U-duality.

Finally, one can speculate about how such a solution in the extended space can be probed by branes.
Totally wrapped branes appear as four-dimensional particles.
An action for particles coupling to the ExFT vector $\mathcal{A}_\mu{}^M$ was studied in \refref{Blair:2017gwn} and takes the form
\be
S = \int \dd \tau \, \frac{1}{2 e } \left( g_{\mu\nu} \dot{X}^\mu \dot{X}^\nu + \gM_{MN} ( \dot{Y}^M + \Aa_\mu{}^M \dot{X}^\mu  + V^M ) (\dot{Y}^N + + \Aa_\mu{}^N \dot{X}^\mu + V^N ) \right) \,,
\ee
where $e$ is a worldline einbein and $V^M$ is the auxiliary worldline one-form which gauges the shift symmetry in the dual directions on choosing a physical spacetime. 
Here we see an effective combination of the extended 56-dimensional space with the 4-dimensional external space, combining the external metric $g_{\mu\nu}$, the extended internal metric $\MM_{MN}$ and the vector potential ${\AAA_\mu}^M$ as an apparent 60-dimensional metric
\begin{equation}
\HH_{\hat{M}\hat{N}} = 
\begin{pmatrix}
g_{\mu\nu} + {\AAA_\mu}^M{\AAA_\nu}^N\MM_{MN} & {\AAA_\mu}^M\MM_{MN} \\
\MM_{MN}{\AAA_\nu}^N & \MM_{MN}
\end{pmatrix} \,.
\end{equation}
For the ExFT monopole/wave solution, this extended extended metric takes a block diagonal form:
\begin{equation}
\HH_{\hat{M}\hat{N}} = 
\begin{pmatrix}
H^{1/2} \HH_{AB}^\mathrm{wave} & 0 \\
0 & H^{-1/2} \HH_{\bar{A}\bar{B}}^\mathrm{mono}
\end{pmatrix}
\label{eq:60metric}
\end{equation}
where the top left block is $29 \times 29$ dimensional, and is simply the metric of a wave with 27 transverse dimensions and the bottom right block is $31 \times 31$ dimensional, 	is the metric of a monopole, also with 27 transverse dimensions
\begin{align}
\HH_{AB}^\mathrm{wave} &= 
\begin{pmatrix}
H-2 & H-1 & 0 \\
H-1 & H & 0 \\
0 & 0 & \delta_{27}
\end{pmatrix} , \\
\HH_{\bar{A}\bar{B}}^\mathrm{mono} &= 
\begin{pmatrix}
H(\delta_{ij} + H^{-2}A_iA_j) & H^{-1}A_i & 0 \\
H^{-1}A_j & H^{-1} & 0 \\
0 & 0 & \delta_{27}
\end{pmatrix} .
\end{align}
In \refeqn{eq:60metric} it is interesting to see that there is a natural split into a block diagonal form simply by composing the fields \`a la Kaluza-Klein. These two blocks come with prefactors of $H^{\pm1/2}$ with opposite power, so the geometry will change distinctly between large and small $r$. 

As we approach the core of the solution, where $r$ is small, $H$ becomes large and the wave geometry dominates. Far away for large $r$, $H$ will be close to one (and thus $A_i$ vanishes) and neither the monopole nor wave dominates. Thus one would imagine asymptotically either description is valid and the different choices related through a duality transformation. Note then the wave solution dominates in the small $r$ region core region. This smoothes out the singular behaviour near the brane core.

\subsection{Superalgebras}
\label{salgebras} 

\subsubsection{Idea}

In M-theory, the type II superalgebras in ten dimensions are lifted to the unique eleven-dimensional superalgebra.
\begin{align}
\{ Q_\alpha, Q_\beta \}= P_\mu (C\Gamma^\mu)_{\alpha \beta} + Z^{\mu \nu} (C \Gamma_{\mu \nu})_{\alpha \beta} + Z^{\mu_1\dots\mu_5}(C \Gamma_{\mu_1..\mu_5})_{\alpha \beta} \,.
\label{11dsusyalg}
\end{align}
Here the supercharges $Q_\alpha$ are 32-component Majorana spinors, with spinor index $\alpha$ and $\mu$ is the eleven-dimensional spacetime index. $P$ and the $Z$s are the momentum and central charges respectively.
In the IIA algebra, the ten-dimensional central charge associated to the D0 brane becomes identified with the momentum in the eleventh dimension. 
By comparing the  BPS state equation
\begin{align}
P_0{}^2=|Z|^2  \label{BPS}\,,
\end{align}
to the equation for a null wave:
\begin{align}
P_0{}^2=|\vec{P}|^2  \, ,
\end{align}
we see the D0 brane is the  null wave in M-theory. This  exactly realises the Kaluza-Klein paradigm where the electrically charged particles are identifed with the Kaluza-Klein momentum.
Now at the level of superalgebras it is the central charge, $Z$ being identified with $P_{11}$.
From the M-theory perspective the D0 brane is massless. 
Following our previous exploration of generalised and extended geometry, we would like to now reinterpret all the central charges as arising from momenta in extra novel dimensions.

To do so, we will examine the superalgebra with {\it{generalised}} coordinates from the extended geometry as follows (for now consider $d<6$):
\begin{align}
X^I=(x^\mu, y_{\mu  \nu}, y_{\mu_1..\mu_5} ) \, . \label{extracoords}
\end{align}
We will seek to remove all central charges and postulate a simple superalgebra in which purely generalised momentum appears in the anticommutator of the supercharges:
\begin{align}
\{Q_\alpha, Q_\beta \} = (C\Gamma_I)_{\alpha \beta} P^I  \, .  \label{susynocentral}
\end{align}
We will then examine the representation theory for this superalgebra. The {\it{massless}} (in the generalised sense) representations will obey a quadratic constraint on the momenta,
\begin{align}
Y^{IJ}{}_{PQ} P_I P_J =0 \,  , \quad P^IP^J  \delta_{IJ}=0 \, .
\end{align}
 The first quadratic constraint is the {\it{weak constraint}} that was determined from demanding closure of the algebra of local symmetries  \cite{Berman:2011cg,Coimbra:2011ky,Berman:2012vc} and the second condition is just the {\it{generalised}} massless condition. 

These generalised massless states will be the 1/2 BPS states. The section condition determines a generalised {\it{lightcone}} structure. The 1/2 BPS states lie on the lightcone and the other states with less supersymmetry lie in the interior of the lightcone.

\subsubsection{Example: $\Gfour$}

As an example, let's consider the case of $\Gfour$.
We will here work with the Lorentzian coset $\frac{\Gfour}{\HfourLor}$, and ignore the seven-dimensional Euclidean external space. 

We will construct spinors of the double cover of the local group, $\HfourLor$,  with spinor index $\alpha=1,\dots,4$. 
The associated gamma matrices forming the Clifford algebra for $\HfourLor$ are denoted:
 \begin{align}
 (\Gamma^i){}^{\alpha}{}_ {\beta} \ , \qquad  i=1,\dots,5 
\end{align}
We denote the charge conjugation matrix by $(C)_{\alpha \beta}$, with inverse $(C^{-1})^{\alpha \beta}$, with which we can lower and raise spinor indices respectively through left multiplication.

From these $\Gamma$ matrices we can form a representation of the global $\Gfour$ using the set of antisymmetrised products of the $\Gamma^i$ matrices:
\begin{align}
(\Gamma^I)^\alpha{}_\beta   \equiv (\Gamma^{[ij]})^\alpha{}_\beta \, , \qquad I=1,\dots, 10
\end{align}
which are in the {\bf{10}} of $\Gfour$. To compare with the usual supersymmetry algebra we can decompose into $SO(1,3)$ $\Gamma$ matrices:
\begin{align}
\Gamma^I=(\Gamma^{[a5]},\Gamma^{[cd]})  \, . \label{sl5gammas}
\end{align}
Similarly, the generalised momentum decomposes as:
\begin{align}
P_I=P_{[ij]}=(P_{a5}, \frac{1}{2} \epsilon_{abcd} Z^{cd}) \, .
\end{align}
We make the obvious identification of $P_{a5}=P_a$ as momenta in the usual four dimensional spacetime and the set $\{Z^{cd}\}$ as momenta in the novel extended directions.
Now we form a supersymmetry algebra using this set of generalised gamma matrices, $\{ \Gamma^I \}$, the set of generalised momenta $P_I$ and the supercharges, $Q_\alpha$. No central charges are required; the bosonic sector has only the generators of the generalised Poincar\'e group.
Thus the complete superalgebra is given by:
\begin{align}
\{Q_\alpha,Q_\beta\}= (C\Gamma^I)_{\alpha \beta} P_I \, ,     \qquad [Q_\alpha,P_I]=0   \, , \qquad [P_I,P_J]=0 \, .
\end{align}
Elementary states are irreducible representations of the Poincar\'e algebra and we may use the Casimirs of the algebra to label the representation. In this case it is the generalised Poincar\'e algebra that will be relevant to classify the states of the theory through its Casimirs.

We will proceed exactly as in the usual superalgebra case when one wishes to examine the massless representations i.e. where the quadratic Casimir of momentum vanishes, and show that they form ``short multiplets''.
As usual, the vacuum expectation value of the $\{ Q_\alpha, Q_\beta \}$ commutator, \refeqn{susynocentral} is positive definite and vanishes in a BPS state.
We calculate the square of this commutator.
This produces (suppressing spinor indices):
\be
\begin{split}
( \Gamma^I C^{-1} )\, (C \Gamma^J) P_I P_J & = 
2(   \eta_{ab} P^a P^b - \frac{1}{2}  Z_{ab} Z_{cd} (\eta^{ac}\eta^{bd}-\eta^{ad}\eta^{bc})   ) \mathbb{I}   
 \\ &\quad +  4 P^a Z_{ab} \Gamma^b  
\\ &\quad + 2 Z_{ab} Z_{cd}   \Gamma^{abcd}  \, . 
\end{split}
\label{susyaxiom}
\ee
For a BPS state, (the expectation value of) this vanishes.
Demanding that this is zero, we need each line on the right-hand side to vanish separately (as the matrices on the right-hand side are independent basis elements for four-by-four matrices).
This means the constraints in terms of the four dimensional momenta and central charges are:
\begin{align}
P_aP^a = \frac{1}{2}Z_{ab} Z^{ab} \, , \qquad P^a Z_{ab}= 0   \, ,  \qquad \,  Z_{ab} Z_{cd} \epsilon^{abcd} =0 \, .
\end{align}
The first term is the standard BPS condition \refeqn{BPS} requiring the mass be equal to the central charge and the second two equations are the quadratic constraints required for the state to be 1/2 BPS as calculated in \refref{Obers:1998fb} by essentially the same calculation.

In terms of the $\Gfour$ generalised momenta, $P_{[ij]}$ these equations become:
\bea
P_{[ij]} P^{[ij]}= P_I P^I=0 \,  , \\  \label{bpssl5}
\epsilon^{ijklm} P_{[ij]} P_{[kl]} =0  \,  . \label{scsl5}
\eea
The first equation \refeqn{bpssl5} implies that the state is massless from the point of view of the extended space Poincar\'e algebra. Thus in extended geometry the usual BPS states are massless. Supersymmetry works because the massless multiplet of SO(2,3) has the same number of degrees of freedom as the massive multiplet in SO(1,4).

The second equation \refeqn{scsl5} is precisely the {\it{physical section condition}} that we need to impose so that the local symmetry algebra of the extended geometry i.e. the algebra of generalised Lie derivatives, closes. 

Thus we see that from studying the representation of the supersymmetry algebra we reproduce the quadratic constraints on the generalised momenta. The foundation of this calculation has essentially already appeared in the literature in the context of U-duality multiplets for 1/2 BPS states \cite{Obers:1998fb}. What this calculation shows is the connection between: 1/2 BPS states in four dimensions, these states in the 10 dimensional extended space, the spinors of SO(2,3), the local Lorentz group of the extended space and their Clifford algebra, and representations of $\Gfour$ the global symmetry of the extended space. 

Finally, we now rewrite the supersymmetry algebra of the generalised space with no central charges  \refeqn{susynocentral}  in terms of a four dimensional quantities i.e. four-dimensional momenta and central charges as follows:
\begin{align}
\{Q_\alpha, Q_\beta \} = (C\Gamma^II)_{\alpha \beta} P_I  = (C  \Gamma^a \Gamma^5)_{\alpha \beta} P_a + \frac{1}{2} \eta_{abcd} ( C \Gamma^{ab})_{\alpha \beta}   Z^{cd}  \, . \label{5to4}
\end{align}
Now we wish to think of this in $\mathrm{SO}(1,3)$ language so that we can reinterpret the spinors as being Dirac spinors of $\mathrm{SO}(1,3)$. Obviously, the $\mathrm{Spin}(2,3)$ spinors also can be thought of as $\mathrm{Spin}(1,3)$ Dirac.
Crucially, the charge conjugation matrix will be different because of the presence of two time like directions in $\mathrm{SO}(2,3)$. Thus the four dimension $\mathrm{SO}(1,3)$ charge conjugation matrix which we denote by $C_{(4)}$ will be related to the $\mathrm{SO}(2,3)$ charge conjugation matrix by:
\begin{align}
C=C_{(4)} \Gamma^5
\end{align}
with $(\Gamma^5)^2= -1$.
We can then insert this into \refeqn{5to4} to give:
\bea
\{Q_\alpha, Q_\beta \} = (C\Gamma_I)_{\alpha \beta} P^I  &=& (C_{(4)} \Gamma^5 \Gamma^a \Gamma^5)_{\alpha \beta} P_a \nonumber
+\frac{1}{2} \eta_{abcd}  C_{(4)} \Gamma^5 \Gamma^{ab}_{\alpha \beta}   Z^{cd}  \nonumber \\
&=& (C_{(4)} \Gamma^a )_{\alpha \beta} P_a + (  C_{(4)}  \Gamma_{cd})_{\alpha \beta}   Z^{cd} \, ,
\eea
where we have used the elementary properties of four dimensional $\Gamma$ matrices:
\begin{align}
\{\Gamma^5,\Gamma^a \}=0 \, \, { \quad \rm{and}} \quad \Gamma^5 \Gamma^{ab} =- \frac{1}{2}\eta^{abcd} \Gamma_{cd} \, .
\end{align}
Thus we have seen how the usual four-dimensional supersymmetry algebra may be lifted to an algebra with global $\Gfour$ symmetry and no central charges. All the central charges arise from momenta in the new extended directions. The section condition of the extended geometry arises from considering the massless representations of this algebra.

For any dimension $d$, there is a coset $G/H$ structure with which one carries out the following process:

\begin{enumerate}

\item Construct spinor representations of $H$ and form the associated Clifford algebra, with with generators, $\Gamma^i$.

\item  A representation of $G$ may then be found from the set of antisymmetrised products of $\Gamma^i$ to give $\Gamma^I$.

\item The momenta and central charges may be combined to form a representation of $G$ which we lable as the generalised momenta $P_I$.

\item The superalgebra is then written in terms of the generalised momenta $P_I$ and the set of generalised gamma matrices $\Gamma^I$.

\item  Then, \refeqn{susyaxiom} is saturated for the {\it{massless}} representation which provides constraints on $P_I$.

\end{enumerate}

As a final comment, the reader might be puzzled by the fact that the section condition (in terms of the generalised momenta) is obeyed by 1/2 BPS objects, while generic solutions which are not necessarily supersymmetric do of course exist as solutions of SUGRA and hence DFT/ExFT. 
This simply reflects the fact that the extended coordinates are in correspondence with the winding modes of the 1/2 BPS objects themselves, which are the fundamental brane states of string and M-theory (as we will discuss further in the next subsection).

\subsection{Winding modes, central charges and extended momenta}
\label{windingcentral}

Let us now see how to connect the above discussions to the interpretation of the extra coordinates themselves.
First consider toroidal backgrounds where many issues are clearer. Let us begin with double field theory. The new coordinates that we introduced were $\tilde{x}_\mu$ are conjugate to string winding modes. That is the translation generators in $\tilde{x}_\mu$ are the generalised momenta $\tilde{P}^\mu$. This generalised momentum is the string winding around the $x^\mu$ direction. For a string described by worldsheet coordinates $X^\mu (\tau,\sigma)$ the winding is given by 
\begin{equation}
Q^\mu_w = \int \partial_\sigma X^\mu d \sigma \, . 
\end{equation}
This is also the central charge in the supersymmetry algebra. In $\mathbb{R}^d$ this charge would be formally infinite (as would be the energy) but the tension is still finite and so the solution is well defined. 

For exceptional field theory there are the momenta are conjugate to all possible brane windings, which in turn are related to the central charges in the supersymmetry algebra. For a membrane described by worlvolume coordinates 
$X^\mu(\tau, \sigma^1,\sigma^2)$, the membrane central charge is given by
\begin{equation}
Z^{\mu \nu}= \int \partial_{\sigma^i} X^\mu \partial_{\sigma^j} X^\nu \epsilon ^{0ij} d^2 \sigma \, ,
\end{equation}
which again is the membrane winding mode wrapping the $X^\mu$ and $X^{\nu}$ directions.
This membrane winding or central charge is equivalent to the generalised momenta $P^{[\mu \nu]}$ which is conjugate to novel coordinates $y_{\mu\nu}$ introduced in the exceptional field theory.
In fact a study of the equations of motion of the membrane world volume naturally introduces the winding mode current
\begin{equation}
j^{\mu \nu}(\tau, \sigma^i)= \partial_{\sigma^i}  X^\mu \partial_{\sigma^j} X^\nu \epsilon ^{0ij} \,
\end{equation}
and couples it in combination with the momentum $\partial_\tau X^\mu$ to the $\Gfour$ metric \cite{Duff:1990hn,Berman:2010is} though such approaches to the membrane theory have issues \cite{Duff:2015jka}.
Similar expressions also express the fivebrane winding current and relate to the central charge $Z^{\mu_1 \dots \mu_5}$.

The astute reader may worry that if the background is not toroidal then winding number is not conserved. This is is not necessarily a problem. Ordinary momentum is not conserved in curved space since there is a potential. The same is true with winding space and this novel momentum (conjugate to winding) need not be a conserved quantity in curved space.

\subsection{Worldvolume actions} 
\label{branewvol}

Having discussed DFT/ExFT and now their brane solutions, we should for completeness briefly discuss the realisation of worldvolume actions for branes probing the extended geometries of these theories. 
This is best developed for the so-called doubled string, where the worldsheet theory involves the doubled $2d$ target space coordinates which must satisfy a (twisted) self-duality constraint halving the degrees of freedom such that (locally) the formulation is equivalent to the standard worldsheet theory \cite{Duff:1989tf, Tseytlin:1990nb,Tseytlin:1990va, Siegel:1993bj, Siegel:1993th, Siegel:1993xq, Hull:2004in, Hull:2006va}. A doubled chiral world sheet theory was studied recently in the context of alpha prime corrections in \refref{Hohm:2013jaa}.

There is a satisfyingly complete story here where the worldsheet beta functional calculation for the doubled string gives the background field equations of DFT \cite{Berman:2007xn,Berman:2007yf,Avramis:2009xi,Copland:2011yh, Copland:2011wx}, while the zero mode fluctuations around the DFT wave/string solution give back the doubled worldsheet \cite{Berkeley:2014nza}. The string partition function for the doubled string has also been investigated in \refref{Berman:2007vi,Tan:2014mba}. Various other aspects of the world sheet theory in doubled space have been investigated in \refref{Blair:2013noa,DeAngelis:2013wba,Hatsuda:2014aza,Hatsuda:2014qqa,Bandos:2015cha,Hatsuda:2015cia,Chatzistavrakidis:2015vka,Driezen:2016tnz,Chatzistavrakidis:2018ztm,Hatsuda:2018tcx,Nikolic:2019wza,Hatsuda:2019xiz}.

The issue for other branes and for U-duality in general is that duality maps between branes of different dimensions (an exception being the 5-branes of type II theories, for which see \refref{Blair:2017hhy}).
This, and other issues, means that developing a similar theory of, say, an exceptional membrane is problematic \cite{Duff:1989tf, Duff:2015jka}.

One can perhaps deal with this by rethinking complete $\Edd$ covariance. 
From the point of view of the ExFT tensor hierarchy, we have a multiplet of $p$-branes coupling to the generalised gauge field in the representation $R_{p+1}$. 
To write the natural Wess-Zumino coupling one has to introduce a ``charge'' $q \in \bar R_{p+1}$, with $\mathcal{L}_{\text{WZ}} \sim q \cdot \mathcal{C}_{p+1} + \dots$ where $\mathcal{C}_{p+1} \in R_{p+1}$ is the ExFT $(p+1)$-form pulled back to the worldvolume.
The remaining terms denoted by ellipsis must be determined by gauge invariance and will involve lower rank ExFT $q$-forms with $q < p+1$.
In particular, however, we have the gauge transformation $\delta \mathcal{C}_{p+1} = \hat \partial \lambda$ with $\lambda \in R_{p+2}$, under which the lower rank forms are invariant. 
This motivates the imposition of a charge constraint:
\be
q \otimes  \partial\big|_{\bar R_{p+2}} = 0 
\label{magic}
\ee
such that $q \cdot ( \hat \partial \lambda) = 0$.
On choosing a solution of the section condition, we then have to solve this constraint, and this will select the allowed $p$-branes in the 10- or 11-dimensional theory.
Both the choice of section condition solution and the choice of $q$ break the global $\Edd$. 
On the other hand, when isometries are present we can view the resulting brane action as describing the $\Edd$ multiplet of $p$-branes in $n$-dimensions obtained by wrapping the 10- or 11-dimensional brane spectrum on a torus.
Elements of the above discussion are present in the approaches of \refref{Hatsuda:2012vm,Hatsuda:2013dya,Linch:2015fya,Linch:2015qva,Linch:2016ipx,Sakatani:2016sko,Sakatani:2017xcn,Arvanitakis:2017hwb,Sakatani:2017vbd,West:2018lfn,Arvanitakis:2018hfn,Blair:2019tww,Sakatani:2020umt} describing branes in (various versions of) exceptional geometry. 

For example, consider the ``exceptional string'' of \refref{Arvanitakis:2017hwb, Arvanitakis:2018hfn}, for $\Gfour$ where the ExFT two-form is $\mathcal{B}_{\mu\nu \fM}$.
The string charge is $q^{\fM}$ and the constraint \refeqn{magic} is just $q^{\fN} \partial_{\fM \fN} = 0$. 
In an M-theory solution of the section condition we have $\partial_{i5} \neq 0$, $i=1,\dots,4$ and one must have $q^{\fM} = 0$: there are no strings in 11-dimensions.
In IIA, we have $\partial_{45} = 0$, $\partial_{i5}\neq0$ for $i=1,2,3$ and now $q^{4} \neq 0$ picks out the fundamental string.
In IIB, with $\partial_{ij} \neq 0$, $i=1,2,$ we have an $\mathrm{SL}(2)$ doublet of non-zero charges, $q^\alpha = (q^4,q^5) \neq 0$, picking out the F1 string and the D1 brane.

%% file: sec7NonGeometryFromDuality.tex
In this section, we will describe simple examples of backgrounds which go beyond supergravity, and into the realm of non-geometry. 
We will firstly review non-geometric backgrounds defined using T-duality (for more details on which we recommend the review \refref{Plauschinn:2018wbo}) before showing how this generalises to U-duality.
After this initial encounter with such T- and U-folds, we will describe examples of so-called exotic branes in string theory and M-theory, and discuss their interpretation.
Throughout, we will see that the natural language with which to describe such non-geometries is provided by DFT and ExFT.

\subsection{Toy models of T-folds} 
\label{toyT}

We will describe a pedagogical example introduced in the context of work on string theory flux compactifications \cite{Dasgupta:1999ss,Kachru:2002sk,Hellerman:2002ax,Dabholkar:2002sy,Flournoy:2004vn,Shelton:2005cf}. 
Here it was realised that when one considers compactifications on geometries carrying flux of the $p$-form gauge fields, acting with duality transformations can lead to configurations which are ``non-geometric'' meaning that T- or U-duality are needed to globally define the space. 

This example is based on the NSNS sector and so can be efficiently described using aspects of double field theory, which we described in section \ref{genKK_DFT}.
In particular, we encode the physical geometry using the generalised metric, $\cH_{MN}$, or equivalently a generalised vielbein $E^A{}_M$.
The latter parametrises the coset $O(d,d)/O(d) \times O(d)$ element, and we have $\cH_{MN} = E^A{}_M E^B{}_N \delta_{AB}$, $E^A{}_M E^B{}_N \eta_{AB} = \eta_{MN}$ where $\eta_{AB}$ agrees numerically with $\eta_{MN}$ (this can be changed by choice of basis).
Conventionally we choose a geometric parametrisation involving a metric $g_{ij} = e^a{}_i e^b{}_j \delta_{ab}$ and $B$-field:
\be
E^A{}_M = \begin{pmatrix} e^a{}_i & 0 \\ B_{ij} e^j{}_a & e_a{}^i \end{pmatrix}\,,\quad
\cH_{MN} = 
\begin{pmatrix}
g_{ij} - B_{ik} g^{kl} B_{lj} & B_{ik} g^{jk} \\ 
- g^{ik} B_{kj} & g^{ij} 
\end{pmatrix} \,.
\label{EHnormal}
\ee
Given a choice of generalised vielbein, we can define an algebra 
\be
\mathcal{L}_{E_A} E^M{}_B = - F_{AB}{}^C E^M{}_C \,,
\label{oddfluxes}
\ee
of the type \refeqn{genpar} that we saw previously in section \ref{genSS} when we discussed generalised Scherk-Schwarz reductions.
The ``generalised fluxes'' $F_{AB}{}^C$ defined here will generally not be constant.
However if for some choice of globally defined generalised vielbein they are constant, then this choice gives a generalised parallelisation and can be used as the starting point of a generalised Scherk-Schwarz reduction.
Our example will admit such possibilities.

Note that after lowering an index via $F_{ABC} = F_{AB}{}^D \eta_{DC}$ we have $F_{ABC} = F_{[ABC]}$. 
For the parametrisation \refeqn{EHnormal}, the non-zero components of the generalised fluxes $F_{ABC}$ are: 
\be
F_{ab}{}^c =  - 2 e_j{}^c e^i{}_{[a|} \partial_i e^j{}_{|b]}\,,\quad
F_{abc}  = 3 e^i{}_a e^j{}_b e^k{}_c \partial_{[i} B_{jk]} \,.
\label{geofluxes}
\ee
Inspired by \refeqn{geofluxes}, let's give names to the components of $F_{AB C}$:
\be
F_{abc} = H_{abc} \,,\quad F_{ab}{}^c = f_{ab}{}^c \,,\quad
F_{a}{}^{bc} = Q_{a}{}^{bc} \,,\quad 
F^{abc} = R^{abc} \,.
\label{namefluxes}
\ee
Here $f_{ab}{}^c$ and $H_{abc}$ have natural geometric interpretations.  
The question is then what sort of backgrounds give rise to the $Q$- and $R$-fluxes.
By starting with a simple background consisting of a three-torus with $H$-flux, we will use T-duality to generate non-geometric backgrounds giving rise to these.
This duality chain may be summarised as:
\be
\begin{array}{clclclc}
 & T_c & & T_b & &  T_a &  \\
H_{abc} & \longrightarrow & f_{ab}{}^c & \longrightarrow & Q_{a}{}^{bc}&  \longrightarrow & R^{abc} \\
T^3 \,\text{with flux} & & \text{Twisted torus} & & \text{T-fold} & & \text{T-fold}\\
& &  & & \text{(global)} & & \text{(local)}\\
\end{array}
\label{fluxchain}
\ee
This is the harmonic oscillator of non-geometry. 
It does not in itself define a genuine solution of supergravity, but can be embedded in or upgraded to a solution in a number of ways.
In section \ref{exoticbranes} below, we will describe a duality chain starting with the NS5 brane which shares the key features of this educational example.
Alternatives are possible, e.g. one can embed the backgrounds appearing in \refeqn{fluxchain} in string theory by fibering them over a line interval \cite{Chaemjumrus:2019ipx,Chaemjumrus:2019wrt,Chaemjumrus:2019fsw}.

\subsubsection*{The three-torus with $H$-flux}

We take the metric and $B$-field to be:
\be
\begin{split}
ds^2 & = (dy^1)^2 + (dy^2)^2 + (dy^3)^2 \,,\\
B & = {\tfrac{h}{2\pi} y^3} dy^1 \wedge dy^2 \,,
\end{split}
\label{toyH}
\ee
where the constant $h \in \mathbb{Z}$. We assume that $y^1, y^2, y^3$ are all $2\pi$ periodic.
For $y^3 \rightarrow y^3 + 2\pi$, the $B$-field shifts by a gauge transformation: $B \rightarrow B + h dy^1 \wedge y^2$.
This can be viewed as a $B$-shift transformation realised as a geometric $O(2,2) \subset O(d,d)$ T-duality.
In terms of the $O(2,2)$ generalised metric describing $(g_{ij}, B_{ij})$ with $i,j=1,2$, we have
\be
\cH_{MN}(y^3 + 2\pi) = (\mathcal{P}_B)_M{}^P (\mathcal{P}_B)_N{}^Q \cH_{PQ} (y^3) \,,
\ee
where the monodromy matrix $\mathcal{P}_B \in O(2,2)$ is
\be
(\mathcal{P}_B)_M{}^N = \begin{pmatrix} 
\delta_i{}^j & \lambda_{ij} \\
0 & \delta^i{}_j 
\end{pmatrix} \,,
\quad
\lambda_{ij} = h \begin{pmatrix} 0 & - 1 \\ 1 & 0 \end{pmatrix} \,.
\label{Hmono}
\ee
The field strength of the two-form is constant,
\be
H_3 = \tfrac{h}{2\pi} dy^1 \wedge dy^2 \wedge dy^3 \,,
\ee
and so this space can be said to carry $H$-flux, $H_{abc} = $ constant.

\subsubsection*{The twisted-torus with $f$-flux}

We can T-dualise the background \refeqn{toyH} on the $y^1$ direction. This leads to the twisted torus:
\be
\begin{split}
ds^2 & = ( d y^1 - {\tfrac{h}{2\pi} y^3} dy^2 )^2  + (dy^2)^2 + (dy^3)^2 \,,
\end{split}
\label{toyF}
\ee
with vanishing $B$-field.
For $y^3 \rightarrow y^3 + 2\pi$, we identify the coordinates as $y^1 \rightarrow y^1 + h y^2$, which is the eponymous ``twist''.
We can view this coordinate patching as a geometric $\mathrm{GL}(2) \subset O(2,2) \subset O(d,d)$ transformation.
This is an example of group manifold, admitting a globally well-defined vielbein via the left-invariant one-forms
\be
u^1 = d y^1 - \tfrac{h}{2\pi} y^3 dy^2 \,,\quad u^2 = dy^2 \,,\quad u^3 = dy^3
\,,
\ee
obeying the Maurer-Cartan equation \be
du^a = \frac{1}{2} f_{bc}{}^a u^b \wedge u^c\,,\quad
f_{23}{}^1 = \tfrac{h}{2\pi} \,,
\ee
defining the geometric flux $f_{ab}{}^c$.

\subsubsection*{The global T-fold with $Q$-flux}

Next, we T-dualise on the direction $y^2$.
This leads to the following space:
\be
\begin{split}
ds^2& = \frac{1}{1+\left({\tfrac{h}{2\pi} y^3}\right)^2} \left( ( d y^1 )^2  + (d y^2)^2 \right) + (dy^3)^2 \,,\\
B_{12} & = - \frac{{\tfrac{h}{2\pi} y^3}}{1+\left({\tfrac{h}{2\pi} y^3}\right)^2}\,.
\end{split}
\label{toyQ}
\ee
For $y^3 \rightarrow y^3 + 2 \pi$ there is no global geometric transformation (combination of $\mathrm{GL}(2)$ coordinate transformations and $B$-field gauge transformations) that makes this space well-defined.
Instead, the metric and $B$-field transform under a non-geometric $O(d,d)$ transformation. 
Hence \refeqn{toyQ} is a non-geometric background, or T-fold.
The global patching is most simply described using the $O(2,2)$ generalised metric which obeys:
\be
\cH_{MN}(y^3 + 2\pi) = (\mathcal{P}_\beta)_M{}^P (\mathcal{P}_\beta)_N{}^Q \cH_{PQ} (y^3) \,,
\ee
where the monodromy matrix $\mathcal{P}_\beta \in O(2,2)$ is
\be
(\mathcal{P}_\beta)_M{}^N = \begin{pmatrix} 
\delta_i{}^j & 0 \\
\lambda^{ij} & \delta^i{}_j 
\end{pmatrix} \,,
\quad
\lambda^{ij} = h \begin{pmatrix} 0 & - 1 \\ 1 & 0 \end{pmatrix} \,.
\label{Qmono}
\ee
In a doubled description of this background, this non-geometric transformation identifies the physical coordinates $y^1$ and $y^2$ with their T-dual coordinates $\tilde y^1$ and $\tilde y^2$ as $y^1 \rightarrow y^1 + h \tilde y^2$, $y^2 \rightarrow y^2 - h \tilde y^1$. Hence one cannot globally distinguish the physical and dual coordinates.

In a T-fold, the metric and $B$-field are not good variables with which to describe the background.
We should instead use the doubled description, or else define new spacetime variables.
Comparing the form of the transformation \refeqn{Qmono} with that of the $H$-flux background, \refeqn{Hmono}, suggests to not use a $B$-field, but instead an alternative parametrisation of the generalised metric based on a dual metric $\tilde g_{ij} = \tilde e^a{}_i \tilde e^b{}_j \delta_{ab}$ and a \emph{bivector} $\beta^{ij} = - \beta^{ji}$:
\be
E^A{}_M = \begin{pmatrix} 
\tilde e^a{}_i & \beta^{ij} \tilde e^a{}_j \\
 0 & \tilde e^i{}_a 
 \end{pmatrix} \,,
 \quad
\mathcal{H}_{MN} = 
\begin{pmatrix}
\tilde g_{ij} & - \tilde g_{ik}\beta^{kj} \\
\beta^{ik} \tilde g_{kj} & \tilde g^{ij} - \beta^{ik} \tilde g_{kl} \beta^{lj}
\end{pmatrix}\,.
\label{EHbiv}
\ee
Equating the generalised metrics of \refeqn{EHnormal} and \refeqn{EHbiv} (the doubled vielbein are related by a local $O(d) \times O(d)$ transformation) gives the transformation from what we might call the $B$-field frame to the bivector frame, which can be most compactly expressed as: 
\be
g+B=(\tilde g^{-1} + \beta)^{-1}\,.
\ee
This can be viewed as a field redefinition in supergravity, leading to what is called $\beta$-supergravity \cite{Andriot:2011uh,Andriot:2012wx, Andriot:2013xca, Blumenhagen:2013aia}.

For the T-fold \refeqn{toyQ}, we find
\be
\begin{split} 
\widetilde{ds^2} & = (dy^1)^2 + (dy^2)^2 + (dy^3)^2 \,,\\
\beta^{12} & = \tfrac{h}{2\pi} y^3 \,,
\end{split}  
\label{toyQbi}
\ee
which makes it apparent that the non-geometric transformation \refeqn{Qmono} is a shift of the bivector.
Writing the background as in \refeqn{toyQbi} in a sense ``undoes'' the effect of the two T-dualities we have performed, as functionally it is the same as the original background \refeqn{toyH} with the indices shifted.

We can compute the fluxes for the frame \refeqn{EHbiv} using the general definition \refeqn{oddfluxes}, finding for the components:
\be
\begin{split}
H_{abc} & = 0 \,,\\
f_{ab}{}^c & = - 2 e^c{}_j e^{i}{}_{[a|} \partial_i e^j{}_{|b]} \,,\\
Q_{a}{}^{bc} & = e^i{}_a e^{[b}{}_j e^{c]}{}_k \left( \partial_i \beta^{jk} +  2 \beta^{kl} e^d{}_i \partial_l e^j{}_d \right) \,,\\ 
R^{abc} & = - 3 e^{[a}{}_j e^{b}{}_{k} e^{c]}{}_l \beta^{i j} \partial_i \beta^{kl} \,.
\end{split}
\ee
This defines the $Q$-flux for the background \refeqn{toyQbi} to be
\be
Q_i{}^{jk} = \partial_i \beta^{jk} \,,
\ee
with $Q_3{}^{12} = \tfrac{h}{2\pi}$.
Hence in this T-fold, working with the alternative generalised vielbein \refeqn{EHbiv} is necessary to produce the expected generalised parallelisation \refeqn{oddfluxes} in which the $Q$-flux appears.

\subsubsection*{The local T-fold with $R$-flux}

Technically, we should not T-dualise on the $y^3$ direction.
A naive application of the Buscher rules would produce:
\be
\begin{split}
ds^2& = \frac{1}{1+\left({\tfrac{h}{2\pi} \tilde y^3}\right)^2} \left( ( d y^1 )^2  + (d y^2)^2 \right) + (dy^3)^2\,, \\
B_{12} & = - \frac{{\tfrac{h}{2\pi} \tilde y^3}}{1+\left({\tfrac{h}{2\pi} \tilde  y^3}\right)^2}\,,
\end{split}
\ee
which depends on $\tilde y^3$, the dual to the physical coordinate $y^3$ of this background ($\tilde y^3$ is what we called $y^3$ in the previous background).
Alternatively, this is described by a flat metric and a bivector $\beta^{12} = \tfrac{h}{2\pi} \tilde y^3$.
A schematic definition of a non-geometric $R$-flux is
\be
R^{ijk} = 3 \tilde \partial^{[i} \beta^{jk]} \,,
\ee
using a derivative $\tilde \partial^i = \frac{\partial}{\partial \tilde y_i}$ with respect to the dual coordinates.
In this case, we clearly have $R^{123} =\tfrac{h}{2\pi}$.
This ``completes'' the duality chain starting with the three-torus with $H$-flux.

We can make sense of such backgrounds within DFT, as the dependence on dual coordinates can be accommodated via generalised Scherk-Schwarz reductions, similarly to the ExFT examples we discussed in section \ref{genSS}.

An important disclaimer here is that although in this toroidal example, non-geometry led to the $Q$- and $R$-fluxes, it is also possible to have such fluxes present in geometric compactifications of ordinary geometries. For instance, compactifications on spheres with fluxes can lead to $Q$-flux. 
When we refer to a flux as being ``non-geometric'' it should therefore be interpreted as such only when the background producing it is also known.

\subsection{Toy models of U-folds} 
\label{toyU}

The generalisation of this toy model to U-duality was written down in \refref{Blair:2014zba}.
We now start with a four-torus carrying flux of the four-form field strength of the M-theory three-form.
We will then act with U-dualities in the torus directions. The transformation we use to generate new backgrounds is that of \refeqn{UIJK}, acting on three directions of this $T^4$ while also, unlike in T-duality, having a non-trivial effect on the remaining eight directions. 
For this reason, we write the full 11-dimensional backgrounds below.

Although it is possible to carry out the dualities solely using $\Gthree$ U-duality transformations, as we are dealing with a four-dimensional torus it is more convenient to use the larger $\Gfour$ duality group.
We will consider backgrounds consisting of a product of a seven-dimensional ``external'' space, with coordinates labelled by $\mu,\nu,\dots$, and a four-dimensional ``internal'' space, with coordinates labelled by $i,j,\dots$. The three-form will only have non-zero components in the internal directions.
We can write these backgrounds as:
\be
\begin{split}
ds^2 &= G_{\mu\nu} dx^\mu dx^\nu + g_{ij} dy^i dy^j \,,\\
C_{(3)} & = \frac{1}{3!} C_{ijk} dy^i \wedge dy^j \wedge dy^k \,,
\end{split}
\ee
where $\mu = 0,5,\dots,10$ indexes the non-compact directions and $i=1,\dots,4$ indexes the toroidal directions.
As we have seen in section \ref{genKK_EFT} and again in section \ref{action5}, the metric $g_{ij}$ and three-form $C_{ijk}$ combine into a unit determinant $5 \times 5$ symmetric matrix, $m_{\cal{M} \cal{N}}$, in the coset $\Gfour / \Hfour$. We decompose the 5-dimensional index as $\mathcal{M} = (i, 5)$ and parametrise it as:
\be
m_{\cal{M} \cal{N}}
=\begin{pmatrix} 
g^{-2/5} g_{ij} & - g^{-2/5} g_{ik} C^k \\
- g^{-2/5} g_{jk} C^k & g^{3/5} + g^{-2/5} g_{kl} C^k C^l 
\end{pmatrix} \,,\quad
C^i \equiv \frac{1}{3!} \epsilon^{ijkl} C_{jkl} \,,
\label{5gm}
\ee
where $\epsilon^{1234}=1$ denotes the alternating symbol and $g \equiv \det g$.
This is equivalent to choosing a generalised vielbein
\be
E^{\fA}{}_{\fM} 
 =\begin{pmatrix}
 e^{-2/5} e^a{}_i & -  e^{-2/5} e^a{}_j C^j\\
0 & e^{3/5} 
\end{pmatrix} \,.
\ee
The generalised metric transforms as $m \rightarrow U m U^T$ for $U \in \Gfour$.
In addition, 
\be
g^{\text{ExFT}}_{\mu\nu} = g^{1/5} G_{\mu\nu} 
\label{5gext}
\ee
is invariant under $\Gfour$.
In principle the 7-dimensional metric $G_{\mu\nu}$ can be an arbitrary $y$-independent metric, its form will play no role in the following (as before this is a toy model and we do not yet concern ourselves with whether it is a valid supergravity solution).

The U-duality transformation \refeqn{UIJK} acting on (say) the $1,2,3$ directions is realised as a permutation of the $\mathcal{M} = 5$ index with the $\mathcal{M}=4$ index:
\be
U_{\cal{M}}{}^{\cal{N}} = \begin{pmatrix}
\delta_i^j - \delta_i^4 \delta^j_4 & \delta^i_4 \\
- \delta_i^4 & 0 
\end{pmatrix} \,,
\label{5UBuscher}
\ee
where $i=1,2,3,4$.
This acts as the identity on the $1,2,3$ components and as an $\mathrm{SL}(2)$ inversion in the $4,5$ components.

\subsubsection*{The four-torus with $F$-flux}

The background with four-form flux is: 
\be
\begin{split}
ds^2 & =  G_{\mu\nu} dx^\mu dx^\nu  + \delta_{ij} dy^i dy^j \,,\\
C_{123} & = - \tfrac{h}{2\pi} y^4 \,.
\end{split}
\label{toyUF}
\ee
For $y^4\rightarrow y^4+2\pi$, the three-form shifts by $-h$, with $h\in \mathbb{Z}$.
The four-form flux is
\be
F_{1234} = \tfrac{h}{2\pi} \,.
\ee

\subsubsection*{The global U-fold with $Q$-flux}

We perform a U-duality transformation on the $123$ directions. This U-duality reduces to a pair of Buscher T-dualities. It therefore ``skips a step'' in the T-duality chain, and immediately produces a non-geometric background:
\be
\begin{split}
ds^2& = \left( 1 +  \left( \tfrac{h}{2\pi} y^4 \right)^2 \right)^{1/3} ( G_{\mu\nu} dx^\mu dx^\nu + (dy^4)^2)
\\ & \qquad
+ \left(1 +  \left( \tfrac{h}{2\pi} y^4 \right)^2\right)^{-2/3}( (dy^1)^2 + (dy^2)^2 + (dy^3)^2) \,,\\
C_{123} & =   \frac{\tfrac{h}{2\pi} y^4 }{1+ \left( \tfrac{h}{2\pi} y^4 \right)^2} \,.
\end{split}
\label{toyUQ}
\ee
This is a U-fold. 
For $y^4 \rightarrow y^4+ 2\pi$ the $\Gfour$ generalised metric transforms under a non-geometric U-duality transformation:
\be
m_{\cal{M} \cal{N}} (y^4+2\pi) 
= ( {U}_\Omega )_{\cal{M}}{}^{\cal{P}}
 ( {U}_\Omega )_{\cal{N}}{}^{\cal{Q}}
 m_{\cal{P}\cal{Q}}(y^4)\,,
\ee
with
\be
( {U}_\Omega )_{\cal{M}}{}^{\cal{N}}
= 
\begin{pmatrix}
 \delta_i{}^j & \Omega_i  \\
 0 & 1 
\end{pmatrix} \,,
\quad
\Omega_i = h \delta_i^4 \,. 
\ee
Now, instead of describing this geometry using a three-form, we can introduce a totally antisymmetric trivector $\Omega^{ijk}$ via the alternative parametrisation
\be
m_{\cal{M} \cal{N}} 
= 
\begin{pmatrix}
\tilde g^{-2/5} \tilde g_{ij} + \tilde g^{3/5} \Omega_i \Omega_j & \tilde g^{3/5} \Omega_i \\
 \tilde g^{3/5} \Omega_i & \tilde g^{3/5}
\end{pmatrix} 
\,,\quad
\Omega_i \equiv \frac{1}{3!} \epsilon_{ijkl} \Omega^{jkl}\,,
\ee
corresponding to the alternative generalised vielbein
\be
E^{\fA}{}_{\fM} 
 =\begin{pmatrix}
 \tilde e^{-2/5} \tilde e^a{}_i & 0\\
\tilde e^{3/5} \Omega_i  & \tilde e^{3/5} 
\end{pmatrix} \,.
\ee
The background \refeqn{toyUQ} in these terms is:
\be
\begin{split}
ds^2& = G_{\mu\nu} dx^\mu dx^\nu + \delta_{ij} dy^i dy^j \,,\\
\Omega^{123} & =  -\tfrac{h}{2\pi} y^4  \,.
\end{split}
\label{toyUQtri}
\ee
We see that for $y^4 \rightarrow y^4 + 2\pi$, we have a non-geometric U-duality transformation producing a shift of the trivector, $\Omega^{123} \rightarrow \Omega^{123} - h$.
Furthermore, we can define an M-theory Q-flux, of the form (note there will be additional contributions in a background with a non-trivial metric \cite{Blair:2014zba})
\be
Q_l{}^{ijk} = \partial_l \Omega^{ijk} \,.
\ee
The U-duality chain is analogous to the T-duality one:
\be
F_{1234} \xrightarrow{U_{123}} Q_4{}^{123} \,.
\ee

\subsubsection*{The twisted-torus with $f$-flux}

Now let's consider a twisted torus in M-theory. This is the same twisted torus that we saw before, times an additional circular direction. Let's write the metric as
\be
ds^2 = G_{\mu\nu} dx^\mu dx^\nu + \left( dy^4 - \tfrac{h}{2\pi}y^2 d y^1 \right)^2 + (dy^1)^2 + (dy^2)^2 + (dy^3)^2 \,,
\label{toyUf}
\ee
with
\be
f_{12}{}^4 = \tfrac{h}{2\pi} \,.
\ee

\subsubsection*{The local U-fold with $R$-flux}

We want to U-dualise the background \refeqn{toyUf} on three of the four toroidal directions.
The isometry directions are $y^1, y^3, y^4$. U-duality on these directions in fact produces a twisted torus of the same form. This is uninteresting.
In order to find a non-geometric background, we should instead consider what would happen if we apply the U-duality rules to the directions $y^1, y^2, y^3$, despite the fact that $y^2$ is not an isometry. 
The result of doing so is:
\be
\begin{split}
ds^2 & = \left( 
1 + \left( \tfrac{h}{2\pi} \tilde y^2 \right)^2
\right)^{1/3} \left( G_{\mu\nu} dx^\mu dx^\nu + (dy^1)^2 \right) 
\\ & \qquad
+ \left( 
1 + \left( \tfrac{h}{2\pi} \tilde y^2 \right)^2
\right)^{-2/3} \left(
(dy^2)^2 + (dy^3)^2 + (dy^4)^2
\right)\,,\\
C_{234} & = - \frac{\tfrac{h}{2\pi} \tilde y^2 }{1 + \left( \tfrac{h}{2\pi} \tilde y^2 \right)^2} \,.
\end{split}
\label{toyUR}
\ee
This geometry now depends on $\tilde y^2$ (which is what we called $y^2$ in \refeqn{toyUf}), which is a dual coordinate.
Hence this is a locally non-geometric U-fold.

In the $\Gfour$ ExFT, we have four physical coordinates $y^i$ along with six dual coordinates conjugate to M2 brane wrappings, denoted by $y_{ij} = - y_{ji}$.
Together these transform in the $\mathbf{10}$ of $\Gfour$, which is the antisymmetric representation.
Let $Y^{\cal{M} \cal{N}} = - Y^{\cal{N}\cal{M}}$ denote these extended coordinates, with $Y^{i 5} = y^i$ and $Y^{i j} = \frac{1}{2} \epsilon^{ijkl} y_{kl}$. 

Here we have $Y^{\cal{M} \cal{N}}_{\text{U-fold}} = (U^{-T})^{\cal{M}}{}_{\cal{K}}  (U^{-T})^{\cal{N}}{}_{\cal{L}} Y^{\cal{K} \cal{L}}_{\text{Torus}}$.
So the physical coordinates appearing in \refeqn{toyUR} are $y^1_{\text{U-fold}} = - y^{14}_{\text{Torus}}$, $y^2_{\text{U-fold}} = - y^{24}_{\text{Torus}}$, $y^3_{\text{U-fold}} = - y^{34}_{\text{Torus}}$, $y^4_{\text{U-fold}} = y^{4}_{\text{Torus}}$, while the dual coordinate $\tilde y^2 \equiv y^2_{\text{Torus}} = y^{24}_{\text{U-fold}}$. The latter is then the membrane winding coordinate $y_{31}$ in the U-fold background.

To establish a definition of a non-geometric flux, we use the parametrisation in terms of a trivector:
\be
\begin{split}
\widetilde{ds^2} & = G_{\mu\nu} dx^\mu dx^\nu + \delta_{ij} dy^i dy^j \,,\\
\Omega^{234} & = \tfrac{h}{2\pi} y_{31} \,,
\end{split}
\label{toyURtri}
\ee
and define an M-theory R-flux (again, note there will be additional contributions in a background with a non-trivial metric \cite{Blair:2014zba})
\be
R^{i;jklm} \equiv  4 \partial^{i[j}  \Omega^{klm]}\,,
\ee
using derivatives with respect to the M2 winding coordinates.
For the background \refeqn{toyURtri}, this evaluates to
\be
R^{3;1234} = \tfrac{h}{2\pi}\,.
\ee

%% file: sec8ExoticBranes.tex
\subsection{Exotic branes}
\label{exoticbranes}

The ``standard'' branes of string theory and M-theory are those that couple to the $p$-form gauge fields of supergravity.
In string theory, in the NSNS sector, this is the fundamental string itself plus its electromagnetic dual, the NS5 brane, while in the RR sector these are the D-branes. 
In M-theory, we have the M2 and the M5.
If a compact direction is present in spacetime, we can include the Kaluza-Klein monopole (magnetically charged under the KK vector, as we have seen in section \ref{kksolns}), and pp-waves as their electromagnetic duals. The Kaluza-Klein monopole (or KKM for short) is then the T-dual of the NS5 brane, while the fundamental string solution in supergravity is T-dual to pp-waves. 
Meanwhile, the M-theory Kaluza-Klein monopole unites both the IIA D6 brane and Kaluza-Klein monopole. 

Wrapping this brane spectrum on tori, and acting with duality transformations, points to the existence of further ``exotic'' branes, as was realised long ago in \refref{Elitzur:1997zn, Blau:1997du, Hull:1997kb,Obers:1997kk,Obers:1998fb}.

What makes these branes exotic? 
Firstly, they cannot exist in decompactified 10- or 11-dimensional spacetime, but require (possibly many) compact isometry directions to be present.
In this, they are direct generalisations of the Kaluza-Klein monopole.
Secondly, their tensions depend on the string coupling to powers of $g_s^{-n}$ with $n \geq 2$. 
Hence, they are very non-perturbative.
Thirdly, their supergravity solutions -- which can again be obtained via duality -- are not globally well-defined but are \emph{non-geometric}.
Here duality transformations are needed to make sense of the solution globally. 

We will now describe some important examples of such exotic branes.
We will follow very similar steps to the toy models of the previous section (one can think of the solutions below as being fully backreacted versions of the toy models we have written down above).

\subsubsection{Example: the $5_2^2$}

We begin with the duality chain that connects the NS5 brane to the Kaluza-Klein monopole and then to the exotic $5_2^2$ brane.

\subsubsection*{The NS5 brane}

The NS5 brane is the SUGRA solution carrying magnetic $H$-flux, and thus is analogous to the torus with $H$-flux of section \ref{toyT}.
The NS5 solution is:
\be
\begin{split}
ds^2 & = dx^2_{012345} + f dx_{6789}^2 \,,\\
H_{abc} & = \epsilon_{abcd} \delta^{de} \partial_e f \,,\\
e^{-2\phi} & = f^{-1} \,.
\end{split} 
\label{NS5}
\ee
The fields depend on the four transverse coordinates $x^a = (x^6,x^7, x^8, x^9)$ via the function
\be
f(x^a) = 1 + \frac{q_{\text{NS5}}}{r^2}\,,\quad
r  =\sqrt{ (x^6)^2 + (x^7)^2 + (x^8)^2 + (x^9)^2 } \,,
\ee
which is harmonic except at the origin $r=0$ where the brane is situated (it extends in the remaining six directions). 
We have condensed our notation for the line element to write $dx^2_{01\dots n} \equiv -(dx^0)^2 + (dx^1)^2 + \dots (dx^n)^2$ for the Minkowski metric in Cartesian coordinates (and similarly for the Euclidean).
This solution is magnetically charged under the $B$-field, as measured by integrating the field strength $H_{(3)}$ over a transverse three-sphere at infinity surrounding the brane,
\be
Q_{\text{mag}} \sim \int H_{(3)} \sim q_{\text{NS5}} \,.
\ee
The transverse space of the solution \refeqn{NS5} is $\mathbb{R}^4$.
In order to T-dualise, we need a (translational) isometry. 
We can obtain a form of the NS5 solution with a dualisable isometry by compactifying the $x^9$ direction and smearing the brane on this direction.
Technically this can be performed by arraying an infinite number of NS5 branes at intervals of $2\pi R_{9}$ along the $x^9$ direction and resumming. 
The resulting smeared solution is:
\be
\begin{split}
ds^2 & = dx^2_{012345} + f \left( dx_{678}^2 + (dx^9)^2 \right) \,,\\
H_{ij9} & = \epsilon_{ijk} \delta^{kl} \partial_l f \,,\\
e^{-2\phi} & = f^{-1} \,.
\end{split} 
\label{NS5smeared}
\ee
where now $x^i = (x^6, x^7, x^8)$ labels the non-compact transverse coordinates, and the harmonic function becomes:
\be
f(x^i) = 1 + \frac{q_{\text{KKM}}}{r}\,,\quad
r  \equiv \sqrt{ (x^6)^2 + (x^7)^2 + (x^8)^2  } \,,
\label{fsmeared}
\ee
where we anticipate the next brane we will come to by denoting $q_{\text{KKM}} = \tfrac{q_{\text{NS5}}}{2R_9}$.
We can pick a gauge such that the $B$-field has components $B_{i9} \neq 0$, $B_{ij} = 0$.

\subsubsection*{The Kaluza-Klein monopole (KKM)}

Now we can T-dualise \refeqn{NS5smeared} on the $x^9$ direction. This gives the string theory Kaluza-Klein monopole solution, which is purely metric and is analogous to the ``twisted torus'' with geometric flux of section \ref{toyT}.
The KKM solution is:
\be
ds^2 = dx^2_{012345} + f dx^2_{678} + f^{-1} ( dx^9 + A_i dx^i )^2 \,,
\label{KKM}
\ee
where the field strength of the ``Kaluza-Klein vector'' $A_i$ obeys
\be
F_{ij}  = 2 \partial_{[i} A_{j]} = \epsilon_{ijk} \delta^{kl} \partial_l f \,,
\ee
with $f$ as in \refeqn{fsmeared}. 

Again, there are 6 worldvolume directions. However we make a distinction between the three non-compact transverse directions $x^6,x^7,x^8$ and the other direction $x^9$, which is called the \emph{special isometry} direction. The transverse space of the KKM is then $\mathbb{R}^3 \times S^1$, and in an entirely non-compact background this solution does not exist.

In order to dualise further, we can compactify the $x^8$ direction and smear.
This is a less well-defined procedure, because the result is of codimension-2 and the resulting harmonic function is a logarithm.
Hence the solution is not asymptotically flat. This is not really a problem: in order to deal with the divergence in the harmonic function we need to view the codimension-2 solution as being a local approximation to a configuration involving objects of opposite charge which have the effect of cancelling out the flux of the apparently problematic solution, and allow for well-defined asymptotics. We will see examples of how this can be done later on.

The smeared KKM can be written as:
\be
ds^2 = dx^2_{012345} + f dx^2_{67} + f (dx^8)^2 + f^{-1} ( dx^9 + A_i dx^i )^2 \,,
\label{KKMsmeared}
\ee
where $F_{ij}$ has components 
\be
F_{\alpha 8}  = 2 \partial_{[\alpha} A_{8]} = - \epsilon_{\alpha \beta} \delta^{\beta \gamma} \partial_\gamma f \,,
\ee
with $x^\alpha =(x^6,x^7)$ denoting the remaining non-compact transverse directions, and
\be
f = - \tilde{q} \log r \,,\quad r = \sqrt{(x^6)^2 + (x^7)^2} \,.
\ee
Here after the resummation we have denoted $\tilde{q} = \tfrac{q_{\text{KKM}}}{\pi R_8} = \tfrac{q_{\text{NS5}}}{2\pi R_8 R_9}$.
If we adopt polar coordinates $(x^6,x^7) = ( r \cos \theta, r \sin \theta)$ then we can take $A_8 = -\tilde{q}\theta$, $A_\alpha=0$.

\subsubsection*{The $5_2^2$ brane} 

T-dualising the smeared KKM \refeqn{KKMsmeared} on $x^8$ we arrive at the following background:
\be
\begin{split}
ds^2 & = dx^2_{012345} + f (dr^2 + r^2 d\theta^2) + \frac{f}{f^2 + (\theta \tilde{q})^2} ( (dx^8)^2 + (dx^9)^2 ) \,,\\
B_{89} & = - \frac{\theta \tilde{q}}{f^2 + (\theta \tilde{q})^2} \,,\\
e^{-2\phi} & = \frac{f^2 + (\theta \tilde{q})^2}{f} \,.
\end{split}
\label{522}
\ee
This is non-geometric; it is a T-fold. It is analogous to the solution with $Q$-flux in the toy model.
As we encircle the brane in the two-dimensional non-compact space, $\theta \rightarrow \theta +2 \pi$, and the metric and $B$-field transform into themselves by a non-geometric $O(2,2)$ transformation acting as a bivector shift. 
This can be seen by writing the generalised metric describing the components of the metric and $B$-field in the $(x^8,x^9)$ directions:
\be
\begin{split}
\mathcal{H}_{MN}& = 
\begin{pmatrix} f^{-1} I_2 & - f^{-1} {\theta} \tilde{q} \epsilon
\\ f^{-1} {\theta} \tilde{q} \epsilon & ( f + f^{-1} ({\theta} \tilde{q})^2 )I_2
\end{pmatrix} 
= 
\begin{pmatrix} 
I_2 & 0 \\ 
{\theta} \tilde{q}\epsilon & I_2 
\end{pmatrix}
\begin{pmatrix}
f^{-1} I_2 & 0 \\ 0 & f I_2
\end{pmatrix}
\begin{pmatrix}
I_2 & - {\theta} \tilde{q} \epsilon \\ 
0 & I_2
\end{pmatrix}
\end{split}
\ee
where
\be
\epsilon = \begin{pmatrix} 0 & 1 \\ -1 & 0 \end{pmatrix} \,.
\ee
The factorisation shows, using the parametrisation \refeqn{EHbiv}, that in terms of a bivector, this background is: 
\be
\begin{split}
ds^2 & = dx^2_{012345} + f (dr^2 + r^2 d\theta^2) + f ( (dx^8)^2 + (dx^9)^2 ) \,,\\
\beta^{89} & =\theta \tilde{q} \,,\\
e^{-2\phi} & = f^{-1} \,.
\end{split}
\label{522bi}
\ee
The dilaton here is determined by the invariance of the $\sqrt{g}e^{-2\phi}$ measure.
We see clearly that the bivector shifts for $\theta \rightarrow \theta + 2\pi$, and that one could define a $Q$-flux with components $Q_\theta{}^{89} = \tilde{q}$.

This brane is known as the $5_2^2$ brane, using the notation explained in section \ref{branename} below.
It can be thought of as a generalisation of the KKM requiring \emph{two} special isometry directions to be defined. For that reason, it could be called a generalised Kaluza-Klein monopole.

As with the duality chain starting with the three-torus with H-flux, we now reach a point where we have no further isometries in which to take T-duality.
A T-duality on the angular direction $\theta$ would formally generate a locally non-geometric solution, analogous to the $R$-flux configuration we saw previously. 
We could alternatively have started by smearing on three transverse directions, rather than two, generating a sequence of codimension-1 solutions (domain walls) leading after three T-dualities to a further codimension-1 exotic brane.
However, we are going to concentrate mostly codimension-2 exotic branes in this review.

\subsubsection{Example: the $5^3$}

This is an example of an exotic brane that is present in M-theory.
We will present it via a U-duality chain that can be viewed as the uplift of the NS5 to $5_2^2$ transformation in type IIA. 

\subsubsection*{The M5 brane} 

The M5 brane solution in SUGRA is:
\be
\begin{split}
ds^2 & = f^{-1/3} dx^2_{012345} + f^{2/3} dx_{6789 \#}^2 \,,\\
F_{abcd} & = \epsilon_{abcde} \delta^{ef} \partial_f f\,.\\
\end{split} 
\label{M5}
\ee
The fields depend on the five transverse coordinates $x^a = (x^6,x^7, x^8, x^9, x^{\#})$ via the function
\be
f(x^a) = 1 + \frac{q_\text{M5}}{r^3}\,,\quad
r  =\sqrt{ (x^6)^2 + (x^7)^2 + (x^8)^2 + (x^9)^2 + (x^{\#})^2 } \,,
\ee
which is harmonic except at the origin $r=0$ where the brane is situated (it extends in the remaining six directions). 
This solution measures carries magnetic charge of the 11-dimensional three-form, as measured by integrating the field strength $F_{(4)}$ over a transverse four-sphere at infinity surrounding the brane,
\be
Q_{\text{mag}} \sim \int F_{(4)} \sim q_{\text{M5}} \,.
\ee
The transverse space of the solution \refeqn{M5} is $\mathbb{R}^5$.
In order to U-dualise, we compactify on three transverse directions, say the $x^8, x^9, x^{10}$ directions.
This immediately gives us a solution of codimension-2, with transverse space $\mathbb{R}^2 \times T^3$.
We again defer dealing with issues with divergences and write this as:
\be
\begin{split}
ds^2 & = f^{-1/3}(r) dx^2_{012345} + f^{2/3} (r) (dr^2 + r^2 d\theta^2) + f^{2/3}(r) dx^2_{89 \#} \,,\\
C_{8 9 \#} & = - \tilde q_\text{M5} \theta \,,
\end{split} 
\label{M5smeared}
\ee
where $\tilde q = q_{\text{M5}} / 2 \pi^2 R_{8} R_9 R_{\#}$.
Here we switched to polar coordinates in the non-compact transverse directions, $(x^6, x^7)=(r\cos \theta,r \sin \theta)$.
The harmonic function is now:
\be
f(r) =  - \tilde q \log r \,.
\ee

\subsubsection*{The $5^3$ brane} 

U-duality acting on the $89\#$ directions gives the background:
\be
\begin{split}
ds^2 & = K^{1/3}(r,\theta) \left( f^{-1/3} (r)  dx^2_{012345} + f^{2/3}(r)   (dr^2 + r^2 d\theta^2) \right) 
\\ & \qquad + f^{2/3}(r)K^{-2/3}(r,\theta) dx^2_{89 \#} \,,\\
C_{8 9 \#} & = K^{-1}(r,\theta)\tilde q \theta \,,
\end{split} 
\label{53}
\ee
where
\be
K(r,\theta) =  f^2(r)  + (\tilde q \theta)^2 \,.
\ee
That this is a U-fold is most easily seen by re-expressing the solution in terms of a trivector, using for instance the alternative $\Gfour$ generalised metric parametrisation described in section \ref{toyU}. This gives:
\be
\begin{split}
{ds}^2 & =  f^{-1/3} (r)  dx^2_{012345} + f^{2/3}(r)   (dr^2 + r^2 d\theta^2)  + f^{2/3}(r) dx^2_{89 \#} \,,\\
\Omega^{8 9 \#} & =-\tilde q \theta \,,
\end{split} 
\label{53tri}
\ee
which makes clear the U-duality monodromy as we encircle the brane $\theta \rightarrow \theta+2\pi$.

To match the discussion in section \ref{toyU}, we could further consider a U-duality transformation of the M-theory Kaluza-Klein monopole along a non-isometric direction, and so generate a local U-fold. The details are not particularly illuminating, so we will move on to explain more of the general features of exotic branes.

\subsection{Exotic brane multiplets and non-geometric uplifts}
\label{exoticmult}

\subsubsection{Brane tension notation}
\label{branename}

We are going to describe notation which gives the name of the brane in terms of the expression for its tension after wrapping all spatial and special isometry directions on a (sufficiently large) torus. 
Let's start with the NS5 brane. 
We will wrap the $12345$ worldvolume directions and the $89$ transverse directions on a $T^7$. The NS5 brane tension in 10-dimensional Minkowski space is
\be
T_{\text{NS5}} = \frac{1}{(2\pi)^5} \frac{1}{g_s^2 l_s^6}\,,
\ee
and the tension of the wrapped brane is
\be
T_{\text{NS5}} = \frac{R_1 \dots R_5}{g_s^2 l_s^6} \,,
\label{TNS5wrapped}
\ee
where $R_i$ denotes the radius of the direction $i$. 
We see that the tension \refeqn{TNS5wrapped} depends \emph{linearly on 5 radii} and on the string coupling $g_s$ to power of \emph{minus 2}. We encode these numbers by labelling the NS5 as the $5_2$ brane.

Now T-dualise on the $9$ direction to turn this wrapped NS5 into a wrapped KKM
We obtain
\be
T_{KKM} = \frac{R_1 \dots R_5 (R_9)^2}{g_s^2 l_s^8} \,,
\label{KKMwrapped}
\ee
which depends \emph{linearly on 5 radii}, \emph{quadratically on 1 radius} and on the string coupling $g_s$ to power of \emph{minus 2}. We encode these numbers by labelling the KKM as the $5_2^1$.

Now T-dualise on the $8$ direction to turn this wrapped KKM into wrapped $5_2^2$ exotic brane, with
\be
T_{5_2^2} = \frac{R_1 \dots R_5 (R_8R_9)^2}{g_s^2 l_s^{10}} \,,
\label{KKMwrapped}
\ee
which depends \emph{linearly on 5 radii}, \emph{quadratically on 2 radii} and on the string coupling $g_s$ to power of \emph{minus 2}. 

In general, we can encode these brane tensions as:
\be
b_n{}^{(\dots , d,c) }
\ee
signifying that the tension formula for the fully wrapped brane depends linearly on the radii of $b$ directions, quadratically on the radii of $c$ directions, cubically on the radii of $d$ directions, and so on; and further depends on $g_s$ to the power of minus $n$.
For branes in M-theory, where there is no string coupling, we merely drop the $n$ subscript.
Thus the M5 would simply be denoted as the $5$ brane. 

With this notation, the fundamental string F1 is written $1_0$ as its tension is $T= \frac{1}{2\pi l_s^2}$, while a D$p$-brane is denoted $p_1$. 
We will see many more examples below.

\subsubsection{Brane multiplets} 

We will find exotic branes by constructing the U-duality orbits of the standard branes wrapped on a torus, by focusing on the those branes that are codimension-2 (or less) in the non-compact directions. 
It is convenient to illustrate this by compactifying down to three dimensions and considering the wrapped branes (whose spatial worldvolumes and special isometry branes wrap the torus) which appear as particle states in the three-dimensional non-compactified directions.

Let's take for example type IIB compactified on $T^7$. 
We can start with the tension formula for any given brane of the theory wrapped on this torus -- for example the NS5 as above -- and then act with T- and S-duality using the rules $T: R \rightarrow l_s^2 /R$, $g_s \rightarrow l_s g_s/R$ and $S: g_s \rightarrow 1/g_s$, $l_s \rightarrow g_s^{1/2}l_s$ to generate a multiplet of the U-duality group $\Geight$.

These basic duality transformations then generate 240 states. 
We list these in table \ref{tab:ex248B}, displaying the brane, its conventional name and its definition according to the notation we have introduced in the previous subsection.
The numbers in parenthesis indicate how many states arise from each brane. 
This almost fills up the 248-dimensional adjoint representation of $\Geight$.
We therefore have to add a further 8 states, which appear when we act with more general $\Edd$ transformations \cite{Obers:1998fb}. We will discuss this in more detail momentarily. 

\begin{table}[ht]
\tbl{The 240 states of the Weyl orbit of $\Geight$ in type IIB, with multiplicities under $\mathrm{SL}(7)$ in parentheses.}
{
\centering
\begin{tabular}{c|ccc|ccc|ccc|ccc}
$g_s^0$ & P & & (7) & F1 & $\red{1_0}	$ & (7) \\
$g_s^{-1}$ & D1 & $\red{1_1}$ & (7) & D3 & \red{$3_1$} & (35) & D5 & $\red{5_1}$ & (21) & D7 & {$\red{7_1}$} & (1) \\
$g_s^{-2}$ & NS5 & $\red{5_2}$ & (21) & KKM & \red{$5_2^1$} & (42) &  & $\red{5_2^2}$ & (21) \\
$g_s^{-3}$ &  & $\red{1_3^6}$ & (7) & & \red{$3_3^4$} & (35) &  & $\red{5_3^2}$ & (21) &  & {$\red{7_3}$} & (1) \\
$g_s^{-4}$ & & \red{$0_4^{(1,6)}$} & (7) & & \red{$1_4^6$} & (7) 
\end{tabular}
}
\label{tab:ex248B}
\end{table}

In table \ref{tab:ex248B}, there is one state corresponding to a totally wrapped D7 brane, which we denote $7_1$. The S-dual of this brane is the $7_3$.
Bound states of the D7 and its S-dual are known as $(p,q)$ 7-branes. These branes are \emph{already codimension-2 in 10 dimensions}, and are characterised by their monodromy under the $\mathrm{SL}(2)$ of S-duality.
They already share many of the features of the non-geometric $5_2^2$ solution, and indeed from this point of view the D7 can be considered the least exotic ordinary brane, or the most ordinary exotic brane. 

If we start with type IIA string theory, or with the branes of 11-dimensional M-theory, we find the same multiplet of codimension-2 branes, but now with different (dual) higher-dimensional interpretations.
The IIA and M-theory multiplets are shown in tables \ref{tab:ex248A} and \ref{tab:ex248M} respectively.

\begin{table}[ht]
\tbl{The 240 states of the Weyl orbit of $\Geight$ in type IIA, with multiplicities under $\mathrm{SL}(7)$ in parentheses.}
{
\centering
\begin{tabular}{c|ccc|ccc|ccc|ccc}
$g_s^0$ & P & & (7) & F1 & $\red{1_0}	$ & (7) \\
$g_s^{-1}$ & D0 & $\red{1_1}$ & (1) & D2 & \red{$2_1$} & (21) & D4 & $\red{4_1}$ & (35) & D6 & {$\red{6_1}$} & (7) \\
$g_s^{-2}$ & NS5 & $\red{5_2}$ & (21) & KKM & \red{$5_2^1$} & (42) &  & $\red{5_2^2}$ & (21) \\
$g_s^{-3}$ & & $\red{0_3^7}$ & (1) & & \red{$2_3^5$} & (21) &  & \red{$4_2^3$} & (35) &  & \red{$6_3^1$} & (1) \\
$g_s^{-4}$ & & \red{$0_4^{(1,6)}$} & (7) & & \red{$1_4^6$} & (7) 
\end{tabular}
}
\label{tab:ex248A}
\end{table}

\begin{table}[ht]
\tbl{The 240 states of the Weyl orbit of $\Geight$ in M-theory, with multiplicities under $\mathrm{SL}(8)$ in parentheses.}
{
\centering
\begin{tabular}{ccc|ccc|ccc}
 P & & (8) & & & &&& \\
 M2 & $\red{2}$ & (28) & M5 & \red{$5$} & (56) & KKM & \red{$6^1$}& (56)  \\
  & $\red{5^3}$ & (56) &  & \red{$2^6$} & (28) &  & \red{$0^{(1,7)}$}& (8)  \\
\end{tabular}
}
\label{tab:ex248M}
\end{table}

We do not have to necessarily stop at codimension-2.
We could consider codimension-1 or even codimension-0 branes, and similarly apply duality to fill up even larger multiplets of further exotic branes.
We will comment further on such classifications below.
	
The branes generated above by acting with basic U-duality transformations filled out 240 of the 248 states of the adjoint of $\Geight$.
In order to explain the missing eight states, which are apparently missed in this process, we need a somewhat technical digression.
Here we are attempting to generate a representation of the adjoint of $\Edd$ (in the particular case $d=8$ but in fact this applies for any $d$).
The basic U-duality transformations (formed from combinations of T- and S-dualities) act in the adjoint of $\Edd$ as the so-called \emph{Weyl subgroup}, which is generated by reflections in hyperplanes orthogonal to roots. As such, they preserve the length of roots. 
The group $\Edd$ is simply-laced, i.e. all roots have the same length except the $d$ roots corresponding to the Cartan subalgebra.
This means that if we start with a brane state corresponding to a root of definite length, we can only reach $\dim \Edd - d$ states, excluding those in the Cartan.
The latter appear if we then act with more general transformations. We can generate the full group by identifying alongside the set of Weyl reflections a set of Borel transformations, which we can pick for instance as the transformations that in supergravity would act as constant shifts of the $p$-form gauge fields (and of the dual graviton and other mixed symmetry tensor fields, for $d \geq 7$).

We also need to be a little more sophisticated in our discussion of what the duality group means in terms of its physical interpretation. 
For the bulk of this review, we have been working with the continuous groups $\Edd(\mathbb{R})$ which appear as the continuous global symmetries of dimensionally reduced supergravity.
We also have the basic U-dualities forming the Weyl subgroup of $\Edd(\mathbb{Z})$, which are singled out by the fact they preserve the mass spectrum. Finally, we can also consider the automorphism group of the lattice of brane charges. 

Let us discuss these and how they differ. First, it is assumed that Dirac quantisation holds for the brane charges. 
This immediately breaks the continuous global symmetry $\Edd(\mathbb{R})$ of the classical supergravity to its arithmetic subgroup $\Edd(\mathbb{Z})$. 
How does this relate to the mass spectrum of brane solutions? 
The orbit of the Weyl subgroup leaves the mass of the branes in the orbit invariant. 
This acts both on the charges and on the moduli (scalar fields) to leave the mass of these states invariant. This is what one often thinks about as duality, and most studies of duality concentrate on these transformations. 
In addition, there are Borel transformations which together with the Weyl transformations generate the full group $\Edd(\mathbb{Z})$. 
Under Borel transformations, the mass of a given brane state changes. These are sometimes then called spectral flow operators as they appear to change the spectrum. 
One may then worry whether they are not true duality transformations. This would be at odds with the statement that $\Edd(\mathbb{Z})$ is the duality group. 

A detailed study has been made of this difference in \refref{Cremmer:1997xj} for simpler duality groups where it was shown that the lattice of brane charges is invariant under the full duality group acting non-linearly, making use of the subgroup of the conventional duality group which preserves the scalar moduli plus the so-called trombone symmetry $\mathbb{R}^+$, which produces a homogeneous scaling of the action and hence is a symmetry of the equations of motion. When the dust settles, it is the full $\mathbb{R}^+ \times \Edd$ group that leaves the lattice of brane charges invariant. It is this that we consider the duality group of the theory. The Weyl orbit is privileged in that it doesn't require 
trombone rescalings.

These subtleties are relevant whenever one encounters the brane states of codimension 2 in the $n$-dimensional external (non-compact) space. 
So for $\Geight$, these are the particle states appearing in the $R_1$ representation, while for $\Gseven$ these are the string states in the $R_2$ representation, and so on. 
More details about the orbits in different dimensions, the decomposition under T-duality, and many more features of these basic U-duality orbits can be found in the review \cite{Obers:1998fb}.

\subsubsection{Non-geometric uplifts}

What are we to make of these additional branes that we did not originally discover in 10 or 11 dimensions?
The following argument \cite{deBoer:2010ud, deBoer:2012ma} tells us that we should associate them to higher-dimensional non-geometric backgrounds. 

From the three-dimensional point of view, these wrapped branes all appear as particles, which are codimension-2 and hence couple magnetically to scalars. 
The compactification of (let's say) 10-d type IIB SUGRA to three dimensions gives scalars in the coset $\Geight/\Height$, parametrised by a generalised metric $\gM_{MN}(g_{ij}, B_{ij},\phi, C_0, C_{ij}, \dots)$. 
These scalars include the RR zero-form $C_0$. 
In \emph{ten} dimensions, the D7 brane is a codimension-2 state which is magnetically charged under this scalar, and characterised by a monodromy $C_0 \rightarrow C_0 +1$ as we encircle the position of the D7 in the two-dimensional transverse space.

From the point of view of the compactification, this monodromy is an element of $\mathrm{SL}(2) \subset \Geight$. 
Acting with $\Geight$ transformations leads to the full multiplet shown in table \ref{tab:ex248B}, in which the D7 monodromy transforms into more general $\Geight$ monodromies for each particular brane listed. 
These more general $\Geight$ monodromies when reinterpreted in terms of 10 dimensions act non-trivially and, generically, non-geometrically on the metric and form fields. Thus the higher-dimensional uplift of the exotic states in \ref{tab:ex248B} will be non-geometric.
Furthermore, we can characterise an arbitrary particle-like state in three dimensions via some $\Geight$ monodromy, and generically this must give something non-geometric in 10-dimensions.

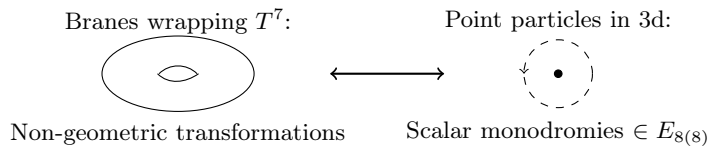
\begin{figure}[ht]
\centering

\begin{tikzpicture}
\draw (4,2.2) node {\footnotesize {Branes wrapping $T^7$:}};
\draw (4,0.7) node {{\footnotesize Non-geometric transformations}};

\draw (4,1.5) ellipse (1cm and 0.5cm);

\draw (4.25,1.5) to [out=135,in=45] (3.75,1.5);
\draw (4.27,1.5) to [out=215,in=-45] (3.73,1.5);
\draw [thick,<->] (6,1.5) -- (7.5,1.5);

\draw (9,2.2) node {\footnotesize {Point particles in $3$d:}};
\draw (9,0.7) node {\footnotesize {Scalar monodromies $\in \Geight$}};

\filldraw (9,1.5) circle (0.05cm);

\draw [dashed,->] (9.45,1.5) arc (0:180:0.45cm);
\draw [dashed] (9.45,1.5) arc (0:-180:0.45cm);

\end{tikzpicture}

\caption{Uplifting exotic branes}
\label{fig:uplift}
\end{figure}

Given these non-geometric uplifts, we have to then make sense of them in string theory. 
If we allow patching by duality symmetries in the full string theory, then clearly these are allowed configurations even though they do not make geometric sense in supergravity.
To resolve the latter problem, one needs an extended geometry such as the doubled or exceptional geometry we have been discussing in this review.

However, there are a number of other issues with these exotic branes. 
They are highly non-perturbative: their tension (or mass) depends on the string coupling as $\sim$ $g_s^{-3}$, $g_s^{-4}$.
They backreact very strongly on spacetime, being of codimension-2 they are characterised by a logarithmic harmonic function, as we have seen, and so the metric diverges asymptotically.
One can interpret this divergence as being due to the fact that the mass of a such a brane is not localised at $r=0$ but spread over spacetime
However, what this is telling is you is that you should \emph{not} extend the solution for an exotic brane to all of spacetime.
The configurations such as we found for the $5_2^2$, equation \refeqn{522}, should be treated as approximate descriptions valid near the brane, and in order to construct well-defined backgrounds (up to non-geometric properties) we should replace the bad asymptotics with something else.
Put differently, codimension-2 exotic brane states should not exist as standalone objects in string theory, but should always -- and indeed, \emph{will} always -- appear as part of more involved configurations of multiple branes.

One way to construct asymptotically well-defined configurations is to mirror the construction of 7-brane solutions in type IIB \cite{Greene:1989ya, Gibbons:1995vg}.
Seven-branes in type IIB are codimension-2 and are characterised by a monodromy of the IIB S-duality group $\mathrm{SL}(2)$, being magnetically charged under the type IIB axio-dilaton which transforms under $\mathrm{SL}(2)$:
\be
\tau = C_0 + i e^{-\phi} \,,\quad
\tau \rightarrow \frac{a\tau+b}{c\tau+d} \,,\quad 
\begin{pmatrix} a & b \\ c& d \end{pmatrix} \in \mathrm{SL}(2) \,.
\ee
Using complex coordinates $z=re^{i\theta}$ for the non-compact two-dimensional transverse space, the metric and axio-dilaton describing a single D7 is:
\be
\begin{split}
ds^2 & = dx^2_{01234567} + \mathrm{Im}\,\tau dz d \bar z\,,\quad \tau(z)  = \frac{1}{2\pi i} \log z \,.
\end{split}
\label{D7allalone}
\ee
From this we immediately see that there is a monodromy for $\theta \rightarrow \theta+2\pi$ corresponding to $\tau \rightarrow \tau +1$, i.e. the shift $C_0\rightarrow C_0+1$. 

The background \refeqn{D7allalone} describes the seven-brane solution near a brane at the origin.
As $\mathrm{Im}\,\tau \sim \log r$, this solution should not be extended to all of spacetime.
In order to have a globally well-defined background sourced by seven-branes, it is necessary to consider $N$ such branes.
The solution for $N$ branes at the positions $z=z_i$ is:
\be
ds^2  = dx^2_{01234567} + \mathrm{Im}\,\tau \frac{|\eta(\tau)|^4}{\prod_{i=1}^N |z-z_i|^{1/6} }dz d \bar z\,,
\quad 
\tau(z) = j^{-1} \left( \frac{P(z)}{Q(z)} \right) \,.
\label{D7N}
\ee
The Dedekind $\eta$ function $\eta(\tau) = e^{\frac{\pi i \tau}{12}} \prod_{n=1}^\infty \left( 1 - e^{2\pi i n \tau} \right)$
ensures that the metric is $\mathrm{SL}(2)$ invariant. 
The axio-dilaton is defined using a pair of polynomials, $P(z), Q(z)$, such that the roots of $Q(z)$ are the positions $z=z_i$ of the branes, and via the modular invariant $j$-function, admitting the expansion $j(\tau) = e^{-2\pi i \tau} + 744 + (196883+1) e^{2\pi i\tau} + \dots$.
In order to get a well-behaved geometry (without conical defects), one needs exactly $N=24$ seven-branes, in which case the transverse space compactifies into an $S^2$. This is a consequence if you like of the very strong backreaction of the codimension-2 branes.
The total monodromy encircling all 24 seven-branes must be trivial for the geometry to be well-defined. The individual branes can have arbitrary $\mathrm{SL}(2)$ monodromy, and this means that even if we can view a particular brane as being a D7, the others will be even more non-perturbative.


In F-theory \cite{Vafa:1996xn}, one interprets $\tau$ as the complex structure of an auxiliary zero area two-torus.
The fibration of this two-torus over spacetime can be described in terms of elliptic curves, allowing one to use techniques from algebraic geometry.
The torus fibration degenerates at singularities, which are interpreted as the positions of seven-branes, and monodromies can be associated to such degenerations.
The full configuration again allows access to non-perturbative physics.


The seven-brane, or F-theory, approach can be applied to exotic branes. 
In the simplest case, one can focus on monodromies which fall within $\mathrm{SL}(2)$ subgroups of larger dualities groups, and realise these via the very same construction.
For instance \cite{Hellerman:2002ax,McOrist:2010jw,Lust:2015yia}, consider type II string theory compactified on a two-torus with coordinates $x^8, x^9$.
The duality group is in this case is $O(2,2) \sim \mathrm{SL}(2) \times \mathrm{SL}(2)$ (plus some $\mathbb{Z}_2$ factors).
One of these $\mathrm{SL}(2)$ factors acts on the genuine complex structure $\tau$ of the torus on which we reduce, while the other acts on the combination $\rho = B_{89} +i \det g$, where $\det g$ denotes the determinant of the metric on the torus.

The monodromies of the NS5, KKM and $5_2^2$ can be encoded as transformations of $\tau$ and $\rho$:
\be
\text{NS5}: \rho \rightarrow \rho + 1 \,,\quad
\text{KKM}: \tau \rightarrow \frac{\tau}{-\tau+1}\,,\quad
5_2^2 : \rho \rightarrow \frac{\rho}{-\rho +1} \,.
\ee
We can build a globally defined configuration involving N brane with monodromies in $\tau$, and $\tilde N$ with monodromies in $\rho$, with $N+\tilde N= 24$ via:
\be
\begin{split}
ds^2 & = dx^2_{012345} +  \frac{\mathrm{Im}\,\tau |\eta(\tau)|^4}{\prod_{i=1}^N |z-z_i|^{1/6} }\frac{\mathrm{Im}\,\rho|\eta(\rho)|^4}{\prod_{i=1}^{\tilde N} |z-\tilde z_i|^{1/6} }dz d \bar z\,,\\
\tau(z)& = j^{-1} \left( \frac{P(z)}{Q(z)} \right) \,,\quad \rho(z)	 = j^{-1} \left( \frac{\tilde P(z)}{\tilde Q(z)} \right) \,.
\end{split}
\label{ns247}
\ee
This suggests an F-theory like description using a semi-auxiliary $T^2_\tau \times T^2_\rho$.

Various approaches have been attempted to generalise this picture to more general mondromies, involving both cases where the auxiliary space is a torus $T^n$, $n>2$, \cite{Kumar:1996zx, Liu:1997mb, Curio:1998bv,Vegh:2008jn,Achmed-Zade:2018rfc} and where it is a $K3$ \cite{Martucci:2012jk, Braun:2013yla, Candelas:2014jma, Candelas:2014kma} (dubbed ``G-theory''). 

The general idea is to assoicate (a subset of) the scalar moduli -- which couple magnetically to codimension-2 branes, and transform under monodromies -- to geometric moduli of an auxiliary space fibred over a physical base space. 
Degenerations of this fibration correspond to the locations of (generically exotic branes).

Away from simple torus fibrations, we have less control over the realisations of this idea.
One might ask, what space, if any, realises $\Geight/\Height$? 

The relationship to doubled and exceptional geometry also warrants further attention.
ExFT is at heart a reformulation of supergravity, or at least is supergravity-like.
It may be that F-theory is to type IIB supergravity as a more general F-theory of arbitrary monodromies is to ExFT.
Although it has been proposed to view ExFT as providing an action principle for F-theory \cite{Hohm:2015xna, Berman:2015rcc, Chabrol:2019kis}, and it is clear that the extended geometry of ExFT is much in the same spirit as the auxiliary torus of F-theory (one can view the generalised metric as the equivalent of the complex structure of this torus), how one might reproduce and generalise mathematical successes of F-theory within the ExFT framework is not currently known.   

\subsection{Exotic supertubes}
\label{stubes} 

We will now briefly describe another way in which exotic branes should appear in string and M-theory, via a brane polarisation effect, as suggested in \cite{deBoer:2010ud,deBoer:2012ma}.
Again, the purpose is to explain how to overcome the bad asymptotics of the codimension-2 nature of the solutions.
The non-geometric nature of the solutions remains and should be resolved using doubled or exceptional geometry.

\subsubsection{The supertube effect}

The supertube effect \cite{Mateos:2001qs} is a spontaneous polarisation of a bound state of branes, in which the original branes ``puff up'' into a (generically) higher-dimensional brane ``dipole''.
By a brane dipole, we mean a configuration in which the brane charge is non-zero locally, but globally zero, i.e. the net brane charge cancels out. 
The configuration forming the brane dipole ``worldvolume'' is supported by (angular) momentum (preventing it from shrinking to zero size) and carries the original brane charges ``dissolved'' in the worldvolume of the dipole brane.
Schematically, we denote this polarisation process by:
\be
\text{Brane} + \text{Brane} \rightarrow \text{brane}^\prime + \text{(angular) momentum} \,.
\ee
The original branes are denoted with upper case letters, while the resulting dipoles will be denoted with lower case letters.
Remarkably, supertubes retain the preserved supersymmetry of the original brane pair.

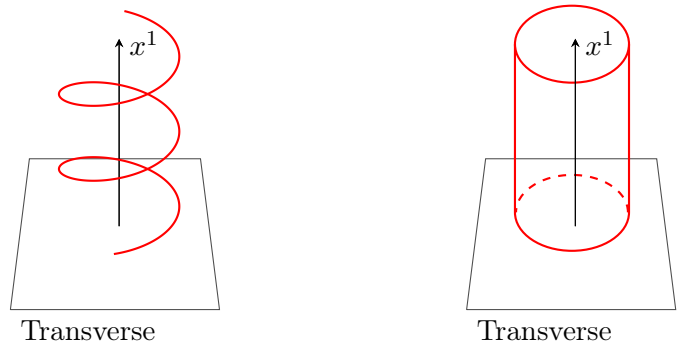
\begin{figure}[ht]
\centering

\begin{tikzpicture}

	\coordinate (corner) at (2,1);
	\coordinate (corner1) at (4.5,3);
	\draw [black!70] (corner) -- (4.75,1) -- (corner1) -- (2.25,3) -- (corner) node [below right,black] {Transverse};


	\begin{scope}[xshift = 0cm, yshift=0.5cm]
	\begin{axis} [
	view={5}{50},
	xlabel=$x$,
	ylabel=$y$,
	zlabel=$x^1$,
	xtick=\empty,
	ytick=\empty,
	ztick=\empty,
	xmin=-1,
	xmax=1,
	ymin=-1,
	ymax=1,
	axis z line=middle,
	axis x line=none,
	axis y line=none,
	z label style={at={(axis description cs:0.5,0.76)}},
	]
	\addplot3 [thick, red, domain=pi:6*pi, samples = 500, samples y=0] ({0.25*sin(deg(-x))}, {0.25*cos(deg(-x))}, {x});
	\end{axis}


	\end{scope}

\begin{scope}[xshift=6cm]
	\coordinate (corner) at (2,1);
	\coordinate (corner1) at (4.5,3);
	\draw [black!70] (corner) -- (4.75,1) -- (corner1) -- (2.25,3) -- (corner) node [below right,black] {Transverse};

	\coordinate (cyl) at (4.15,3);
	\draw (cyl) node [above,cylinder,red,thick,draw,minimum height=3.25cm, aspect=4,minimum width=1.5cm,rotate=90] (A) {};
	  \draw[dashed,thick,red]
		let \p1 = ($ (A.after bottom) - (A.before bottom) $),
			\n1 = {0.5*veclen(\x1,\y1)-\pgflinewidth},
			\p2 = ($ (A.bottom) - (A.after bottom)!.5!(A.before bottom) $),
			\n2 = {veclen(\x2,\y2)-\pgflinewidth}
	  in
		([xshift=-\pgflinewidth] A.before bottom) arc [start angle=0, end angle=180,
		x radius=\n1, y radius=\n2];

	\begin{scope}[xshift = 0cm, yshift=0.5cm]
	\begin{axis} [
	view={5}{50},
	xlabel=$x$,
	ylabel=$y$,
	zlabel=$x^1$,
	xtick=\empty,
	ytick=\empty,
	ztick=\empty,
	xmin=-1,
	xmax=1,
	ymin=-1,
	ymax=1,
	axis z line=middle,
	axis x line=none,
	axis y line=none,
	z label style={at={(axis description cs:0.5,0.76)}},
	]
	\addplot3 [thick,white, domain=0:2, samples = 100, samples y=0]     ({cos(x)}, {cos(x)}, {sin(x)});
	\end{axis}
	\end{scope}

\end{scope}
\end{tikzpicture}

\caption{Left: the F1-pp supertube. Right: the original D2 supertube.}
\label{fig:stubes}

\end{figure}

A simple example of a supertube polarisation is formed when adding momentum in the longitudinal direction of a fundamental string. 
As a string only has transverse directions, it cannot carry this longitudinal momentum except by puffing up into the transverse directions, forming a helix configuration with non-zero angular momentum. 
If the original longitudinal direction is the $x^1$ direction, then we denote this process by:
\be
\text{F1}(1) + \text{P}(1) \rightarrow \text{f1}(\psi) + p(\psi) \,,
\ee
where $\psi$ denotes the curve in the transverse directions along which the string extends.
The statement that this is a dipole is that there is no net F1 charge in the transverse directions i.e. no string winding along $\psi$. Rather than contract to the origin of the transverse directions, angular momentum supports the string which (instantaneously) forms a helix shape as illustrated in figure \ref{fig:stubes}.

Assuming this is a type IIB string, the supertube can be S- and then T-dualised to
\be
\text{D0} + \text{F1}(1) \rightarrow \text{d2}(1\psi) + p(\psi) \,.
\ee
This is the original supertube configuration of Mateos and Townsend \cite{Mateos:2001qs}. 
It is a polarised state interpreted as the puffing up a D0 and F1 into a D2 brane configuration, in which the D2 brane configuration extends along the $x^1$ direction and in a curve $\psi$ in the directions transverse to $x^1$. This is shown in figure \ref{fig:stubes}, in which the curve $\psi$ is the circle projected onto the transverse plane. 

The charges corresponding to the F1 and D0 appear via the flux of the D2 worldvolume Born-Infeld field strength, with the electric components corresponding to F1 charge, the magnetic components to D0 charge, and these fluxes carrying the angular momentum $p(\psi)$ which supports the configuration from collapse.

\subsubsection{Exotic supertubes}

\begin{figure}[ht]
\centering

\begin{tikzpicture}
\begin{scope}[xshift=0cm, yshift=0cm]
\draw [->] (0,0) -- (3,0) node [right] {$x^2$};
\draw [->] (0,0) -- (0,2) node [above] {$x^3$};
\draw [->] (0,0) -- (-0.5,-0.5) node [below] {$x^1$};

\draw [thick, blue] (1.75,1) arc (180:360:0.5cm);

\draw [thick,red] (0.5,-0.5) to [out=-45,in=-95] (2.5,1) to [out=85,in=-45] (1.25,2) to [out=135,in=35] (0.75,1.75) to [out=215,in=134] (0.5,-0.5); 
\draw [red] (2,2.25) node {\small Curve $\psi$: $\vec{x} = \vec{F}(v)$};

\draw [thick, blue,->] (2.75,1) arc (0:180:0.5cm);
\draw [thick, blue] (2.75,1) node [right] {$c$};

\end{scope}
\end{tikzpicture}

\caption{An exotic supertube. Traversing the curve $c$ leads to a non-geometric monodromy.}
\label{fig:exoticstube}

\end{figure}
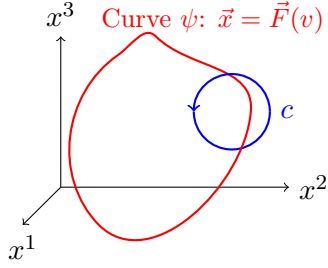

It was pointed out in \refref{deBoer:2010ud,deBoer:2012ma} that further T-dualities lead to supertube configurations in which ordinary branes polarise into exotic branes.
For instance, one can obtain:
\be
\text{D4}(6789) + \text{D4}(4589) \rightarrow 5_2^2 (4567\psi,89) +p(\psi) \,.
\ee
An explicit supergravity solution for this exotic supertube can be written down.
It is a solution of type IIA SUGRA on $\mathbb{R}^{1,3}_{0123} \times T^6_{456789}$.
We denote the transverse non-compact directions by $\vec{x} = (x^1,x^2,x^3)$.
The NSNS part of the solution is:
\be
ds^2 = 
-\frac{1}{\sqrt{f_1 f_2}} ( dt - A_i {dx^i})^2 
+ \sqrt{f_1 f_2} {dx_{123}^2}
+ \sqrt{\frac{f_1}{f_2}} dx_{45}^2 + \sqrt{\frac{f_2}{f_1}} dx_{67}^2 
+ \frac{\sqrt{f_1f_2}}{f_1 f_2 + {\gamma}^2} dx_{89}^2
\ee
\be
B_{89} = \frac{{\gamma}}{f_1f_2 + {\gamma}^2}\,, \quad e^{2\phi} = \frac{\sqrt{f_1f_2}}{f_1f_2+ {\gamma}^2}\,,
\ee
and it also carries non-trivial RR fields, $C_3 \neq 0, C_5 \neq 0$, $C_1 =C_7 = 0$, see \refref{deBoer:2012ma} for the details.
The harmonic functions appearing are:
\be
f_1 = 1 + \frac{Q_1}{L} \int_0^L \frac{dv}{|\vec{x} - \vec{F}(v)|} \,,\quad
f_2 = 1 + \frac{Q_1}{L} \int_0^L \frac{dv |\dot{\vec{F}}(v)|}{|\vec{x} - \vec{F}(v)|}  \,,
\label{exstf}
\ee
where $\vec{x} = \vec{F(v)}$ describes the position of the supertube in the transverse space, via the profile function $\vec{F}$. 
We also have the one-form
\be
A_i = - \frac{Q_1}{L} \int_0^L \frac{dv \,\dot{F}_i}{|\vec{x} - \vec{F}(v)|} \,,
\ee
in terms of which the function $\gamma$ appearing in the solution is given by
\be
\partial_i \gamma = \epsilon_{i}{}^{jk} \partial_j A_k  \,.
\ee 
This function is therefore not single-valued.
Consider a curve $c$ which has a non-trivial intersection with the supertube curve $\vec{x} = \vec{F}(v)$, as in figure \ref{fig:exoticstube}.
Parametrising the curve $c$ by $w$, we have
\be
\begin{split}
\oint_c d \gamma 
=\frac{Q_1}{L} \int dw\, \int dv \, \epsilon_{ijk} \frac{\dot{x}^i(w)  \dot{F}^j(v)(x^k(w)-F^k(v))}{|\vec{x}(w) - \vec{F}(v)|^3}
 =\frac{4\pi Q_1 n}{L} \,,
\end{split}
\ee
using the fact that the integral in the first line is Gauss' linking integral measuring the linking number $n$ of the two curves (the number of times that $c$ winds around the supertube curve).
Hence
\be
\gamma \rightarrow \gamma + \frac{4\pi Q_1}{L} \,,
\ee
as we go once through the curve.
This leads to a non-geometric monodromy transforming the fields in the $89$ directions, and makes this an exotic supertube.
Crucially though, as can be seen from the harmonic functions \refeqn{exstf}, the solution is asymptotically flat, and has no global exotic monodromy.

The importance of the arguments of \refref{deBoer:2010ud, deBoer:2012ma} is the following.
As the polarisation of branes into supertubes is inevitable, hence the polarisation of branes into exotic supertubes is inevitable.
Therefore exotic branes will generically appear in string theory in situations where multiple branes are present and will polarise.

Supertubes play an important role in the microstate geometry program, where they provide supergravity realisations of microstates of black holes, and so account for some portion of the entropy of black holes in string theory.
It has been proposed in \refref{deBoer:2010ud,deBoer:2012ma} that for three- and four-charge black holes that non-geometric microstates may be required to obtain the correct entropy.

\subsection{Winding localisation}
\label{windingmode}

Let's revisit the smeared NS5 on $\mathbb{R}^3 \times S^1$.
The metric is
\be
ds^2 = dx^2_{012345} + f ( dx^2_{678} + d\theta^2) \,,
\label{NS5smearedrevisit}
\ee
with
\be
f(r) = 1 + \frac{q^\prime}{r} \,,\quad r = \sqrt{ (x^6)^2 + (x^7)^2 + (x^8)^2 } \,.
\label{fsmearedrevisit}
\ee
We previously obtained this solution by a smearing procedure starting with the honest localised NS5 solution on $\mathbb{R}^4$.
Alternatively, we could have used the solution to the Laplace equation on $\mathbb{R}^3 \times S^1$:
\be
f(r,\theta) = 1 + \frac{q^\prime}{r} \frac{\sinh r}{\cosh r -\cos \theta} \,.
\label{flocalised}
\ee 
The NS5 solution \refeqn{NS5smearedrevisit} with the localised harmonic function \refeqn{flocalised} should represent the true NS5 brane solution with transverse space $\mathbb{R}^3 \times S^1$, localised at $r=0, \theta=0$.
How should we interpret the smeared solution, which is what appears naturally as the T-dual partner of the KKM? 
The answer is believed to be that the string worldsheet theory in the smeared background quantum corrects into the full localised solution, after taking into account the presence of worldsheet instanton configurations. Such a possibility was first suggested in a paper by Gregory, Harvey and Moore \cite{Gregory:1997te} and then evidence for this came much later from an impressive calculation by Tong \cite{Tong:2002rq}.

The T-dual version of the worldsheet instanton calculation was studied by Harvey and Jensen in \refref{Harvey:2005ab,Jensen:2011jna} and suggests that the Kaluza-Klein monopole\footnote{The further generalisation to the $5_2^2$ brane was done in \refref{Kimura:2013zva} and similar effects for general branes in DFT studied in \refref{Kimura:2018hph}} can be thought of as being localised in winding space:
\be
ds^2 = dx^2_{012345} + f dx^2_{678} + f^{-1} (  d\theta^2 + A_i dx^i) \,,
\label{KKMrevisit}
\ee
\be
f(r,\tilde\theta) = 1 + \frac{q^\prime}{r} \frac{\sinh r}{\cosh r -\cos  \tilde \theta} \,.
\label{flocalisedKKM}
\ee 
As pointed out in \refref{Jensen:2011jna}, this should be interpreted in doubled geometry.
Such a background does not violate the section condition, as it depends on the dual coordinate $\tilde\theta$ (the $\theta$ of the localised NS5 \refeqn{NS5smearedrevisit}), but not also on the physical coordinate.  As a solution in DFT it is fine but there is no supergravity interpretation. This is not a problem from the point of view of the string worldsheet theory. 

To understand how such localisation comes about let us look at some more details. The full transverse space of the monopole is an $S^3$. It is written as a Hopf fibration of $S^1$ over $S^2$ with $\theta$ the coordinate of the Hopf fibre. Crucially, the first homotopy group of $S^3 $ is trivial and so there are no string winding modes on this space. Locally though one can think of the winding modes around the $S^1$ Hopf fibre and construct a current corresponding to this winding mode. Examining the worldsheet theory, one can show that the there are classical unwinding solutions. These are solutions to the string equations of motion that can alter the winding number. They are like Lorentzian world sheet instantons in that they are world sheet cylinders where the winding number changes from the past circle to the future circle. In the path integral one should sum over these solutions. These configurations act as a sources for the winding modes and so strings in that background feel a potential in winding space. To put it even more simply, because there are no conserved winding numbers of strings in the KK background, the KK winding number can be changed and so momentum in the $\tilde{\theta}$ direction is not conserved (momentum in winding mode space is equivalent to winding charge). This means there is some potential in that direction which is captured by the Laplace equation with delta function source.

\subsection{Discussion} 
\label{exoticexft}

\subsubsection*{Exotic branes in exceptional field theory} 

We have seen that duality predicts the existence of exotic branes, which are characterised by non-trivial T- or U-duality monodromies as one encircles the position of the brane.
These are generalisations of 7-branes in type IIB, which have S-duality monodromies, and of Kaluza-Klein monopoles, which require the presence of compact ``special isometry'' directions. 
The spacetime backgrounds sourced by these branes are thus non-geometric. 

One of the original motivations for treating physical and dual coordinates together was to provide a ``geometry for non-geometry'' \cite{Hull:2004in}. 
It is clear that if one takes the extended geometry of DFT or ExFT seriously, then it is a natural home for non-geometries.  

Let's review the state of affairs in ExFT.
Given some particular brane solution (geometric or non-geometric) there are multiple ways to realise it as an ExFT solution, depending on how one chooses to orient the worldvolume, transverse and special isometry directions between the $n$-dimensional spacetime and the extended geometry.
The standard branes couple electrically to the tensor hierarchy gauge fields from $R_1$ to $R_{8-d}$, the latter corresponding to those of codimension $3$ in the $n$-dimensional space, including the Kaluza-Klein monopole.
The first exotic branes appear coupled electrically to the adjoint-valued tensor hierarchy gauge field in $R_{9-d}$, and magnetically to the generalised metric. 
Being magnetically charged under $n$-dimensional scalars, they are characterised by monodromies of the latter, leading to their non-geometric nature when interpreted from an 11- or 10-dimensional perspective.
Branes of lower codimension appear at the next level in the tensor hierarchy. 

One can also use exceptional field theory to speculate about configurations which are locally non-geometric i.e. configurations where the background depends on dual coordinates (without violating the section condition by not also depending on the actual physical coordinates of the spacetime), as in the $R$-flux examples of sections \ref{toyT} and \ref{toyU} and in the winding mode localisation example discussed above, see for instance \refref{Bakhmatov:2016kfn,Kimura:2018hph}.

Note that we can write down descriptions of exotic branes that remain exotic within the exceptional field theories themselves. For example a large spectrum of exotic branes were constructed in the $\Gseven$ exceptional theory where the non-geometric monodromy acts non-trivially on the external spacetime as well as in the extended geometry \cite{Berman:2018okd}: hence this is a configuration which is exotic in $\Gseven$ as it really makes use of a $\Geight$ monodromy.
By making the extended geometry as large as possible we can fully geometrise such transformations using higher rank $\Edd$ transformations.

Ultimately, this leads to brane solutions in orbits of the infinite-dimensional extensions of the $\Edd$ sequence, which have proven a useful starting point for obtaining and classifying orbits of branes, see for example \refref{Englert:2003py,West:2004st,Cook:2004er,West:2004kb,Englert:2007qb,Riccioni:2006az,Riccioni:2007au,Cook:2008bi, Kleinschmidt:2011vu}.
The goal of obtaining a complete classification of all possible exotic brane states has been pursued also via the formulation of ``wrapping rules'' controlling the brane spectrum on a torus \cite{Bergshoeff:2006gs,Bergshoeff:2010xc,Bergshoeff:2011zk,Bergshoeff:2011mh,Bergshoeff:2011ee,Bergshoeff:2011se,Bergshoeff:2012ex,Bergshoeff:2012jb,Bergshoeff:2017gpw}, while classifications taking exceptional field theory as a starting point include \refref{Bakhmatov:2017les,Fernandez-Melgarejo:2018yxq,Berman:2018okd,Fernandez-Melgarejo:2019mgd, Fernandez-Melgarejo:2019pvx}.
All these approaches provide access to large orbits of non-geometric branes and also describe the exotic generalisations of $p$-form gauge fields to which they couple.
These are mixed symmetry tensor potentials, carrying sets of antisymmetric indices, related to unusual dualisations of ordinary $p$-forms and of the graviton.
Within exceptional field theory, these appear as components of the tensor hierarchy gauge fields at high levels.

\subsubsection*{Other exotic features}
\label{exoticfurther}

Below, we will list a sample of works on non-geometric backgrounds to illustrate some of the other interesting features associated to the appearance of such backgrounds in string and M-theory.
Some of these have been explicitly connected to DFT or ExFT, while others offer entirely different angles and approaches which have not (yet!) been explicitly discussed in this context.

\emph{Other versions of non-geometry.} 
Non-geometric brackgrounds can be described as topological interfaces on the worldsheet, see for example \refref{Satoh:2015uia}.
Mirror symmetry has been shown to be related to fibre wise T-duality, so one can use mirror symmetry to construct so called ``mirrorfolds" where mirror symmetry is used to patch together local solutions.\cite{Kawai:2007nb}.

\emph{A ``mysterious'' duality'' with del Pezzo surfaces.}
An intriguing observation relates the moduli space of M-theory to the moduli spaces of del Pezzo surfaces \cite{Iqbal:2001ye}, where the usual 1/2 BPS branes are given by spheres in the del Pezzo. In a very interesting recent work Kaidi \cite{Kaidi:2019dqr} has constructed the relevant cycles corresponding to exotic branes and has showed they are given by higher genus curved in the del Pezzo. This then predicts more novel exotic branes.

\emph{Non-commutativity and non-associativity.} 	
The physics of strings and branes probing non-geometric backgrounds leads to some interesting features.
Particularly noteworthy is the evidence for non-commutativity and non-associativity of closed strings in T-fold backgrounds \cite{Lust:2010iy,Blumenhagen:2010hj,Blumenhagen:2011ph,Condeescu:2012sp,Mylonas:2012pg,Andriot:2012vb,Bakas:2013jwa,Blair:2014kla,Bakas:2015gia}, and of membranes in U-fold backgrounds \cite{Gunaydin:2016axc,Kupriyanov:2017oob,Lust:2017bgx, Lust:2017bwq}. 

\emph{Non-geometric engineering.} Recently S- and U-folds have been used to construct novel four dimensional $\mathcal{N}=3$ SUSY field theories \cite{Garcia-Etxebarria:2015wns,Aharony:2016kai}. This works by having branes in backgrounds where we make some non-trivial (and non-perturbative) duality identification, such as a D3 brane in an S-fold or an M5 brane in a U-fold \cite{Garcia-Etxebarria:2016erx}. The S- or U-folding in the background means the supersymmetry on the brane gets broken and the coupling gets fixed at order one leading to a non-perturbative field theory with an unusual amount of supersymmetry.
These constructions can be extended (via the double copy) to produce new variants of supergravity with $\mathcal{N}=7$ supersymmetry, again by making use of duality twists \cite{Ferrara:2018iko}. 

\emph{Worldvolume actions for exotic branes.}
The worldvolume actions for exotic branes can be determined by starting with the known worldvolume theories of standard branes, and applying duality transformations.
This procedure was used in \refref{Eyras:1999at} for the $6_3^1$, $7_3$ (the S-dual of the D7) and some other codimension-1 exotic branes, and later in \refref{Chatzistavrakidis:2013jqa,Kimura:2014upa} for the $5_2^2$ and \refref{Kimura:2016anf} for the $5^3$.

\emph{Electric non-geometry.} 
Throughout this review, electromagnetic duality has played an important role in the background.
One might then wonder what would be the electromagnetic dual of an exotic brane. 
Let's consider the $5_2^2$, for example.
We saw that one way to obtain this brane was to start with the NS5, magnetically charged under the $B$-field, T-dualise to the KKM, which is magnetically charged under KK vector, and then T-dualise again to the $5_2^2$, which we could think of as being magnetically charged under the bivector. 
An electromagnetic dual of this chain would start with the fundamental string, electrically charged under $B$-field and then T-dualise along the spatial worldsheet direction to the pp-wave, which is electrically charged under the KK vector. 
To go one step further, we should consider a timelike duality\cite{Welch:1994qm} (studied in general by Hull in \refref{Hull:1998vg, Hull:1998ym}) and obtain a novel solution which can be viewed as being electrically charged under the bivector.
This solution is that of a fundamental string with negative tension:
\be
\begin{split}
ds^2 & = \tilde H^{-1} ( - dt^2 + dz^2 ) + d\vec{x}_8{}^2 \,,\\
B_{tz} & = \tilde H^{-1} - 1 \,,\\
e^{-2\Phi} & = \tilde H \,,
\end{split}
\ee
where the harmonic function $\tilde H = 1 - \frac{h}{r^6}$ with $r\equiv |\vec{x}_8|$ differs from that of the usual string solution by the appearance of a minus rather than a plus sign. When $\tilde H = 0$, there is a naked singularity, furthermore the solution can be checked to have negative ADM mass. Such solutions have been re-analysed in \refref{Dijkgraaf:2016lym}  where it is argued that backgrounds corresponding to negative tension branes may still make sense in string theory, for instance other (mutually supersymmetric) branes can safely probe beyond the singularity.
In DFT terms, this is borne out by the non-singularity of the generalised metric:
\be
\cH_{MN} = \begin{pmatrix} \tilde H - 2 & 0 & 0 & \tilde H-1 \\ 0 & 2 - \tilde H & \tilde H -1 & 0 \\ 0 & \tilde H -1 &  -\tilde H & 0 \\ \tilde H - 1  & 0 &0 & \tilde H \end{pmatrix}
\ee
(here we ignore the transverse directions). The only issue is that at $\tilde H = 0$ we cannot interpret the bottom-right block as an inverse spacetime metric. One can either interpret the singularity then as a locus where the generalised metric admits a so-called non-Riemannian parametrisation (to be discussed in the next section), or else switch parametrisations to that involving a dual metric and bivector, in which case one will obtain the bivector analogue of the usual fundamental string solution with $\beta^{tz} = H^{-1} -1$. This should then be interpreted as some sort of instantonic solution electrically charged under $\beta^{ij}$ \cite{Sakatani:2014hba}.
Although their status within the full string theory remains unclear, it seems that from the DFT perspective (discussed further in \refref{Blair:2016xnn}) these type of solutions are better behaved than expected.

%% file: sec9NonRieGeometry.tex
\subsection{Non-Riemannian backgrounds}

So far we have seen how exceptional field theory can beautifully encode the metric and $p$-form fields of ordinary supergravity. 
We have seen how there can be backgrounds which although locally may be described by supergravity there are global obstructions, leading to a plethora of exotic branes and backgrounds. 
We have also seen how there can be all manner of exotic or non-geometric fluxes which we can encode with local fields such as the Q-flux, or if we wish to be adventurous the R-flux. We will now consider the question of an even more unusual backgrounds which will have no description in terms of the familiar local fields. 
There will be none of the usual supergravity fields, no metric, no $p$-forms. Why would such backgrounds exist?

The logic is as follows. Normally, one writes the action for ExFT or DFT in terms of the generalised metric and we then pick a gauge that allows us to choose a parametisation of the generalised metric, and in doing so introduce the local supergravity fields (or perhaps those involving bivectors or similar, which are related by field redefinitions). Instead one may write down the theory as a constrained system whereby the generalised metric obeys an algebraic constraint determined by the appropriate group of the ExFT or DFT. For example, in DFT one may work with a generalised metric $\cH_{MN}$ which obeys the constraint equation:
\begin{equation}
\cH_{MN}= \eta_{MK} \cH^{KL} \eta_{LN}  \, ,
\label{oddcomp}
\end{equation}
with $\eta_{MN}$ the usual $O(d,d)$ matrix.
We know that we can solve this constraint by writing the generalised metric as an element of the $O(d,d)/O(d) \times O(d)$ coset. Doing so introduces the usual form of the generalised metric, \refeqn{Hfirstlook}, that we are by now so familiar with. The constraint we have introduced means that the equations of motion derived from the action for $\cH_{MN}$ obeys a projection equation as discussed in section \ref{genKK_DFT}, namely:
\begin{align}
P_{MN}{}^{KL} \frac{\delta S}{\delta {\cH}^{KL}} =0 \,,
\quad
P_{MN}{}^{KL} 
= \frac{1}{2}(\delta_{M}^{(K} \delta_N^{L)}- \cH_{MP} \eta^{P(K}\eta_{NQ}\cH^{L)Q})\,.
\label{projeom}
\end{align}
We can now ask however are there other solutions to the constraint equation? And indeed by inspection we can see that one obvious solution stands out
\begin{equation}
\cH_{MN}=\eta_{MN} \, .  \label{mequaleta}
\end{equation}
That is we can make the DFT metric be equal to the $O(d,d)$ metric!
This has been reported and discussed at length in works by Park and collaborators \cite{Lee:2013hma, Ko:2015rha,Morand:2017fnv,Cho:2018alk,Cho:2019ofr,Cho:2019npq}. Furthermore, the DFT equations of motion are then trivially solved as the projector appearing in \refeqn{projeom} automatically vanishes. 
Note, the $O(d,d)$ metric is rigid, there are no moduli and thus no fields.
We call such a background \emph{non-Riemannian} since it does not contain a Riemannian metric. 

One might then ask, is this is just a rather trivial vacuum and can we just ignore it? There are several aspects to this. First in DFT one could  choose the number of dimensions one wish to solve this way with the remaining dimensions having the usual coset parameterisation in terms of metric and B-field. This then gives a fascinating compactification where there are {\bf{no moduli}} from the compact space.

Before returning to the properties of the specific background \refeqn{mequaleta}, let us discuss the general picture of non-Riemannian parametrisations of the generalised metric.
As shown in \refref{Morand:2017fnv}, solutions of the $O(d,d)$ compatibility condition \refeqn{oddcomp} can be classified based on whether the $d \times d$ block $\cH^{\mu\nu}$ of the generalised metric is degenerate. This block normally corresponds to the inverse spacetime metric.
When it is degenerate, one can introduce alternative non-Riemannian parametrisations in which both $\cH^{\mu\nu}$ and its zero vectors appear.\footnote{Alternatively, in some cases one could pass to a parametrisation with a bivector \cite{Malek:2013sp}, though this is not possible for the case \refeqn{mequaleta}.}
This degenerate metric structure is reminiscent of approaches to non-relativistic geometries, in which the prototypical situation would be that one has a degenerate rank $D-1$ ``metric'' with a zero vector. The latter is used to define the direction of time, and on orthogonal spatial hypersurfaces the degenerate metric restricts to a normal Euclidean metric.
Indeed, this allows one to embed various non-relativistic limits of string theory \cite{Gomis:2000bd, Danielsson:2000gi, Andringa:2012uz, Harmark:2017rpg} into doubled or exceptional geometry \cite{Ko:2015rha,Park:2016sbw, Cho:2019ofr,Park:2016sbw,Berman:2019izh,Blair:2019qwi,Blair:2020ops}, which is rather surprising when you consider that this was never an intended goal of these approaches.

From a worldsheet perspective the zero vectors of the degenerate ``metric'' play a rather interesting role. 
One can classify the non-Riemannian parametrisations by a pair of integers $(n,\bar n)$ such that the total number of zero vectors is $n+\bar n$ (and the trace of the generalised metric is $2(n-\bar n)$). Then what happens is that $n$ target space coordinates become chiral and $\bar n$ become anti-chiral, or more precisely one obtains chiral/antichiral $\beta\gamma$ systems. In effect there is a decoupling of left- and right-movers and this is what makes the target space non-Riemannian.

So, crucially, there is a sensible string worldsheet description of these backgrounds. If there were not one might just suspect they were pathological solutions so it is very important to see that these backgrounds are fully quantum consistent string backgrounds. The quantum consistency of strings in such backgrounds was already investigated in \refref{Kim:2007pc,Kim:2007hb} and more recently in \refref{Lindstrom:2020oow}. Thus it is already remarkable that DFT contains more string consistent backgrounds than are described by supergravity.

Once we have seen the consistency of the worldsheet theory for strings in such a non-Riemannian background it is natural to ask what the ExFT generalisation is. We wish to emphasize the fact that the string theory is consistent so that we have faith that non-Riemannian backgrounds are good even if they have no local supergravity description.

\subsection{Non-Riemannian Exceptional Field theory}

We will seek the analogue of equation \refeqn{mequaleta} for exceptional field theory following \refref{Berman:2019izh}. The first challenge is that most exceptional groups are not equipped with a symmetric tensor of the right dimension to allow a similar equation. There is one case that stands out. In $\Geight$, the $R_1$ representation is the adjoint, and the generalised metric for $\Geight$ is a symmetric $248 \times 248$ matric which obeys $\gM_{MK} \gM_{NL} \gM_{PQ} f^{KLQ} = - f_{MNP}$. This is the analogue of the condition \refeqn{oddcomp}, and in particular it is used to ensure that the ExFT action is invariant under generalised diffeomorphisms \cite{Hohm:2014fxa}.
One can construct such an object for $\Geight$ using the $\Geight$ Killing metric. 
Recall, the Killing metric is constructed from the structure constants of the algebra as follows:
\begin{equation}
\kappa^{MN}= \frac{1}{60} f^{MP}{}_Q f^{NQ}{}_P  \, .
\end{equation}
The projector for the $\Geight$ equations of motion is given by \cite{Berman:2019izh}:
\begin{equation}
P_{MN}{}^{KL} = \frac{1}{60} \left( \delta^{(K}_M \delta^{L)}_N + \gM_{MN} \gM^{KL} - \gM_{MQ} Y^{Q(K}{}_{RN} \gM^{L) R}\right) \, .
\end{equation}
It is then a simple exercise to see that:
\begin{equation}
\gM^{MN}=-\kappa^{MN}  \label{E8nonrie}
\end{equation}
solves
\begin{equation}
P_{MN}{}^{KL} (\gM)=0  \, .
\end{equation}
Thus, the relation \refeqn{E8nonrie} defines our $\Geight$ non-Riemannian space.
It is a solution of the $\Geight$ ExFT equations of motion for the internal space that does not admit any supergravity parameterisation in terms of a metric or p-forms on the internal space.

One may then ask what is the $\Geight$ ExFT with this choice of generalised metric taken for the internal space. The theory then becomes a rather interesting theory which appeared first in \refref{Hohm:2018ybo}, with the following action:
\be
S = \int d^3x d^{248} Y   \sqrt{|g|} R[g] +\int d^3x \epsilon^{\mu\nu\rho} \langle \mathbb{A}_\mu, \partial_\nu \mathbb{A}_\rho - \frac{1}{3} \mathbb{A}_\nu \circ \mathbb{A}_\rho \rangle
\label{truncaction} 
\ee	
where we have rewritten the Chern-Simons term of the $\Geight$ ExFT (see section \ref{action8}), involving the $\Geight$ ExFT gauge fields $\mathbb{A}_\mu \equiv ( \mathcal{A}_{\mu}{}^M, \mathcal{B}_{\mu M})$, in terms of a ``Leibniz algebra'' product, with
\be
\mathbb{A}_\mu \circ \mathbb{A}_\nu \equiv ( \mathcal{L}_{{\cal A}_\mu} {\cal A}_\nu{}^M , \mathcal{L}_{{\cal A}_\mu} \mathcal{B}_{\nu M} + \mathcal{B}_\nu^N \partial_M (f_{N}{}^P{}_{Q} \partial_P \mathcal{A}_\mu{}^Q + \mathcal{B}_{\mu N}))\,,
\ee
obeying $x \circ (y \circ z) = ( x \circ y) \circ z + y \circ ( x \circ z)$, and
\be
\langle \mathbb{A}_\mu, \mathbb{A}_\nu \rangle = \int d^{248} Y \left( \mathcal{A}_{(\mu}{}^M \mathcal{B}_{\nu) M} - f^M{}_{NK} \mathcal{A}_{(\mu}{}^N \partial_M \mathcal{A}_{\nu)}{}^K \right)\,.
\ee
Given that the 3d gravity of the external sector is also topological, one has a fully topological sector of the theory. Remarkably then one has something quite unique. The $\Geight$ ExFT has a topological phase given by the solution to the metric constraint \refeqn{E8nonrie} and a fully geometric phase given by the generalised metric parametrised as usual in terms of the metric and $p$-form fields of eleven dimensional supergravity. At the classical level which phase we are in is a choice of background, essentially spontaneous symmetry breaking. One would hope that at the level of the path integral different phases are selected as one changes some order parameter.

In addition to this the existence of a new vacuum disconnected from the usual set of vacua in supergravity has implications for the vacuum energy.  In some form of minisuperspace quantisation one would construct a wavefunction over the space of vacua with the action providing an effective potential. The presence of a disconnected vacuum (with no moduli) would be like a delta function potential. The wavefunction over vacua would then have some support on this delta function which would lower the overall vacuum energy \cite{bermatapp}.

%% file: sec10conclusions.tex
After many years of development, and after overcoming numerous stumbling blocks, we are now at the stage where the reformulation of supergravity theory with manifest U-duality has been achieved. 
The theory is complete for all the finite exceptional groups and there has been significant progress for the infinite dimensional case of $E_9$ \cite{Bossard:2017aae,Bossard:2018utw} as well as attempts \cite{Bossard:2019ksx} at the ultimate $E_{11}$ theory proposed by West and collaborators \cite{West:2001as} (but so far relatively little progress on the intermediate case of $E_{10}$  \cite{Damour:2002cu}).  Understanding how different ExFTs with different groups are related is still not completely understood, despite the provisional work in \refref{Berman:2019efr}.
The development of the infinite dimensional cases will no doubt continue and one imagines that the various technical difficulties will be overcome as we understand how different ExFTs are embedded in each other.

We have touched on several of the key applications of ExFT (and of its predecessor DFT) in this review.
We described for instance the connection with gauged supergravities, reductions and consistent truncations in section \ref{genSS} and in section \ref{corrections} the power of these formalisms in describing higher-derivative corrections to supergravity.
Here the subjects have proven their technical utility, and we expect they will continue to produce exciting new results in these directions.

The presentation taken in this review has been quite ahistoric, in that we did not spend a great deal of time first discussing DFT. DFT was in fact first given as a truncation of closed string field theory and it is string field theory that lies beneath many of the structures of DFT. 
The proper justification for the truncation from closed string field theory to DFT has been suggested by Sen \cite{Sen:2016qap} to arise as a generalised Wilson effective action via a procedure of integrating out degrees of freedom from the string field theory path integral.
Indeed the role of T-duality in the context of closed string field theory, studied in \refref{Kugo:1992md}, provides an insight into many of the structures in DFT. 

What perhaps is unexpected is how exceptional field theory follows the structures of DFT. String field theory is a second quantised theory that is perturbative in the string coupling. M-theory is a meant to be the theory that is defined nonperturbatively in the string coupling and indeed its low energy effective theory does not even share the same dimension as the perturbative string. The M-theory low energy effective action, eleven-dimensional supergravity, has of course a different massless field content from its ten-dimensional counterparts the IIA and IIB supergravities that are the perturbative string effective actions. It is thus something of a mystery that the structures in DFT get replicated for exceptional field theory given that exceptional field theory is in some sense the nonperturbative  version of DFT. This observation has proved to be a useful guide in exceptional field theory and has led to the maxim, {\it{what happens in DFT happens for ExFT}}. Physically though we are far from understanding why this is. 

There are of course some outstanding questions. 
An example where there are difficulties in moving to exceptional field theory as compared to DFT has been in precisely understanding the geometry of the extended space. 
We have described how the theory has a local symmetry described infinitesimally by the generalised Lie derivative. For a normal Lie algebra the finite transformations (at least those connected to the identity) are obtained by exponentiating the infintesimal generators. For the case of DFT the role of the section constraint makes this process technically involved, and there have been a series of works examining the finite transformations in DFT \cite{Hohm:2012gk,Park:2013mpa, Berman:2014jba, Hull:2014mxa, Naseer:2015tia, Rey:2015mba,Chaemjumrus:2015vap}. However, for exceptional field theory the local finite transformations are not known. The global structure of exceptional spaces is also not known and even for DFT there are various global issues \cite{Papadopoulos:2014mxa,Papadopoulos:2014ifa,Hassler:2016srl,Howe:2016ggg,Alfonsi:2019ggg}. Getting to grips with these local and global questions is crucial to understanding what possible exotic spaces there may be. 
The exotic branes presented here are easily seen through the application of dualities but there is no reason to suppose that these are the only exotic spaces that ExFTs allow. It is clear that one needs to develop the appropriate topological classification for the spaces of exceptional field theory and at the moment this is not known.

For DFT in a background that admits a complex structure, there is a symplectic strucutre and subsequently one may form a description of the space as a para-Hermitian manifold \cite{Vaisman:2012ke,Vaisman:2012px,Marotta:2018myj,Mori:2019slw,Marotta:2019eqc,Carow-Watamura:2020xij,Ikeda:2020lxz}. This approach goes far in describing global issues in DFT. In this framework, the doubled spacetime is a foliated manifold for which the polarization (or solution of the section condition) determines the physical spacetime as a quotient space. This is in contrast to the idea of the physical spacetime as a submanifold (as such ``section condition'' may be somewhat of a misnomer). From the string worldsheet perspective this is the {\it{doubled but gauged}} approach \cite{Lee:2013hma,Basile:2019pic}. 

The approach relies on two things that exceptional field theory is missing. M-theory is not a theory of membranes (or indeed fivebranes) so even though one may have worldvolume actions for branes in exceptional backgrounds as discussed in section \ref{branewvol}, brane actions do not play the same role in M-theory as they do in string theory. They are not fundamental objects whose quantisation may be used to define the theory as one does with strings. 
(In spite of this situation, some first steps towards generalising the para-Hermitian structures to exceptional geometry based on versions of U-duality covariant brane actions, have been recently taken in \refref{Sakatani:2020umt}, see also the comments in \refref{Arvanitakis:2018hfn}.)
Secondly, the para-Hermitian approach relies on similarities to symplectic geometry and geometric quantisation. 
It is curious how DFT has many similarities to geometric quantisation as discussed in \refref{Berman:2019biz}. Extending that approach to ExFT would be fascinating but as yet has not been developed. To do so would be a fascinating exercise: beyond being an application of ExFT it would be a geometric approach to quantum nonperturbative physics.


One further geometric topic we did not discuss in detail in this review was the construction of generalised notions of connections and curvatures.
This can indeed be done, but the situation is more subtle than in ordinary geometry, where the Levi-Civita connection is the unique torsion-free metric compatible connection, leading to the Ricci scalar which gives the Einstein-Hilbert action.
In double or exceptional geometry, one can define generalised connections that annihilate the generalised metric and preserve the appropriate $O(d,d)$ or $\Edd$ structures, but these are not unique \cite{Siegel:1993xq,Siegel:1993th,Jeon:2010rw, Jeon:2011cn,Hohm:2011si, Coimbra:2011nw,Coimbra:2011ky, Park:2013gaj, Cederwall:2013naa,Aldazabal:2013mya,Godazgar:2014nqa}.
This leads to obstructions when defining unambiguous curvature tensors.
By taking projections and contractions, the only unique covariant objects reduce to the generalised Ricci tensor and generalised Ricci scalar which appear in the Lagrangian and equations of motion of the generalised metric.

For example, in (pure) DFT, the generalised Ricci tensor and scalar vanish by the equations of motion, and it can further be shown that there is no $O(d,d)$ covariant quantity that reduces to the square of the Riemann tensor \cite{Hohm:2011si}. This means it is unclear whether or how one can characterise singularities in DFT (leading, for instance, to subtleties when trying to study black holes in doubled geometry \cite{Arvanitakis:2016zes}), and is perhaps expressing something important about the nature of how $\alpha^\prime$ corrections must be incorporated in this string theoretic geometry.
 An absence of singularities may of course also be a virtue! 
DFT and ExFT also offer novel resolutions of geometric singularities where an apparent naked singularity in spacetime can be interpreted as a breakdown in the parametrisation of the generalised metric, signalling the need to switch from a Riemannian to a non-Riemannian parametrisation \cite{Blair:2016xnn, Berman:2019izh}, as we described briefly at the end of section \ref{exoticfurther} in the context of putative electromagnetic duals of exotic branes. 

A further idea that would be very interesting to lift from the doubled to the exceptional setting is that of generalised dualities.
The usual T- and U-dualities apply in backgrounds with abelian isometries.
In string theory backgrounds with a non-abelian group of isometries, one can perform a transformation known as non-abelian T-duality\cite{delaOssa:1992vci} which generates a second string theory background, which generically will have no isometries. This can be further generalised to Poisson-Lie duality \cite{Klimcik:1995dy,Klimcik:1995ux} which can be used to transform between sigma models in two backgrounds, neither of which have isometries.
The secret to these generalised T-dualities is that there exists an underlying algebraic structure based on an even-dimensional Lie algebra known as the Drinfeld double. 
This algebra naturally comes with an $O(d,d)$ invariant metric, and turns out to have a natural description in terms of doubled geometry, with Poisson-Lie transformations (which contain abelian and non-abelian T-duality as special cases) acting as non-constant $O(d,d)$ transformations in DFT \cite{Hassler:2017yza,Demulder:2018lmj,Bugden:2019vlj,Sakatani:2019jgu,Catal-Ozer:2019hxw}.
In the absence of a clear worldvolume understanding of U-duality it has recently been proposed to use the natural ExFT generalisations of this framework to construct the algebras and geometries that would appear in a non-abelian or Poisson-Lie version of U-duality \cite{Sakatani:2019zrs,Malek:2019xrf,Sakatani:2020iad}. This seems to be a promising approach to use the power of ExFT to find new types of dualities in M-theory.

There is still much to do in fully understanding  the role of exotic branes. From the string perspective, exotic branes should apparently dominate the theory at strong coupling since their naive tension makes them the lightest states at strong coupling. The reason why this might be naive is shown when we look at M-theory and the IIA string. The NS5 has tension $g_s^{-2}$, the D-branes $g_s^{-1}$, and the fundamental string $g_s^0$ yet at strong coupling it is not that the NS5 dominates. The NS5 and D4 are described by the M5 and the fundamental string and D2 combine to the M2. The M2 and M5 have the same dimensionless tension and there is no hierarchy of scales at strong coupling, see \cite{Berman:2007bv} for a review. One would imagine a similar thing happening with the exotic branes in exceptional field theory, but this has as yet not been accomplished. 
The physics of the production of exotic supertubes from ordinary brane configurations, which as suggested in \refref{deBoer:2010ud, deBoer:2012ma} may contribute to black hole microstate counting, is another area in which has not yet begun to be fully explored.

From a perhaps more mathematical perspective, a promising new technical development, which may ultimately shed light on many of the above questions, has been to see the role of $L_{\infty}$ algebras \cite{Hohm:2017pnh,Cederwall:2018kqk,Cederwall:2018aab,Cagnacci:2018buk,Arvanitakis:2018cyo,Grewcoe:2020ren}, Leibniz gauge theories \cite{Bonezzi:2019ygf} and differential graded Lie algebras \cite{Bonezzi:2019bek} in exceptional field theories.
This shows that underlying the exceptional geometry of ExFT are powerful (and, indeed, applicable much generally than just in this subject) algebraic structures, whose role in the formulation has perhaps not yet been fully appreciated.

Finally, we turn to cosmology. It is here where the ideas look most likely to have physical application in the understanding the geometry of the universe at small scales where string gas dynamics dominate the physics \cite{Brandenberger:2017umf,Brandenberger:2018bdc,Bernardo:2019pnq,Lescano:2020smc}.
For cosmological applications, one can also obtain more control over higher-derivative expansions leading to proposals for duality invariant cosmology at all orders in $\alpha^\prime$ \cite{Hohm:2019jgu} in which one can obtain de Sitter vacua \cite{Hohm:2019ccp} (follow-ups include \cite{Krishnan:2019mkv,Wang:2019kez,Bernardo:2019bkz, Bernardo:2020zlc}).
We might also mention that non-geometric backgrounds may play a role in finding de Sitter vacua as demonstrated in \refref{Dibitetto:2015bia}, where seven-dimensional de Sitter solutions were found by compactifying on highly non-geometric spaces. This is one of the lessions; we may need to include what may be thought of as exotic non-geometric spaces to obtain physically interesting spaces in lower-dimensional theories. 

Another unexplored possibility is to explore how the thermodynamics of strings are high temperatures would be captured by DFT -- in fact it was one of the original motivations of \refref{Tseytlin:1990va} where doubled geometry was introduced to understand strings beyond Hagedorn temperature \cite{Atick:1988si}. Even more fascinting would be to see if exceptional field theory can help capture the thermodynamics of M-theory.

We began the review by quoting Kaluza's request for assistance from extra dimensions. The extra dimensions, well beyond Kaluza's imagination, introduced by double and expectional field theory have provided much needed help in reorganising string and M-theory but we have yet to see the full role these new dimensions will play.

%% file: sec_appendix.tex
\section{Fixing the coefficients of the $\Gfour$ ExFT}
\label{fixingsl5}

Here, we will give a sense of the type of calculation involved in verifying the invariance of the ExFT invariance under $n$-dimensional diffeomorphisms, and how this fixes the coefficients in the action.
We will work with the case of $\Gfour$ that we described in section \ref{action5}.
We start with the topological term written as an integral over one dimension higher is:
\be
S_{\text{top}} 
= \kappa \int \dd^8 x\,\dd Y \epsilon^{\mu_1\dots\mu_8}\left(
\frac{1}{4} \hat\partial \Fc_{\mu_1 \dots \mu_4} \p \Fc_{\mu_5 \dots \mu_8}
- 4 \Fa_{\mu_1 \mu_2} \p ( \Fb_{\mu_3\mu_4\mu_5} \p \Fb_{\mu_6\mu_7\mu_8 })
\right)
\ee
where the coefficients have been chosen so that its variation is a total derivative:
\be
\begin{split}
\delta S_{\text{top}} = 2 \kappa
\int \dd^8 x\,\dd Y \epsilon^{\mu_1\dots\mu_8}\ \D_{\mu_1} \Big(&
- 4 \delta \Aa_{\mu_2} \p ( \Fb_{\mu_3\mu_4\mu_5} \p \Fb_{\mu_6\mu_7\mu_8}) 
\\ &
- 12 \Fa_{\mu_2 \mu_3} \p ( \Delta \Ab_{\mu_4\mu_5} \p \Fb_{\mu_6\mu_7\mu_8})
\\ & \qquad + ( \hat \partial \Delta \Ac_{\mu_2 \mu_3 \mu_4} ) \p \Fc_{\mu_5 \dots \mu_8}
\Big) \,.
\end{split}
\ee
Using the definition of $\p$, this is:
\be
\begin{split}
\delta S_{\text{top}} =  2\kappa
\int \dd^8 x\,\dd Y \epsilon^{\mu_1\dots\mu_8}\ \D_{\mu_1} \Big(&
+  2 \delta \Aa_{\mu_2}{}^{\fM \fN}  \Fb_{\mu_3\mu_4\mu_5 \fM} \Fb_{\mu_6\mu_7\mu_8 \fN} 
\\ &
+ 6 \Fa_{\mu_2 \mu_3}{}^{\fM \fN}  \Delta \Ab_{\mu_4\mu_5 \fM} \Fb_{\mu_6\mu_7\mu_8\fN}
\\ & \qquad +  \partial_{\fN \fM} \Delta \Ac_{\mu_2 \mu_3 \mu_4}{}^{\fN} \Fc_{\mu_5 \dots \mu_8}{}^{\fM}
\Big) \,.
\end{split}
\ee
The kinetic term for the one- and two-forms are
\be
S_1 = - \frac{1}{8} \int \dd^7 x\,\dd Y \sqrt{|g|} m_{\fM \fP} m_{\fN \fQ} \Fa_{\mu\nu}{}^{\fM \fN} \Fa^{\mu\nu \fP \fQ} \,,
\ee
\be
S_2 =  - \frac{1}{12} \int \dd^7 x\,\dd Y\sqrt{|g|} m^{\fM \fN} \Fb_{\mu\nu\rho \fM} \Fb^{\mu\nu\rho}{}_{\fN} \,.
\ee
Recall that
\be
\begin{split}
\delta \Fa_{\mu\nu}{}^{\fM \fN} & = 2 \D_{[\mu} \delta \Aa_{\nu]}{}^{\fM \fN} + \frac{1}{2} \epsilon^{\fM \fN \fP \fQ \fK} \partial_{\fP\fQ} \Delta \Ab_{\mu\nu \fK} \,,\\
\delta \Fb_{\mu\nu\rho \fM} & = 3 \D_{[\mu} \Delta \Ab_{\nu\rho]\fM} - \frac{3}{4} \epsilon_{\fM \fN \fP \fQ \fK} \delta \Aa_{[\mu}{}^{\fN \fP} \Fa_{\nu\rho]}{}^{\fQ \fK}
+ \partial_{\fN \fM } \Delta \Ac_{\mu\nu\rho}{}^{\fN}\,,
\end{split}
\ee
and hence the field equation for $\Ac_{\mu\nu\rho}$ is
\be
\partial_{\fN \fM} \left(
\frac{1}{6} \sqrt{|g|} m^{\fM \fP} \Fb^{\mu\nu\rho}{}_{\fP} - 2 \kappa \epsilon^{\mu\nu\rho \sigma_1\dots \sigma_4} \Fc_{\sigma_1\dots\sigma_4}{}^{\fM} 
\right)=0 \,.
\ee
Under external diffeomorphisms, we take
\be
\begin{split}
\delta_\xi \Aa_\mu{}^{\fM \fN} &= \xi^{\sigma} \Fa_{\sigma \mu}{}^{\fM \fN} + m^{\fM \fP} m^{\fN \fQ} g_{\mu\nu} \partial_{\fP \fQ} \xi^\nu \,,\\
\Delta_\xi \Ab_{\mu\nu \fM}& = \xi^\sigma \Fb_{\sigma \mu\nu \fM} \,,\\
\Delta_\xi \Ac_{\mu\nu\rho}{}^{\fM}  &= - \frac{1}{12 \cdot 4!\cdot 3! \cdot \kappa} \sqrt{|g|} \xi^\sigma \epsilon_{\sigma \mu \nu \rho \lambda_1 \lambda_2\lambda_3}m^{\fM \fN} \Fb^{\lambda_1\lambda_2\lambda_3}{}_{\fN} \,.
\end{split}
\ee
The complete variation of the topological term is then:
\be
\begin{split}
\delta_\xi S_{\text{top}} =  2\kappa
\int \dd^8 x\,\dd Y \epsilon^{\mu_1\dots\mu_7} \Big(&
+  2 \xi^{\sigma} \Fa_{\sigma \mu_1}{}^{\fM \fN} \Fb_{\mu_2\mu_3\mu_4 \fM} \Fb_{\mu_5\mu_6\mu_7 \fN} 
\\ &
+ 2m^{\fM \fP} m^{\fN \fQ} g_{\mu_1\sigma} \partial_{\fP \fQ} \xi^\sigma  \Fb_{\mu_2\mu_3\mu_4 \fM} \Fb_{\mu_5\mu_6\mu_7 \fN} 
\\ &
+ 6\xi^\sigma  \Fa_{\mu_1 \mu_2}{}^{\fM \fN}  \Fb_{\sigma \mu_3 \mu_4 \fM} \Fb_{\mu_5\mu_6\mu_7 \fN}
\Big)
\\ &\hspace{-3cm}
+ \frac{1}{6}  \int \dd^8 x\,\dd Y
 \partial_{\fN \fM} \left(  \sqrt{|g|} \xi^\mu m^{\fN \fP} \Fb^{\nu\rho\sigma}{}_{\fP} \right)
 \Fc_{\mu\nu\rho\sigma}{}^{\fM}
 \,.
\end{split}
\label{deltaxiStop}
\ee
The first and third lines here cancel via the Schouten identity.
For the second line, we write
\be
\begin{split}
\epsilon^{\mu_1\dots\mu_7} g_{\mu_1\sigma} \partial_{\fP \fQ} \xi^\sigma  \Fb_{\mu_2\mu_3\mu_4 \fM} \Fb_{\mu_5\mu_6\mu_7 \fN} 
& = -|g| \epsilon_{\sigma \lambda_1 \dots \lambda_6} \partial_{\fP \fQ} \xi^\sigma \cH^{\lambda_1 \lambda_2\lambda_3}{}_{\fM} \cH^{\lambda_4 \lambda_5\lambda_6}{}_{\fN} 
\end{split}\,.
\ee
Then we integrate by parts to obtain the final expression 
\be
\begin{split}
\delta_\xi S_{\text{top}} =  8\kappa
\int \dd^8 x\,\dd Y \sqrt{|g|}  &
\xi^\sigma \epsilon_{\sigma \mu_1 \dots \mu_6} \partial_{\fP \fQ}  \left(\sqrt{|g|} m^{\fM \fP} \cH^{\mu_1 \mu_2 \mu_3}{}_{\fM} \right) m^{\fN \fQ} \cH^{\mu_4 \mu_5 \mu_6}{}_{\fN} 
\\ &\hspace{-2cm}
+ \frac{1}{6}  \int \dd^8 x\,\dd Y
 \partial_{\fN \fM} \left(  \sqrt{|g|} \xi^\mu m^{\fN \fP} \Fb^{\nu\rho\sigma}{}_{\fP} \right)
 \Fc_{\mu\nu\rho\sigma}{}^{\fM}
 \,.
\end{split}
\label{deltaxiStopfinal}
\ee
Next, we consider
\be
\begin{split}
\delta_\xi \cH_{\mu\nu\rho\fM} & = 
L_\xi \cH_{\mu\nu\rho \fM}
\\ &\qquad
 - \frac{3}{4} \epsilon_{\fM \fN \fP \fQ \fK}
 m^{\fN \fN^\prime} m^{\fP \fP^\prime} g_{\sigma[\mu|}  \partial_{\fN^\prime \fP^\prime}  \xi^\sigma\Fa_{|\nu\rho]}{}^{\fQ \fK} 
\\&\qquad -\xi^\sigma \partial_{\fM \fN} \Fc_{\sigma\mu\nu\rho}{}^{\fN} 
- \frac{1}{12 \cdot 4!\cdot 3! \cdot \kappa}  \partial_{\fP \fM } \left( \sqrt{|g|} \xi^\sigma \epsilon_{\sigma \mu \nu \rho \lambda_1 \lambda_2\lambda_3}m^{\fP \fQ} \Fb^{\lambda_1\lambda_2\lambda_3}{}_{\fQ}\right)
\end{split}
\label{LH5}
\ee
after using the Bianchi identity for $\cH_{\mu\nu\rho\fM}$ to simplify the expression.
The final two lines can be viewed as the anomalous variation of this field strength under the $7$-dimensional diffeomorphisms.
The very final line gives:
\be
\begin{split} 
\delta_\xi^{\text{anom}} S_2 & \supset 
 \int \dd^7x\,\dd Y \sqrt{|g|}
 \frac{1}{6}  m^{\fM \fN} \Fb^{\mu\nu\rho}{}_{\fM} \xi^\sigma \partial_{\fN \fP} \Fc_{\sigma\mu\nu\rho}{}^{\fP}
\\ &
+ \frac{1}{6} \frac{1}{12 \cdot 4!\cdot 3! \cdot \kappa}\int \dd^7x\,\dd Y  \partial_{\fP \fM} \left( \sqrt{|g|} m^{\fM \fN} \cH^{\mu\nu\rho}{}_{\fN} \right)
\sqrt{|g|} \xi^\sigma \epsilon_{\sigma \mu \nu \rho \lambda_1 \lambda_2\lambda_3}m^{\fP \fQ} \Fb^{\lambda_1\lambda_2\lambda_3}{}_{\fQ}
\end{split}
\ee
after integrating by parts.
The first line here combines with the second line of \refeqn{deltaxiStopfinal} into a total derivative, and the second line cancels against the first line of \refeqn{deltaxiStopfinal} if
\be
 \frac{1}{6} \frac{1}{12 \cdot 4!\cdot 3! \cdot \kappa} = 8 \kappa
  \Rightarrow \kappa = \pm \frac{1}{12 \cdot 4!}  \,.
\ee
The choice of sign is immaterial and we pick the plus sign.
Next, we can consider 
\be
\delta_\xi \Fa_{\mu\nu}{}^{\fM \fN} = L_\xi \Fa_{\mu\nu}{}^{\fM \fN}
+ \frac{1}{2} \epsilon^{\fM \fN \fP \fQ \fK} \partial_{\fP \fQ} \xi^\sigma \Fb_{\sigma\mu\nu \fK}
+ 2 \D_{[\mu|} ( m^{\fM \fP} m^{\fN \fQ} \partial_{\fK \fL} \xi^\sigma g_{|\nu]\sigma})\,.
\ee
It is straightforward to check that the contribution to the variation of the kinetic term for this field strength arising from the second term here cancels against the remaining piece coming from the anomalous variation of $\cH_{\mu\nu\rho \fM}$, i.e. the second line in \refeqn{LH5}.
This fixes the coefficients of $\mathcal{L}_1$ and $\mathcal{L}_2$ relative to each other.
With further work, the third term here can be shown to cancel against a term coming from the variation of the Einstein-Hilbert term, $R_{\text{ext}}(g)$, for the metric $g_{\mu\nu}$.
This fixes the relative coefficients of $\mathcal{L}_1$ and $R_{\text{ext}}(g)$. Finally, further anomalous variations from $R_{\text{ext}}(g)$, the kinetic term for the generalised metric and the internal part of the Lagrangian all conspire to cancel against each other and fix all relative coefficients (up to the overall scale).
We refer the diligent reader to the original literature to check the precise details.


\section{$O(d,d)$ double field theory}

In section \ref{genKK_DFT}, we explained the double field theory which describes the full 10-dimensional NSNS sector action using $O(10,10)$ generalised diffeomorphisms, a generalised metric and a generalised dilaton.
We can also formulate double field theory with $n=10-d$ ``undoubled'' coordinates, $x^\mu$, and $2d$ doubled coordinates, $Y^M$.
The field content of this theory is:\cite{Hohm:2013nja}
\be
\left\{ 
g_{\mu\nu} , \cH_{MN}, \gendil, \Aa_{\mu}{}^M , \Ab_{\mu\nu} 
\right\}\,.
\ee
As well as the $n$-dimensional metric, $O(d,d)$ generalised metric, $\cH_{MN}$, and generalised dilaton, $\gendil$, there is a particularly simple tensor hierarchy, consisting solely of the one-form $\Aa_\mu{}^M$ and $O(d,d)$ singlet two-form, $\Ab_{\mu\nu}$, which transform under generalised diffeomorphisms and gauge transformations as:
\be
\begin{split}
\delta \Aa_\mu{}^M = \D_{\mu} \Lambda^M - \eta^{MN}\partial_N \lambda_\mu \,,\quad
\Delta \Ab_{\mu\nu} & = \eta_{PQ} \Lambda^P \Fa_{\mu\nu}{}^Q  + 2 \D_{[\mu} \lambda_{\nu]} \,.
\end{split}
\label{BgaugeODD}
\ee
As in section \ref{thier}, we can define the corresponding generalisations of the wedge product and exterior derivative.
If $A_1,A_2 \in R_1 = \mathbf{2d}$ and $B \in R_2 = \mathbf{1}$, 
\be
( \hat \partial B)^M = \eta^{MN}\partial_N B\,,\quad
A_1 \p A_2 = \eta_{MN} A_1^M A_2^N \,.
\ee
The formulae \refeqn{FABC} may then be used for the field strengths.
In this case, it turns out that the field strength $\cH_{\mu\nu\rho}$ of the two-form is already fully covariant without the need to introduce a three-form, so the tensor hierarchy terminates.

The remaining fields transform under generalised diffeomorphisms as follows.
The generalised metric and dilaton transform as a rank 2 tensor and a scalar of weight one, exactly as in section \ref{genKK_DFT}.
The $n$-dimensional metric is simply a scalar:
\be
\delta_\Lambda g_{\mu\nu} = \Lambda^N\partial_N g_{\mu\nu} \,.
\ee
It is then straightforward to write down terms which are invariant under generalised diffeomorphisms, and under the further tensor hierarchy gauge transformations. 
In order to fix the action uniquely, we again use transformations under external diffeomorphisms.
In terms of an external vector field $\xi^\mu$ and the usual $n$-dimensional Lie derivative $L_\xi$ (defined with $\mathcal{D}_\mu$ in place of $\partial_\mu$), we have:
\be
\begin{split}
\delta_\xi g_{\mu\nu} & = L_\xi g_{\mu\nu} \,,\\
\delta_\xi \Aa_\mu{}^M & = \xi^\nu \Fa_{\nu\mu}{}^M  + \cH^{MN}  g_{\mu\nu} \partial_N \xi^\nu \,,\\
\Delta_\xi \Ab_{\mu\nu} & = \xi^\rho \Fb_{\mu\nu\rho} \,,\\
\delta_\xi \cH_{MN} & = L_\xi \cH_{MN} \,,\\
\delta_\xi \gendil & = L_\xi \gendil\,,
\end{split}
\ee
The action for the full double field theory invariant under all these tranformations is then:
\be
\begin{split} 
S = 
 \int \dd^n x \, \dd Y \,\sqrt{|g|} e^{-2\gendil} \big(& R_{\text{ext}}(g)
+ 4 \mathcal{D}_\mu \gendil \mathcal{D}^\mu \gendil
 + \frac{1}{8} \mathcal{D}_\mu \cH_{MN} \mathcal{D}^\mu \cH^{MN} 
\\ & 
 - \frac{1}{4} \cH_{MN} \Fa_{\mu\nu}{}^M \Fa^{\mu\nu N} 
- \frac{1}{12} \Fb_{\mu\nu\rho} \Fb^{\mu\nu\rho}
\\ &
+ \frac{1}{4} \cH^{MN} ( \partial_M g_{\mu\nu} \partial_N g^{\mu\nu} + \partial_M \ln |g| \partial_N \ln |g| )
\\ &
+ \mathcal{R}_{\text{DFT}}(\cH, \gendil) 
\big)\,.
\end{split}
\label{DFTsplit}
\ee